%% file: hoofd.tex
\documentclass[titlepage,11pt]{book}
\usepackage{amsmath, amssymb}
\newtheorem{theo}{Theorem}
\newtheorem{exie}{Example}
\newtheorem{deffie}{Definition}
\usepackage{epsfig}

\usepackage{color}

\def\Red#1{{\color{red}#1}}
\def\Green#1{{\color{darkgreen}#1}}
\def\Blue#1{{\color{blue}#1}}

\definecolor{red}{rgb}{1,0,0}
\definecolor{blue}{rgb}{0,0,1}
\definecolor{green}{rgb}{0,1,0}
\definecolor{black}{rgb}{0,0,0}
\definecolor{yellow}{rgb}{1,1,0}
\definecolor{mdwblue}{rgb}{0.2,0.2,0.6}
\definecolor{gray}{rgb}{0.7,0.7,0.7}
\definecolor{darkgreen}{rgb}{0.2,0.7,0.2}

\begin{document}

\pagenumbering{roman}

\tableofcontents
\pagenumbering{arabic}
\include{preface}
\include{inleiding}
\include{symblijst}
\include{hoofdstuk1}

\include{symmetry}

\include{gromov}
\include{ls}

\include{conclusions}
\appendix

\include{AppendixA}
\include{appendB}

\include{appendF}
\include{appendG}

\include{appendI}

\include{bib}
\include{appendJ}
\end{document}

%% file: preface.tex
\section{\mbox Preface}
Before we lift off, I would like to grasp the occasion to say
thanks to some people who have drawn my attention towards this
subject.  In the first place, this work would never have been
accomplished without my promotor for many reasons; the most
important one being his attitude to let me explore this
fascinating subject by myself while bearing in mind the penalty
that it might actually never get finished.  No less important was
his classical relativist's scepticism which has served this work
well in a twofold way: from one side it challenged me to prove to
him that what I had done really was the good way to approach the
subject and on the other hand it forced me to construct my work
with a rigour which is not entirely mine.  Also, his willingness
to get me into contact with specialists in a field which was not
entirely his own was of no lesser importance.  I guess my
agitating personality certainly has cost him some grey hairs, but
I think it is safe to say that we also had some great deal of fun
together.  Thank you Norbert and I hope to hear of you later
again. \\* \\* I am also indebted to Rafael Sorkin and Luca
Bombelli, whose warm reception in Syracuse and Mississippi
contributed a great deal to my understanding of the subject.
These visits in $2002$ and $2003$ were periods in which I could
present my own work and learn a lot about theirs, which was the
really interesting stuff.  I could learn and discuss physics first-hand
in the coffeeshop, during the walk home, at the evening
supper and sometimes even at $2$ am.  Thanks Luca and Rafael. \\*
\\* The next person who has had his own input into this work is my
copromotor Frans Cantrijn.  The most surprising thing for me was
that Frans actually never kicked me out of his office or told me
to stop phoning him when I was asking his advice or when I wanted
to talk to somebody about some (crazy) ideas.  Frans has always
been prepared to listen with patience and to comment very
carefully.  Frans, thanks again for all the nice conversations we
had.  \\* \\*
Also, I would like to thank Fay Dowker and Renate Loll for the short but pleasant visits in London and Utrecht. \\* \\*
Last but not least there is my wife Monika, my
family in Herentals and my friends and collegues Willy Arts,
Theodore Kolokolnikov, Luc Hoegaerts, Lode Wylleman, Soeren
Krausshar and Benny Malengier who have made my life pleasant
during the sometimes frustrating work.  Also, I express my gratitude to the head of the department, Roger Van Keer, for the interest shown in my work.  Thanks to all of them and
kisses to Monika.

%% file: inleiding.tex
\section{\mbox {Introduction}}
The problem, which I shall discuss in this thesis, can be situated in the area of mathematical physics.
It has to do with a theoretical challenge which I find extremely fascinating: the merging of general relativity (GR) with quantum mechanics.  The latter is a theory which is defined on a background, i.e., a differentiable manifold with a slicing necessary to produce a differentiable time function.  The challenge is how to define the background quantum?  What are the ``quanta'' that build the geometry of spacetime, which looks like a manifold with a Lorentzian tensor on scales larger than the Planck scale?  If one formulates the question of quantum general relativity like this, then it is de facto necessary to reformulate quantum theory from the start.  As usual, most approaches to the problem are far more conservative than that. \\*
\\*
Some fifteen years ago, some conjectures about the background quantum
were made by R. Sorkin.  The one which I believe to be the best candidate, is the ``causal set'' approach.
This thesis is not about causal sets in the first place, neither
about the quantum dynamics causal sets have to obey.  However, this work deals with the kinematics of (quantum) gravity, i.e., we study the space of ``generalised spacetimes'' which might occur in a path integral formulation of quantum gravity.  As is by now common wisdom in the quantum gravity
community, summing over classical spacetimes with isometric initial
and final data surfaces just won't work \footnote{See Donaldson \cite{Donaldson}.} and therefore it is necessary to consider more general structures. \\*
\\*
The first chapter of this thesis is an attempt to provide a good introduction to
general relativity and might even be interesting for the
specialist.  Throughout this chapter, the representation of the
diffeomorphism group is highlighted and the discussion is focused
around this topic. \\* \\* In the second chapter, I present some
rather technical properties of the diffeomorphism group which I
needed for my first paper \cite{Noldus}.  It may seem to be
paradoxical to include this paper only in an appendix and not in the main body of this thesis.  The reason
for doing so is that the theory presented there is not as elegant
as the Gromov Hausdorff one which I developed afterwards and
moreover, the obtained results are still incomplete.  However, I
believe this chapter is a must for everyone working in the field:
the material presented has immediate value for any quantisation
program aiming at representing the Lie algebra of the
diffeomorphism group as essentially self adjoint operators on some
Hilbert space and regrettably these results are not as well known
as they should be.  \\* \\* The third and fourth chapter of this
thesis deal with Lorentzian Gromov Hausdorff theory.  Like the
well known Gromov Hausdorff theory, this theory is formulated for spaces $\mathcal{M}$ equipped with a \emph{Lorentz} distance $d$ which satisfies the following properties: 
\begin{itemize}
\item $d(p,p) = 0$ for all $p \in \mathcal{M}$,
\item $d(p,q) > 0$ implies $d(q,p) = 0$ (antisymmetry),
\item if $d(p,r)d(r,q) > 0$ then $d(p,q) \geq d(p,r) + d(r,q)$ (reverse triangle inequality).
\end{itemize}
Hence, a Lorentz distance defines a partial order, and order is
something foreign to metrics!  Moreover, the spheres of fixed
radius are hyperbolae and hence not \emph{a priori} compact in a
non compact spacetime.  The reader might guess that these
properties make the Lorentzian convergence theory look very
different from its metric counterpart although there are also many
similarities.   \\* \\* 
The most important object occuring in chapter $3$ is the so called strong metric $D$ which is defined as $$D(p,q) = \sup_{r \in \mathcal{M}} \left| d(p,r) + d(r,p) - d(r,q) - d(q,r) \right| $$ and a Lorentz space is a set $\mathcal{M}$ with a Lorentz distance $d$ defined on it such that $(\mathcal{M},D)$ is a compact metric space.  In the same chapter, I also construct a metric $d_{GH}$ as well as a stronger uniformity (GGH) on the space of all Lorentzian interpolating spacetimes.  Also, a suitable limit space is constructed by using the strong metric $D$.  At the end of this chapter, some quantitative control mechanisms on convergence are introduced and they turn out to be very important as illustrated by many results in chapter $4$.  In the latter, I repeat the definition of a Lorentz space and it will be shown that $d_{GH}$ as well as GGH have natural extensions as metric and quantitative Hausdorff uniformity respectively on the moduli space $\mathcal{LS}$ of all Lorentz spaces.  $(\mathcal{LS},GGH)$ is then investigated in some detail: for example a study of causality on Lorentz spaces is initiated (on which only the chronology relation is given) and properties like path metricity of $d$ are proven to be stable under convergence assuming the validity of some of the control mechanisms mentioned before.  Also, some criteria for precompactness of a subset in $\mathcal{LS}$ are investigated and these could be seen as metric conditions which imply that the geometry of the respective (equivalence classes of) Lorentz spaces is ``bounded'' in some sense.  In my opinion, it would be desirable to develop general algorithms which construct classes of discrete Lorentz spaces of ``bounded geometry'' if possible.  The reason for this claim is that there are probably too many generic causal sets and restriction to a subset of bounded geometry would seriously improve one's chances to construct a \emph{finite} quantum dynamics.  Hence, this last section of chapter $4$ might have much more importance than it appears to have by its rather formal presentation. All these results have been published in three further papers in Classical and Quantum Gravity \cite{Noldus1, Noldus2,
Noldus3}. \\* \\*
I end this introduction by making a small comment about terminology.  In the literature, one does not really make the distinction between a Lorentzian manifold $(\mathcal{M},g)$ and a spacetime which could be defined as an equivalence class under active diffeomorphisms of the former.  I shall follow this convention when we are talking about ``spacetime'' and it should be clear from the context which interpretation is meant.

%% file: symblijst.tex
\section{\mbox {Explanatory list of symbols}}
\begin{verse}
\hspace{13pt} \textsl{The list has to be understood as ``symbol(s) \ - \  short explanation(s) \ - \ references \ - \ $x$'' where the optional $x$ takes the values $B,C, \ldots$  For example \\* \vspace{5pt} \hspace{13pt} $\mathcal{J}^{+}$ \ - \ Future null infinity in an asymptotically flat or weakly asymptotically simple and empty spacetime \ - \ ~\cite{Wald} p.276, ~\cite{Hawking1} p.222 \ - \ B \\* \vspace{5pt} \hspace{13pt}
means that $\mathcal{J}^{+}$ is shortly explained in appendix B and that a full treatment can be found in Hawking and Ellis and the book of Wald.  If there is no option available it simply means that the reader has to look it up in the literature.}
\end{verse}
\begin{itemize}
\item $\mathcal{U}, \mathcal{V}$ \ - \  Sets and in particular
convex normal neighbourhoods \ - \ ~\cite{Hawking1} p.34,
~\cite{Kriele} p. 127 \item $\mathcal{M}, \tilde{\mathcal{M}}$ \ -
\ 4 dimensional spacetimes \item $g, \tilde{g}, h$ \ - \
Lorentz metrics \item $|g| = - det(g_{\alpha \beta})$ \item $\nabla$ \ - \
Levi Civita connection of a model $(\mathcal{M},g)$ \ - \
~\cite{Hawking1} p. 30, ~\cite{Wald} p. 30, ~\cite{Nakahara} p.
209, ~\cite{Kriele} p. 130 \item $\Sigma$ \ - \ 3
dimensional (spacelike) hypersurface \ - \ ~\cite{Hawking1} p. 44
\item $\gamma_{ab}$ \ - \ Riemannian metric on a 3 dimensional
manifold, first fundamental form on a spacelike hypersurface
$\Sigma$\ - \ ~\cite{Hawking1} p. 45 \item $D$ \ - \ Levi Civita
connection of a 3 dimensional model $(\Sigma , \gamma_{ab})$ \ - \
~\cite{Hawking1} p. 47, ~\cite{Wald} p. 257, ~\cite{Thiemann} p.
33, ~\cite{Kriele} p. 275 \item $K_{ab}$ \ - \ Second fundamental
form (intrinsic curvature) of a spacelike hypersurface $\Sigma$ \
- \ ~\cite{Hawking1} p.45, ~\cite{Wald} p. 230 \item
$I^{+}(\mathcal{U}), I^{-}(\mathcal{U})$ \ - \  Chronological
future, past of a set $\mathcal{U}$ \ - \ ~\cite{Hawking1} p. 182;
~\cite{Kriele} p. 354 , ~\cite{Wald} p. 190 \item
$J^{+}(\mathcal{U}), J^{-}(\mathcal{U})$ \ - \ Causal future, past
of a set $\mathcal{U}$ \ - \ ~\cite{Hawking1} p. 183, ~\cite{Wald}
p. 190, ~\cite{Kriele} p. 359 \item $E^{+}(\mathcal{U}),
E^{-}(\mathcal{U})$ \ - \ Future, past horismos of a set
$\mathcal{U}$\ - \ ~\cite{Hawking1} p. 184, ~\cite{Kriele} p. 403
\item $D^{+}(\mathcal{U}), D^{-}(\mathcal{U}), D (\mathcal{U} )$ \
- \ Future, past and full Cauchy developments of a set
$\mathcal{U}$ \ - \ ~\cite{Hawking1} p. 201, ~\cite{Wald} p. 200,
~\cite{Kriele} p. 376 \item $I^{\pm}_{\mathcal{V}},
J^{\pm}_{\mathcal{V}}, E^{\pm}_{\mathcal{V}},
D^{\pm}_{\mathcal{V}}, D_{\mathcal{V}}$ \ - \ Have the same
meaning as before but now one has to restrict the spacetime to
$\mathcal{V} \subset \mathcal{M}$ \item $C^{r, \alpha} \quad
\alpha=+ \text{ or } \emptyset$ \ - \ Class of $r$ times
differentiable real valued functions on $\mathcal{M}$ with $r$'th
derivatives which are continuous if $\alpha = \emptyset$ or
Lipschitz if $\alpha = +$. \item $A(p,q) \quad p,q \in
\mathcal{M}$ \ - \ The Alexandrov set corresponding to the
points $p,q$, $A(p,q) = J^{+}(p) \cap J^{-}(q)$ \item
$\mathcal{C}(p,q)$ \ - \ The set of all causal curves with initial
end point $p$ and final endpoint $q$ \ - \ ~\cite{Hawking1} p.
208, ~\cite{Wald} p. 206 \item $Diff(\mathcal{M})$ \ - \ The group
of ($C^{\infty}$) diffeomorphisms on a manifold $\mathcal{M}$.
\item $L Diff(\mathcal{M})$ \ - \ The Lie algebra of
$Diff(\mathcal{M})$. \item $Diff_{0} \mathcal{M}$ \ - \ The
principal component of $Diff(\mathcal{M})$.
\end{itemize}

%% file: hoofdstuk1.tex
\chapter{\mbox {Classical general relativity}}
\begin{verse}
\hspace{13pt} \textsl{I discuss the theory of relativity from three different perspectives.  I start off with the Lagrangian formulation, then pass to the Hamiltonian one and finish with the less known and quite old metric formulation of Synge.  It is the last formulation which serves as motivation for the kinematical point of view taken in this thesis.}
\end{verse}

\section{The Lagrangian formulation}
General relativity, as found in most textbooks, starts from the
assumption that all \emph{geometrical} data are encoded in the
pair $(\mathcal{M},g)$, where $\mathcal{M}$ is a connected, four
dimensional, Hausdorff $C^{\infty}$ manifold carrying a $C^{r,
\alpha}$ Lorentz metric $g$.  It is obvious that in order to
obtain an evolution equation for the ``gravitational field'', one
needs a notion of derivative.  The covariant derivative $\nabla$
\emph{must} somehow (i.e. by some covariant recipe) be
uniquely determined by the geometrical data.  One takes $\nabla$
to be the unique torsion free connection which makes the metric
$g$ covariantly constant, i.e. such that $\nabla_{\alpha} g_{\beta \gamma} = 0$.  $\nabla$
is well known as the \emph{Levi Civita} connection.  Given two pairs $(\mathcal{M},g)$ and $(\mathcal{\tilde{M}},\tilde{g})$, we
say that $(\mathcal{\tilde{M}},\tilde{g})$ is an $C^{r,\alpha}$
extension of $(\mathcal{M},g)$ if there exists a
$C^{r,\alpha}$ imbedding $\mu: \mathcal{M} \rightarrow
\mathcal{\tilde{M}}$, such that $\mu_{*}g = \tilde{g}$ on $\mu(
\mathcal{M})$ - such a $\mu$ is also called a $C^{r,\alpha}$
diffeomorphism.  A model $(\mathcal{M},g)$ is called
$C^{r,\alpha}$ \emph{inextensible} if $(\mathcal{M},g)$ has no $C^{r,\alpha}$ extension.  The relation
``is an extension of'' is a partial order relation and since every
ordered sequence of models $((\mathcal{M}_{\nu},g_{\nu}))_{\nu \in
I}$, where $I$ is an ordered net, has an upper bound by taking the
directed limit of the $(\mathcal{M}_{\nu},g_{\nu})$, one can
conclude by Zorn's lemma that every $C^{r,\alpha}$ model has an
inextendible $C^{r,\alpha}$ extension (which is highly non
unique).  In this thesis, as far as manifolds are concerned, we shall deal with $C^{2}$ extendible tuples $(\mathcal{M},g)$ with
boundary, athough, as mentioned in ~\cite{Hawking1}, it is
sufficient to consider $C^{1,+}$ models in order for the field
equations to be defined in a distributional sense and still
guarantee the local existence and uniqueness of geodesics.
\\*
\\*
In physics we want to speak about the direction of time and therefore we shall always assume that spacetime is time orientable, i.e., there exists a nowhere vanishing smooth timelike vectorfield $k^{\alpha}$ on $\mathcal{M}$.  As pointed out in ~\cite{Hawking1} a non orientable model $(\mathcal{M},g)$ always admits a connected, time orientable, double covering space $(\tilde{\mathcal{M}},g)$ constructed by doubling the points where an orientation cannot be consistently defined ~\cite{Hawking1} (p. 181).  Since the direction of time can be perceived by us through observation of the \emph{time irreversible} dynamics of collective phenomena as predicted by statistical (quantum) mechanics, one would prefer the orientation to coincide with a thermodynamic arrow of time.  Suppose one has a \emph{local} foliation (in any convex normal neighborhood $\mathcal{U}$) of spacelike hypersurfaces $\Sigma_{t}$ where $g_{\alpha \beta} \big( \frac{ \partial}{\partial t} \big)^{\alpha} k^{\beta} < 0$, $\big( \frac{ \partial}{\partial t} \big)^{\alpha} \big( \frac{ \partial}{\partial t} \big)_{\alpha} = -1$ and $0 \leq t \leq \epsilon$, then the previous statement says that for any evolving isolated, thermodynamic system $\Delta_{t} \subset \Sigma_{t}$, one has that $\frac{\delta S_{t}}{ \delta t} \geq 0$, i.e., the entropy $S_{t}$ increases in time with respect to a congruence of observers moving on the integral curves of the vectorfield $\big( \frac{ \partial}{\partial t} \big)^{\alpha}$.  It is clear that, locally, a thermodynamic arrow of time always exists but it is however not obvious that it can be continuously defined everywhere.  Moreover, the relationship between such arrow of time and for example the arrow defined by the expansion of the universe is obscure.  We shall therefore drop the condition that entropy should be increasing with respect to the direction $k^{\alpha}$ and in fact, this assumption will play no role in the sequel. \\*
\\*
Now we come to a very delicate point which even puzzled Einstein for quite some time.  Two time oriented models $(\mathcal{M},g)$ and $(\tilde{\mathcal{M}},\tilde{g})$ are equivalent iff they are diffeomorphic.  An equivalence class of models is considered to be a \emph{single} physical model, i.e., two diffeomorphic models are regarded as physically equivalent.  This means that one can identify observers 1, 2 with world lines $\gamma_{1}$ respectively $\gamma_{2}$ in models $(\mathcal{M}_{1},g^{1})$ respectively $(\mathcal{M}_{2},g^{2})$, and phyical measurements ${\Psi_{1}}^{\alpha_{1} \ldots \alpha_{k}}_{\textrm{phy} \quad \beta_{1} \ldots \beta_{l}}$, ${\Psi_{2}}^{\alpha_{1} \ldots \alpha_{k}}_{\textrm{phy} \quad \beta_{1} \ldots \beta_{l}}$ made by these observers if and only if there exists a diffeomorphism $\psi : \mathcal{M}_{1} \rightarrow \mathcal{M}_{2}$ such that $\psi \circ \gamma_{1} = \gamma_{2}$, $\psi_{*}g^{1} = g^{2}$ and $\psi_{*} ({\Psi_{1}}^{\alpha_{1} \ldots \alpha_{k}}_{\textrm{phy} \quad \beta_{1} \ldots \beta_{l}}) = {\Psi_{2}}^{\alpha_{1} \ldots \alpha_{k}}_{\textrm{phy} \quad \beta_{1} \ldots \beta_{l}}$.  It will become clear later that this point of view constitutes the main problem in quantizing the theory of general relativity.  Einstein came first to part of this conclusion by his famous ``hole argument''.  Imagine a spacetime $\mathcal{M}$ with an open set $\mathcal{U} \subset \mathcal{M}$ on which the energy momentum tensor $T_{\alpha \beta}$ vanishes identically.  Let $\psi$ be a diffeomorphism which differs only from the identity on (a subset of) $\mathcal{U}$ and let $g$ be a solution of the Einstein equations.  Then $\psi_{*}g$ is a solution of the \emph{same} equations so he concluded that $g$ and $\psi_{*}g$, which can be identified by an active shift, \emph{have} to represent the same physical metric.  The above principle, which one calls ``general covariance'', leads a vast majority of scientists to the conclusion that points have \emph{no} physical significance in general relativity.  However, as Komar and Bergmann argued, there is a way to assign \emph{physical}\footnote{Scalars constructed from the metric.} coordinates to points of a \emph{generic} spacetime.  For example, in a vacuum spacetime such coordinates exist only if it is a Petrov type I solution (four different eigenvalues of the Weyl tensor).  It is clear that both approaches are very different from each other in the sense that the former is an exclusively kinematical statement while the latter has a \emph{dynamical} aspect since the form of matter and the intial data present determine, through the Einstein equations, the Bergmann Komar coordinates.  The debate between both sides is still alive today and it is almost impossible to pick one side on grounds that are limited to the classical theory itself.  In the more general framework of discrete ``spacetimes'', the first principle has an obvious generalisation while the latter has not.  Therefore, in this thesis, we shall adopt the convention that points in a ``spacetime'' have no physical meaning.  Notice also that both approaches make a physical statement about a Lorentz manifold $(\mathcal{M},g)$ while some people attach a physical interpretation to the points of the manifold themselves \emph{without} any recourse to the metric tensor field.  This last attitude is in my opinion unacceptable since it conflicts with the spirit of general relativity and brings us back to the Newtonian point of view. \\* \\*
We now study some of the topological implications the existence of a Lorentz metric might have.  The existence of a Lorentz metric restricts the topology of the underlying manifold in some cases.  A classical result proven by Steenrod says that an even dimensional, \emph{compact} manifold \emph{without boundary} carrying a Lorentz metric must have Euler number $\chi$ zero (~\cite{Steenrod} p. 207).  In four dimensions, this implies that, since,
$$ \chi(\mathcal{M}) = \sum_{n=0}^{4} (-1)^{n} B_{n}$$
and $B_{n} = B_{4-n}$ for $0 \leq n \leq 2$ :
$$2B_{0} - 2B_{1} + B_{2} = 0.$$
Hence, $B_{1} > 0$, which means that the manifold is not simply connected\footnote{$B_{i}$ are the Betti numbers, namely the dimension of the i'th cohomology group.}.  However, this result is not of interest to us since as mentioned before, we are interested in Lorentz manifolds with boundary which can be simply connected.  Let $\mathcal{S}_{1}$ and $\mathcal{S}_{2}$ be $n-1$ dimensional manifolds, possibly with boundary.  A connected, $n$ dimensional manifold $\mathcal{M}$ with boundary is called \emph{interpolating} between $\mathcal{S}_{1}$ and $\mathcal{S}_{2}$ if there exists a further $n-1$ dimensional(possibly empty) manifold $\mathcal{T}$ such that $\partial \mathcal{M} = \mathcal{S}_{1} \cup \mathcal{S}_{2} \cup \mathcal{T}$.  If $\mathcal{T}$ is not empty, then it is required that $\mathcal{T} \cap \mathcal{S}_{i} = \partial \mathcal{S}_{i}$ $(i=1,2)$, $\partial \mathcal{S}_{1}$ and $\partial \mathcal{S}_{2}$ are diffeomorphic, and $\mathcal{T} \equiv \partial{S}_{1} \times \left[0,1\right]$.  $\mathcal{M}$ is an interpolating spacetime iff there exists a smooth Lorentz metric on $\mathcal{M}$ such that $\mathcal{S}_{1}$ and $\mathcal{S}_{2}$ are spacelike and $\mathcal{T}$ is timelike or empty.  If $\mathcal{S}_{1}$ and $\mathcal{S}_{2}$ are \emph{closed} (compact and without boundary), then a \emph{compact} interpolating manifold $\mathcal{M}$ is called a \emph{cobordism} and $\mathcal{S}_{1}$ and $\mathcal{S}_{2}$ are cobordant (mind: $\mathcal{T} = \emptyset$). Let $\mathcal{M}$ be an interpolating cobordism between $\mathcal{S}_{1}$ and $\mathcal{S}_{2}$, then the following results are of interest.  If $n$ is even, then $\mathcal{M}$ is an interpolating spacetime iff $\chi ( \mathcal{M}) = 0$.  In case $n$ is odd, $\mathcal{M}$ is an interpolating spacetime iff $\chi (S_{1}) = \chi (S_{2})$.  Using surgery methods, one obtains that for $n >2$ even, there exists \emph{always} an interpolating spacetime between any $\mathcal{S}_{1}$ and $\mathcal{S}_{2}$ if there exists an interpolating cobordism between them\footnote{This is always the case for $n=4,7$.}.  For $n$ odd, such interpolating spacetime exists only iff $\chi (S_{1}) = \chi(S_{2})$ given the existence of an interpolating cobordism\footnote{This is always true for $n=3,7$.  The general condition is that the Stiefel-Whitney numbers must be the same, and if orientation is important, then the Pontryagin numbers must be the same too.}.  More details can be found in \cite{Borde} and references therein.  We shall come back in more detail to the kinematical aspects of general relativity later on.     \\*
\\*
Now we turn to the conditions the equations governing the matter fields must satisfy.  First of all, one has the \emph{requirement} that for every convex normal neighborhood $\mathcal{U}$, a ``signal'' in $\mathcal{U}$ can be sent between two points $p$ and $q$ \emph{if and only if} $p$ and $q$ can be joined by a causal curve lying entirely in $\mathcal{U}$.  Stated in a more mathematical language the \emph{local causality} requirement says that a \emph{physical} field $\Psi^{\alpha_{1} \ldots \alpha_{k}}_{\textrm{phy} \quad \beta_{1} \ldots \beta_{l}}$ must satisfy the condition that for any convex normal neighborhood $\mathcal{U}$ and spacelike hypersurface $ \Sigma \subset \mathcal{U}$ and for any  $\mathcal{V} \subset D^{+}( \Sigma) \cap \mathcal{U} $ one has that the value of $\Psi^{\alpha_{1} \ldots \alpha_{k}}_{\textrm{phy} \quad \beta_{1} \ldots \beta_{l}}$ on $\mathcal{V}$ is uniquely determined by the values of the field and its derivatives up to some finite order (usually first order) on $J^{-}_{\mathcal{U}} (\mathcal{V}) \cap \Sigma$.  Moreover the field on $\mathcal{V}$ is not uniquely defined by some initial data on any retract of $J^{-}_{\mathcal{U}} (\mathcal{V}) \cap \Sigma$.  One could introduce a stronger notion of \emph{causality} by not demanding that the hypersurface $\Sigma$ be a part of a convex normal neighborhood: it is usually this form of causality which shows up in theorems about the Cauchy problem in general relativity ~\cite{Wald}, ~\cite{Hawking1},~\cite{Choquet} but it is too strong for our purposes since the unique solution of the Cauchy problem, given suitable initial data, is always a globally hyperbolic space-time.
As argued in ~\cite{Hawking1}, the local causality postulate enables one to determine the null cone $N_{p}$ for each point $p$ and hence one can measure the metric up to a conformal factor.  \newline
The second postulate states that locally energy and momentum are conserved.  One expresses this postulate in terms of the symmetric energy momentum tensor $T_{\alpha \beta}$ as follows:
\begin{itemize}
\item $T_{\alpha \beta}$ vanishes on an open set $\mathcal{V}$ \emph{if and only if} all matter fields vanish on $\mathcal{V}$.  This implies that all fields must have positive energy ~\cite{Hawking1}.  One could reject this restriction and allow for two matter fields (at least one with negative energy) to exist whose energy momentum tensors cancel each other exactly on $\mathcal{V}$.  An argument why one does not allow for negative energies to occur is given later on.
\item $T^{\alpha \beta}$ satisfies the equation
\begin{equation} \label{em}
 \nabla_{\alpha} T^{\alpha \beta} = 0
\end{equation}
\end{itemize}
The second condition is mainly inspired by the results valid in Minkowski space time.  Minkowski space time is maximally symmetric with 4 ``translation'' Killing vectorfields $L_{\alpha}$ (generators of momentum) and 6 ``proper Lorentz'' Killing vectorfields $M_{\alpha \beta}$ (generators of angular momentum) defined by:
\begin{itemize}
\item $L_{\alpha} = \frac{\partial}{\partial x^{\alpha}}$
\item $M_{\alpha \beta} = e_{\alpha}x^{\alpha} \frac{\partial}{\partial x^{\beta}} - e_{\beta}x^{\beta} \frac{\partial}{\partial x^{\alpha}}$ (no summation)
\end{itemize}
where $e_{4} = -1$ and $e_{a} = 1$ otherwhise.  If one interprets $P^{\alpha} = T^{\alpha \beta} L_{\beta}$ as the flux of four momentum in the direction $\frac{\partial}{\partial x^{\alpha}}$and $P^{\alpha}_{\beta} = T^{\alpha \gamma}M_{\gamma \beta}$ as the flux of angular momentum then classical results in Minkowski space say that these currents are conserved in the sense that the integral of the flux over the boundary $\partial \mathcal{D}$ of any compact orientable region $\mathcal{D}$ must vanish.  By Gauss' theorem one has then $\nabla_{\alpha} P^{\alpha} = \nabla_{\alpha}P^{\alpha}_{\beta} = 0$ everywhere.  Using the Killing equation and the symmetry of $T^{\alpha \beta}$ one obtains that $( \nabla_{\alpha} T^{\alpha \beta})L_{\beta} = ( \nabla_{\alpha} T^{\alpha \beta})M_{\beta \gamma} = 0$ which implies that, since all the 10 Killing vectors vectors are lineary independent and $\nabla_{\alpha}T^{\alpha \beta}$ has exactly 4 independent components, $\nabla_{\alpha} T^{\alpha \beta} = 0$ (in principle it would be sufficient to have four independent Killing fields which produce conserved currents to deduce that $\nabla_{\alpha} T^{\alpha \beta} = 0$).  This equation generalizes by the minimal coupling principle from Minkowski space to a general space time as (\ref{em}).  As further explained in ~\cite{Hawking1}, $\nabla_{\alpha} T^{\alpha \beta} = 0$ implies that locally one has an almost conservation of energy, momentum and angular momentum.  Using this, one can show that small isolated bodies move approximately on timelike geodesics independent of their internal constitution provided that the energy density of matter is \emph{non-negative} ~\cite{Dixon}, ~\cite{Fock}.  As a consequence, the second postulate enables one to measure the conformal factor up to a constant multiplying factor. \\*

Now, we come to the field equations for the metric tensor.  Einstein proposed an equation which couples matter and geometry:
\begin{equation*} \label{vvgl} G_{\alpha \beta} = R_{\alpha \beta} - \frac{1}{2}g_{\alpha \beta}R + \Lambda g_{\alpha \beta} = 8 \pi T_{\alpha \beta}
\end{equation*}
where, $G_{\alpha \beta}$ is the Einstein tensor,
$R_{\alpha \beta}$ is the Ricci tensor, $R = R_{\alpha \beta} g^{\alpha \beta}$ is the Ricci
scalar and $\Lambda$ is the cosmological constant.  It has been
proven that the Einstein tensor is the unique tensor, in four
dimensions, which satisfies $\nabla_{\alpha}G^{\alpha \beta} = 0$ and contains
only derivatives up to order 2 of the metric $g_{\alpha \beta}$ \cite{Lovelock}.  One can show that the field equations reduce in the
Newtonian limit and with respect to stationary observers moving on
Killing integral curves with an affine parameter close to the
Killing parameter, to the Poisson equation $\nabla^{2} \phi = 4
\pi \rho$, where $\phi$ is the (Newtonian) gravitational potential
and $\rho$ is the matter density as measured by such observers
\cite{Wein}.  A short, but beautiful ``deduction'' of the Einstein
field equations, starting from the assumption that the Newtonian
limit has to produce the Poisson equation, can be found in
~\cite{Hawking1}.  One can also construct a Lagrangian formulation of
general relativity by demanding that the energy momentum tensor be
derivable from a Lagrangian density $\mathcal{L}_{M}$.  Let
$\mathcal{L}_{M}$ be some scalar (functionally) differentiable
function of a finite set of physical fields $ \Psi^{\alpha_{1} \ldots
\alpha_{k_{i}} }_{i \quad \beta_{1} \ldots  \beta_{l_{i}} }$ and the first
order derivatives $ \nabla_{\gamma} \Psi^{\alpha_{1} \ldots \alpha_{k_{i}}}_{i
\quad \beta_{1} \ldots \beta_{l_{i}}}$ and the metric $g_{\alpha \beta}$.  Denote by
$\mathcal{D}$ a orientable compact, four dimensional region of
$\mathcal{M}$ and let $I(\mathcal{D}) = \int_{\mathcal{D}}
\mathcal{L}_{M} \sqrt{ |g|} d^{4}x $ be the action corresponding
with $\mathcal{L}_{M}$ on $\mathcal{D}$.  Extremizing this action
by perturbing the matter fields $ \Psi^{\alpha_{1} \ldots \alpha_{k_{i}}
}_{i \quad \beta_{1} \ldots  \beta_{l_{i}}} \rightarrow \Psi^{\alpha_{1} \ldots
\alpha_{k_{i}} }_{i \quad \beta_{1} \ldots  \beta_{l_{i}} } + \Delta
\Psi^{\alpha_{1} \ldots \alpha_{k_{i}} }_{i \quad \beta_{1} \ldots  \beta_{l_{i}} }$
and demanding that $\Delta \Psi^{\alpha_{1} \ldots \alpha_{k_{i}} }_{i \quad
\beta_{1} \ldots  \beta_{l_{i}} }$ vanishes on $ \partial \mathcal{D}$,
yields by using Gauss' integration formula the well known Euler
Lagrange equations:
$$ \nabla_{\gamma} \left( \frac{\partial \mathcal{L}_{M}}{\partial \nabla_{\gamma} \Psi^{\alpha_{1} \ldots \alpha_{k_{i}} }_{i \quad \beta_{1} \ldots  \beta_{l_{i}} } } \right) - \frac{\partial \mathcal{L}_{M}}{\partial \Psi^{\alpha_{1} \ldots \alpha_{k_{i}} }_{i \quad \beta_{1} \ldots  \beta_{l_{i}} }} = 0.$$
Variation of $I(\mathcal{D})$ with respect to the metric $g_{\alpha \beta}$ yields, after some mani\-pulation, the result:
$$ \Delta I(\mathcal{D}) = \frac{1}{2} \int_{\mathcal{D}} T^{\alpha \beta} \Delta g_{\alpha \beta} \sqrt{  |g|} d^{4} x$$ where $T^{\alpha \beta}$ is to be interpreted as the energy momentum tensor.  The conservation equations of the energy momentum tensor follow from the \emph{diffeomorphism} invariance of the integral for fields satisfying the Euler Lagrange equations.  More precisely, let $\psi$ be a diffeomorphism which differs from the identity only in the interior of $\mathcal{D}$, then one has:
$$ I(\mathcal{D}) = \int_{\mathcal{D}} \mathcal{L}_{M} \sqrt{ |g|}d^{4}x = \int_{\psi ( \mathcal{D})}  \mathcal{L}_{M} \sqrt{ |g|} d^{4}x = \int_{\mathcal{D}} \psi^{*} ( \mathcal{L}_{M} \sqrt{ |g|} d^{4}x ).$$
Hence if a vectorfield $X$ generates a one parametergroup of diffeomorphisms then one has
$$ \int_{\mathcal{D}} L_{X} ( \mathcal{L}_{M} \sqrt{|g|} d^{4}x ) = 0$$
from which one can easily deduce that $\nabla_{\alpha} T^{\alpha \beta}$ has to be zero.
Now one still needs an action for which the Euler Lagrange equations under variation of the metric, deliver the Einstein tensor.  It is well known that the correspon\-ding Lagrangian density can be written as $\frac{1}{16 \pi}(R - 2 \Lambda )$.  This results in the Einstein Hilbert action:
\begin{equation*} \label{action} I(\mathcal{D}) = \int_{\mathcal{D}} \left( \frac{1}{16 \pi}(R - 2 \Lambda) + \mathcal{L}_{M} \right) \sqrt{ |g|} d^{4}x \, . \end{equation*}
Let me stress that $\nabla_{\alpha}T^{\alpha \beta} = 0$ \emph{because} $\nabla^{\alpha} G_{\alpha \beta} = 0$.  One can show that for any $\mathcal{D}$ the action $I(\mathcal{D})$ with $\mathcal{L}_{M} = 0$ is unbounded as a function of $g$ which is easily seen by transforming $g$ with a constant conformal factor.  This is important for Euclidian quantum gravity, since if the Euclidian action is not positive semidefinite then it is unbounded from below which makes the path integral ill defined.  The Einstein-Hilbert action needs a boundary term, if there is a boundary \cite{Wald}.
\section{The Hamiltonian formulation}
I shall present the canonical theory in the old ADM variables.
The reason why I do not present the connection representation,
which is more suitable for quantisation, is that the latter topic
is not of interest in this thesis.  It is merely my intention to present the classical algebra and to discuss the
``principle'' of canonical quantisation applied to general
relativity, instead of pointing out the technical benefits of the
Ashtekhar variables which would not bring any improvement to this
discussion.  The presentation shall be based upon the old papers
of Kucha\v{r} and Isham (\cite{Isham1} and \cite{Isham2}).  To the
best knowledge of the author, these $1984$ papers are the first
and last ones that dealt with the subject of the representation
of the Lie algebra of the diffeomorphism group in the canonical
formulation of gravity.  Recently however, the implementation of
the diffeomorphism group in a kind of combination of the covariant
and canonical approach to classical relativity was studied by
Savvidou in \cite{Sav1} and \cite{Sav2}.  It shall be
made clear to the reader that, even at the classical level, the
representation of the diffeomorphism group in the canonical
framework is very unsatisfactory, although the dynamical character
of that representation is a bonus over the purely kinematical (and
thus theory independent) approach followed in the rest of this
thesis. \\*  \\* The Hamiltonian formulation of gravity starts
from a globally hyperbolic spacetime with a fixed foliation.  Let
$\psi :\Sigma \times \mathbb{R} \rightarrow \mathcal{M}$ be a
diffeomorphism which describes the foliation and denote by $n^{\mu}$ the unit normal vectorfield
to the surfaces $\psi \left( \Sigma \times \left\{ t \right\}
\right)$.  By the Frobenius theorem there exist
$C^{\infty}$-functions $h$ and $f$ such that $n_{\mu} = h f_{,
\mu}$ locally.  The vectorfield with components
$$T^{\mu} = \frac{\partial \psi^{\mu} (x,t)}{\partial t} = N(\psi(x,t))n^{\mu} + N^{\mu}(\psi(x,t))$$
determines timelike integral curves which one can think of as world lines of observers.  The positive function $N$ is called the ``lapse'' and the vectorfield $N^{\mu}$ is the shift vectorfield, which satisfies $N^{\mu}g_{\mu \nu}n^{\nu} = 0$ everywhere.  The configuration space is formed by the first fundamental form (or just the induced Riemannian metric) $$\gamma_{\mu \nu} = g_{\mu \nu} + n_{\mu} n_{\nu}.$$   The second fundamental form (or shape operator) is defined by
$$ K_{\mu  \nu } = \gamma_{\mu}^{ \alpha} \gamma_{\nu}^{ \beta} \nabla_{\alpha} n_{\beta}.$$
In the last equation, all indices are raised or lowered by means of the metric $g_{\mu \nu}$.  It is easy to derive that the extrinsic shape operator is symmetric.  Also, Arnowitt, Deser and Misner showed that the $n+1$ dimensional vacuum Einstein Hilbert action can be cast in the form:
$$I\left[\gamma_{ab}, \dot{\gamma}^{ab}, N, N^{a}\right] = \int_{\mathbb{R}} dt \int_{\Sigma} d^{n}x \sqrt{det(\gamma)} N \left[ R^{n} + K_{ab}K^{ab} - K^{2} \right], $$
where $R^{n}$ is the Ricci scalar associated to $\gamma_{ab}$.  Note that when $n=1$, then $I$ vanishes identically.  The canonical momentum is computed to be
$$ \pi^{ab} = \frac{\partial I}{\partial \dot{\gamma}_{ab}} = \sqrt{det(\gamma)}\left[ K^{ab} - \gamma^{ab} K \right] $$
Notice that the conjugate momentum $\pi^{ab}$ is a smooth contravariant $(2,0)$ tensor density with weight\footnote{A quantity $T^{a_{1}\ldots a_{k}}_{b_1 \ldots b_l}$ is a tensor density of weight $m$ iff it transforms under $x^{a} \rightarrow \tilde{x}^a$ as $$\tilde{T}^{a_{1}\ldots a_{k}}_{b_1 \ldots b_l} = det\left( \frac{\partial \tilde{x}^{a}}{\partial x^{b}}\right)^{-m} \frac{\partial \tilde{x}^{a_1 }}{\partial x^{c_1}} \ldots \frac{\partial \tilde{x}^{a_k }}{\partial x^{c_k}} \frac{\partial x^{d_1} }{\partial \tilde{x}^{b_1}} \ldots \frac{\partial x^{d_l }}{\partial \tilde{x}^{b_l}} T^{c_1 \ldots c_k}_{d_1 \ldots d_l}.$$} 1.
Defining the densities,
\begin{itemize}
\item $H_{a} = -2D_{b} \pi_{a}^{b}$, the so called spatial diffeomorphism constraint,
\item $H = - \sqrt{det(\gamma)} R^{n} +  \frac{1}{ \sqrt{det(\gamma)} } \left[ \pi_{ab}\pi^{ab} - \frac{1}{n-1} \pi^{2} \right]$, the Hamiltonian constraint,
\end{itemize}
where $D$ is the covariant derivative for the metric $\gamma_{ab}$ and $\pi^{2} = \pi^{ab} \pi_{ab}$, we are able to cast the action in the following form
$$ I\left[\gamma_{ab}, \pi^{ab}, N, N^{a}\right] = \int_{\mathbb{R}} dt \int_{\Sigma} d^{n}x \left( \pi^{ab}\dot{\gamma}_{ab} - NH - N^{a}H_{a} \right). $$
Since one can vary the lapse and shift ``rather'' independently\footnote{One needs that $N^{2} - N^{a}\gamma_{ab}N^{b} > 0$.  I should note that this restriction is \emph{not} strictly necessary from the mathematical point of view and it is not used by Kucha\v{r} and Isham.  I add it however since otherwise the canonical formalism wouldn't make much physical sense to me.} of $\gamma_{ab}$ and $\pi^{ab}$ one arrives at the constraint equations
$$ H=H_{a}=0 $$
The Hamiltonian $H(N,N^{a})$ equals
$$H(N,N^{a}) = H(N) + H(N^{a}) = \int_{\Sigma} d^{n}x N(x)H(x) + \int_{\Sigma}d^{n}x N^{a}(x)H_{a}(x) $$
and the (unprojected) constraints satisfy the Dirac algebra
$$ \left\{ H(M,M^{a}), H(N,N^{a}) \right\} = H(M^{a}N_{,a} - N^{a}M_{,a}, \left[M^{c},N^{c}\right]^{a} + \gamma^{ab}(MN_{,b}-NM_{,b})).$$
The question around $1980$ was how diffeomorphism invariance was implemented in the Dirac algebra.  Clearly, the mapping $N^{a} \rightarrow H(N^{a})$ is an \emph{anti}homomorphism\footnote{By choice of the Lie bracket on the Lie algebra $L Diff(\Sigma)$ of $Diff(\Sigma)$.} from $L Diff(\Sigma)$ to the Poisson bracket algebra on the space of functionals on the phase space $(\gamma_{ab},\pi^{ab})$.  However, the appearance of $\gamma^{ab}$ in the above formula shows that the representation cannot be extended to $L Diff(\mathcal{M})$.  The solution Kucha\v{r} and Isham provided to this problem is to make the embedding dynamical and as such extend the configuration space (with respect to some gauge fixed one parameter family of embeddings). \\* \\*
Denote (as in \cite{Isham2}) by $Y:\Sigma \times \mathbb{R} \rightarrow \mathcal{M}$ the \emph{fixed} foliation (gauge) and define $\sigma : \mathcal{M} \rightarrow \Sigma$ and $\tau : \mathcal{M} \rightarrow \mathbb{R}$ by $Y^{-1}(p) = (\sigma(p),\tau(p))$.  Define $Emb_{Y}( \Sigma, \mathcal{M})$ as the set of all embeddings $X:\Sigma \rightarrow \mathcal{M}$ which can be continuously deformed into one of the leaves $Y_{t}$ by transporting the points of $X(\Sigma)$ over the curves of the congruence $\left\{ Y_{x} | x \in \Sigma \right\}$.  We shall take as configuration space\footnote{$Riem \Sigma$ is the vector space of all Riemannian metric tensors on $\Sigma$.} $Emb_{Y}(\Sigma, \mathcal{M}) \times Riem \Sigma$, where $(X,\gamma)$ determines a spacetime metric $g$ through the following conditions defined by $Y$:
\begin{itemize}
\item $g^{-1}(d\tau, d\tau) = -1$
\item $g^{-1}(d\tau, \sigma^{*}(\rho)) = 0$ for all $\rho \in T_{*}\Sigma$
\item $g_{X(x)}(X_{*}v(x),X_{*}w(x)) = \gamma_{x}(v(x),w(x))$ for all vector fields $v,w \in T^{*}\Sigma$
\end{itemize}
which are equivalent to
\begin{itemize}
\item $g_{Y(x,t)}(\dot{Y}_{x}(t),\dot{Y}_{x}(t)) = -1$
\item $g_{Y(x,t)}(\dot{Y}_{x}(t), Y_{t \,*}V(x,t)) = 0$ for all $V(x,t) \in T^{*} (\Sigma \times \left\{t \right\})$
\item $g_{X(x)}(X_{*}v(x),X_{*}w(x)) = \gamma_{x}(v(x),w(x))$ for all vector fields $v,w \in T^{*}\Sigma$.
\end{itemize}
Explicitly,
\begin{eqnarray*} g_{\alpha \beta}(X(x)) & = & \gamma_{cd}(x) \tilde{\sigma}^{-1 \, c}_{,a}(x) \tilde{\sigma}^{-1 \, d}_{,b}(x) \sigma^{a}_{,\alpha}(X(x)) \sigma^{b}_{,\beta}(X(x)) + \\ & & \tilde{\tau}_{,c}(x)\tilde{\sigma}^{-1 \, c}_{,a}(x) \tilde{\tau}_{,d}(x) \tilde{\sigma}^{-1 \, d}_{,b}(x) \sigma^{a}_{,\alpha}(X(x)) \sigma^{b}_{,\beta}(X(x)) - \\ & & \tau_{,\alpha}(X(x)) \tau_{,\beta}(X(x)) \end{eqnarray*}
where $\tilde{\sigma} = \sigma \circ X$ belongs to $Diff(\Sigma)$
and $\tilde{\tau} = \tau \circ X$ is an element of
$C^{\infty}(\Sigma , \mathbb{R})$.  The second formulation of the
gauge conditions shows that we also have to make sure that
$X(\Sigma)$ is \emph{nowhere} tangent to $\dot{Y}_{x}(t)$ since
then $\gamma_{ab}$ is not positive definite and we do not allow
signature changes.  It is important to notice that due to the
specific form of the gauge conditions $g_{\alpha \beta}$ is a
functional of the embedding $X$ and the first fundamental form
$\gamma_{ab}$ and \emph{not} of its conjugate momentum.  On the
other hand, it should be noted that a globally hyperbolic metric
$g$ can, in general, only be cast into the above form by action
of an (active) diffeomorphism $\psi \in Diff_{0}(\mathcal{M})$ in
a \emph{neighbourhood} of the compact time slab $Y_{t}$.  Define
the Lie group $\mathcal{G}$ of spacetime diffeomorphisms $\psi = Y
\circ \alpha \circ Y^{-1}$ where $\alpha \in Diff(\Sigma \times
\mathbb{R})$ is of the form $\alpha(x,t) = (\beta(x),f(x,t))$ with
$\beta(x) \in Diff(\Sigma)$ and $f(x, \cdot) \in
Diff(\mathbb{R})$.  Then it is easy to see that a diffeomorphism
$\psi$ defines a transformation on $Emb_{Y}(\Sigma, \mathcal{M})$
by left multiplication iff $\psi \in \mathcal{G}$. The previous
considerations lead us to the following important conclusions:
\begin{itemize}
\item  for any vectorfield $X \in T^{*} \mathcal{M}$, we have that the flow map of $X$ (interpreted as an element of $T^{*} Emb_{Y}(\Sigma , \mathcal{M})$) is well defined on $Emb_{Y}(\Sigma, \mathcal{M})$ for sufficiently small times iff $X$ generates a one parametergroup of diffeomorphisms in $\mathcal{G}$.
\item  the action of the Lie algebra of $Diff(\mathcal{M})$ is \emph{not} the active one because of the gauge fixing.
\end{itemize}
The rest of this section is devoted to showing how the action of $L Diff(\mathcal{M})$ can be implemented on the extended phase space with canonical coordinates $(X^{\alpha}, P_{\alpha}, \gamma_{ab}, \pi^{ab})$.  For this purpose, Kucha\v{r} and Isham continue by constructing a new action principle which is dynamically equivalent to the Einstein Hilbert action on $(X^{\alpha}, P_{\alpha}, \gamma_{ab}, \pi^{ab})$.  From the ADM action, one computes that the conjugate momenta $P_{\alpha}$ of the embedding variables $X^{\alpha}$ equal
$$P_{\alpha}(x) = H(x)n_{\alpha}(x) - H_{a}(x) X^{,a}_{\alpha}(x) $$
This leads to the constraints
$$ \Pi_{\alpha} := P_{\alpha} - Hn_{\alpha} + H_{a}X^{,a}_{\alpha} = 0 $$
By introducing the \emph{Lagrange multipliers}\footnote{It must be obvious to the reader why we use our old notation of the lapse and shift here.} $N^{\alpha}$, we obtain a Lagrangian
$$ I\left[ X^{\alpha}, P_{\alpha}, \gamma_{ab}, \pi^{ab} , N^{\alpha} \right] = \int_{\mathbb{R}} dt \int_{\Sigma}d^{n}x (P_{\alpha} \dot{X}^{\alpha} + \pi^{ab} \dot{\gamma}_{ab} - N^{\alpha} \Pi_{\alpha} ) $$
The main algebraic result is that the projected quantities $\Pi = n^{\alpha} \Pi_{\alpha}$, $\Pi_{a} = X^{\alpha}_{,a} \Pi_{\alpha}$ satisfy the Dirac algebra and that the Poisson brackets between the unprojected constraints vanish strongly\footnote{That means identically.  Poisson brackets vanish weakly if they are zero on the constraint surface.}, i.e.,
$$ \left\{\Pi_{\alpha}(x), \Pi_{\beta}(x') \right\} = 0. $$
Using this, one can show that the following mapping $\Pi$
$$ \Pi(V) :=  \int_{\sigma} d^{n}x V^{\alpha}(x) \Pi_{\alpha}(x) $$ from $L Diff(\mathcal{M})$ to the Poisson bracket algebra of $C^{\infty} ( T^{*} Emb_{Y}( \Sigma , \mathcal{M} ) \times T^{*} Riem \Sigma , \mathbb{R})$ is in fact a homomorphism.  Hence, the conclusion is that in the canonical framework, $LDiff(\mathcal{M})$ is faithfully represented, but ``most'' vectorfields are not integrable on the extended phase space which implies that $Diff(\mathcal{M})$ is only ``partially'' realised.  \\* \\*  We now turn to a brief discussion of the metric formulation of relativity by J. L. Synge.
\section{The World function formulation}
In this section, I give the basic equations underlying the world
function formulation of relativity.  It is \emph{not} my intention
to give a precise and complete account of this subject, but rather
this section is meant as a motivation for the kinematical point of
view which I take in the rest of this thesis.  In the previous
section, I came to the conclusion that the canonical formulation
of a generally covariant theory (gravity in particular) does
\emph{not} possess a representation of the spacetime diffeomorphism
group.  However, the advantage of the canonical approach is that
the dynamical variables are encoded in the representation of $L
Diff(\mathcal{M})$ through the constraints.  But this should be
taken with some care since this small victory carries with itself
a huge cost since the Dirac Hamiltonian is ``pure'' constraint and
hence there is no quantum evolution of the wave function of the
universe\footnote{This is one of the so called problems of time.}.
Apart from this remark, some people doing canonical quantum
gravity think of diffeomorphism invariance as having a dynamical
character since the representation of $L Diff(\mathcal{M})$ in the
Poisson bracket algebra on $C^{\infty} ( T^{*} Emb_{Y}( \Sigma ,
\mathcal{M} ) \times T^{*} Riem \Sigma , \mathbb{R})$ has only a
sensible interpretation as an active shift of the intial data on
the (theory dependent) constraint surface.  This does not make
much sense to me since the ``on shell character of the
representation'' teaches me everything about the representation
itself and nothing about the principle of diffeomorphism
invariance.  At the Lagrangian level, active diffeomorphism
invariance clearly is a \emph{theory independent} property and one
should not be surprised that imposing a gauge ``breaks'' this
symmetrygroup in the canonical framework.   This, in my point of
view, implies that the canonical language does not do justice to
generally covariant theories and neither does the principle of
canonical quantisation.  \\* \\* One could go one step further and
ask for a formulation of general relativity in which active
diffeomorphism invariance is clearly represented and moreover, the
main kinematical object does not depend on the differentiable
structure at all.  The latter is important if one does not believe
spacetime is a differentiable manifold but, say, a discrete
set\footnote{Arguments for discreteness will be given in the next
section.}.  Thanks to L. Bombelli, I heard about the formulation
of Synge which I shall describe now.  The world function
$\Omega_{g} : \mathcal{M} \times \mathcal{M} \rightarrow
\mathbb{R}$ for a spacetime $(\mathcal{M},g)$ is defined as
follows:
$$\Omega_{g}( p,q) = \frac{1}{2} \epsilon L_{g}^{2} (p,q) $$ where $\epsilon = \pm 1$ depending on whether $p$ and $q$ are spacelike or timelike separated respectively\footnote{As a matter of fact, I prefer to define the World function using the Lorentz distance $d_{g}$.  The result that $g_{\alpha \beta}$ can be retrieved from coincidence limits of second partial order derivatives of the World function then remains true, since knowledge of all metric components in the timelike directions determines the entire metric tensor.}.  $L_{g}$ stands for the geodesic length between the points $p$ and $q$ if it is defined\footnote{Notice that this definition has huge global problems since then it is possible for two points to be connected by many geodesics.  In fact, timelike related points could even be connected by an infinite number of spacelike geodesics.  But we shall simply disregard these problems in this section.  Remark that in the absence of such problems $L_{g}$ coincides with $d_{g}$ for timelike related pairs.}.  To get a feeling for the definition and some properties of it, I make a small study of 4 dimensional Minkowski spacetime $(\mathbb{R}^{4} , \eta )$.  Obviously, $$\Omega_{\eta}(x^{\mu},y^{\nu}) = \frac{1}{2} (x^{\alpha} - y^{\alpha})\eta_{\alpha \beta}(x^{\beta} - y^{\beta}).$$
We have the following formulas:
\begin{itemize}
\item $\eta^{\alpha \beta} \frac{\partial \Omega_{\eta}(x^{\mu},y^{\nu})}{\partial x^{\alpha}} \frac{\partial \Omega_{\eta}(x^{\mu},y^{\nu})}{\partial x^{\beta}} = 2 \Omega_{\eta} (x^{\mu},y^{\nu}) $ and the same formula is valid when one derives to the $y^{\nu}$ coordinates.
\item $\lim_{x \rightarrow y} \Omega_{\eta} (x^{\mu}, y^{\nu}) = 0$
\item $\lim_{x \rightarrow y} \frac{\partial \Omega_{\eta}(x^{\mu}, y^{\nu})}{\partial x^{\alpha}} = 0$ and the same formula is valid when one derives with respect to the $y^{\nu}$ coordinates.
\item $\lim_{x \rightarrow y} \frac{\partial^{2} \Omega_{\eta}(x^{\mu}, y^{\nu})}{\partial x^{\alpha} \partial x^{\beta}} = \eta_{\alpha \beta}(y^{\nu})$
\end{itemize}
It is easy to see from this example why the ``Lorentzian geodesic energy function'' rather than the ordinary Lorentz distance is used, since derivatives of the latter have bad coincidence properties.  Assuming that $L_{g}$ is well defined and that $g$ is of sufficiently high differentiability class, the above properties remain true in the general case.  It is important to notice that the last coincidence limit in the general case is computed \emph{without} knowledge of the connection since:
\begin{eqnarray*} g_{\alpha \beta}(q) & = & \lim_{p \rightarrow q} \nabla_{p^{\beta}}\partial_{p^{\alpha}} \Omega_{g}(p,q) \\
& = & \lim_{p \rightarrow q} \partial_{p^{\beta}} \partial_{p^{\alpha}} \Omega_{g}(p,q) - \left( \lim_{p \rightarrow q} \Gamma_{\beta \alpha}^{\mu}(p) \right) \left( \lim_{p \rightarrow q} \partial_{p^{\mu}} \Omega_{g}(p,q) \right) \\
& = & \lim_{p \rightarrow q} \partial_{p^{\beta}} \partial_{p^{\alpha}} \Omega_{g}(p,q)
\end{eqnarray*}
All tensors of importance (Riemann curvature and Weyl tensor) can be written in terms of coincidence limits of derivatives of the world function.  In particular the Einstein equations are fourth order partial differential equations of $\Omega_{g}$.  For more details, the reader is referred to the beautiful classic \cite{Synge} and to the references therein.  It is also important to know that to the best knowledge of the author \emph{no} set of intrinsic conditions on $\Omega$ is known such that the coincidence limits of second partial derivatives exist \emph{and} give rise to a Lorentzian tensor.  Again, I stress that this section is merely meant as motivation for the further use of the observable $d_{g}$ as basic kinematic variable in the rest of this thesis.  In all fairness, I do not believe that the dynamics for $\Omega_{g}$ will give us any hint for classical dynamics of generalised structures such as discrete \textsl{Lorentz spaces}.  Classical dynamics on the subset of causal sets has already been studied on basis of general physical principles and we shall come back to this in the next section.        
\section{Quantum theory of gravity}
After this rather personal birds eye view of classical relativity, I allow myself in my young enthousiasm to contemplate a bit about possible directions for quantum gravity.  The reader should be aware that this section is an outline of my \emph{contemporary} laundry list which is more constructed by exclusion guided by the utter wish not to break some generalised form of active diffeomorphism invariance at the ``quantum level'', rather than by knowledge of a \emph{mathematically} worthy alternative for the contemporary state of canonical quantum gravity.  I think I explained very carefully what I think are the shortcomings in the canonical program:
\begin{itemize}
\item no \emph{active} spacetime diffeomorphism invariance
\item a classical phase space which is a priori dependent on the gauge fixing (choosing a particular class of observers)
\item no representation of the spacetime diffeomorphism group (apart from the subgroup $\mathcal{G}$)
\item no nontrivial evolution operators at the quantum level (since the Hamiltonian is pure constraint)
\end{itemize}
One also encounters a measurement problem of how to express observables in a spacetime language since the knowledge of how to do so would require one to be able to integrate the field equations from the initial data (in some gauge). \\* \\*
I guess the basic question is whether spacetime is continuous or discrete.  There are many indications which speak for discreteness such as the divergences in Quantum Field Theory which have to be ``cured'' by making an ultraviolet cutoff (which implies breaking of Lorentz invariance), the singularities in classical relativity and more recently the results emerging from canonical quantum gravity concerning the discreteness of the spectra of \emph{spacelike} geometric operators like (geodesic) length, area and volume.  The main result here is that these operators have a \emph{positive} smallest eigenvalue\footnote{As argued before, it is reasonable to expect that good results concerning the spacelike sector emerge from canonical quantum gravity.}.  The reader can find a more extended expos\'e in \cite{Sorkin3}.  \\* \\*
If spacetime really is discrete, what does this discrete substratum look like?  The answer R. Sorkin provides to this question is the causal set $(\mathcal{P},\prec)$ approach, where $\mathcal{P}$ is a finite or countable set, $\prec$ is a partial order relation and $p \prec q$ should be interpreted as ``$p$ is in the past of $q$''.  $(\mathcal{P},\prec)$ has a natural interpretation as a measure space by simply counting the number of elements.  For a relation between a causal set and a spacetime the reader may consult \cite{Bombelli3} and references therein\footnote{In this thesis, we shall study Lorentz spaces and as the reader shall see in chapter four, most causal sets have a natural interpretation as Lorentz space.}.  In the weakest sense, a growth process of a causal set is a sequence $(\mathcal{P}_{n} , \prec_{n})_{n \in \mathbb{N}}$ of causal sets such that $\mathcal{P}_{n+1} = \mathcal{P}_{n} \cup \left\{p_{n+1} \right\}$,  $p \prec_{n} q$ iff $p \prec_{n+1} q$ for all $p,q \in \mathcal{P}_{n}$ and there exists no $p \in \mathcal{P}_{n}$ such that $p_{n+1} \prec_{n+1} p$.  Classical dynamics of a causal set is then defined as the \emph{labelled} markovian growth process where the transition probabilities have to satisfy the physical requirements of \emph{discrete} general covariance and Bell causality.  Discrete general covariance is just labelling invariance (the discrete analogue of coordinate invariance) and Bell causality states that the transition probabilities $P(\mathcal{P}_{n} \rightarrow \mathcal{P}_{n+1})$ only dependend upon the past set of $p_{n+1}$ in $\mathcal{P}_{n}$.  A full classification of these \emph{physical} stochastic growth dynamics has been given by D. Rideout in \cite{Rideout}.  Denote by $\tilde{\Omega}$ the set of labelled, infinite causal sets which can be seen as the set of all growth processes.  Let $\tilde{a}(n)$ be a labelled causal set with $n+1$ elements.  We define the cylinder set $cyl(\tilde{a}(n))$ as the set of all elements $\tilde{x} \in \tilde{\Omega}$ such that $\tilde{x}$ coincides with $\tilde{a}(n)$ on the first $n+1$ elements.  As usual, this type of stochastic dynamics induces a measure on the set of cylinder sets which can be uniquely extended to the regular\footnote{Regular means that there exists a countable family of elements in $\tilde{\mathcal{R}}$ which is separating and generating.} $\sigma$-algebra $\tilde{\mathcal{R}}$ generated by the cylinder sets.  Physical predicates then correspond to labelling invariant subsets of $\tilde{\mathcal{R}}$.  Define the equivalence relation $\sim$ on $\tilde{\Omega}$ by $\tilde{x} \sim \tilde{y}$ iff $\tilde{x}$ and $\tilde{y}$ are different labellings of the same partial order.  Let $\Omega = \left\{ \left[ \tilde{x} \right]_{\sim} | \tilde{x} \in \tilde{\Omega} \right\}$ and $\mathcal{R} = \tilde{\mathcal{R}}/_{\sim}$, then the question is how to provide a physical interpetation of the elements of $\mathcal{R}$.  The stem sets $st(\left[\tilde{a}(n)\right]) = \left[ cyl(\tilde{a}(n)) \right]$ are the unlabelled versions of the cylinder sets and have a direct physical meaning\footnote{The physical meaning of $st(\left[\tilde{a}(n)\right])$ is the probability that $\left[\tilde{a}(n)\right]$ occurs.}.   Let $\mathcal{S} \subset \mathcal{R}$ be the set of all stem sets. The main result in \cite{Sorkin4} states that the complement of the $\sigma$-algebra $\mathcal{R}(\mathcal{S})$, generated by $\mathcal{S}$ in $\mathcal{R}$, has measure zero for all physical growth dynamics.  So far, causal set theory has no quantum dynamics and quantum measurement theory yet. \\* \\* Let us now come back to some questions on the laundry list like: ``is topology, dimension fixed or dynamical?''.  First of all, there is a philosophical reason which forces us to choose the latter : why would a causal growth process happen \emph{a priori} on some fixed and God given four dimensional continuum which ``affects without being affected''?  My main motivation for claiming that topology and dimension are dynamical, lies in the realisation that two Lorentz spaces are approximately the same iff \emph{all} measurements of geodesic timelike distances can almost be identified.  If this were not true then measurements of timelike intervals (the only ones we make!) wouldn't make much sense, since small measurement errors could indeed jeopardize our Weltanschaung seriously from a \emph{global} point of view.  I stress the word ``all'' since it is easy to give examples of causally discontinuous spacetimes where a small, local perturbation (w.r.t. some class of observers) of the metric could have drastic global consequences.  If one takes this to be the main guideline, then properties like topology and dimension are highly unstable as could be expected \emph{a priori} from a discrete theory.  Hence, we arrive at the conclusion that in a good kinematical framework one needs some \emph{isometry}\footnote{This is our generalized form of active diffeomorphism invariance.} invariant notion of comparison between discrete and continuum Lorentz spaces for the purpose of either taking a continuum limit or constructing an active \emph{isometry} invariant measure.   This work is almost entirely accomplished in this thesis.

%% file: symmetry.tex
\chapter{\mbox {Symmetry}}
\begin{verse} 
\hspace{13pt} \textsl{One of the first things you do in a gauge theory is to study the symmetry group of the model.  There is a huge amount of literature available on the diffeomorphism group of a differentiable manifold.  The intention of this chapter is to give an outline of the results which I needed in my first paper.  It is unnecessary to say that such survey is quite technical and incomplete.  I opted for using a minimal but necessary amount of functional analytic concepts, which unavoidably leads to a presentation which is perhaps too compressed.  Moreover, I do not believe that a perfect knowledge of this subject (which might take quite some time of study) is necessary in order to understand the most important subtleties.  Nevertheless, I hope that the ambitious reader can find what he or she might be looking for in the references listed below.} 
\end{verse}
\section{Functional analytic preliminaries.}
The aim of this section is to give a quick introduction to some basic concepts from functional analysis.  The reader who wishes to gain a better understanding of the subject is reffered to \cite{Yosida} and \cite{Michor} and references therein.  
\begin{deffie} 

\begin{itemize}
\item A \emph{locally convex} space $E$ is a (real) vector space with a Hausdorff topology which is generated by a family of seminorms.  A seminorm $p:E \rightarrow \mathbb{R}$ is defined by the following properties: $p(x) \geq 0$, $p(x+y) \leq p(x) + p(y)$ and $p(\lambda x) = \left| \lambda \right| p(x)$.  
\item A subset $A$ in a vector space is called absorbing iff $\bigcup_{r >0}rA$ is the whole space.  
\item The Minkowski functional $q_{A} : x \rightarrow \inf \left\{ t > 0 : x \in tA \right\}$ of a convex absorbing set is a convex function.
\item A set $A$ in a vector space is called abolutely convex iff $\lambda x + \mu y \in A$ for all $x,y \in A$ and $\left|\lambda \right| + \left| \mu \right| \leq 1$ 
\end{itemize} $\square$
\end{deffie}   
It is clear that in a locally convex space $0$ has a basis of neighorhoods consisting of absolutely convex sets.  Also, scalar multiplication and addition are continuous by definition.  Conversely, one can show that a vector space with a Hausdorff topology having these previous properties is necessarily a locally convex space, by proving that the Minkowski functionals associated to the absolutely convex neighbourhoods of the zero vector generate the topology.  An important concept is the one of \emph{Fr\'echet space}:
\begin{deffie}
A Fr\'echet space is a complete, locally convex space with a metrizable topology or, equivalently, with a countable basis of seminorms.
\end{deffie}       
There is a natural way to construct Banach spaces out of locally convex spaces by taking quotients.  Let $E$ be a locally convex vector space and suppose $U$ is a convex neighbourhood of $0$.  Denote by $\widetilde{E_{U}}$ the completed quotient of $E$ with the Minkowski functional of $U$ as norm and let $Pr_{U} : E \rightarrow \widetilde{E_{U}} : x \rightarrow \left[ x \right]_{U}$ be the natural projection\footnote{The construction of this quotient goes a follows: $N(U) = \left\{x \in E : q_{U}(x) = 0 \right\}$ is a vector space and obviously $E/N(U)$ is Hausdorff with respect to the Minkowski functional $q_{U}$.}.  Suppose that $V \subset U$ and $V,U$ are convex neighbourhoods of $0$, then $q{_U}(x) \leq q_{V}(x)$ for all $x \in E$.  Hence $N(V) \subset N(U)$ and there exists a natural linear projection of $\widetilde{E_{V}} \rightarrow \widetilde{E_{U}}$ which, with a slight abuse of notation, can be written as $Pr_{U} \circ (Pr_{V})^{-1}$.  A linear operator $T:E \rightarrow F$ between Banach spaces is called nuclear or \emph{trace class} iff $T$ can be written in the form
$$T(x) = \sum_{j=1}^{\infty} \lambda_{j} x_{j}(x) y_{j} $$ 
where $x_{j} \in E'$, $y_{j} \in F$ with $\left\|x_{j} \right\| \leq 1$ and $\left\| y_{j} \right\| \leq 1$ and $(\lambda_{j})_{j} \in \ell^{1}$ \footnote{$E'$ denotes the topological dual of $E$, i.e., $E'\subset E^{*}$ is the subspace of continuous linear functionals from $E$ to $\mathbb{R}$.}.  The trace\footnote{One can however define the trace in a more familiar form by: 
$$ Tr(T) = \sup_{S \in E \otimes F'} \frac{\left| Tr(TS) \right|}{\left\| S \right\|}$$ where the norm on $E \otimes F'$ is defined as $$ \left\| S \right\| = \inf \sum_{i} \left\| s_{i}\right\|_{E}\left\| t_{i}\right\|_{F'} $$ where the infimum is taken over all representations $\sum_{i} s_{i} \otimes t_{i}$ of $S$.} of $T$ is then defined as 
$$Tr(T) = inf \sum_{j =1}^{\infty} \left| \lambda_{j} \right|$$ 
where the infimum is taken over all such representations.  It is well known that the topological dual of the compact operators $\mathcal{K}(\mathcal{H})$ on a separable Hilbert space $\mathcal{H}$ equals the $W^{*}$ algebra of the bounded operators $\mathcal{B}(\mathcal{H})$ on $\mathcal{H}$.  Moreover, the topological dual of $\mathcal{B}(\mathcal{H})$ is the space of all trace class operators.  \\* \\*
Now we are in position to define a nuclear locally convex space:
\begin{deffie}
A locally convex space $E$ is nuclear if for any absolutely convex neighbourhood $U$ of $0$, there exists another such neighbourhood $V \subset U$ such that the canonical projection $Pr_{U} \circ Pr^{-1}_{V}:\widetilde{E_{V}} \rightarrow \widetilde{E_{U}}$ is trace class.  
\end{deffie}   
Before we start of with the real material, we still need to recall the notion of projective limit and some stability result.  In general, let $J$ be a set directed by a partial order relation $\leq$ and let $(E_{i})_{i \in J}$ be a family of topological vector spaces.  The direct product space $\prod_{i \in J} E_{i}$ is equipped with the coarsest topology which makes all the projections $\pi_{j}: \prod_{i \in J} E_{i} \rightarrow E_{j}$ continuous.  Assume further that we are given for any $j \leq k$ a continuous linear map $T_{j,k}: E_{k} \rightarrow E_{j}$ such that 
\begin{itemize}
\item $T_{j,j} = Id_{E_{j}}, \quad \forall j \in J$
\item $T_{i,k} = T_{i,j} \circ T_{j,k}$ if $i \leq j \leq k$
\end{itemize}                       
The object $(E_{j},T_{j,k})_{(J, \leq)}$ is called a \emph{projective system} and the subspace $F \subset \prod_{i \in J} E_{i}$ of all sequences $(x_{j})_{j \in J}$ such that $T_{j,k}(x_{k}) = x_{j}$ is called the \emph{projective limit}. The stability result we need is that the projective limit of a projective system of nuclear Fr\'echet spaces is a nuclear Fr\'echet space ~\cite{Abbati,Yosida,Dubinsky}.  The advantage of working with Fr\'echet spaces is that, as the reader will see, some different notions of differentiability coincide.       
\section{The manifold structure of the diffeomorphism group}
We begin by putting a manifold structure on the space of smooth mappings between two manifolds $\mathcal{M}$ and $\mathcal{N}$ in the case of compact $\mathcal{M}$.  After this we will give the key modifications to noncompact $\mathcal{M}$.  
\begin{theo}
Let $\mathcal{M}$ and $\mathcal{N}$ be ordinary $C^{\infty}$ manifolds with $\mathcal{M}$ compact.  Then $
C^{ \infty} (\mathcal{M},\mathcal{N}) $ has the structure of a $ C^{ \infty }_{c} $
manifold.  The local model at $ f \in C^{ \infty } ( \mathcal{M}, \mathcal{N} ) $ is
given by the nuclear Fr\'echet space $ C^{ \infty }_{f} ( \mathcal{M} ,T^{*}\mathcal{N} )
$
\end{theo} \cite{Kriegl2} First we recall when a map $f$ is $ C^{ \infty }_{c}$.
Let $E, F$ be locally convex Hausdorff linear spaces and let $W
\subset E$ be open, then $f : W \rightarrow F $ is $ C^{1}_{c} $
if and only if there exists a mapping $ Df : E \rightarrow
L(E,F) $ such that $$ \lim_{ t \rightarrow 0 } \frac{ f(x + tv) -
f(x) }{t} = Df(x)v, \quad \forall v \in E, \quad x \in W, \quad t
\in \mathbb{R}  $$ exists, and the mapping $ W \times E \rightarrow
F : (x,v) \rightarrow Df(x)v$ is (jointly) continuous\footnote{It is easy to see that a $C^{1}_{c}$ function is continuous by proving that in a complete locally convex space the following formula makes sense \cite{Kriegl2}:
$$f(x+v)-f(x) = \int_{0}^{1}ds \, Df(x+sv)v.$$ Hence 
$$p(f(x+v)-f(x)) \leq \int_{0}^{1}ds \, p(Df(x+sv)v)$$ for any seminorm $p$.}.  The set of all $C^{1}_{c} $ mappings is a linear space.  The space $ C^{k}_{c} $
is defined recursively: a map $f$ is $C^{k}_{c} $ if $ D^{k-1}f
: W \times E^{k-1} \rightarrow L^{k}(E^{k},F)$\footnote{$L^{k}(E^{k},F)$ is the set of multilinear mappings from $E^{k}$ to $F$.} is $C^{1}_{c}$. Finally $$
C^{ \infty }_{c} = \bigcap_{k \geq 1 } C^{ k }_{c} $$ Now for
Fr\'echet spaces this concept of smoothness is equivalent to the
following: a map $ f : W \subset E \rightarrow F $ is smooth if
and only if every $ C^{ \infty }_{c} $ curve in $W$ is mapped to a
$ C^{ \infty }_{c} $ curve in $F$. Many results which we will
state now, can be generalized to a larger class of locally convex
Hausdorff linear spaces ~\cite{Balanzat}.  Let $  W \subset
\mathbb{R}^{n} $, then the topology on $ C^{ \infty } ( W, F)$,
where $F$ is a normed linear space, is defined by the following countable family $
\rho_{m,K }$ of seminorms. Let $m \in \mathbb{N}_{0} $, $ p =(
p_{1},p_{2}, \ldots , p_{n} ) \in \mathbb{N}^{n}$ and $ |p| =
\sum_{i=1}^{n} p_{i} $, then for any compact $K \subset W$
$$ \rho_{m,K} (f) = sup_{ |p| \leq m } (  sup_{ x \in K } \parallel
D^{p}f(x) \parallel ) $$ where $$ D^{p} = \frac{ \partial^{ |p|
}}{ \partial x_{1}^{ p_{1}} \partial x_{2}^{ p_{2}} \ldots
\partial x_{n}^{ p_{n}} }.$$  It is well known that this family of
seminorms makes $ C^{ \infty } ( W , F ) $ into a nuclear
Fr\'echet space  ~\cite{Dubinsky,Yosida}.  Now let $\mathcal{M}$ be any
ordinary manifold and let $( U ,\phi )$ be a chart, then the map
$$ \phi^{*} : C^{ \infty } ( U, F) \rightarrow C^{ \infty }( \phi
( U ) , F ) : f \rightarrow f  \circ \phi^{-1} $$ is a linear
isomorphism and induces on $ C^{ \infty } ( U, F) $ the structure
of a nuclear Fr\'echet space.  Because $\mathcal{M}$ is second countable
there exists a countable covering of charts $ ( U_{ k } , \phi_{ k
} ) $ such that one can construct the following restriction maps
$$ C^{ \infty } ( \mathcal{M}, F ) \rightarrow C^{ \infty } ( U_{k}, F ):f \rightarrow f_{|U_{k}}.$$
It then follows that
$$ C^{ \infty } ( \mathcal{M}, F ) = \lim_{k} C^{ \infty } ( U_{k}, F  )$$
where this limit is a projective limit of nuclear Fr\'echet spaces
and, hence, it is a nuclear Fr\'echet space
~\cite{Abbati,Yosida,Dubinsky}. This topology is known as the
Schwartz topology.  One can also introduce jet bundles in order to
characterise the Schwartz topology: for this the reader is referred to
~\cite{Kriegl2}.
\\* \\* Let $ exp : U \subset T^{*}\mathcal{N} \rightarrow \mathcal{N} $ be the
exponential map associated to a Riemannian metric $h$ on $\mathcal{N}$ , $
U$ an open neighbourhood of the zero section on which the
exponential map is defined and $ \pi_{N} : T^{*}\mathcal{N} \rightarrow \mathcal{N} $ the
canonical projection.  One can choose $U$ such that
$$ \nu \equiv ( \pi_{\mathcal{N}} , exp ): U \rightarrow \mathcal{N} \times \mathcal{N} : v \rightarrow ( \pi_{\mathcal{N}} (v) ,
exp_{ \pi_{\mathcal{N}} (v)}(v) ) $$ is a diffeomorphism from $U$ to a
neighbourhood $V$ of the diagonal in $\mathcal{N} \times \mathcal{N}$.  Now we can
define the local model $C^{ \infty }_{f} ( \mathcal{M}, T^{*}\mathcal{N} ) $ of $C^{ \infty } ( \mathcal{M}, \mathcal{N} )$ at $f$:
$$ C^{ \infty }_{f} ( \mathcal{M}, T^{*}\mathcal{N} ) = \{ g \in C^{ \infty } (
\mathcal{M}, T^{*}\mathcal{N} ) | \quad \pi_{\mathcal{N}} \circ g = f \} $$ $C^{ \infty }_{f} ( \mathcal{M},
T^{*}\mathcal{N} ) $ is a linear space which can be identified with $ Sec( E ) $
in the bundle $( E , \pi_{E} ,\mathcal{M} )$ where $E$ is defined as
follows:
$$ E = \{ (x,v) | v \in T^{*}\mathcal{N}_{f(x)} \} $$
The topology on $E$ is defined by the open sets $$ O_{W,U} = \{
(x,v_{f(x)}) | \, x \in W \, \text{and} \, v_{f(x)} \in U
\} $$ where $W$ is open in $\mathcal{M}$ and $U$ is open in $T^{*}\mathcal{N}$ with $f(W)
\subset \pi_{\mathcal{N}} ( U ) $. The differential structure is defined by
the charts $ ( O_{W,U} , ( \chi , pr_{2} \circ \psi  )) $ where $
( W ,\chi ) $, $ ( U , \psi  ) $ are charts in $ \mathcal{M} $, respectively
$ T^{*}\mathcal{N} $, with $ \psi( U ) \subset V \times F $ and $ pr_{2} $ is
the projection on the second factor.  We clearly can endow the
$C^{ \infty } $ sections on $E$ with the Schwartz topology, so
this induces on $C^{ \infty }_{f} ( \mathcal{M}, T^{*}\mathcal{N} ) $ the structure of a
nuclear Fr\'echet space.
\\ We can now define the charts on $ C^{ \infty } ( \mathcal{M}, \mathcal{N} )
$.  Define $$ U_{f} = \{ g \in C^{ \infty } ( \mathcal{M}, \mathcal{N} ) | \,
(f(x),g(x)) \in V, \, \forall x \in \mathcal{M} \}.$$ This induces the
map:
$$ u_{f} : U_{f} \rightarrow C^{ \infty }_{f} ( \mathcal{M} , T^{*}\mathcal{N} ) $$ with $ u_{f} (g) (x) =
exp_{f(x)}^{-1} ( g(x)) =  ( \nu^{-1} \circ ( f \times g )) ( x )
$. \\ $u_{f}$ maps $U_{f}$ bijectively to $ \{ s \in C^{ \infty
}_{f} ( \mathcal{M} , T^{*}\mathcal{N} ) | s(x) \in U \} $ which is by definition open in
the Schwartz topology because $\mathcal{M}$ is compact.  Now the inverse of
$u_{f}$ is given by:
$$ u_{f}^{ -1 } ( s )(x)  =  exp_{f(x)}( s(x) ) $$ so $ u_{f}^{-1}
(s) = ( pr_{2} \circ \nu ) \circ s $.  The atlas on $ C^{ \infty }
( \mathcal{M} , \mathcal{N} ) $ is hence defined as $$ \{ ( U_{f} , u_{f} ) | f \in
C^{ \infty } ( \mathcal{M}, \mathcal{N} ) \} $$  One determines transition maps
and proves that they are $ C^{ \infty }_{c} $.  It is clear that
if $ U_{f} \cap U_{ g } \neq \emptyset $ that $ u_{f} \circ
u_{g}^{ -1 } : u_{g} ( U_{f} \cap U_{ g } ) \rightarrow u_{f} (
U_{f} \cap U_{ g } ) $ is given by :
$$ (( u_{f} \circ u_{g}^{ -1 } )( s ))(x)  = u_{f(x)}( exp_{g(x)}
( s(x) )) = exp_{f(x)}^{-1}( exp_{g(x)} ( s(x) )) $$ or $$ u_{f}
\circ u_{g}^{ -1 } = exp_{f}^{-1} \circ exp_{g} $$  This
transition map is $ C^{ \infty }_{c} $ if and only if they map
smooth curves to smooth curves, which can be shown to be the case.  It is easy to see that
the differential structure is independent of the chosen Riemannian
metric so we arrive at our result.  \\* \\*  It is worthwhile noticing that
$ exp $ is defined everywhere on $T^{*}\mathcal{N}$ for compact $\mathcal{N}$ and that every two
points can be connected by a unique $h$ - geodesic (this is a
special case of the Hopf - Rinow theorem) ~\cite{Jost}.
\\* \\* The construction in the noncompact case is essentialy the same and the first result we are interested in is stated in the theorem below
\begin{theo} \cite{Kriegl2}
 $Diff(\mathcal{M})$ is an open $C^\infty _c$ submanifold of $C^\infty _{\mathcal{F} \mathcal{D}}
(\mathcal{M},\mathcal{M})$, composition and inversion are smooth.  The Lie algebra of
the smooth infinite dimensional Lie group $Diff(\mathcal{M})$ is the
convenient\footnote{The definition of a convenient locally convex space requires some work and the interested reader can find it in \cite{Kriegl2}.} vector space $\mathfrak{X} _{cpt}(\mathcal{M})$ of all smooth vector
fields on $\mathcal{M}$ with compact support, equipped with the negative of
the usual Lie bracket and with the $\mathcal{D}$ topology.  The
exponential mapping $EXP : \mathfrak{X}_{cpt}(\mathcal{M}) \to Diff(\mathcal{M})$ is the
flow mapping to time 1, and it is smooth.  A definition of the $
\mathcal{F} \mathcal{D} $ topology can be found in
~\cite{Kriegl2}.  In the case of compact $\mathcal{M}$ the $
\mathcal{F} \mathcal{D} $ topology coincides with the Schwartz topology.
\end{theo}
So far it seems that everything is in agreement with what we are used to in the finite dimensional case.  However, a beautiful result of Grabowski learns us that $EXP$ is not locally surjective near the identity diffeomorphism.  Before we state the formal result let us give a simple but illustrative example which can be found in the literature. \\* \\*
\textit{\textbf{Example}}
We construct an infinite family of diffeomorphisms on the circle $S^{1}$ wich contains the identity diffeomorphism as a limit point in the Schwartz topology on $Diff(S^{1})$ and which does not belong to $EXP(\mathfrak{X} _{cpt}(S^{1}))$.  As a starter we prove that any fixed point free diffeomorphism on $S^{1}$ belonging to the image of $EXP$ is conjugated to a translation $R_{a} : s \rightarrow s + a \mod 2 \pi$, where $s$ is the usual arclength.  Denote the vectorfield by the scalar $X(s)$ and choose any point $x_{0}$ on the circle.  The differential equation for the flow looks like
\begin{eqnarray}
 \dot{s}(t) & = & X(s(t)) \\
s(0) & = & x_{0} 
\end{eqnarray}
Hence if X is nowhere zero then 
$$ t = \int_{x_{0}}^{\psi_{t} (x_{0})} \frac{ds}{X(s)} $$
where $\psi_{t}$ denotes the one parametergroup of diffeomorphisms generated by $X$.  So in particular one has that
$$ 1 = \int_{x_{0}}^{(EXP(X))(x_{0})} \frac{ds}{X(s)} $$
Define $h(x) = \theta \int_{0}^{x} \frac{ds}{X(s)}$ where $\theta = (\int_{0}^{2 \pi} \frac{ds}{X(s)})^{-1}$ then $h(x + 2 \pi) - h(x) = 1$ and we have that $(h \circ EXP(X))(x_{0}) = R_{\theta}(h(x_{0}))$.  To confirm this last equation we calculate
$$(h \circ EXP(X))(x_{0}) = \theta \int_{0}^{(EXP(X))(x_{0})} \frac{ds}{X(s)} = \theta \left( 1 +  \int_{0}^{x_{0}} \frac{ds}{X(s)} \right) = R_{\theta}(h(x_{0}))$$
So we can write $EXP(X) = g \circ g$ where $g = h^{-1} \circ R_{\frac{\theta}{2}} \circ h$ belongs to $Diff(S^{1})$.  Consider for large $n$ the following family of fixed point free diffeomorphisms 
$$ f_{n} ( s) = s + \frac{1}{2^{n-1}} \sin^{2} ( \frac{n s}{2}) + \frac{2 \pi}{n} \mod 2 \pi$$
It is easy to see that $f_{n}$ has just one periodic orbit of period $n$ namely $\{ \frac{2 k \pi}{n} | k : 0 \ldots n-1 \}$.  Now for $n$ even, it is easily seen that $f_{n}$ is not in the range of $EXP$.  Suppose it is, then $f_{n} = g \circ g$ but then $g$ has a periodic orbit of period $2n$, which splits into two disjoint orbits of period $n$ for $g \circ g$ which gives the necessary contradiction.                          $\square$
\\*
\\*
We now state the result due to Grabowski ~\cite{gra}.
\begin{theo}
In $Diff(\mathcal{M})$ there exists a smooth curve through the identity which contains a pathconnected free subgroup of $2^{\aleph_{0}}$ generators which meets the exponential map only in the identity.
\end{theo} 
An equally surprising result is that the exponential map is not even locally injective on $Diff(\mathcal{M})$ when the dimension of $\mathcal{M}$ is larger than 1 or when $\mathcal{M}$ equals $S^{1}$.  \\*
\\*
These results would severely jeopardize the program of canonical quantisation if we would not have the following result.
\begin{theo}
$EXP(\mathfrak{X} _{cpt}(\mathcal{M}))$ generates the identity component of the diffeomorphism group, $Diff_{0} (\mathcal{M})$\footnote{Private communication with P.W. Michor.}.  
\end{theo}           
The reader who wants to get some feeling for this result might study the slightly weaker proposition by Michor which states that the subgroup generated by $EXP(\mathfrak{X} _{cpt}(\mathcal{M}))$ is dense in $Diff_{0} (\mathcal{M})$ \cite{Michor1}.  For the more general result, it was shown by Epstein that for a connected $\mathcal{M}$,  $Diff_{0}(\mathcal{M})$ is a simple group and Thurston has proven that the subgroup generated by $EXP(\mathfrak{X} _{cpt}(\mathcal{M}))$ is normal.  Hence, the result follows. \\* \\*
Gravity is a theory with a \emph{localizable} symmetry group, that is the Lie algebra of $Diff(\mathcal{M})$ consists of sections of the tangent bundle $T^{*} \mathcal{M}$ with compact support.  One has proven in general that Lagrangian theories with localizable symmetries give rise to a singular Legendre transform ~\cite{abba}, that is the theory is a constraint theory.  Now, it is a general belief that a localizable symmetry group gives rise to \emph{first class} constraints and in the case of gravity it was explicitly shown in the previous chapter that this is the case.

%% file: gromov.tex
\chapter{Lorentzian Gromov Hausdorff theory}
\begin{verse}
\hspace{13pt} \textsl{In this chapter, I introduce the
Gromov-Hausdorff metric and generalised Gromov-Hausdorff
uniformity.  I construct limit spaces and study some properties of
them.  The whole construction is focused around the strong metric
which is introduced in the second section.}
\end{verse}
\section{Introduction}
This chapter introduces Lorentzian, Gromov-Hausdorff theory.
First, we shall focus on compact interpolating spacetimes and
construct a metric on the moduli space of isometry classes of such
objects.  Next, we try to characterise the closure of this space
by studying limit spaces of Cauchy sequences.  However, in order
for such a limit space to be well defined, it is necessary to
introduce a tighter definition of closeness which we call the
generalised, Gromov-Hausdorff uniformity (GGH).   While studying
limit spaces of GGH Cauchy sequences of compact, interpolating
spacetimes, a natural metric arises.  This metric, which I call
the strong metric, enables us to define the abstract notion of
Lorentz space.  The next chapter is devoted to the study of
the GGH convergence theory of Lorentz spaces.  \\*  \\* The reader
shall be introduced to many new concepts, which will prove their
value in the derivation of some important results.  Many of these
concepts, which have no metric analogue, make clear that the
Lorentzian theory is deeper, richer and more complicated than its
metric counterpart.  Every new concept is motivated by many
examples and counterexamples of weaker notions which just do not
work, but are equivalent to the stronger one on the subspace of
interpolating spacetimes.  As such, the author hopes that his
theory will appear to the reader as being self evident.

\section{Compact interpolating spacetimes.}  \label{Coit}

The aim of this section is to make first steps in the construction
of a convergence theory for Lorentz manifolds, which constitute
the geometrical playground of general relativity.   To start with,
we generalise the Lipschitz notion of distance between two compact
metric spaces to a notion of distance between two globally
hyperbolic, compact, Lorentz manifolds with spacelike boundary.
It turns out that we can only measure a distance between
conformally equivalent structures, and in this sense this chapter
is only a warm-up. The most important result is a Lorentzian
analogue for the Ascoli-Arzela theorem which guarantees
convergence to isometry.  The most important lesson, however, is
that we have found a class of mappings that is rich enough to
enable us to compare conformally equivalent structures and which
is small enough to keep a good control over it.  \\* Next, we
introduce a Lorentzian notion of Gromov-Hausdorff distance
$d_{\textrm{GH}}$.  I will give some examples to show that the
Lorentzian theory is rather different from the ``Riemannian'' one.

\subsection{A Lipschitz distance} \label{ld}
Our aim is to define a Lorentzian analogue of the classical ``Riemannian'' Lipschitz distance between (locally) compact (pointed) metric spaces.  Let $(X,d_{X})$ and $(Y,d_{Y})$ be two compact metric spaces and $f : X \rightarrow Y$ be a bi-Lipschitz mapping, i.e., there exist numbers $ 0 < \alpha < \beta $ such that $$ \alpha d_{X} (x,y) \leq d_{Y} (f(x),f(y)) \leq \beta d_{X} (x,y) \quad \forall x,y \in X $$
The minimal such $\beta$ is the \emph{dilatation} $\textrm{dil}(f)$ of $f$ and the maximal such $\alpha$ the \emph{co-dilatation} of $f$ ( or the inverse of the dilatation of $f^{-1}$ if $f^{-1}$ exists).  The Lipschitz distance $d_{L}(X,Y)$ between $X$ and $Y$ is the infimum over all bi-Lipschitz homeomorphisms of the expression: $$ \left| \ln (\textrm{dil}(f)) \right|  + \left| \ln( \textrm{dil}(f^{-1}) ) \right| $$
The key result is that $d_{L}(X,Y) = 0$ iff $(X,d_{X})$ is isometric to $(Y,d_{Y})$, which is a direct consequence of the Ascoli-Arzela theorem \cite{Yosida}.
\begin{theo} (\textbf{Ascoli-Arzela})
Let $X$ and $Y$ be second countable, locally compact spaces, moreover $(Y,d_{Y})$ is assumed to be metrically complete.  Assume that the sequence $\left\{ f_{n} \right\}$ of mappings $f_{n} : X \rightarrow Y $ is equicontinuous such that the sets $ \bigcup_{n} \left\{f_{n}(x) \right\}$ are bounded with respect to $d_{Y}$ for every $x \in X$.  Then there exists a continuous mapping $f : X \rightarrow Y$ and a subsequence of $\left\{ f_{n} \right\}$ which converges uniformly on compact sets in $X$ to $f$.
\end{theo}
Let us now point out some analogy as well as some discrepancy with the Lorentzian case.  For now we restrict ourselves to spacetimes, i.e., pairs $(\mathcal{M},g)$ where  $\mathcal{M}$ is a $C^{\infty}$, paracompact, Hausdorff manifold and $g$ is a Lorentzian metric tensor on it, such that $(\mathcal{M},g)$ is time orientable.  Abstract Lorentz spaces will be defined later on in analogy with Seifert and Busemann \cite{Busemann}.  We will make now a convention in terminology which is not standard in the literature, but is somehow necessary to keep the discussion clear.
\begin{deffie} \label{lrd}
Let $X$ be a set, a Lorentz distance is a function $d: X \times X \rightarrow \mathbb{R}^{+} \cup \left\{ \infty \right\} $ which satisfies for all $x,y,z \in X$:
\begin{itemize}
\item $d(x,x) = 0$
\item $d(x,y) > 0$ implies $d(y,x) = 0$ (antisymmetry)
\item if $d(x,y)d(y,z) > 0$ then $d(x,z) \geq d(x,y) + d(y,z)$ (reverse triangle inequality).
\end{itemize}
\end{deffie}
It is well known that every chronological spacetime determines a
canonical Lorentz distance by defining $d_{g} (x,y)$ as the
supremum over all lengths of future oriented causal curves from
$x$ to $y$ if such curves exist and zero otherwise.  One has that
$d_{g}$ is continuous and finite if $(\mathcal{M},g)$ is globally
hyperbolic and, vice versa, that if $d_{g}$ is continuous then
$(\mathcal{M},g)$ is causally continuous.  More equivalences
between properties of $d_{g}$ and causality restrictions can be
found in \cite{Beem}.  We shall only be interested in globally
hyperbolic spacetimes since the continuity of $d_{g}$ is a
desirable property if one wants to work out a comparison theory
according to Lipschitz.  Note immediately that a compact globally
hyperbolic spacetime does not exist unless we consider manifolds
with a boundary.  We assume the boundary is spacelike and that
$\mathcal{M}$ and $\mathcal{N}$ are locally extendible across
their boundary.  All the results in section \ref{Coit} are also
valid when an extra timelike or null boundary is added.  The
motivated reader is invited to construct proofs similar to those
in this section.  In sections \ref{construction} and \ref{GHvGGH},
all results in this section shall be proven in a much greater
generality.  Remark first that every point of the boundary is
contained in a neighbourhood $\mathcal{U}$ which is diffeomorphism
equivalent to a hypercube in $\mathbb{R}^{n}$ which is closed on
one face and otherwise open.  By local extendibility I mean that
there exists an isometric embedding of $(\mathcal{U}, g_{|
\mathcal{U}})$ in a open spacetime $(\mathcal{V} , g_{ |
\mathcal{V}})$ such that the image of $\mathcal{U}$ has compact
closure in $\mathcal{V}$.  We stress that this does not correspond
to the usual notion of \emph{causal} local extendibility.\\* First we have to contemplate which
mappings between two spacetimes need to be considered for
comparison.  For that purpose, let $(\mathcal{M},g)$ and
$(\mathcal{N},h)$ denote globally hyperbolic spacetimes.  A
mapping $f : \mathcal{M} \rightarrow \mathcal{N}$ is said to be
timelike Lipschitz if and only if it has bounded timelike
dilatation $\textrm{tdil} (f)$, which is defined as the smallest
number $\beta$ such that
$$ d_{h} ( f(x),f(y)) \leq \beta d_{g} (x , y), \quad \forall x,y \in \mathcal{M}. $$
The above construction for the ``Riemannian'' case suggests that
we consider timelike bi-Lipschitz homeomorphisms.  However, a
slight generalisation of a classical result for homotheties
teaches us that a surjective timelike bi-Lipschitz map is
automatically a homeomorphism (see Appendix B).  Indeed, a result
of Hawking, Mc Carthy and King \cite{Hawking2}, proves that such
mapping is a $C^{\infty}$ conformal diffeomorphism.  One might be
concerned that such maps are too restrictive in the sense that
they only allow conformally equivalent spacetimes to be compared
but as mentioned before, this section is meant as a warm-up to get
used to the techniques needed for the next section.  The next
logical step is formulating and proving a Lorentzian version of a
suitably modified Ascoli-Arzela theorem.
\begin{theo} \label{asc1} (\textbf{Lorentzian Ascoli-Arzela}) Let $f_{n} : \mathcal{M} \rightarrow \mathcal{N}$ be onto timelike bi-Lipschitz mappings such that $\bigcup_{n} \left\{ f_{n} (x) \right\}$ and $\bigcup_{n} \left\{ f^{-1}_{n} (y) \right\}$ are precompact\footnote{A subset $A$ of a topological space $X$ is precompact iff the closure of $A$ is compact.} in $\mathcal{N}$, respectively $\mathcal{M}$, for all $x \in \mathcal{M}$ and $y \in \mathcal{N}$.  Moreover, let $(c_{n})_{n \in \mathbb{N}}$ be a descending sequence ($c_{n} < 1$) converging to zero such that $\textrm{tdil} (f_{n}) \leq 1+c_{n}$ and $\textrm{tdil} (f_{n}^{-1}) \leq \frac{1}{1 - c_{n}}$; then there exists a subsequence $(n_{k})_{k \in \mathbb{N}}$ and an isometry $f$ such that $f_{n_{k}}$ converges to $f$ pointwise.
\end{theo}  We recall that $x \prec y$ means that $x$ is in the causal past of $y$ and $x \ll y$ indicates that $y$ is in the chronological future of $x$.  Note also that we did not specify that $\ll$ is a partial order relation induced by a metric tensor $g$ since this would unnecessarily complicate the notation.  It should be clear from the context by which metric the particular partial order relation is defined.  We shall also omit the notational distinction between $d_{g}$ and $d_{h}$.    \\*
\textsl{Proof}: \\* Let $\mathcal{C}$ be a countable dense subset
of $\mathcal{M}$.  By a diagonalisation argument, one obtains a
subsequence $\left\{ f_{n_{k}} \right\}$ such that $f_{n_{k}} ( p
) \stackrel{k \rightarrow \infty}{\rightarrow} f(p), \quad \forall
p \in \mathcal{C}$. Let $r$ be any interior point of $\mathcal{M}$
which is not in $\mathcal{C}$, we show now that the definition of
$f$ can be extended to $r$ such that $\lim_{k \rightarrow \infty}
f_{n_{k}} ( r ) = f(r)$.  Let $\mathcal{U}$ be a causally convex
normal neighbourhood of $r$ and choose a point $p_{1} \in
\mathcal{C} \cap I^{-}(r) \cap \mathcal{U}$ close enough to $r$.
Let $\gamma$ be the unique timelike geodesic from $p_{1}$ through
$r$ and define $\tilde{p}_{i} , \tilde{q}_{i} \in \gamma $ by
$d(\tilde{p}_{i}, r) = \frac{d(p_{1},r)}{i} $ and $ d(r,
\tilde{q}_{i}) = \frac{d(p_{1},r)}{i}$.  Hence $d( \tilde{p}_{i},
\tilde{p}_{i+j}) = \frac{j d(p_{1},r) }{i(i+j)} = d(
\tilde{q}_{i+j}, \tilde{q}_{i})$ and $
d(\tilde{p}_{i},\tilde{q}_{i}) = \frac{ 2 d(p_{1},r)}{i} $. Define
now sequences $p_{i} , q_{i} \in \mathcal{C}$ such that
$\tilde{p}_{i} \ll p_{i} \ll \tilde{p}_{i+1}$, $ \tilde{q}_{i+1}
\ll q_{i} \ll \tilde{q}_{i}$ with the exception that $p_{1} =
\tilde{p}_{1}$.  We shall now prove the following claims:
\begin{itemize}
\item  the sequence $( f(p_{i}))_{i \in \mathbb{N}_{0}}$ is contained in the compact set $A(f(\tilde{p}_{1}) , f(q_{1}))$\footnote{$A(p,q) = \{ r | p \prec r \prec q \}$.} and has exactly one accumulation point $f_{\uparrow} (r)$ which turns out to be a limit point.
\item $f_{\uparrow} (r)$ is independent of the choice of $(p_{i})_{i \in \mathbb{N}}$
\end{itemize}
The first claim is an easy consequence of the observation that for all $i$ one has that :

\begin{eqnarray*}
d(f(\tilde{p}_{1}) , f(p_{i})) & = & \lim_{k \rightarrow \infty} d(f_{n_{k}} ( \tilde{p}_{1}) , f_{n_{k}} (p_{i})) \\
& = & d(\tilde{p}_{1} ,p_{i}), \\
\end{eqnarray*}
where we used the continuity of $d$ in the target space and the
property of the convergence of the timelike dilatation and
co-dilatation of the mappings $f_{n}$.  The above also proves that
$ d(f(p_{i}) , f(p_{i+j})) = d(p_{i}, p_{i+j})$ and hence $
f(p_{i}) \ll f(p_{i+j})$.  This in turn implies that any
accumulation point of the sequence $( f(p_{i}))_{i \in
\mathbb{N}_{0}}$ must lie to the future of all $ f(p_{i})$, hence
it is a limit point which must be unique.  \\* The second claim
follows from the observation that if $\hat{p}_{i}$ is another such
sequence with corresponding $\hat{f}_{\uparrow} (r)$ then one has
that $$p_{i} \ll \hat{p}_{i+1} \ll p_{i+2} \ll  \hat{p}_{i+3} \ll
\ldots \ll r. $$ Hence $$0 < d(f(\hat{p}_{i+1}), f_{\uparrow} (r)
) \leq d(f(p_{i}), f_{\uparrow} (r) ) -
d(f(p_{i}),f(\hat{p}_{i+1})) $$ However, the first term on the
rhs.\ converges to zero for $i \rightarrow \infty$ and the second
term is estimated by $d(f(p_{i}),f(\hat{p}_{i+1})) \leq \frac{2
d(\tilde{p}_{1},r)}{i(i+2)}$.  Hence $\hat{f}_{\uparrow} (r) \in
E^{-} ( f_{\uparrow} (r) )$.  The reverse is proven similary and
this concludes the second claim. \\*
\\*  The same result is of course also true for $p$ replaced by $q$, and we denote the corresponding accumulation point by $f_{\downarrow} (r)$.   Note that $d(f(\tilde{p}_{1}),f_{\uparrow}(r)) = d(\tilde{p}_{1},r)$, $d(f_{\downarrow}(r), f(q_{1})) = d(r, q_{1})$ and $ d(f(\tilde{p}_{1}), f(q_{1})) = d(\tilde{p}_{1},q_{1})$, which all follow from continuity of $d$ and the present properties of $f$.   $f_{\uparrow}(r) = f_{\downarrow}(r)$ follows from the observation that changing $q_{1}$ by a point in the future of $q_{1}$, so that we can come arbitrary close to $\tilde{q}_{1}$, does not change the point $f_{\downarrow}(r)$.  For the same reasons as before, such a sequence of points $q_{1}$ will define a sequence $f(q_{1})$ which converges to a point, say, $f_{\uparrow}(\tilde{q}_{1})$ in the future of all points $f(q_{1})$.  Hence due to continuity we have that $d(f_{\downarrow}(r), f_{\uparrow}(\tilde{q}_{1})) = d(r, \tilde{q}_{1})$ and $d(f(\tilde{p}_{1}), f_{\uparrow}(\tilde{q}_{1})) = d(\tilde{p}_{1}, \tilde{q}_{1})$.  But this implies that $d(f_{\downarrow}(r),f_{\uparrow}(r)) = 0$ and more strongly $f_{\uparrow}(r) = f_{\downarrow}(r)$, otherwise by ``rounding off the edges'' we could find a timelike curve with length larger than $d(f(\tilde{p}_{1}), f_{\uparrow}(\tilde{q}_{1}))$, which is a contradiction.  It is now easy to see that $f_{n_{k}} (r) $ converges to $f(r)$, since for every $i$ we can find a $k_{0}$ such that for all $k \geq k_{0}$ one has that $f(p_{i}) \ll f_{n_{k}} ( p_{i+1} ) \ll f(r) \ll f_{n_{k}}(q_{i+1}) \ll f(q_{i}) $, which implies (because of the properties of $f_{n_{k}}$) that $f(p_{i}) \ll f_{n_{k}}(r) \ll f(q_{i}) $.  This concludes the proof when $r$ is an interior point, since the open Alexandrov sets $int(A(f(p_{i}),f(q_{i})))$ form a basis for the topology around $f(r)$.
The case when $r$ is a past boundary point is rather different, since then we cannot squeeze the point $r$ anymore in an Alexandrov set (the case of a future boundary point is identical).  Obviously, $f_{n_{k}}(r)$ belongs to the past boundary of $\mathcal{N}$.  Let $\gamma$ be the unique geodesic segment orthogonal to the past boundary in $r$, and choose the sequences $(\tilde{q}_{i})_{i \in \mathbb{N}}$ and $(q_{i})_{i \in \mathbb{N}}$ as before.  Then, we can find a subsequence $f_{n_{k_{l}}}$ such that $f_{n_{k_{l}}}(r) \stackrel{l \rightarrow \infty}{\rightarrow} f(r)$, where $f(r)$ belongs to the past boundary and $f_{n_{k_{l}}} (\gamma_{|\left[ r, \tilde{q}_{1}\right]}) \rightarrow f(\gamma_{|\left[ r, \tilde{q}_{1}\right]})$ in the $C^{0}$ topology of curves.  It is easy to see that $f(\gamma_{|\left[ r, \tilde{q}_{1}\right]})$ is the unique geodesic segment in $\mathcal{N}$ orthogonal to the past boundary in  $f(r)$.  But in this case, we have that $$f(q_{i}) \gg f_{n_{k}}(q_{i+1}) \gg f_{n_{k}}(r),$$ and since the $I^{-}(f(q_{i}))$ form a basis for the topology around $f(r)$, we have that $\lim_{k \rightarrow \infty} f_{n_{k}}(r) = f(r)$, which concludes the proof.  It is not difficult to see that $f$ is continuous by construction.  As a matter of fact, we should still prove that $f$ is onto.  Performing the same construction for $f_{n_{k}}^{-1}$ we find (by eventually taking a subsequence) a limit mapping $f^{-1}$.  We now
show that $f^{-1} \circ f = id_{\mathcal{M}}$, $f \circ f^{-1} = id_{\mathcal{N}}$.  We shall prove the former, the proof of the latter is identical.  Suppose there exists an interior point $x$ such that $\lim_{k \rightarrow \infty} f_{n_{k}}^{-1} \circ f (x)  \neq x$, then there exist points $p_{1} , p_{2} , p_{3} , q_{1}, q_{2} , q_{3}$ such that
$$ p_{1} \ll p_{2} \ll p_{3} \ll f^{-1} \circ f(x) \ll q_{3} \ll q_{2} \ll q_{1}$$ and $x \notin A(p_{1} , q_{1})$.  Then for $k$ big enough:
\begin{itemize}
\item $p_{3} \ll f_{n_{k}}^{-1} \circ f(x) \ll q_{3}$
\item $f_{n_{k}} ( p_{1} ) \ll f(p_{2}) \ll f_{n_{k}} ( p_{3})$
\item $f_{n_{k}}(q_{3}) \ll f(q_{2}) \ll f_{n_{k}} ( q_{1}), $
\end{itemize}
hence
$$ f_{n_{k}} (p_{1}) \ll f(p_{2}) \ll f(x) \ll f(q_{2}) \ll f_{n_{k}} ( q_{1}), $$ but $f_{n_{k}}(x) \notin A(f_{n_{k}} (p_{1}) , f_{n_{k}} (q_{1}))$, which implies that $f(x)$ cannot lie between $f(p_{2})$ and $f(q_{2})$, which is a contradiction.  Hence $f^{-1} \circ f$ equals the identity on the interior of $\mathcal{M}$, and therefore it equals the identity everywhere since it is continuous. The conclusion that $f$ is an isometry follows from the discussion in Appendix B.
 $\square$
\\*
\\*
Remark first that in the proof of the theorem we needed the requirement that $\bigcup_{n} \left\{ f^{-1}_{n} (y) \right\}$ is precompact in $\mathcal{M}$ for all $y \in \mathcal{N}$ only to guarantee the surjectivity and hence the smoothness of $f$.  It is also possible to get the following stronger result:
\begin{theo}
Let $\alpha < 1 < \beta , \quad f_{n} : \mathcal{M} \rightarrow \mathcal{N}$ be as in Theorem \ref{asc1} with the difference that $\textrm{tdil}(f_{n}) \leq \beta$ and $\textrm{tdil}(f_{n}^{-1}) \leq \frac{1}{\alpha}$.  Then, there exists a subsequence $f_{n_{k}}$ and an $f$ such that $f_{n_{k}}$ converges pointwise to $f$.  Moreover one has that $\textrm{tdil}(f) \leq \beta$ and $\textrm{tdil}(f^{-1}) \leq \frac{1}{\alpha}$.
\end{theo}
\textsl{Proof}: \\*
Let $\mathcal{C}$ and $\mathcal{D}$ be countable dense subsets in $\mathcal{M}$ and $\mathcal{N}$ respectively.  By a diagonalization argument we find a subsequence $f_{n_{k}}$ such that $f_{n_{k}} ( p )$ converges to $f(p)$ and $f_{n_{k}}^{-1}(q)$ converges to $f^{-1}(q)$ for all $p \in \mathcal{C}$ and $q \in \mathcal{D}$ respectively.
Suppose $r$ is an interior point and let $\gamma$, $(p_{i})_{i \in \mathbb{N}}$ and $(\tilde{p})_{i \in \mathbb{N}}$ be as before.  Take $q \in \mathcal{D}$ arbitrarily close in the chronological future of $f_{\uparrow}(r)$, we have then that     \begin{eqnarray*}
d(r,f^{-1}(q)) & = & \lim_{i \rightarrow \infty} d(p_{i}, f^{-1} (q) ) \\
& = & \lim_{i \rightarrow \infty} \lim_{k \rightarrow \infty} d(p_{i}, f_{n_{k}}^{-1} (q) ) \\
& \geq & \frac{1}{\beta} \lim_{i \rightarrow \infty} \lim_{k \rightarrow \infty} d(f_{n_{k}} (p_{i}), q) \\
& \geq & \frac{d(f_{\uparrow}(r), q)}{\beta}. \\
\end{eqnarray*}
Hence $r \ll f^{-1}(q)$.  Take now $q_{1}, q_{2} \in  \mathcal{D}$ such that $f_{\uparrow}(r) \ll q_{1} \ll q_{2}$ with $q_{2}$ arbitrarily close to $f_{\uparrow}(r)$.  Choose $i>0$, then for $k$ sufficiently large one has
$$ r \ll f_{n_{k}}^{-1}(q_{1}) \ll f^{-1} (q_{2}) $$
and
$$ f(p_{i}) \ll f_{n_{k}} ( p_{i+1} ) \ll f_{\uparrow}(r). $$
Hence
$$ f(p_{i}) \ll f_{n_{k}}( p_{i+1}) \ll f_{n_{k}} ( r ) \ll q_{1}, $$
which proves $\lim_{k \rightarrow \infty} f_{n_{k}} (r) = f_{\uparrow}(r)$.  \\*
Let $r$ be a point of the ``past'' boundary (the future situation is dealt with identically).  Let $\gamma$ be a distance maximizing geodesic with past endpoint $r$ and let $(\tilde{q}_{i})_{i \in \mathbb{N}}, (q_{i})_{i \in \mathbb{N}}$ be sequences of points as before where now the ``futuremost'' point $\tilde{q}_{1}$ is sufficiently close to $r$ and $q_{1}$ can be chosen equal to $\tilde{q}_{1}$.  Without loss of generality, we can assume that $\tilde{q}_{1} \ll q \in \mathcal{D}$ such that $J^{-} (q)$ is compact.  For $k$ sufficiently large we have that
$$ f_{n_{k}} ( q ) \gg f(q_{1}) \gg f(q_{2}) \gg \ldots $$
Since $f_{n_{k}}$ is continuous $J^{-}(f_{n_{k}}(q)) = f_{n_{k}} ( J^{-} (
q))$ is compact, therefore the sequence $(f(p_{i}))_{i \in \mathbb{N}}$ has an accumulation point $f_{\downarrow}(r)$, which is as usual also a limit point.  Suppose $f_{\downarrow}(r)$ is not on the past boundary, then we can find a point $p \in \mathcal{D}$ such that $ p \ll f_{\downarrow} (r)$.  The calculation above shows that $f^{-1}(p) \ll r$, which is impossible.  Hence $f_{\downarrow}(r)$ belongs to the past boundary.  Since the past light cones $I^{-}(f(p_{i}))$ constitute a local basis for the topology around $f_{\downarrow}(r)$, the result follows.
\\*
The other conclusions of the theorem are obvious. $\square$ \\*
\\*
Having this theorem in the pocket, the theorem which guarantees convergence to isometry follows immediately:
\begin{theo}
Let $(\mathcal{M},g)$ and $(\mathcal{N},h)$ be compact globally hyperbolic spacetimes with boundary, then $d_{L} ((\mathcal{M},g),(\mathcal{N},h)) = 0$ iff $(\mathcal{M},g)$ and $(\mathcal{N},h)$ are isometric.
\end{theo}
The notion of Lipschitz distance however is too severe and does not give rise to a rich comparison theory since there is too much geometric control.  A result of Defrise-Carter ~\cite{Defrise} shows that\footnote{The assumption in the paper of Defrise-Carter that the group needs to be finite dimensional, is not necessary.} the group of \emph{local} conformal isometries of a four dimensional Lorentz manifold is, with two exceptions, a group of isometries.  By this, I mean that for every spacetime $(\mathcal{M},g)$ having such a group, which is not conformally equivalent to Minkowski or a plane-wave spacetime with parallel rays, there exists a conformal factor $\Omega$ such that this group \emph{only} constitutes of isometries for the spacetime $(\mathcal{M}, \Omega g)$.  In Minkowski spacetime,  there is a $15$-dimensional group of proper conformal transformations\footnote{Generators consist of the $10$ Poincare transformations, $1$ dilatation and $4$ accelerations.} and in a plane-wave spacetime with parallel rays only a $6$ or $7$ dimensional group of homotheties\footnote{$5$ respectively $6$ generators form an isometry group, and $1$ generator forms a dilatation.}.  Hence, there are ``not many'' infinitesimal conformal isometries, and there are even fewer which can be integrated.  Note that the result of Defrise-Carter does not mention anything about \emph{discrete} conformal isometries.  However, the results of this section are still very important, since:
\begin{itemize}
\item we shall be forced to generalise this Lipschitz theory to \emph{abstract} globally hyperbolic Lorentz spaces, which will be done in chapter \ref{lsp}.
\item the proofs give a hunch how to prove convergence to isometry in case the family of mappings gets enlarged, such as will happen in the next section.
\end{itemize}
For a complete Riemannian manifold which is not locally flat, Kobayashi and Nomizu have proven that there are no homotheties which are not isometries.  As stated before, this is not true in the Lorentzian case as the next plane wave spacetime shows \cite{Beem}.
\begin{exie} \label{ex1}
\end{exie}
Consider $\mathbb{R}^{3}$ with the metric $ds^{2} = e^{xz}dxdy + dz^{2}$.  $(\mathbb{R}^{3}, ds^{2})$ is not flat and the mappings $\phi_{t}(x,y,z) = (e^{t}x,e^{-3t}y,e^{-t}z)$ are proper homotheties with factor $e^{-2t}$. $\square$  \\*
\\*
Moreover, we will see in the next subsection that for every compact globally hyperbolic spacetime $(\mathcal{M},g)$ there exists a ``Riemannian'' metric $D_{\mathcal{M}}$ such that all $d_{g}$ isometries are $D_{\mathcal{M}}$ isometries.  Suppose $(\mathcal{M},g)$ has a nontrivial Lie algebra of conformal isometries and is not a compact piece cut out of Minkowski or a plane-wave spacetime (with parallel rays), then there exists a ``Riemannian'' metric $\tilde{D}_{\mathcal{M}}$ such that ``most'' (apart from possible discrete conformal isometries) $g$-conformal isometries are $\tilde{D}_{\mathcal{M}}$ isometries\footnote{We know there exists a global conformal factor $\Omega$ such that essentially all $g$ conformal isometries are $\Omega g$ isometries, hence the claim follows.}.
\subsection{A Gromov-Hausdorff distance} \label{GHD}
We recall the notion of Gromov-Hausdorff distance in the ``Riemannian'' case.  For this purpose define the Hausdorff distance $d_{H}$ between subsets $U,V$ of a metric space $(X,d_{X})$ as
$$ d_{H} (U,V) = \inf \{\epsilon | U \subset B(V, \epsilon), V \subset B(U, \epsilon) \}$$
where $B(U, \epsilon) = \{ x \in X | \exists a \in U : d_{X}(x,a) < \epsilon \}$.
Gromov had around $1980$ the following idea \cite{Gromov} :  consider two compact metric spaces $(X,d_{X})$ and
$ ( Y, d_{Y})$, define a metric $d$ on the disjoint union $X \sqcup Y$ to be \emph{admissible} iff the restrictions of $d$ to $X$ and $Y$ equal $d_{X}$ and $d_{Y}$ respectively.  Then $$d_{GH} ( (X,d_{X}),(Y,d_{Y})) = \inf \{ d_{H} (X,Y) | \textrm{all admissible metrics on} X \sqcup Y \}.$$
In other words the Gromov-Hausdorff distance between two metric spaces is the infimum over all Hausdorff distances between $X$ and $Y$ in $X \sqcup Y$ with respect to metrics which extend the given metrics on $X$ and $Y$.  Suppose $d$ is an admissible metric on $X \sqcup Y$; then there exist mappings $f : X \rightarrow Y$, $ g :Y  \rightarrow X$ such that $d (x,f(x)) \leq d_{H}(X,Y)$ and $d(y,g(y)) \leq d_{H}(X,Y)$ for all $x \in X$, $y \in Y$ respectively.
The triangle inequality and the properties of $d$ imply that :
\begin{eqnarray}
\left| d_{Y} ( f(x_{1}) , f(x_{2})) - d_{X} ( x_{1} , x_{2} ) \right| & \leq & 2 d_{H}(X,Y)  \\
\left| d_{X} ( g(y_{1}) , g(y_{2})) - d_{Y} ( y_{1} , y_{2} ) \right| & \leq & 2 d_{H}(X,Y)  \\
d_{X} ( x , g \circ f (x) ) & \leq & 2 d_{H}(X,Y) \\
d_{Y} (y , f \circ g (y )) & \leq & 2 d_{H}(X,Y)
\end{eqnarray}
Observe that the last two inequalities imply that in the limit for $d_{H} (X , Y)$ approaching to zero, $f$ becomes invertible.  But for compact metric spaces, invertibility also follows from the observation that in the limit for $d_{H} (X , Y)$ going to zero, $f$ and $g$ become distance-preserving maps.  Hence $g \circ f$ and $f \circ g$ are distance-preserving maps on $X$ and $Y$ respectively.  The compactness assumption then implies that they are both bijections and, as a consequence, so are $f$ and $g$.  We shall first prove a similar result in the Lorentzian case.
\begin{theo}
Let $f : \mathcal{M} \rightarrow \mathcal{M}$ be continuous and Lorentz distance preserving on the interior of $\mathcal{M}$; then $f$ maps the interior onto itself.
\end{theo}
\textsl{Proof}:
Remark that an interior point is mapped by a distance-preserving map to an interior point.  Suppose $p$ is an interior point not in $f(\overset{\circ}{\mathcal{M}})$, then there exists a neighborhood $\mathcal{U}$ of $p$ for which $f(\overset{\circ}{\mathcal{M}}) \cap \mathcal{U} = \emptyset$.  For suppose not, then we can find a sequence $r_{n} \stackrel{n \rightarrow \infty}{\rightarrow} r$ in $\overset{\circ}{\mathcal{M}}$ such that $f(r_{n}) \stackrel{n \rightarrow \infty}{\rightarrow} p$.  Hence, $r$ is not an interior point and without loss of generality we can assume it belongs to the future boundary.  But then $f(\overset{\circ}{\mathcal{M}}) \cap I^{+} ( p) = \emptyset$, otherwise there would exist an interior point to the future of all $r_{n}$, which is impossible.\\*
Hence, we may assume that there exist points $r \ll p \ll s$ such that $f(\overset{\circ}{\mathcal{M}}) \cap I^{+}(r) \cap I^{-}(s) = \emptyset$ and $d_{g}(r,p) = d_{g} ( p,s) > 0$.   Since $f^{k} (p) \notin I^{+}(r) \cap I^{-}(s)$ for all $k$, we get that $f^{k}(p) \notin  I^{+}(f^{l}(r)) \cap I^{-}(f^{l}(s))$ for all $k \geq l$.  By taking a subsequence if necessary, we can assume that $f^{n} ( p) \stackrel{n \rightarrow \infty}{\rightarrow} \tilde{p}$, $f^{n} ( r) \stackrel{n \rightarrow \infty}{\rightarrow} \tilde{r}$, $f^{n} ( s) \stackrel{n \rightarrow \infty}{\rightarrow} \tilde{s}$.  Hence $\tilde{r} \ll \tilde{p} \ll \tilde{s}$, but this is impossible since this implies that for $n$ big enough $\tilde{p} \in I^{+}(f^{n}(r)) \cap I^{-}(f^{n}(s))$.  $\square$ \\*
\\*
Let us now make the following definition.
\begin{deffie}
(\textbf{Lorentzian Gromov-Hausdorff} ) We call $(\mathcal{M},g)$ and $(\mathcal{N},h)$ $\epsilon$-close iff there exist mappings $\psi : \mathcal{M} \rightarrow \mathcal{N}$, $\zeta : \mathcal{N} \rightarrow \mathcal{M}$, not necessarily continuous, such that
\begin{eqnarray}
\left| d_{h} ( \psi ( p_{1} ), \psi (p_{2})) - d_{g} (p_{1} , p_{2} ) \right| & \leq & \epsilon \quad \forall p_{1}, p_{2} \in \mathcal{M} \\
\left| d_{g} ( \zeta ( q_{1} ) ,\zeta (q_{2})) - d_{h} ( q_{1} ,q_{2} )\right| & \leq & \epsilon \quad \forall q_{1}, q_{2} \in \mathcal{N}.
\end{eqnarray}
The Gromov-Hausdorff distance $d_{GH}((\mathcal{M},g),(\mathcal{N},h))$ is defined as the infimum over all $\epsilon$ such that $(\mathcal{M},g)$ and $(\mathcal{N},h)$ are $\epsilon$-close.

\end{deffie}
We show that the previous theorem implies that $d_{GH}$ is a metric.
\begin{theo} \label{GHD1}
$d_{GH}((\mathcal{M},g),(\mathcal{N},h)) = 0$ iff $(\mathcal
{M},g)$ and $(\mathcal{N},h)$ are isometric.
\end{theo}
\textsl{Proof}: \\*
Suppose we are given sequences $(\psi_{n})_{n \in \mathbb{N}}, (\zeta_{n})_{n \in \mathbb{N}}$ of, possibly discontinuous, maps which make $(\mathcal{M},g)$ and $(\mathcal{N},h)$ $\frac{1}{n}$-close.
Let $\mathcal{C}$ and $\mathcal{D}$ be countable dense subsets of $\mathcal{M}$ respectively $\mathcal{N}$.  Take subsequences $(\psi_{n_{k}})_{k \in \mathbb{N}}$ and $(\zeta_{n_{k}})_{k \in \mathbb{N}}$ such that
\begin{itemize}
\item $\psi_{n_{k}} ( p ) \stackrel{k \rightarrow \infty}{\rightarrow} \psi(p)$ for all $p \in \mathcal{C}$
\item $\zeta_{n_{k}} ( q ) \stackrel{k \rightarrow \infty}{\rightarrow} \zeta(q)$ for all $q \in \mathcal{D}$

\end{itemize}
Obviously $d_{h} ( \psi ( p ) , \psi ( \tilde{p} )) = d_{g} ( p , \tilde{p})$ for all $p, \tilde{p} \in \mathcal{C}$ and $ d_{g} ( \zeta ( q ) , \zeta ( \tilde{q} )) = d_{h} ( q , \tilde{q})$ for all $q, \tilde{q} \in \mathcal{D}$, which is an easy consequence of the global hyperbolicity and the limiting properties of the sequences $(\psi_{n_{k}})_{k \in \mathbb{N}}$ and $(\zeta_{n_{k}})_{k \in \mathbb{N}}$. \\*
We shall now prove that the limit map $\psi$ exists and is distance-preserving.
Let $r$ be an interior point of $\mathcal{M}$ and take sequences $(\tilde{p}_{i})_{i \in \mathbb{N}}$ , $(\tilde{q}_{i})_{i \in \mathbb{N}}$, $(p_{i})_{i \in \mathbb{N}}$ and $(q_{i})_{i \in \mathbb{N}}$ in $\mathcal{M}$ as before.
In exactly the same way as in the proof of theorem \ref{asc1}, we obtain that $\psi_{ \uparrow } (r ) = \psi_{ \downarrow } (r)$.  Also $\psi(r) = \lim_{k \rightarrow \infty} \psi_{n_{k}} (r ) $ since for arbitrary $i$ we can find a $k_{0}$ such that $\forall k \geq k_{0}$
\begin{itemize}
\item $\frac{1}{k} < \min \{ d(p_{i + 1},r) , d(r , q_{i+1}) \}$
\item $\psi(p_{i}) \ll \psi_{n_{k}} ( p _{i+1} ) \ll \psi_{n_{k}} ( q_{i+1} ) \ll \psi (q_{i} )$
\end{itemize}
hence
$$ \psi (p_{i}) \ll \psi_{n_{k}} ( p_{i+1} ) \ll \psi_{n_{k}} (r) \ll \psi_{n_{k}} ( q_{i+1} ) \ll \psi ( q_{i} ) $$
which proves the case.  From this it is easy to prove that $\psi$ is continuous on the interior points. \\*
In exactly the same way one constructs a continuous limit map $\zeta$ on the interior of $\mathcal{N}$.  \\*
The previous theorem now shows that $\psi$ and $\zeta$ are distance preserving homeomorphisms from the interior of $\mathcal{M}$ to $\mathcal{N}$ and from the interior of $\mathcal{N}$ to $\mathcal{M}$ respectively.  Using this, it is not difficult to show that one can continuously extend $\psi$ to the boundary so that $\lim_{k \rightarrow \infty} \psi_{n_{k}} ( r ) = \psi(r)$ for every boundary point $r$.  Hence the result follows.$\square$   \\*
\\*
Furthermore, it is obvious that $d_{GH}$ is symmetric and satisfies the triangle inequality.  We will now discuss some properties of $d_{GH}$.  Let us start with an obvious one which is similar to the ``Riemannian'' case.
\begin{theo} \,
$d_{GH} ((\mathcal{M},g),(\mathcal{N},h)) \leq \max \{\textrm{tdiam}(\mathcal{M}), \textrm{tdiam} (\mathcal{N}) \}$ where \\* $\textrm{tdiam}( \mathcal{M} )$ denotes the timelike diameter, i.e.,
$$ \textrm{tdiam}( \mathcal{M} ) = \max_{p,\tilde{p} \in \mathcal{M}} d_{g} ( p ,\tilde{p} ).$$
\end{theo}
We shall now give an example that might feel strange in the beginning for people used to Riemannian geometry, although the result itself is what one should expect from Lorentzian geometry.
\newpage
\begin{exie} \label{ex2}
\end{exie}
As mentioned before, an occasion will present itself for abstraction of the concept of an interpolating spacetime.  Therefore, we need to contemplate about the topology (which coincides with the manifold topology for interpolating spacetimes) we are going to equip such abstract spaces with.  This example shows that such topology is necessarily determined by the Lorentz distance.  Since suppose not, then one could imagine that a Riemannian manifold is a Lorentz space in which every point is null-connected with itself and not causally related to any other point (imagine that the Riemannian manifold serves as a spacelike Cauchy surface in a globally hyperbolic spacetime).  Then the previous theorem shows that \emph{any} two Riemannian manifolds are a distance zero apart since their timelike diameters are zero.  This is clearly undesirable, however this situation can be cured easily.  The reader should notice that the Alexandrov topology on a Riemannian manifold is trivial and therefore this space is topologically equivalent to a point.  Hence, if one only wants to consider ``rich'' topological spaces, one should control the timelike diameter, i.e., it must be bounded away from zero as the next example shows.  \\* Consider cylinders $C_{T} = S^{1} \times \left[ 0 , T \right]$ with Lorentz metric $ds^{2} = - dt^{2} + d \theta^{2}$.  A Gromov-Hausdorff limit for $T \rightarrow 0$ is a point.
$\square$
\\*
\\*
Now, we shall show that we can construct a metric $D_{\mathcal{M}}$ such that every $d_{g}$ isometry is a $D_{\mathcal{M}}$ isometry.  For any compact interpolating spacetime, this metric shall be constructed from the Lorentz distance $d_{g}$ alone, which is in contrast to the usual extra assumption of a preferred class of observers in the major part of the literature.  Such a preferred class of observers is for example given if the energy momentum tensor satisfies the type I weak energy condition ~\cite{Hawking1}, i.e., determines a preferred timelike eigenvectorfield.  However, our approach is purely geometrical and dynamical aspects related to the Einstein equations are not considered.  This is moreover the only sensible strategy if
\begin{itemize}
\item one wants to construct a theory of \emph{vacuum} quantum gravity
\item one considers spacetime \emph{not} to be a manifold.  What would the analogue be of the Einstein-Hilbert action on something like a causal set ~\cite{Sorkin} or a spin network ~\cite{Thiemann}, anyway?
\end{itemize}
$D_{\mathcal{M}}$ will also play a crucial part in the
construction of the limit space of a Cauchy sequence of compact
interpolating spacetimes \cite{Noldus1}.  For reasons which will
become clear in section \ref{ls}, $D_{\mathcal{M}}$ will be
referred to as the strong metric.
\begin{deffie} \label{Str}
Let $(\mathcal{M},g)$ be a compact interpolating spacetime.  The strong  metric $D_{\mathcal{M}}$ is defined as
$$ D_{\mathcal{M}} ( p , q ) = \max_{r \in \mathcal{M}} \left| d_{g}(p,r) + d_{g}(r,p) - d_{g}(q,r) - d_{g}(r,q) \right| $$
$\square$
\end{deffie}
\begin{theo} $D_{\mathcal{M}}$ is a metric on $\mathcal{M}$ for any compact interpolating spacetime $(\mathcal{M},g)$.
\end{theo}
\textsl{Proof}: \\*
Clearly, $D_{\mathcal{M}}(p,q) = 0$ iff $p=q$.  $D_{\mathcal{M}}$ is symmetric by definition, so we are only left to prove the triangle inequality.  Choose $p,q \in \mathcal{M}$ and let $r$ be such that $$D_{\mathcal{M}}(p,q) = \left| d_{g}(p,r) + d_{g}(r,p) - d_{g}(r,q) - d_{g}(q,r) \right|.$$ Then for all $z \in \mathcal{M}$,
\begin{eqnarray*} D_{\mathcal{M}}(p,q) & \leq & \left| d_{g}(p,r) + d_{g}(r,p) - d_{g}(r,z) - d_{g}(z,r) \right| + \\ & & \left| d_{g}(z,r) + d_{g}(r,z) - d_{g}(r,q) - d_{g}(q,r) \right| \\ & \leq & D_{\mathcal{M}}(p,z) + D_{\mathcal{M}}(z,q) \end{eqnarray*} $\square$  \\* \\*  The reader should also notice that the strong metric could be defined on any set (with max replaced by sup) equipped with a Lorentz distance.  This remark will lead to the notion of Lorentz space \cite{Noldus2}. $\square$
\\*
\\*
We end this section with a theorem which is an amalgamation of elementary properties of the strong metric.
\begin{theo} \label{amal} This theorem is an amalgamation of results concerning the strong metric.
\begin{itemize}
\item{a)} $d_{g}$ is continuous in the strong topology.
\item{b)} The Alexandrov topology\footnote{A \emph{basis} for the Alexandrov topology is given by $I^{+}(p), I^{-}(p), I^{+}(p) \cap I^{-}(q)$ for all $p,q \in \mathcal{M}$.} is weaker than the strong topology.
\item{c)} If $\mathcal{M}$ is a compact globally hyperbolic spacetime then, the manifold, strong, and Alexandrov topology coincide.
\item{d)} The $\epsilon$-balls of the metric $D_{\mathcal{M}}$ are causally convex, i.e., if $p \ll q$ and $p,q \in B(r, \epsilon)$ for some $r \in \mathcal{M}$, then $I^{+}(p ) \cap I^{-}(q) \subset B(r , \epsilon)$.
\end{itemize}
$\square$
\end{theo}
\textsl{Proof}:  \\*
\begin{itemize}
\item{a)} Choose $p,q \in \mathcal{M}$, $\epsilon > 0$, $r \in B( p , \frac{\epsilon}{2})$, $s \in B( q , \frac{ \epsilon}{2} )$; then
$$ \left| d_g(p,q) - d_g(r,s) \right| \leq \left| d_g(p,q) - d_g(p,s) \right| + \left| d_g(p,s) - d_g(r,s) \right| < \epsilon $$
\item{b)} Since $d_g$ is continuous in the strong topology, $d_g(r,\cdot ),d_g( \cdot ,r): \mathcal{M} \rightarrow \mathbb{R}^{+}$, are also continuous in the strong topology for all $r \in \mathcal{M}$.  Hence $d_g(r, \cdot )^{-1} ( \left( 0 , +\infty \right) )$ and $ d_g( \cdot ,r)^{-1} (\left( 0 , +\infty \right) )$ are open in the strong topology, which implies that the Alexandrov topology is weaker than the strong one.
\item{c)} Since $D_{\mathcal{M}}$ is continuous in the manifold topology\footnote{Let $X$ be a compact topological space and suppose $f: X \times X \rightarrow \mathbb{R}$ is a continuous function.  Then, $F: X \rightarrow \mathbb{R}: x \rightarrow \max_{y \in X} f(x,y)$ is continuous.}, the strong topology is weaker than the manifold topology.  But the Alexandrov topology is weaker than the strong topology and coincides with the manifold topology on a globally hyperbolic spacetime, hence all topologies coincide.
\item{d)} Follows from the definition.
\end{itemize}
$\square$
\section{Limit spaces} \label{ls}
\subsection{Introduction}
In this section, I construct a limit space of a Cauchy sequence of globally hyperbolic spacetimes.  In subsection \ref{construction}, I work gradually towards a construction of the limit space.  I prove that the limit space is unique up to isometry.  I also show that, in general, the limit space has quite complicated causal behaviour.
\\* \\*
I start from a slight modification of the Gromov-Hausdorff metric
introduced in \ref{GHD} (Ref. \cite{Noldus}) and probe for a
suitable construction of ``the'' limit space of a Cauchy sequence
of spacetimes.  The modification (``convergence to
invertibility'') seems necessary to me and I shall come back to
the difference between the Gromov-Hausdorff distance  $d_{GH}$
(GH) and the \emph{generalised, Lorentzian Gromov-Hausdorff uniformity} (GGH)
later on.
\begin{deffie}
(\textbf{Generalised Lorentzian Gromov-Hausdorff uniformity}) We
call $(\mathcal{M},g)$ and $(\mathcal{N},h)$ $(\epsilon , \delta)$-close iff there exist mappings $\psi : \mathcal{M} \rightarrow
\mathcal{N}$, $\zeta : \mathcal{N} \rightarrow \mathcal{M}$ such
that
\begin{eqnarray}
\left| d_{h} ( \psi ( p_{1} ), \psi (p_{2})) - d_{g} (p_{1} , p_{2} ) \right| & \leq & \epsilon \quad \forall p_{1}, p_{2} \in \mathcal{M} \\
\left| d_{g} ( \zeta ( q_{1} ) ,\zeta (q_{2})) - d_{h} ( q_{1} ,q_{2} )\right| & \leq & \epsilon \quad \forall q_{1}, q_{2} \in \mathcal{N}
\end{eqnarray}
and
\begin{eqnarray} D_{\mathcal{M}}(p,\zeta \circ \psi(p)) & \leq & \delta  \\
D_{\mathcal{N}}(q, \psi \circ \zeta (q)) & \leq & \delta
\end{eqnarray}
for all $p \in \mathcal{M}$ and $q \in \mathcal{N}$. $\square$
\end{deffie}
\textbf{Remarks}: \\*
I show that for any Lorentzian distance $d$, the property $$ \left| d(p,r) + d(r,p) - d(q,r) - d(r,q) \right| < \epsilon, \, \forall r \in \mathcal{M},$$ is equivalent to $\left| d(p,r) - d(q,r) \right| < \epsilon$ and $\left| d(r,p) - d(r,q) \right| < \epsilon$ for all $r \in \mathcal{M}$. \\* \\*
\textsl{Proof}: \\*
$\Rightarrow )$ We prove only the first inequality, the second being analogous.  Without loss of generality we may assume that $d(p,r) > 0$.  Suppose $\epsilon > d(p,r) > 0$ and $d(q,r) \geq 0$ then $\left| d(p,r) - d(q,r) \right| < \epsilon $.  Suppose $d(p,r) \geq \epsilon$ then $d(q,r) > 0$ since if $d(r,q) >0$ then $d(p,q) > \epsilon$, hence
$$ \left| d(p,q) + d(q,p) - 2d(q,q) \right| = d(p,q) > \epsilon $$
which is a contradiction.  But if $d(q,r) >0$, then again the first inequality follows. \\*
$\Leftarrow )$ Suppose that for some $r$
$$ \left| d(p,r) + d(r,p) - d(q,r) - d(r,q) \right| \geq \epsilon $$
Without loss of generality (all other cases are symmetric) we can
assume that $d(p,r) > \epsilon$, hence $d(q,r) > 0$ otherwise
$\left| d(p,r) - d(q,r) \right| > \epsilon $ which is excluded by
assumption.  But in this case
$$ \left| d(p,r) + d(r,p) - d(q,r) - d(r,q) \right| = \left| d(p,r) - d(q,r) \right| < \epsilon $$ which is a contradiction.  $\square$ \\*  \\*
\label{rara} Using the above remark I show that Gromov-Hausdorff
$(\epsilon, \delta)$ closeness has all the properties of a
uniformity\footnote{An introduction to uniformities can be found
in Appendix C.}.  Obviously, the notion is symmetric by definition
and we are left to prove the generalised triangle inequality.
Hence, suppose that $(\mathcal{M}_{1} , g_{1})$ and
$(\mathcal{M}_{2} , g_{2} )$ are $(\epsilon_{1}, \delta_{1} )$-close, assume also that $(\mathcal{M}_{2} , g_{2})$ and
$(\mathcal{M}_{3} , g_{3} )$ are $(\epsilon_{2} , \delta_{2} )$-close, then $(\mathcal{M}_{1} , g_{1})$ and $(\mathcal{M}_{3} ,
g_{3} )$ are $( \epsilon_{1} + \epsilon_{2} , \delta_{1} +
\delta_{2} + 2 \max \left\{ \epsilon_{1} , \epsilon_{2} \right\}
)$-close.  \\* \\* \textsl{Proof}: Let $\psi_{i} : \mathcal{M}_{i}
\rightarrow \mathcal{M}_{i+1}$ and $\zeta_{i} : \mathcal{M}_{i+1}
\rightarrow \mathcal{M}_{i}$ be mappings which make
$(\mathcal{M}_{i} , g_{i})$, $(\mathcal{M}_{i+1} , g_{i+1} )$ and
$(\epsilon_{i}, \delta_{i} )$-close for $i=1,2$.  Then we have for
all $r,p \in \mathcal{M}_{1}$ that:
\begin{eqnarray*} \left| d_{g_{1}} ( \zeta_{1} \circ \zeta_{2} \circ \psi_{2} \circ \psi_{1} (p) , r ) - d_{g_{1}} ( p ,r ) \right|  \leq  & \left| d_{g_{2}} ( \psi_{1} ( p ) , \psi_{1} (r)) - d_{g_{1}} ( p ,r ) \right| + \end{eqnarray*}
\begin{eqnarray*} & & \left| d_{g_{1}} ( \zeta_{1} \circ \zeta_{2} \circ \psi_{2} \circ \psi_{1} (p) , r ) - d_{g_{2}} ( \psi_{1} \circ \zeta_{1} \circ \zeta_{2} \circ \psi_{2} \circ \psi_{1} ( p ) , \psi_{1} (r)) \right| + \\ & & \left| d_{g_{2}} ( \psi_{1} \circ \zeta_{1} \circ \zeta_{2} \circ \psi_{2} \circ \psi_{1} (p) , \psi_{1} (r) ) - d_{g_{2}} ( \zeta_{2} \circ \psi_{2} \circ \psi_{1} ( p ) , \psi_{1} (r)) \right| + \\ & &
\left| d_{g_{2}} ( \zeta_{2} \circ \psi_{2} \circ \psi_{1} ( p ) , \psi_{1} (r)) - d_{g_{2}} ( \psi_{1} ( p ) , \psi_{1} (r)) \right|
\end{eqnarray*}
Obviously, this implies that $$ \left| d_{g_{1}} ( \zeta_{1} \circ \zeta_{2} \circ \psi_{2} \circ \psi_{1} (p) , r ) - d_{g_{1}} ( p ,r ) \right|  \leq  2 \epsilon_{1} + \delta_{1} + \delta_{2}.$$  Making the same estimate for
$$ \left| d_{g_{3}} ( \psi_{2} \circ \psi_{1} \circ \zeta_{1} \circ \zeta_{2} ( q ), s ) - d_{g_{3}} ( q ,s ) \right| $$ the result follows.  $\square$ \\* \\*
Let $(\mathcal{M}_{i} , g_{i})_{i = 1}^{n}$ be spacetimes such that $(\mathcal{M}_{i},g_{i})$ and $(\mathcal{M}_{i+1} , g_{i+1})$ are $(\epsilon_{i} , \delta_{i})$ close.  Then, in the same spirit as above, $(\mathcal{M}_{1} ,g_{1})$ and $(\mathcal{M}_{n} , g_{n} )$ are $(\sum_{i=1}^{n-1} \epsilon_{i}, \\* \sum_{i=1}^{n-1} \delta_{i} + 2 \sum_{i = 2}^{n-2} \epsilon_{i} + 2 \max \left\{ \epsilon_{1} , \epsilon_{n-1} \right\})$ close.  \\* \\*  As a consequence of theorem \ref{GHD1} (Ref. ~\cite{Noldus}), $(\mathcal{M},g)$ and $(\mathcal{N},h)$ are isometric iff they cannot be distinguished by the the modified Gromov-Hausdorff uniformity.  Note also that it is impossible to make \emph{directly} a metric out of the $(\epsilon , \delta)$ Gromov-Hausdorff closeness, since if $\delta = f( \epsilon )$ where $f$ is a continuous function, one would obtain for $\epsilon_{2} < \epsilon_{1}$ that
$$ f(\epsilon_{1}) + f(\epsilon_{2}) + 2 \epsilon_{1} \leq f(\epsilon_{1} + \epsilon_{2}) $$
which is impossible.
\\*
\\*
The natural way to proceed now is to consider a generalised
Gromov-Hausdorff Cauchy sequence of spacetimes, construct a
``reasonable'' limit space and finally deduce some properties of
it.  This is the work done in the second and third subsection. The
presentation of this material is conservative, in the sense that,
ab initio, the main goal is to construct a decent limit space
using the Alexandrov topology.  This approach, however, turns out
not to work since the candidate limit space is in general not
$T_{2}$\footnote{A topological space $(X, \tau)$ is $T_{0}$ iff
for any two points there exists an open neighbourhood which
contains one of them but not both.  $(X, \tau)$ is $T_{1}$ iff for
any two points there exist neighbourhoods which do not contain the
other point.  A topological space is $T_{2}$ or Hausdorff iff for
any two points there exist disjoint neighbourhoods.}, the Lorentz distance $d$ is not continuous, nor is the
limit space compact in this topology.  This part of section
\ref{construction} has its merits nevertheless, since the problems
occurring bring alive important notions such as the timelike
capacity, $\mathcal{TC}(\mathcal{M})$, and the timelike continuum
$\mathcal{TCON}(\mathcal{M})$.  These notions express that in a
limit space of a Cauchy sequence of interpolating spacetimes, the
definition of the causal relation from the chronological one is
brought into jeopardy.  Hence, it is not clear whether two
chronologically related points can be connected by a causal curve
(geodesic).  \\*
\\*
Since I want the limit $d$ to be continuous and the $T_{2}$ separation property to be valid (at least on the interior), the strong metric, which is already present in the definition of the GGH uniformity itself, becomes important.  This metric turns out to have great technical potential, as shown in theorems \ref{two}, \ref{three} and \ref{four}.  R. Sorkin pointed out to me that D. Meyer already had a similar idea, although for different purposes\cite{Meyer}.  In section \ref{prop}, I study some examples which indicate what kind of problems show up in a possible definition of the causal relation.  It is also shown that for points $p \ll q$, the connecting geodesic (if it exists) is in general causal and not everywhere timelike.  Since the strong metric, $D_{\mathcal{M}}$, turned out to be such a strong device, I shall examine further properties of it.  For example, it turns out that $D_{\mathcal{M}}$ cannot be a path metric.  Further study of the strong metric and causal curves is left for section \ref{cau}.  There is still a small philosophical remark to be made about $D_{\mathcal{M}}$: it is a globally determined notion of closeness.  This is clear from the definition itself, and the intermezzo at the end of subsection \ref{prop}.
\subsection{Construction of the limit space} \label{construction}
\begin{figure}[h]
\begin{center}
  \setlength{\unitlength}{0.5cm}
\begin{picture}(22,5)

\put(0,0){\line(1,0){4}}
\put(0,0){\line(0,1){4}}
\put(0,4){\line(1,0){4}}
\put(4,0){\line(0,1){4}}

\put(6,0){\line(1,0){2}}
\put(6,0){\line(-1,2){1}}
\put(8,0){\line(1,2){1}}
\put(9,2){\line(-1,2){1}}
\put(8,4){\line(-1,0){2}}
\put(6,4){\line(-1,-2){1}}

\put(11.5,0){\line(1,0){1}}
\put(11.5,0){\line(-3,4){1.5}}
\put(12.5,0){\line(3,4){1.5}}
\put(14,2){\line(-3,4){1.5}}
\put(12.5,4){\line(-1,0){1}}
\put(11.5,4){\line(-3,-4){1.5}}

\put(15,2){$\cdots$}

\put(19,0){\line(1,1){2}}
\put(21,2){\line(-1,1){2}}
\put(19,4){\line(-1,-1){2}}
\put(17,2){\line(1,-1){2}}
\put(17,2){\circle*{0.3}}
\put(21,2){\circle*{0.3}}
\end{picture}
\caption{Illustration of example $3$}
\label{fig0}
\end{center}
\end{figure}
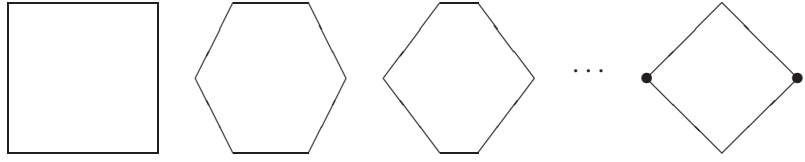

Let $(\mathcal{M}_{i} , g_{i} )_{ i \in \mathbb{N}}$ be a sequence of compact, Lorentzian manifolds such that there exist mappings $\psi^{i}_{i+1} : \mathcal{M}_{i} \rightarrow \mathcal{M}_{i+1}$ and $\zeta^{i+1}_{i} : \mathcal{M}_{i+1} \rightarrow \mathcal{M}_{i}$ making $(\mathcal{M}_{i},g_{i})$ and $(\mathcal{M}_{i+1},g_{i+1})$, $(\frac{1}{2^{i}} ,\frac{1}{2^{i}} )$-close.  If we introduce the following mappings
$$ \psi^{i}_{i+k} = \psi^{i+k-1}_{i+k} \circ \psi^{i+k-2}_{i+k-1} \circ \ldots \circ \psi^{i+1}_{i+2} \circ \psi^{i}_{i+1} : \mathcal{M}_{i} \rightarrow \mathcal{M}_{i+k}$$
$$ \zeta^{i+k}_{i} = \zeta^{i+1}_{i} \circ \zeta^{i+2}_{i+1} \circ \ldots \circ \zeta^{i+k-1}_{i+k-2} \circ \zeta^{i+k}_{i+k-1} : \mathcal{M}_{i+k} \rightarrow \mathcal{M}_{i}$$
then $\psi^{i}_{i+k}$ and $\zeta^{i+k}_{i}$ make $(\mathcal{M}_{i},g_{i})$ and $(\mathcal{M}_{i+k},g_{i+k})$, $(\frac{1}{2^{i-1}} , \frac{3}{2^{i-1}} )$-close.  Consider the set $\mathcal{S}$ of sequences $(x_{i})_{i \in \mathbb{N}}$, $x_{i} \in \mathcal{M}_{i
}$, such that there exists an $i_{0}$ such that for all $i > i_{0}$ one has that $x_{i} = \psi^{i_{0}}_{i} (x_{i_{0}})$.  Hence $x_{i} = \psi^{j}_{i} (x_{j})$ for all $i > j \geq i_{0}$.
We define the following Lorentz distance on $\mathcal{S}$:
$$ d((x_{i})_{i \in \mathbb{N}} , (y_{i})_{i \in \mathbb{N}} ) = \lim_{i \rightarrow \infty} d_{g_{i}} ( x_{i} , y_{i})$$
It is easy to verify that $d$ is indeed well defined as a Lorentz distance.  The resulting partial order is defined by: $(x_{i})_{i \in \mathbb{N}} \ll (y_{i})_{i \in \mathbb{N}}$ iff $d((x_{i})_{i \in \mathbb{N}} , (y_{i})_{i \in \mathbb{N}} )  > 0$.  Before we determine a quotient of the space $\mathcal{S}$, we should tell which topology is defined on it.  The obvious choice is the Alexandrov topology\footnote{We will see later on that this is a rather poor choice.} for which a \textbf{subbasis} is given by the sets:
\begin{itemize}
\item $\mathcal{S}$, $\emptyset$
\item $I^{+} ((x_{i})_{i \in \mathbb{N}})$ and $I^{-} ((x_{i})_{i \in \mathbb{N}})$
\end{itemize}
with $I^{\pm}$ defined by the relation $\ll$. $\square$ \\*
\\*
\textbf{Remark} \\*
I stress the word ``subbasis'' since in general the above sets do \emph{not} constitute a basis of the Alexandrov topology, as will become clear in the examples \ref{ex4} and \ref{ex5}, where specific intersections of sets belonging to the subbasis do not contain any element of the subbasis. $\square$  \\*
\\*
As suggested in section \ref{Coit}, in order to make sure that for any $(x_{i})_{i \in \mathbb{N}}$ there exists a point $(y_{i})_{i \in \mathbb{N}}$ such that $d((x_{i})_{i \in \mathbb{N}} , (y_{i})_{i \in \mathbb{N}} ) > 0$ or $d((y_{i})_{i \in \mathbb{N}} , (x_{i})_{i \in \mathbb{N}} ) > 0$ we have to demand that in the spaces $(\mathcal{M}_{i} ,g_{i})$ every point has a sufficiently long past or a sufficiently long future.  Therefore, we introduce the concept of timelike capacity.
\begin{deffie}
The timelike capacity $\mathcal{TC} (\mathcal{M},g)$ of a spacetime $(\mathcal{M} ,g)$ is defined as
$$ \mathcal{TC}(\mathcal{M},g) = \inf_{p \in \mathcal{M}} \sup_{q \in  \mathcal{M}} ( d_{g} ( p,q) + d_{g} (q,p) ).$$
\end{deffie}
Suppose now that the timelike capacity of the sequence $(\mathcal{M}_{i} ,g_{i})$ is bounded from below, i.e., there exists an $\alpha >0$ such that
$$ \mathcal{TC} ( \mathcal{M}_{i} ,g_{i} ) \geq \alpha \quad \forall i \in \mathbb{N},$$
then for any $(x_{i})_{i \in \mathbb{N}}$ one can find a $(y_{i})_{i \in \mathbb{N}}$ such that the quantity $$d( (x_{i})_{i \in \mathbb{N}} , (y_{i})_{i \in \mathbb{N}} ) + d((y_{i})_{i \in \mathbb{N}} , (x_{i})_{i \in \mathbb{N}} )$$ becomes arbitrarily close to $\alpha$.\footnote{A formal proof of this is the following: choose $\alpha > \epsilon > 0$ and $(x_{i})_{i \in \mathbb{N}} \in \mathcal{S}$. Let $\frac{1}{2^{i_{0}}} < \epsilon$, $i_{0} \in \mathbb{N}$ be such that $x_{j} = \psi_{j}^{i_{0}} (x_{i_{0}})$ for all $j > i_{0}$.  Let $y_{i_{0}} \in \mathcal{M}_{i_{0}}$ be such that $d_{g_{i_{0}}} (x_{i_{0}} , y_{i_{0}} ) + d_{g_{i_{0}}} ( y_{i_{0}} ,x_{i_{0}}) = \alpha$, then $\alpha + \epsilon > d((\psi_{j}^{i_{0}}(y_{i_{0}}))_{j>i},(x_{i})_{i \in \mathbb{N}}) + d((x_{i})_{i \in \mathbb{N}},(\psi_{j}^{i_{0}}(y_{i_{0}}))_{j>i}) > \alpha - \epsilon$.}  If there is no control on the timelike capacity, this clearly needs not be case as is shown in example \ref{ex2}.  We give an illustration of this concept in the next example.
\begin{exie}
\end{exie}
In this example, a Cauchy sequence of globally hyperbolic
interpolating spacetimes is given in which timelike capacity
converges to zero and topology change occurs.  All spacetimes in
picture \ref{fig0} below are compact pieces cut out of two
dimensional Minkowski spacetime and moreover, the fat dots are
assumed to be identified\footnote{The timelike boundaries of all
spaces, except for the limit space, are assumed to be smooth.  The author
apologises that this is not so in the pictures.}.
$\square$ \\* \\*
In what follows, I shall construct a candidate limit space and examine its separation properties in the Alexandrov topology.  It will turn out that the Alexandrov topology is too weak and the strong metric topology will emerge as a natural candidate.  First, we show that one cannot expect any candidate limit space to be Hausdorff in the Alexandrov topology.  Suppose we allow that in the limit the boundary becomes a null surface, then obviously the limit space, equipped with the Alexandrov topology, is at most $T_{0}$.  This is illustrated by the next example.
\begin{exie} \label{ex3}
\end{exie}
Take the ``cylinder universe'' $S^{1} \times \mathbb{R}$ with
metric $- dt^{2} + d\theta^{2}$ and let $p = (0,- \frac{T}{2})$
and $q = (0, \frac{T}{2})$ with $T > 0$. \\* In the notation of
\cite{Beem} let $K^{+}(q,\epsilon) = \{ r | d(q,r) = \epsilon \}$
be the ``future ball'' of radius $\epsilon$ centred at $q$ and
$K^{-}(p, \epsilon) = \{ r | d(r,p) = \epsilon \}$ be the ``past
ball'' centred at $p$.  Consider the spacetimes $(J^{+} ( K^{-}
(p, \epsilon)) \cap J^{-} ( K^{+} ( q, \epsilon )) , -dt^{2} +
d\theta^{2} )$ then the \emph{unique} (up to isometry)
Gromov-Hausdorff limit space for $\epsilon \rightarrow 0$ is
$$(J^{+}(E^{-}(p)) \cap J^{-}( E^{+}(q)), -dt^{2} + d\theta^{2}
)$$ which is $T_{0}$, but not $T_{1}$, in the Alexandrov
topology. $\square$
\\*
\\*
However, this is, as will become clear later on, not only a
\emph{boundary} phenomenon, and in general the interior points of
the $T_{0}$-quotient are not $T_{2}$ separated.  Let us first
characterise the $T_{0}$-quotient of $\mathcal{S}$.
\begin{theo} \label{one}
The $T_{0}$-quotient $T_{0}\mathcal{S}$ of $\mathcal{S} = \{ (x_{i})_{i \in \mathbb{N}} \, | \, \exists i_{0} : \forall i \geq i_{0}, \quad x_{i} = \psi_{i}^{i_{0}}( x_{i_{0}}) \}$ equals $\mathcal{S}/_{\sim}$, where the equivalence relation $\sim$ is defined by: $(x_{i})_{i \in \mathbb{N}} \sim (y_{i})_{i \in \mathbb{N}}$ iff for all $(z_{i})_{i \in \mathbb{N}} \in \mathcal{S}$
\begin{equation} \label{ekkie} d( (x_{i})_{i \in \mathbb{N}} , (z_{i})_{i \in \mathbb{N}} ) + d((z_{i})_{i \in \mathbb{N}} , (x_{i})_{i \in \mathbb{N}} ) = d( (y_{i})_{i \in \mathbb{N}} , (z_{i})_{i \in \mathbb{N}} ) + d((z_{i})_{i \in \mathbb{N}} , (y_{i})_{i \in \mathbb{N}} ) \end{equation}
\end{theo}
\textsl{Proof}: \\* Obviously, if two points $x,y \in \mathcal{S}$
are $T_{0}$-separated then there exists a $z \in \mathcal{S}$ such
that (\ref{ekkie}) is not satisfied.  Suppose $x,y \in
\mathcal{S}$ are not $T_{0}$-separated, but there exists a $z \in
\mathcal{S}$ such that (\ref{ekkie}) is not satisfied.  Then there
are essentially two possibilities, either $d(z,x) > d(z,y)$ and
$d(y,z) = 0$, or $d(x,z) > d(z,y)$ and $d(y,z) =0$.  The latter
case implies that $x$ and $y$ are $T_{0}$-separated by $I^{-}(z)$
which is a contradiction (obviously $(\mathcal{S},d)$ is
chronological).  The former case is proven by noticing that if
$d(z,x) > d(z,y) + \delta$, then for $k$ sufficiently large such
that $\frac{1}{2^{k-1}} < \frac{\delta}{8}$ and $z_{l} =
\psi^{k}_{l} (z_{k})$ , $x_{l} =\psi^{k}_{l} (x_{k})$, $y_{l} =
\psi^{k}_{l} (y_{k})$ for all $l > k$, one has that
$$ d_{g_{k}} (z_{k} , x_{k}) > d_{g_{k}} ( z_{k} , y_{k} ) + \frac{3 \delta}{4} $$
Choose $\gamma$ to be a distance maximising geodesic in
$\mathcal{M}_{k}$ from $z_{k}$ to $x_{k}$ and define the points
$p_{k}$ and $q_{k}$ on $\gamma$ by
$$ d_{g_{k}}(p_{k} , q_{k}) = d_{g_{k}} ( q_{k} , x_{k} ) = \frac{3 \delta}{8} $$
then $p_{k}$ is not in the causal past of $y_{k}$.  Hence, $p = (p_{i})_{i \in \mathbb{N}}$ and $q = (q_{i})_{i \in \mathbb{N}}$, with $p_{i} = \psi_{i}^{k} (p_{k})$ and $q_{i} = \psi_{i}^{k} (q_{k})$ for all $i > k$, satisfy the properties \begin{itemize}
\item $d(p,y) < \frac{\delta}{8}$
\item $d(p,q),d(q,x) > \frac{\delta}{4}$
\end{itemize}
Hence, $q$ is not in the past of $y$, since otherwhise $d(p,y) \geq d(p,q) > \frac{\delta}{4}$ which is a contradiction.  But then, $x$ and $y$ are $T_{0}$ separated by $I^{+}(q)$ which is a contradiction.  $\square$
\\*
\\*
Note that theorem \ref{one} and the remark on page \pageref{rara} imply that the Lorentz distance $d$ on $\mathcal{S}$ induces a Lorentz distance on $T_{0}\mathcal{S}$, also denoted by $d$ (i.e. $d$ is independent of the chosen representative).  Theorem \ref{one} also reveals that the strong (metric) $T_{2}$-quotient of $\mathcal{S}$ equals $T_{0}\mathcal{S}$. \\* \\* In the following, I want to construct the \emph{timelike closure} of the $T_{0}$-quotient of $\mathcal{S}$.  Hence, I should first define timelike Cauchy sequences in $T_{0}\mathcal{S}$.
\begin{deffie}
A sequence $(x^{i})_{i \in \mathbb{N}}$ of points in $T_{0}\mathcal{S}$ is called future timelike Cauchy iff $x^{i} \ll x^{j}$ for all $i < j$ and for $\forall \epsilon > 0, \, \exists i_{0}$ such that for all $k > j \geq i_{0}$
$$ 0 < d(x^{j} , x^{k}) < \epsilon $$
A past timelike Cauchy sequence is defined dually. $\square$
\end{deffie}
Of course, some timelike Cauchy sequences determine the same ``limit point''.  Hence, I need to define when two timelike Cauchy sequences are \emph{equivalent}.
\begin{deffie}
Two future timelike Cauchy sequences $(x^{i})_{i \in \mathbb{N}}$, $(y^{i})_{i \in \mathbb{N}}$ in $T_{0}\mathcal{S}$ are equivalent iff for any $k$ there exists an $i_{0}$ such that $i \geq i_{0}$ implies that $x^{k} \ll y^{i}$ and $y^{k} \ll x^{i}$.  The equivalence relation for two past timelike Cauchy sequences is defined dually.  A future timelike Cauchy sequence $(x^{i})_{i \in \mathbb{N}}$ and a past timelike Cauchy sequence $(y^{i})_{i \in \mathbb{N}}$ are equivalent iff $x^{k} \ll y^{l}$ for all $k,l \in \mathbb{N}$ and there exist no two points $z^{1},z^{2} \in T_{0}\mathcal{S}$ such that
$$ x^{k} \ll z^{1} \ll z^{2}, \quad \forall k \textrm{ and } z^{2} \notin \bigcup_{j \in \mathbb{N}} I^{+} ( y^{j}) $$
or
$$ y^{k} \gg z^{1} \gg z^{2}, \quad \forall k \textrm{ and } z^{2} \notin \bigcup_{j \in \mathbb{N}} I^{-} (x^{j}). $$
$\square$
\end{deffie}
\textbf{Motivation}: \\*
The only point in the above definition which might not be obvious is why there are two points $z^{1},z^{2}$ included in the definition of equivalence between a future timelike Cauchy sequence $(p^{i})_{i \in \mathbb{N}}$ and a past timelike Cauchy sequence $(q^{i})_{i \in \mathbb{N}}$.  Consider a spacetime which is already timelike complete (such as Minkowski spacetime), and let $z^{1}$ be the limit point of the sequence $(p^{i})_{i \in \mathbb{N}}$.  Then, the only conclusion which one can draw from $p^{i} \ll q^{j}$ for all $i,j > 0$ is that $z^{1}$ is null connected to the limit point of the sequence $(q^{i})_{i \in \mathbb{N}}$ which implies that $z^{1} \notin \bigcup_{j \in \mathbb{N}} I^{+} ( q^{j})$.  But the limit point of $(q^{i})_{i \in \mathbb{N}}$ might still coincide with the limit point of $(p^{i})_{i \in \mathbb{N}}$.  Clearly, if there would exist a second point $z^{2}$ satisfying $z^{1} \ll z^{2} \notin \bigcup_{j \in \mathbb{N}} I^{+} ( q^{j})$, then the limit points cannot coincide.  $\square$  
\begin{theo} The previous definition indeed determines an equivalence relation $\sim$. \end{theo} \label{pag72} 
\textsl{Proof}: \\*
Suppose $(p^{i})_{i \in
\mathbb{N}}$, $(q^{i})_{i \in \mathbb{N}}$ and $(r^{i})_{i \in
\mathbb{N}}$ are future timelike Cauchy sequences such that
$(p^{i})_{i \in \mathbb{N}} \sim (q^{i})_{i \in \mathbb{N}}$ and
$(q^{i})_{i \in \mathbb{N}} \sim (r^{i})_{i \in \mathbb{N}}$.  I
show that $(p^{i})_{i \in \mathbb{N}} \sim (r^{i})_{i \in
\mathbb{N}}$.  Choose $k \in \mathbb{N}_{0}$, then there exists an
$i_{0}$ such that $i \geq i_{0}$ implies that $p^{k} \ll q^{i}$.
Also there exists an $i_{1}$ such that $i \geq i_{1}$ implies that
$q^{i_{0}} \ll r^{i}$, hence $p^{k} \ll r^{i}$ for all $i \geq
i_{1}$.  Similarly, one can find an $\hat{i}_{1}$ such that $i
\geq \hat{i}_{1}$ implies that $r^{k} \ll p^{i}$.  Taking the
maximum of $i_{1}$ and $\hat{i}_{1}$ proves the claim.  The case
in which all sequences are past timelike Cauchy is identical.  We
are left to prove the case where one of them is from a different
type than the other two.  I only prove the case where $(p^{i})_{i
\in \mathbb{N}}$ and $(q^{i})_{i \in \mathbb{N}}$ are future
timelike Cauchy and $(r^{i})_{i \in \mathbb{N}}$ is past timelike
Cauchy, the other case being analogous.  I show first that $p^{i}
\ll r^{j}$ for all $i,j >0$.  Suppose there exist $k, l > 0$ such
that $p^{k} \notin I^{-} (r^{l})$ and let $i_{0}$ be such that $i
\geq i_{0}$ implies that $p^{k} \ll q^{i}$.  But then $q^{i}
\notin I^{-} (r^{l})$, which is a contradiction.  Remark that for
all $z \in \mathcal{M}$, $p^{k} \ll z$ for all $k$ iff $q^{l} \ll
z$ for all $l$.    This implies that it is impossible for
$z^{1},z^{2}$ to exist such that $p^{k} \ll z^{1} \ll z^{2}$ for
all $k$ and $z^{2} \notin \bigcup_{j \in \mathbb{N}} I^{+}
(r^{j})$.  Moreover, $\bigcup_{j \in \mathbb{N}} I^{-}(p^{j}) =
\bigcup_{j \in \mathbb{N}} I^{-}(q^{j})$, which implies it is
impossible for $z^{1},z^{2}$ to exist such that $r^{k} \gg z^{1}
\gg z^{2}$ for all $k$ and $z^{2} \notin \bigcup_{j \in
\mathbb{N}} I^{-} (p^{j})$.  $\square$ \\* \\*
I construct now the timelike closure $\overline{T_{0}\mathcal{S}}$ of $T_{0}\mathcal{S}$.  Define $\widetilde{T_{0}\mathcal{S}}$ as the union of $T_{0}\mathcal{S}$ with all timelike Cauchy sequences in $T_{0}\mathcal{S}$.  Define the Alexandrov topology on $\widetilde{T_{0}\mathcal{S}}$ as follows: $\mathcal{O} \subset \widetilde{T_{0}\mathcal{S}}$ in an element of the \textbf{subbasis} if
\begin{itemize}
\item $\mathcal{O} \cap T_{0}\mathcal{S}$ is an element of the afore mentioned subbasis for the Alexandrov topology in $T_{0}\mathcal{S}$
\item A future (past) timelike Cauchy sequence $(p^{i})_{i \in \mathbb{N}}$ in $T_{0}\mathcal{S}$ belongs to $\mathcal{O}$ iff
\begin{itemize}
\item $\mathcal{O} \cap T_{0}\mathcal{S} = I^{+} (q)$ ($I^{-}(q)$) for some $q \in T_{0}\mathcal{S}$ and there exists an $i_{0} \in \mathbb{N}_{0}$ such that $i \geq i_{0}$ implies that $p^{i} \in \mathcal{O} \cap T_{0}\mathcal{S} $, or
\item there exist $r^{1} , r^{2} \in T_{0}\mathcal{S} \cap \mathcal{O}$, $i_{0} \in \mathbb{N}_{0}$ such that $p^{i} \ll r^{2} \ll r^{1}$  ($r^{1} \ll r^{2} \ll p^{i}$) and $p^{i} \in \mathcal{O} \cap T_{0}\mathcal{S}$ for all $i \geq i_{0}$.
\end{itemize}
\end{itemize}
$\overline{T_{0}\mathcal{S}}$ is defined as the $T_{0}$-quotient (in the Alexandrov topology) of $\widetilde{T_{0}\mathcal{S}}$\footnote{It will become clear on page \pageref{page79} that the definition of $\overline{T_{0}\mathcal{S}}$ dependends on the mappings $\psi^{i}_{j}$ and $\zeta^{j}_{i}$ used to construct $\mathcal{S}$.}.
\\*
\\*
\textbf{Remark}:
\\*
Again, the intersection of two sets belonging to the subbasis is in general \emph{not} equal to some union of elements of the subbasis as example \ref{ex4} shows.  After having studied examples \ref{ex4} and \ref{ex5}, the reader should get a taste for the reason why the above definition is constructed in such a delicate way.  $\square$  \\*
\\*
\textbf{Property}: \\*
Two timelike Cauchy sequences are $T_{0}$-separated iff they are inequivalent.  \\*
$\Leftarrow )$ Suppose $(p^{i})_{i \in \mathbb{N}}$ and $(q^{i})_{i \in \mathbb{N}}$ are future timelike inequivalent.  Then, there exists a $k$ and a sequence $(l_{n})_{n \in \mathbb{N}}$ such that $p^{k}$ is not in the timelike past of $q^{l_{n}}$ for any $n \in \mathbb{N}$.  But then $p^{k}$ is not in the past of any $q^{i}$ with $i \geq l_{0}$, which implies that $I^{+} (p^{k})$ contains $(p^{i})_{i \in \mathbb{N}}$ but not $(q^{i})_{i \in \mathbb{N}}$.  Suppose now that $(p^{i})_{i \in \mathbb{N}}$ is future timelike Cauchy and $(q^{i})_{i \in \mathbb{N}}$ is past timelike Cauchy, with $(p^{i})_{i \in \mathbb{N}}$ not equivalent to $(q^{i})_{i \in \mathbb{N}}$.  There are essentially two cases:
\begin{itemize}
\item there exist $k,l$ such that $p^{k}$ is not in the timelike past of $q^{l}$, but then $p^{k}$ is not in the timelike past of all $q^{s}$ for all $s \geq l$.  Hence $I^{+} (p^{k})$ separates $(p^{i})_{i \in \mathbb{N}}$ from $(q^{i})_{i \in \mathbb{N}}$.
\item $p^{k} \ll q^{l}$ for all $k,l \in \mathbb{N}$ but there exist $z^{1}, z^{2} \in \mathcal{M}$ such that, say, $ p^{k} \ll z^{1} \ll z^{2}$ but $z^{2} \notin \bigcup_{l \in \mathbb{N}} I^{+} (q^{l})$; then clearly $I^{-}(z^{2})$ separates $(p^{i})_{i \in \mathbb{N}}$ from $(q^{i})_{i \in \mathbb{N}}$.
\end{itemize}
$\Rightarrow )$ Suppose $(p^{i})_{i \in \mathbb{N}}$ and $(q^{i})_{i \in \mathbb{N}}$ are future timelike equivalent, then every set of the form $I^{+}(r)$ contains an element $p^{i}$ iff it contains an element $q^{j}$ and hence all $p^{s}$ for all $s \geq i$ and $q^{t}$ for all $t \geq j$, and the same property is valid for a finite number of intersections of such sets.  A set of the form $I^{-}(r)$ contains $(p^{i})_{i \in \mathbb{N}}$ iff there exists a $r^{1}$ with $p^{i} \ll r^{1} \ll r$ for all $i \in \mathbb{N}$.  Hence, $q^{i} \ll r^{1}$ for all $i \in \mathbb{N}$ and therefore $(q^{i})_{i \in \mathbb{N}} \in I^{-}(r)$.  Hence, $(p^{i})_{i \in \mathbb{N}}$ and $(q^{i})_{i \in \mathbb{N}}$ are $T_{0}$ equivalent.  Suppose now that $(p^{i})_{i \in \mathbb{N}}$ is future timelike Cauchy and $(q^{i})_{i \in \mathbb{N}}$ is past timelike Cauchy, with $(p^{i})_{i \in \mathbb{N}}$ equivalent to $(q^{i})_{i \in \mathbb{N}}$.  Let $I^{-}(r)$ be an open set containing $(p^{i})_{i \in \mathbb{N}}$.  Then, there exists an $r^{1}$ such that $p^{i} \ll r^{1} \ll r$ for all $i \in \mathbb{N}$.  Consequently, there exists a $j_{0}$ such that $j \geq j_{0}$ implies that $q^{j}$ belongs to $I^{-}(r)$, hence $(q^{j})_{j \in \mathbb{N}} \in I^{-}(r)$.  Hence, every Alexandrov set containing $(p^{i})_{i \in \mathbb{N}}$ contains $(q^{j})_{j \in \mathbb{N}}$.  The symmetric case is proven identically.
$\square$     \\*
\\*
Define the \emph{timelike continuum} of $\overline{T_{0}\mathcal{S}}$ as the subset of all points $r$, such that there exist timelike past \emph{and} future Cauchy sequences in $T_{0}\mathcal{S}$ which are $T_{0}$-equivalent with $r$ in the Alexandrov topology.  \\* \\*
\textbf{Remark}: \\*
It is not true that if future timelike $(p^{i})_{i \in \mathbb{N}}$ and past timelike Cauchy sequences $(q^{i})_{i \in \mathbb{N}}$ converge to $r$ in the Alexandrov topology, then $r$, $(p^{i})_{i \in \mathbb{N}}$ and $(q^{i})_{i \in \mathbb{N}}$ are $T_{0}$ equivalent.  Moreover, $(p^{i})_{i \in \mathbb{N}} \sim (q^{i})_{i \in \mathbb{N}}$ does not imply that $r$ is $T_{0}$ equivalent with $(p^{i})_{i \in \mathbb{N}} \sim (q^{i})_{i \in \mathbb{N}}$.  This is illustrated in example \ref{ex6}. $\square$ \\*
\\*
The next examples show that the Alexandrov topology is too weak.
\\*
Moreover, in general the timelike continuum is a proper subset of $\overline{T_{0}\mathcal{S}}$.  I will also give an example in which every maximal $T_{2}$ subspace of $\overline{T_{0} \mathcal{S}}$ coincides, apart from a few points, with the timelike continuum.
\begin{exie} \label{ex4}
\end{exie}
This example is meant as a technical warm-up, and shows the mildest form of ``exotic'' behaviour.  We study on $S^{1} \times \left[ 0 , 5 \right]$ a family of conformally equivalent\footnote{Two metrics $d_{1}$ and $d_{2}$ on the same underlying space $\mathcal{N}$ are conformally equivalent if $d_{1}(x,y) > 0 \Leftrightarrow d_{2} (x,y) > 0$ for all $x,y \in \mathcal{N}$.} Lorentz metrics $d_{\epsilon}$, whose limit distance $d$ is ``entirely degenerate'' on the strip $\mathcal{U} = S^{1} \times \left[ 1 , 1 + \pi \right]$, i.e., $d(x,y) = 0$ for all $x,y \in \mathcal{U}$.  However, it turns out that all points of $\mathcal{U}$ \emph{are} $T_{2}$ separated. This is a consequence of the ``external'' relations with points outside $\mathcal{U}$ and because the strip $\mathcal{U}$ is not ``thick enough'' (this will be made clear later on).  This stands diametrically opposite to what we are used to from pseudo\footnote{Pseudo in the sense that the ``Riemannian'' metric is allowed to be degenerate.} Riemannian geometry in which this is just impossible because of the triangle inequality \footnote{Let $\mathcal{U}$ be a maximal set in a pseudo ``Riemannian'' space on which the ``Riemannian'' metric $D$ vanishes, then for any exterior point $r$ and $x,y \in \mathcal{U}$ such that $D(r,x)$ or $D(r,y)$ is nonvanishing, one has that $D(r,x) \leq D(r,y) \leq D(r,x)$.}.  In \emph{global} Lorentzian geometry however, the reversed triangle inequality is responsible for this phenomenon.  I am aware that some classical relativists might start objecting to the construction now, but let me convince these people that it is their own, highly Riemannesque, ``local intuition'' which is responsible for this protest.  Moreover this global ``artifact'' has as peculiarity that we are still able to reconstruct the ``faithful'' causal relations even on the strip $\mathcal{U}$ of degeneracy.  In what follows, I construct conformal factors $\Omega_{\epsilon}(t)$ which equal $1$ on $\left[0 , 1 - \epsilon \right] \cup \left[ 1 + \pi + \epsilon , 5 \right]$, $\epsilon$ on $\left[ 1 + \epsilon , 1 + \pi - \epsilon \right]$ and undergo a smooth transition in the strips $\left[ 1 - \epsilon , 1 + \epsilon \right]$ and $\left[ 1 + \pi - \epsilon , 1 + \pi + \epsilon \right]$ respectively.  Such a smooth transition function can be constructed by using the smooth function $f$, defined by $f(x) = 0$ for $x \leq 0$ and $f(x) = \exp^{ - \frac{1}{x^{2}}}$ for $ x > 0$.  Define moreover $\alpha_{\epsilon} (x) = 1 - (1 - \epsilon) \chi (x - \epsilon)$, $\beta_{\epsilon}(x) = \epsilon + ( 1 - \epsilon ) \chi ( x - \epsilon )$ where $\chi$ is the characteristic function defined by $\chi (x ) = 1$ if $x > 0$ and zero otherwise.  Then with $$\psi_{\epsilon} ( x ) = \frac{ \alpha_{\epsilon}(\cdot - 1 + \epsilon ) \ast \left( f(\cdot + \epsilon)f( - \cdot + \epsilon  ) \right) }{ \int_{- \infty}^{+\infty} f(t + \epsilon) f(-t + \epsilon) dt} ( x)$$ and
$$ \zeta_{\epsilon} ( x ) = \frac{\beta_{\epsilon}(\cdot - 1 -\pi + \epsilon) \ast \left( f( \cdot + \epsilon  )f( - \cdot + \epsilon ) \right)  }{ \int_{-\infty}^{+\infty} f(t + \epsilon)f( - t + \epsilon) dt }(x)$$
one can define, for example, $\Omega_{\epsilon}(t) = \psi_{\epsilon} (t) $ for $t \in \left[ 1 - \epsilon , 1 + \epsilon \right]$ and $\Omega_{\epsilon}(t) = \zeta_{\epsilon} (t)$ on $\left[ 1 + \pi - \epsilon , 1 + \pi + \epsilon \right]$. (In the above formula's, $\ast$ denotes the convolution product.)
\begin{figure}
\begin{center}
\mbox{\rotatebox{-90}{\includegraphics[width=4cm,height=6cm]{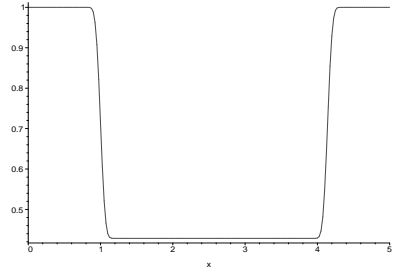}}}
\caption{A plot of $\Omega_{\epsilon}$ for some $\epsilon>0$. }
\label{fig1}
\end{center}
\end{figure}
Consider now the metric tensors
$$ ds^{2}_{\epsilon} = \Omega_{\epsilon}^{2}(t) ( -dt^{2} + d\theta^{2} )$$ and the associated Lorentz distances $d_{\epsilon}$.   On basis of simple geometric arguments, (i.e., without even calculating the geodesics) it is easy to see that $(S^{1} \times \left[0,5 \right], d_{\epsilon})$ is a Gromov-Hausdorff Cauchy sequence (with the maps $\psi_{\delta}^{\epsilon}$ and $\zeta_{\epsilon}^{\delta}$ equal to the identity) converging to the cylindrical space with a Lorentz metric $d$ which is degenerate on the strip $\left[ 1 , 1 + \pi \right]$ (Hint: look for lower and upper bounds of lengths of geodesics).  However, all the points on the strip are $T_{2}$ separated as is shown (partially) in Fig.\ \ref{fig2}.
\begin{figure}
\begin{center}
  \setlength{\unitlength}{1.8cm}
\begin{picture}(8,5)

\put(1,1){\line(1,0){6}}
\put(1,1){\line(0,1){3}}
\put(1,4){\line(1,0){6}}
\put(7,1){\line(0,1){3}}
\put(1,0.8){$0$}
\put(7,0.8){$2\pi$}
\put(0.8,4){$\pi$}

\thicklines
\put(1,3){\line(1,1){1.3}}
\put(7,3){\line(-1,-1){1.7}}
\put(5.3,1.3){\line(-1,1){3}}
\put(2.3,4.4){$q_1$}

\put(2.3,1.3){\Red{\line(1,1){3}}}
\put(2.3,1.3){\Red{\line(-1,1){1.3}}}
\put(5.3,4.3){\Red{\line(1,-1){1.7}}}
\put(5.3,4.4){$q_2$}

\put(1.2,3.2){\Green{\line(1,-1){2.4}}}
\put(3.6,0.8){\Green{\line(1,1){1.1}}}
\put(4.2,3.2){\Green{\line(1,-1){2.4}}}
\put(6.6,0.8){\Green{\line(1,1){0.4}}}
\put(1,1.2){\Green{\line(1,1){0.7}}}
\put(3.6,0.6){$p_1$}
\put(6.6,0.6){$p_2$}

\thinlines
\multiput(1,3)(0.25,0){24}{\line(1,0){0.125}}
\multiput(1,2.6)(0.25,0){24}{\Red{\line(1,0){0.125}}}
\multiput(1,1.2)(0.25,0){24}{\Green{\line(1,0){0.125}}}

\multiput(1.3,3.1)(0.11,-0.11){21}{\Blue{\line(1,1){1.1}}}
\multiput(4.3,3.1)(0.11,-0.11){15}{\Blue{\line(1,1){1.1}}}
\put(5.95, 1.45){\Blue{\line(1,1){1.05}}}
\put(6.06, 1.34){\Blue{\line(1,1){0.94}}}
\put(6.17, 1.23){\Blue{\line(1,1){0.83}}}
\put(1, 2.06){\Blue{\line(1,1){0.27}}}
\put(6.28, 1.12){\Blue{\line(1,1){0.72}}}
\put(1, 1.84){\Blue{\line(1,1){0.38}}}
\put(6.39, 1.01){\Blue{\line(1,1){0.61}}}
\put(1, 1.62){\Blue{\line(1,1){0.49}}}
\put(6.50, 0.9){\Blue{\line(1,1){0.50}}}
\put(1, 1.4){\Blue{\line(1,1){0.60}}}

\put(2.5,2.3){\circle*{0.05}}
\put(2.6,2.3){$r_1$}
\put(5.5,2.3){\circle*{0.05}}
\put(5.6,2.3){$r_2$}

\end{picture}
\caption{Points $r_{1},r_{2}$ in the degenerate area can be Hausdorff separated.}
\label{fig2}
\end{center}
\end{figure}
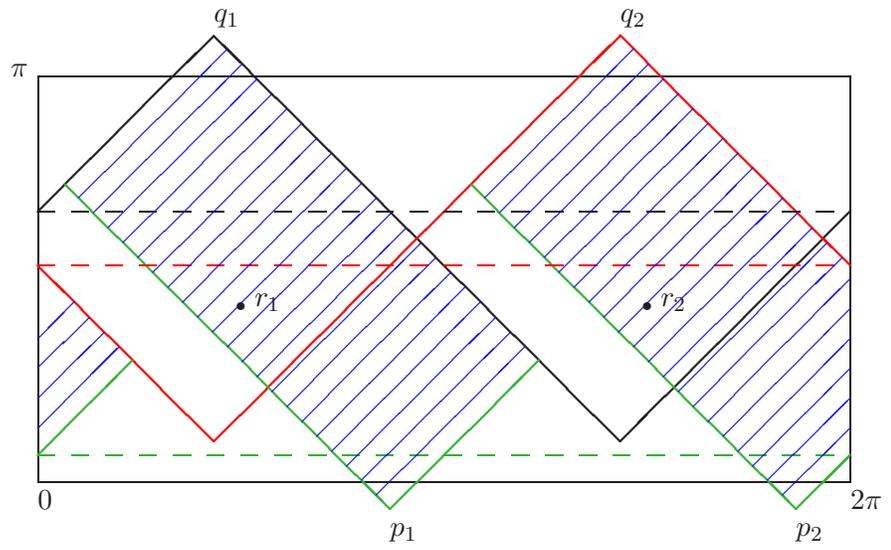
This will not be the case anymore in the next example.  The reader should do the following exercises in order to get used to the ``strange'' things which can happen.
\begin{itemize}
\item Construct an open set in the ``degenerate area'' which does not contain any set of the form $I^{+}(p) \cap I^{-} (q)$ for $p \ll q$, elements of $\overline{T_{0}\mathcal{S}}$.
\item Consider the past timelike Cauchy sequence $((0, 1 + \pi + \frac{1}{n+1}))_{n \in \mathbb{N}}$, construct two \emph{generating} Alexandrov sets (in $\overline{T_{0}\mathcal{S}}$) $I^{+}(p_{i}) \cap I^{-}(q_{i})$, $i=1,2$ such that there exist $z_{i}$ with $p_{i} \ll z_{i} \ll (0,1 + \pi + \frac{1}{n+1})$ for all
$n \in \mathbb{N}$ \emph{but} there exists no point $z \in I^{+}(p_{1}) \cap I^{+}(p_{2}) \cap I^{-}(q_{1}) \cap I^{-}(q_{2})$ such that $z \ll (0, 1 + \pi + \frac{1}{n+1})$ for all $n \in \mathbb{N}$.
\end{itemize}
 $\square$
\begin{exie} \label{ex5}
\end{exie}
This more exotic example makes all points in the ``degenerate area'' only $T_{0}$ and \emph{not} $T_{2}$ separated.
 Consider the cylinder $S^{1} \times \left[ 0 ,8 \right]$ and define conformal factors $\Omega_{\epsilon} (t)$ as follows: $\Omega_{\epsilon}(t)$ equals $1$ on the strips $\left[ 0 ,1 - \epsilon \right]$ and $\left[ 1 + 2 \pi + \epsilon , 8 \right]$, $\epsilon$ on $\left[ 1 + \epsilon , 1 + 2 \pi - \epsilon \right]$, and undergoes smooth transitions on $\left[ 1 - \epsilon , 1 + \epsilon \right]$ and $\left[ 1 + 2 \pi - \epsilon , 1 + 2 \pi + \epsilon \right]$ respectively.  Again, it is not difficult to show that the distances $d_{ \epsilon}$ associated with the metrics
$$ ds^{2}_{\epsilon} = \Omega_{\epsilon}^{2}(t) ( -dt^{2} + d\theta^{2}) $$
form a Gromov-Hausdorff Cauchy sequence, and define a limit distance $d$ (wrt. to the identity mappings $\psi^{\epsilon}_{\delta}$ and $\zeta
^{\delta}_{\epsilon}$ ) which is degenerate on the strip $\left[1 , 1 + 2 \pi \right]$\footnote{$(S^{1} \times \left[ 0 ,8 \right] , d)$ is timelike complete.}.  However, the points with time coordinate $t = 1 + \pi$ are $T_{0}$ equivalent and therefore the $T_{0}$ limit space is topologically a (double) cone with (common) tip $t = 1 + \pi$.  Points in the strip $\left[ 1 , 1 + 2 \pi \right] $ cannot be $T_{2}$ separated, since any Alexandrov set containing such point also contains the tip\footnote{The $T_{1}$ separation property is also not satisfied since no point of the degenerate strip can be $T_{1}$ separated from the tip.} (see Fig.\ \ref{fig3}).
\begin{figure}[h]
\begin{center}
  \setlength{\unitlength}{1cm}
\begin{picture}(8,7)

\put(1,1){\line(1,0){6}}
\put(1,1){\line(0,1){6}}
\put(1,7){\line(1,0){6}}
\put(7,1){\line(0,1){6}}
\put(7.1,1){$t=1$}
\put(7.1,4){$t=1+\pi$}
\put(7.1,7){$t=1+2\pi$}
\put(1,4){\Red{\line(1,0){6}}}

\thicklines
\put(1,6){\line(1,1){1.3}}
\put(7,6){\line(-1,-1){1.7}}
\put(5.3,4.3){\line(-1,1){3}}

\put(1,3.8){\line(1,-1){3}}
\put(4,0.8){\line(1,1){3}}

\thinlines
\put(1,4){\Blue{\line(1,-1){3.1}}}
\put(1.2,4){\Blue{\line(1,-1){3}}}
\put(1.4,4){\Blue{\line(1,-1){2.9}}}
\put(1.6,4){\Blue{\line(1,-1){2.8}}}
\put(1.8,4){\Blue{\line(1,-1){2.7}}}
\put(2,4){\Blue{\line(1,-1){2.6}}}
\put(2.2,4){\Blue{\line(1,-1){2.5}}}
\put(2.4,4){\Blue{\line(1,-1){2.4}}}
\put(2.6,4){\Blue{\line(1,-1){2.3}}}
\put(2.8,4){\Blue{\line(1,-1){2.2}}}
\put(3,4){\Blue{\line(1,-1){2.1}}}
\put(3.2,4){\Blue{\line(1,-1){2}}}
\put(3.4,4){\Blue{\line(1,-1){1.9}}}
\put(3.6,4){\Blue{\line(1,-1){1.8}}}
\put(3.8,4){\Blue{\line(1,-1){1.7}}}
\put(4,4){\Blue{\line(1,-1){1.6}}}
\put(4.2,4){\Blue{\line(1,-1){1.5}}}
\put(4.4,4){\Blue{\line(1,-1){1.4}}}
\put(4.6,4){\Blue{\line(1,-1){1.3}}}
\put(4.8,4){\Blue{\line(1,-1){1.2}}}
\put(5,4){\Blue{\line(1,-1){1.1}}}
\put(5.2,4){\Blue{\line(1,-1){1}}}
\put(5.4,4){\Blue{\line(1,-1){0.9}}}
\put(5.6,4){\Blue{\line(1,-1){0.8}}}
\put(5.8,4){\Blue{\line(1,-1){0.7}}}
\put(6,4){\Blue{\line(1,-1){0.6}}}
\put(6.2,4){\Blue{\line(1,-1){0.5}}}
\put(6.4,4){\Blue{\line(1,-1){0.4}}}
\put(6.6,4){\Blue{\line(1,-1){0.3}}}
\put(6.8,4){\Blue{\line(1,-1){0.2}}}
\put(1,4.1){\Blue{\line(1,0){6}}}
\put(1,4.2){\Blue{\line(1,0){6}}}
\put(1,4.3){\Blue{\line(1,0){6}}}
\put(1,4.4){\Blue{\line(1,0){4.2}}}
\put(1,4.5){\Blue{\line(1,0){4.1}}}
\put(1,4.6){\Blue{\line(1,0){4}}}
\put(1,4.7){\Blue{\line(1,0){3.9}}}
\put(1,4.8){\Blue{\line(1,0){3.8}}}
\put(1,4.9){\Blue{\line(1,0){3.7}}}
\put(1,5){\Blue{\line(1,0){3.6}}}
\put(1,5.1){\Blue{\line(1,0){3.5}}}
\put(1,5.2){\Blue{\line(1,0){3.4}}}
\put(1,5.3){\Blue{\line(1,0){3.3}}}
\put(1,5.4){\Blue{\line(1,0){3.2}}}
\put(1,5.5){\Blue{\line(1,0){3.1}}}
\put(1,5.6){\Blue{\line(1,0){3}}}
\put(1,5.7){\Blue{\line(1,0){2.9}}}
\put(1,5.8){\Blue{\line(1,0){2.8}}}
\put(1,5.9){\Blue{\line(1,0){2.7}}}
\put(1,6){\Blue{\line(1,0){2.6}}}
\put(1.1,6.1){\Blue{\line(1,0){2.4}}}
\put(1.2,6.2){\Blue{\line(1,0){2.2}}}
\put(1.3,6.3){\Blue{\line(1,0){2}}}
\put(1.4,6.4){\Blue{\line(1,0){1.8}}}
\put(1.5,6.5){\Blue{\line(1,0){1.6}}}
\put(1.6,6.6){\Blue{\line(1,0){1.4}}}
\put(1.7,6.7){\Blue{\line(1,0){1.2}}}
\put(1.8,6.8){\Blue{\line(1,0){1}}}
\put(1.9,6.9){\Blue{\line(1,0){0.8}}}
\put(2,7){\Blue{\line(1,0){0.6}}}
\put(2.1,7.1){\Blue{\line(1,0){0.4}}}
\put(2.2,7.2){\Blue{\line(1,0){0.2}}}
\put(7,4.4){\Blue{\line(-1,0){1.6}}}
\put(7,4.5){\Blue{\line(-1,0){1.5}}}
\put(7,4.6){\Blue{\line(-1,0){1.4}}}
\put(7,4.7){\Blue{\line(-1,0){1.3}}}
\put(7,4.8){\Blue{\line(-1,0){1.2}}}
\put(7,4.9){\Blue{\line(-1,0){1.1}}}
\put(7,5){\Blue{\line(-1,0){1}}}
\put(7,5.1){\Blue{\line(-1,0){0.9}}}
\put(7,5.2){\Blue{\line(-1,0){0.8}}}
\put(7,5.3){\Blue{\line(-1,0){0.7}}}
\put(7,5.4){\Blue{\line(-1,0){0.6}}}
\put(7,5.5){\Blue{\line(-1,0){0.5}}}
\put(7,5.6){\Blue{\line(-1,0){0.4}}}
\put(7,5.7){\Blue{\line(-1,0){0.3}}}
\put(7,5.8){\Blue{\line(-1,0){0.2}}}

\end{picture}
\caption{Picture of the candidate limit space.}
\label{fig3}
\end{center}
\end{figure}
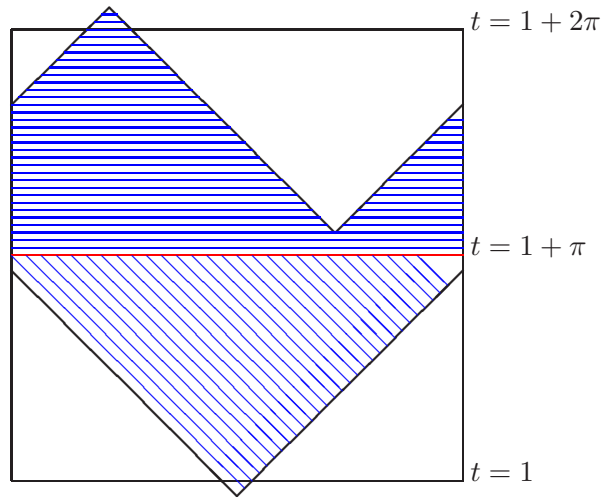
Consider the sequence $((0,1+ \pi + \frac{1}{n+1}) )_{ n \in \mathbb{N}}$. Surprisingly, \emph{every} point in the degenerate strip is a limit point of this sequence in the Alexandrov topology!  However, only the tip is the \emph{unique}, strong limit point.  As proven in theorem \ref{amal} (Ref. \cite{Noldus}, theorem 8), $d$ is continuous in the \emph{strong}, but not in the Alexandrov topology.  This tells us that the strong topology is \emph{more suitable} since one would at least like $d$ to be continuous on ``the'' limit space.  Another example of the rather pathological behaviour of the Alexandrov topology is the following.  Removing the tip from the previous limit space, one obtains a timelike complete, non strongly compact (but compact in the Alexandrov topology) limit space with respect to mappings $\psi_{\delta}^{\epsilon}$ and $\zeta_{\epsilon}^{\delta}$ defined as follows: $\zeta_{\epsilon}^{\delta}$ is simply the identity,  $\psi_{\delta}^{\epsilon}$ however is defined by making a cut at $t = 1 + \pi$; $\psi_{\delta}^{\epsilon} (x,t) = (x,t) $ for $t \in \left[ 0 , 1 \right] \cup \left( 1 + \pi , 8 \right]$ and $( x , 1 + (1 - \epsilon)(t - 1) )$ for $t \in \left[ 1 , 1 + \pi \right]$.  Notice also that in the strong topology, the (first) limit space is compact and Hausdorff.     $\square$  \\*
\\*
These last examples made clear that the strong metric gives rise to a suitable topology on $\overline{T_{0}\mathcal{S}}$.  Moreover, as proven in theorem \ref{amal}, the strong topology coincides with the manifold topology on a compact globally hyperbolic interpolating spacetime\footnote{As a matter of fact, for a general distinguishing spacetime (with or without boundary) with finite timelike diameter, the strong topology is finer as the manifold topology \cite{Meyer} \cite{Kriele1}.}.
The figures in the next example are illustrations of the remark following the definition of the timelike continuum.
\begin{exie} \label{ex6}
\end{exie}
Fig.\ (\ref{fig4}) shows future timelike $(p^{i})_{i \in \mathbb{N}}$ and past timelike Cauchy sequences $(q^{i})_{i \in \mathbb{N}}$ which converge to $r$ in the Alexandrov topology, but $r$, $(p^{i})_{i \in \mathbb{N}}$ and $(q^{i})_{i \in \mathbb{N}}$ are not $T_{0}$ equivalent.
\begin{figure}[h]
\begin{center}
  \setlength{\unitlength}{1.5cm}
\begin{picture}(8,5)

\put(1,1){\line(1,0){6}}
\put(1,1){\line(0,1){3}}
\put(1,4){\line(1,0){6}}
\put(7,1){\line(0,1){3}}
\put(0.8,0.8){$0$}
\put(7,0.8){$2\pi$}
\put(0.8,4){$1$}

\put(2.5,1){\line(1,1){3}}
\put(2.5,4){\line(1,-1){3}}
\put(4,2.5){\circle*{0.1}}
\put(4.1,2.4){$r$}
\put(3,1){\line(1,1){1}}
\put(5,1){\line(-1,1){1}}
\put(3,4){\line(1,-1){1}}
\put(5,4){\line(-1,-1){1}}

\multiput(0,0)(0.2,0.2){6}{\put(3,1){\line(0,1){0.5}}}
\put(2.8,1){\line(0,1){0.3}}
\multiput(0,0)(-0.2,0.2){5}{\put(5,1){\line(0,1){0.5}}}
\put(5.2,1){\line(0,1){0.3}}
\multiput(0,0)(0.2,-0.2){6}{\put(3,4){\line(0,-1){0.5}}}
\put(2.8,4){\line(0,-1){0.3}}
\multiput(0,0)(-0.2,-0.2){5}{\put(5,4){\line(0,-1){0.5}}}
\put(5.2,4){\line(0,-1){0.3}}

\put(4.1,3.8){$(q^i)_{i\in\mathbb{N}}$}
\put(4,3.8){\circle*{0.05}}
\put(4,3.6){\circle*{0.05}}
\put(4,3.4){\circle*{0.05}}
\put(4,3.2){\circle*{0.05}}
\put(4,3.1){\circle*{0.05}}
\put(4,3.0){\circle*{0.05}}
\put(3.95,2.95){\Red{x}}
\put(4.1,1.1){$(p^i)_{i\in\mathbb{N}}$}
\put(4,1.2){\circle*{0.05}}
\put(4,1.4){\circle*{0.05}}
\put(4,1.6){\circle*{0.05}}
\put(4,1.8){\circle*{0.05}}
\put(4,1.9){\circle*{0.05}}
\put(4,2){\circle*{0.05}}
\put(3.95,1.95){\Red{x}}

\end{picture}
\caption{Convergence in the Alexandrov topology is not the same as $T_{0}$ equivalence. }
\label{fig4}
\end{center}
\end{figure}
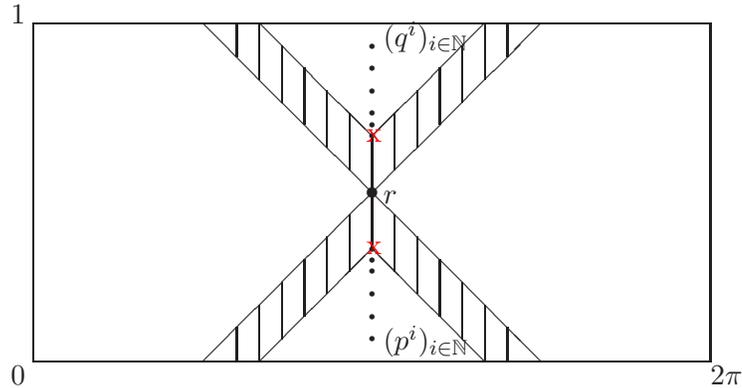
Fig.\ (\ref{fig5}) shows that two equivalent sequences $(p^{i})_{i \in \mathbb{N}} \sim (q^{i})_{i \in \mathbb{N}}$ which converge to $r$ are not necessarily $T_{0}$ equivalent with $r$.
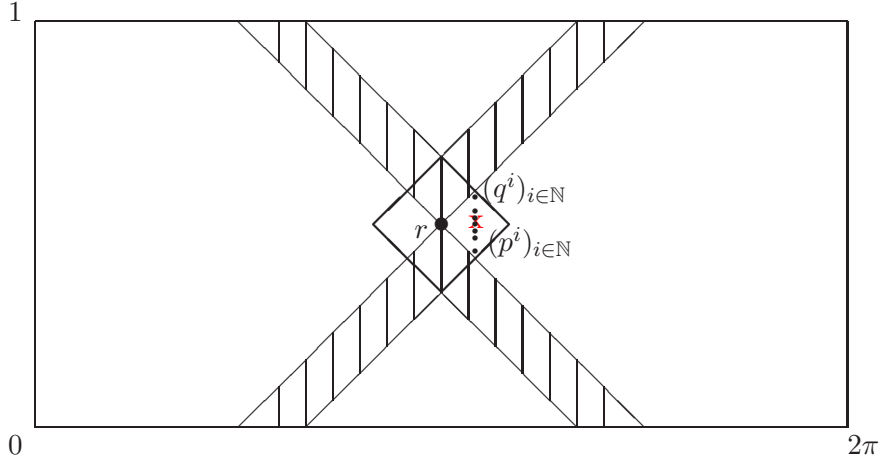
\begin{figure}[h]
\begin{center}
  \setlength{\unitlength}{1.8cm}
\begin{picture}(8,5)

\put(1,1){\line(1,0){6}}
\put(1,1){\line(0,1){3}}
\put(1,4){\line(1,0){6}}
\put(7,1){\line(0,1){3}}
\put(0.8,0.8){$0$}
\put(7,0.8){$2\pi$}
\put(0.8,4){$1$}

\put(2.5,1){\line(1,1){3}}
\put(2.5,4){\line(1,-1){3}}
\put(4,2.5){\circle*{0.1}}
\put(3.8,2.4){$r$}
\put(3,1){\line(1,1){1}}
\put(5,1){\line(-1,1){1}}
\put(3,4){\line(1,-1){1}}
\put(5,4){\line(-1,-1){1}}

\multiput(0,0)(0.2,0.2){6}{\put(3,1){\line(0,1){0.5}}}
\put(2.8,1){\line(0,1){0.3}}
\multiput(0,0)(-0.2,0.2){5}{\put(5,1){\line(0,1){0.5}}}
\put(5.2,1){\line(0,1){0.3}}
\multiput(0,0)(0.2,-0.2){6}{\put(3,4){\line(0,-1){0.5}}}
\put(2.8,4){\line(0,-1){0.3}}
\multiput(0,0)(-0.2,-0.2){5}{\put(5,4){\line(0,-1){0.5}}}
\put(5.2,4){\line(0,-1){0.3}}

\thicklines
\put(4,3){\line(1,-1){0.5}}
\put(4,3){\line(-1,-1){0.5}}
\put(4,2){\line(1,1){0.5}}
\put(4,2){\line(-1,1){0.5}}

\put(4.2,2.47){\Red{x}}
\put(4.3,2.7){$(q^i)_{i\in\mathbb{N}}$}
\put(4.25,2.7){\circle*{0.04}}
\put(4.25,2.6){\circle*{0.04}}
\put(4.25,2.55){\circle*{0.04}}
\put(4.25,2.5){\circle*{0.04}}
\put(4.34,2.3){$(p^i)_{i\in\mathbb{N}}$}
\put(4.25,2.3){\circle*{0.04}}
\put(4.25,2.4){\circle*{0.04}}
\put(4.25,2.45){\circle*{0.04}}

\end{picture}
\caption{Equivalent sequences converging to $r$ are not $T_{0}$ equivalent to $r$.}
\label{fig5}
\end{center}
\end{figure}
The ``universe'' in these pictures is $(S^{1} \times \left[0 ,1\right],d)$, where $d$ is the limit distance defined by a sequence $d_{\epsilon}$.  $d_{\epsilon}$ is constructed from $g_{\epsilon} = \Omega_{\epsilon}^{2}(t) ( -dt^{2} + d\theta^{2})$ where the smooth conformal factor $\Omega_{\epsilon}$ goes, proportionally to $\epsilon$, to zero on the shaded area $\mathcal{D}$ and to $1$ elsewhere.  Hence, $d$ is degenerate on $\mathcal{D}$, but the timelike relations between points $p \in \mathcal{D}$ and $q \in S^{1} \times \left[0 ,1\right] \setminus \mathcal{D}$ are the ones induced by $ds^{2} = -dt^{2} + d\theta^{2}$.
$\square$
\\*
\\*
We shall be mainly interested in the strong topology, but first I
finish with stating a few properties of $d$ in the Alexandrov
topology.  One can show that $d$ is continuous on the timelike
continuum of $\overline{T_{0}\mathcal{S}}$ and that the Alexandrov
topology has the $T_{2}$ property on $\mathcal{TCON}$: a proof can
be found in Appendix D. \label{page64} Also, one can prove that
$(\overline{T_{0}\mathcal{S}},d)$ is a limit space of the sequence
$(\mathcal{M}_{i},g_{i})$, see Appendix D. \\* \\* Let us
summarise our preliminary results: examples \ref{ex4} and
\ref{ex5} show that we have to allow degenerate metrics and that,
moreover, the Alexandrov topology has bad separation properties on
the ``degenerate area''.  The afore mentioned results however show
that the candidate limit space has the required behaviour on the
timelike continuum, i.e., $d$ is continuous and the Alexandrov
topology is $T_{2}$ on $\mathcal{TCON}$.  In the following example, I show that by a
judicious choice of mappings $\psi^{i}_{j}$ and $\zeta^{j}_{i}$,  $\overline{T_{0}\mathcal{S}}$ is \emph{not} compact
in the Alexandrov topology while $\overline{T_{0}\mathcal{S}}$ is
for another set of mappings!  
\begin{exie}
\end{exie}
I show that $\overline{T_{0} \mathcal{S}}$ is dependent upon the mappings $\psi_{\epsilon}^{\delta}$ and $\zeta_{\delta}^{\epsilon}$ used to construct it.  
\begin{figure}[h]
\begin{center}
  \setlength{\unitlength}{1.5cm}
\begin{picture}(8,5)

\put(1,1){\line(1,0){6}}
\put(1,1){\line(0,1){3}}
\put(1,4){\line(1,0){6}}
\put(7,1){\line(0,1){3}}
\put(0.8,0.8){$0$}
\put(7,0.8){$2\pi$}
\put(0.8,4){$1$}

\put(2.5,1){\line(1,1){3}}
\put(2.5,4){\line(1,-1){3}}
\put(4,2.5){\circle*{0.1}}
\put(3.7,2.4){$r$}
\put(3,1){\line(1,1){1}}
\put(5,1){\line(-1,1){1}}
\put(3,4){\line(1,-1){1}}
\put(5,4){\line(-1,-1){1}}

\multiput(0,0)(0.2,0.2){6}{\put(3,1){\line(0,1){0.5}}}
\put(2.8,1){\line(0,1){0.3}}
\multiput(0,0)(-0.2,0.2){5}{\put(5,1){\line(0,1){0.5}}}
\put(5.2,1){\line(0,1){0.3}}
\multiput(0,0)(0.2,-0.2){6}{\put(3,4){\line(0,-1){0.5}}}
\put(2.8,4){\line(0,-1){0.3}}
\multiput(0,0)(-0.2,-0.2){5}{\put(5,4){\line(0,-1){0.5}}}
\put(5.2,4){\line(0,-1){0.3}}

\put(4.1,2.5){\circle*{0.07}}
\put(4.3,2.5){\circle*{0.07}}
\put(4.6,2.5){\circle*{0.07}}
\put(5,2.5){\circle*{0.07}}
\put(5.5,2.5){\circle*{0.07}}
\put(6.1,2.5){\circle*{0.07}}

\end{picture}
\caption{A sequence converging to a point, $r$, which does not belong to $\mathcal{TCON}$.}
\label{seq}
\end{center}
\end{figure}
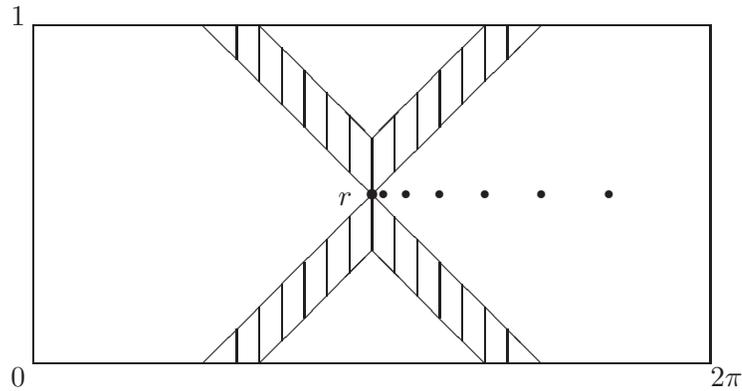
The ``universe'' in picture \ref{seq} is $(S^{1} \times \left[0 ,1\right],d)$, where $d$ is the limit distance defined by a sequence $d_{\epsilon}$.  $d_{\epsilon}$ is constructed from $g_{\epsilon} = \Omega_{\epsilon}^{2}(t) ( -dt^{2} + d\theta^{2})$, where the smooth conformal factor $\Omega_{\epsilon}$ goes, proportionally to $\epsilon$, to zero on the shaded area $\mathcal{D}$ and to $1$ elsewhere.  The mappings $\psi_{\delta}^{\epsilon}$, $\zeta_{\epsilon}^{\delta}$ are the identity.  In this compact (in the Alexandrov and strong topology) limit space we consider a sequence which converges in the Alexandrov topology to the unique point $r$ which does not belong to the timelike continuum.  \\* \\*
Applying a cut procedure as explained at the end of example \ref{ex5}, we can make sure that (only) $r$ does not belong to $T_{0}\mathcal{S}$.  But, then $r \notin \overline{T_{0} \mathcal{S}}$, which implies that by another choice of mappings, we obtain that $\overline{T_{0} \mathcal{S}}$ is not compact in the Alexandrov nor in the strong topology. $\square$ \\* \\*
All this shows, in my opinion, that the Alexandrov
topology is not appropriate and I shall concentrate on the strong
topology from now on. \label{page79}\\* \\* The beautiful thing
about the strong topology is that it is a metric topology, and the
immediate natural question which emerges is whether GGH
convergence of the sequence $(\mathcal{M}_{i},g_{i})$ forces GH
convergence of the compact metric spaces $(\mathcal{M}_{i},
D_{\mathcal{M}_{i}})$.  Recall that the Gromov-Hausdorff distance
between metric spaces $(X,d_{X})$ and $(Y,d_{Y})$ is defined by
$$d_{GH} ( (X,d_{X}),(Y,d_{Y})) = \inf \{ d_{H} (X,Y) |
\textrm{all admissible metrics } d \textrm{ on } X \sqcup Y \}$$
where a metric $d$ on the disjoint union $X \sqcup Y$ is
\emph{admissible} iff the restrictions of $d$ to $X$ and $Y$ equal
$d_{X}$ and $d_{Y}$, respectively.  By $d_{H}$, I denote the
Hausdorff distance associated to $d$.
\begin{theo} \label{two}
Let $(\mathcal{M},g)$ and $(\mathcal{N},h)$ be $(\epsilon , \delta)$-close in the Lorentzian Gromov-Hausdorff sense, then $d_{GH}((\mathcal{M},D_{\mathcal{M}}),(\mathcal{N}, D_{\mathcal{N}})) \leq \epsilon + \frac{3 \delta}{2}$.
\end{theo}
\textsl{Proof}:  \\*
Let $\psi : \mathcal{M} \rightarrow \mathcal{N}$ and $\zeta : \mathcal{N} \rightarrow \mathcal{M}$ be mappings which make $(\mathcal{M},g)$ and $(\mathcal{N},h)$, $(\epsilon , \delta)$-close.  Then, using that $D_{\mathcal{M}} (\zeta \circ \psi (p) , p) , D_{\mathcal{N}} (\psi \circ \zeta (q) ,q ) < \delta$, it is not difficult to derive that
\begin{equation} \label{imp1} \left| D_{\mathcal{N}} ( \psi (p) , \psi (q)) - D_{\mathcal{M}} (p,q ) \right| < 2( \epsilon + \delta), \quad \forall p,q \in \mathcal{M} \end{equation}
and
\begin{equation} \label{imp2} \left| D_{\mathcal{M}} (\zeta (p) , \zeta(q) ) - D_{\mathcal{N}} (p,q) \right| < 2( \epsilon +\delta), \quad \forall p,q \in \mathcal{N}. \end{equation}
I define an admissible metric $D$ on $\mathcal{M} \sqcup \mathcal{N}$ by declaring that $D_{|\mathcal{M}}=D_{\mathcal{M}}$, $D_{|\mathcal{N}}=D_{\mathcal{N}}$ and
\begin{eqnarray*} D(p,q) & = & \min_{r \in \mathcal{M}, s \in \mathcal{N}} \frac{1}{2} \left( D_{\mathcal{M}} (p,r) + D_{\mathcal{N}} ( \psi (r), q) + D_{\mathcal{N}} (q , s ) + D_{\mathcal{M}} (\zeta (s) ,p) \right) \\ & & + (\epsilon + \delta) \end{eqnarray*}
for all $p \in \mathcal{M}$ and $q \in \mathcal{N}$.  It is necessary to check that $D$ satisfies the triangle inequality.  Let $p_{1}, p_{2} \in \mathcal{M}$ and $q \in \mathcal{N}$, then
\begin{eqnarray*}
D(p_{1} , p_{2}) & \leq & \frac{1}{2} ( D_{\mathcal{M}} ( p_{1} , r_{1}) + D_{\mathcal{M}}(r_{1} , r_{2}) + D_{\mathcal{M}} (r_{2} , p_{2}) + D_{\mathcal{M}} ( p_{1} , \zeta (s_{1}) ) + \\ & & D_{\mathcal{M}} (\zeta (s_{1}) , \zeta (s_{2}) ) + D_{\mathcal{M}} ( \zeta (s_{2}) , p_{2} ) ) \\
& \leq & \frac{1}{2} ( D_{\mathcal{M}} ( p_{1} , r_{1}) + D_{\mathcal{N}} ( \psi (r_{1}) , \psi (r_{2}) ) + D_{\mathcal{M}} (r_{2} , p_{2}) + 2(\epsilon + \delta) + \\ & & D_{\mathcal{N}} (s_{1} , s_{2}) + D_{\mathcal{M}} ( \zeta (s_{2}) , p_{2} ) ) + D_{\mathcal{M}} ( p_{1} , \zeta (s_{1}) ) \\
& \leq & \frac{1}{2} ( D_{\mathcal{M}} ( p_{1} , r_{1}) + D_{\mathcal{N}} ( \psi (r_{1}), q) + D_{\mathcal{N}} (s_{1} , q) + D_{\mathcal{M}}(\zeta (s_{1}) , p_{1} )) + \\ & &  \frac{1}{2} (D_{\mathcal{M}}( p_{2} , r_{2} ) + D_{\mathcal{N}} ( \psi (r_{2}) , q ) + D_{\mathcal{N}} (q , s_{2} ) + D_{\mathcal{M}} ( \zeta (s_{2}) , p_{2}) ) + \\ & & 2(\epsilon + \delta)
\end{eqnarray*}
for all $r_{1},r_{2} \in \mathcal{M}$ and $s_{1} ,s_{2} \in \mathcal{N}$.  Hence
$$D(p_{1} , p_{2} ) \leq D(p_{1} ,q) + D(q , p_{2})$$
The other triangle inequalities are proven similarly.  Obviously,
$$D(p ,\psi(p)), D(q, \zeta (q)) \leq \epsilon + \frac{3
\delta}{2},$$ which proves the claim.   $\square$
\\*
\\*
Theorem \ref{two} reveals that any compact limit space (in the strong topology), $(\mathcal{M}^{\textrm{str}},d)$, of a GGH sequence $(\mathcal{M}_{i} ,g_{i})_{i \in \mathbb{N}}$ must be isometric, w.r.t. $D_{\mathcal{M}^{\textrm{str}}}$, to the limit space of the Gromov-Hausdorff sequence $(\mathcal{M}_{i} , D_{\mathcal{M}_{i}})_{i \in \mathbb{N}}$ due to the well known result of Gromov \cite{Gromov},\cite{Petersen}.  One could now proceed as before and define $\mathcal{M}^{\textrm{str}}$ by a completion procedure from the $T_{0}$ quotient of $\mathcal{S}$.  On the other hand, we can, inspired by the previous theorem, construct in a direct way a compact limit space by using the classical Gromov construction.  The reader will easily see that the $T_{0}$ quotient of $\mathcal{S}$ is dense in $\mathcal{M}^{\textrm{str}}$ in the strong topology defined by $D_{\mathcal{M}^{\textrm{str}}}$.
\begin{theo} \label{three}
The Gromov-Hausdorff limit space of the sequence \\* $(\mathcal{M}_{i},D_{\mathcal{M}_{i}})_{i \in \mathbb{N}}$  with a suitably defined Lorentz distance $d$, is a limit space of the sequence $(\mathcal{M}_{i} ,g_{i})_{i \in \mathbb{N}}$.
\end{theo}
\textsl{Proof}:  \\*
Let $\psi_{i+1}^{i} : \mathcal{M}_{i} \rightarrow \mathcal{M}_{i+1}$, $\zeta^{i+1}_{i} : \mathcal{M}_{i+1} \rightarrow \mathcal{M}_{i}$ be as before and denote by $D_{i,i+1}$ the admissible metric on $\mathcal{M}_{i} \sqcup \mathcal{M}_{i+1}$ constructed from $\psi_{i+1}^{i}, \zeta^{i+1}_{i}$ and $D_{\mathcal{M}_{i}}, D_{\mathcal{M}_{i+1}}$ as in the proof of theorem \ref{two}.  \\* Then, $D_{i,i+1} (p_{i} , \psi^{i}_{i+1} (p_{i})), D_{i,i+1} ( p_{i+1} , \zeta^{i+1}_{i} ( p_{i+1} )) \leq \frac{5}{2^{i+1}}$.  The following inequality is crucial:
\begin{eqnarray*}
\left| d_{g_{i+1}} (p_{i+1} ,q_{i+1}) - d_{g_{i}}(p_{i} ,q_{i}) \right| & \leq & \frac{1}{2^{i}} + \left| d_{g_{i+1}} ( p_{i+1} , q_{i+1}) - d_{g_{i+1}} ( \psi^{i}_{i+1} ( p_{i}) , \psi^{i}_{i+1} ( q_{i})) \right| \\ & \leq &
\frac{1}{2^{i}} + D_{\mathcal{M}_{i+1}} ( p_{i+1} , \psi^{i}_{i+1} (p_{i})) + D_{\mathcal{M}_{i+1}} ( q_{i+1}, \psi^{i}_{i+1} ( q_{i})) \\ & \leq & \frac{3}{2^{i-1}} + D_{i,i+1} ( p_{i} , p_{i+1}) + D_{i ,i+1} (q_{i},q_{i+1}).
\end{eqnarray*}
Let $\bigsqcup_{i \in \mathbb{N}} \mathcal{M}_{i}$ be the disjoint union of the $\mathcal{M}_{i}$ and define a metric $D$ on it by declaring that for all $i,k >0$
$$ D(p_{i} , p_{i+k}) = \min_{\left\{p_{i+j} \in \mathcal{M}_{i+j}, j=1 \ldots k-1 \right\} } \left\{ \sum_{j=0}^{k-1} D_{i+j, i+j+1}(p_{i+j} , p_{i+j+1} ) \right\}. $$
Obviously, $D(p_{i} , \psi^{i}_{i+k} (p_{i})),D(p_{i+k} , \zeta^{i+k}_{i} (p_{i+k})) \leq \frac{5}{2^{i}}$ and
$$ \left| d_{g_{i+k}} (p_{i+k} ,q_{i+k}) - d_{g_{i}}(p_{i} ,q_{i}) \right|  \leq \frac{3}{2^{i-2}} + D(p_{i},p_{i+k}) + D(q_{i},q_{i+k}) $$
I construct now the limit space as the ``boundary'' of the completion of $\left( \bigsqcup_{i \in \mathbb{N}} \mathcal{M}_{i} , D \right)$.
Define
$$\widehat{\mathcal{M}} = \left\{ (p_{i})_{i \in \mathbb{N}} | p_{i} \in \mathcal{M}_{i} \textrm{ and } D(p_{i} ,p_{j}) \rightarrow 0 \textrm{ for } i,j \rightarrow \infty \right\}. $$
$\widehat{\mathcal{M}}$ has a pseudometric defined on it
$$ D((p_{i})_{i \in \mathbb{N}} , (q_{i})_{i \in \mathbb{N}}) = \lim_{i \rightarrow \infty} D(p_{i}, q_{i}) $$
and considering the above estimates, the following Lorentz metric
$$ d((p_{i})_{i \in \mathbb{N}} , (q_{i})_{i \in \mathbb{N}}) = \lim_{i \rightarrow \infty} d_{g_{i}}(p_{i} ,q_{i})$$
is also well defined on $\widehat{\mathcal{M}}$.  I show that $D_{\widehat{\mathcal{M}}}$, defined by this $d$ as $$D_{\widehat{\mathcal{M}}} ( (p_{i})_{i \in \mathbb{N}} , (q_{i})_{i \in \mathbb{N}}) = \sup_{r \in \widehat{\mathcal{M}}} \left| d((p_{i})_{i \in \mathbb{N}} , r) + d(r , (p_{i})_{i \in \mathbb{N}}) - d((q_{i})_{i \in \mathbb{N}} , r) - d(r , (q_{i})_{i \in \mathbb{N}}) \right| $$ equals $D$ on $\widehat{\mathcal{M}}$.  Suppose there exist sequences $(p_{i})_{i \in \mathbb{N}}$, $(q_{i})_{i \in \mathbb{N}}$ and $\delta >0$ such that $$D_{\widehat{\mathcal{M}}} ((p_{i})_{i \in \mathbb{N}}, (q_{i})_{i \in \mathbb{N}}) > D((p_{i})_{i \in \mathbb{N}} ,(q_{i})_{i \in \mathbb{N}}) + \delta ;$$
then there exists a sequence $(r_{i})_{i \in \mathbb{N}}$, such that for $k$ big enough:
$$ \left| d_{g_{k}} (p_{k} ,r_{k} ) + d_{g_{k}} (r_{k} , p_{k}) - d_{g_{k}} (q_{k} ,r_{k}) - d_{g_{k}}(r_{k} ,q_{k}) \right| > D_{\mathcal{M}_{k}} ( p_{k} ,q_{k}) + \frac{\delta}{2},$$ which is impossible by definition of $D_{\mathcal{M}_{k}}$.  Hence, suppose that there exist sequences $(p_{i})_{i \in \mathbb{N}}$, $(q_{i})_{i \in \mathbb{N}}$, $\delta >0$ such that $$D_{\widehat{\mathcal{M}}} ((p_{i})_{i \in \mathbb{N}}, (q_{i})_{i \in \mathbb{N}}) + \delta < D((p_{i})_{i \in \mathbb{N}} ,(q_{i})_{i \in \mathbb{N}}).$$
Choose $k > \frac{ ln(\frac{176}{\delta})}{ln(2)}$ big enough such that
$$ \left| D((p_{i})_{i \in \mathbb{N}} ,(q_{i})_{i \in \mathbb{N}}) - D_{\mathcal{M}_{k}}(p_{k} ,q_{k}) \right| < \frac{\delta}{4}  $$
and $D(p_{k}, (p_{i})_{i \in \mathbb{N}}), D(q_{k} , (q_{i})_{i \in \mathbb{N}}) < \frac{5}{2^{k}}$; then it is not difficult to see that the hypothesis implies that
$$ \left| d_{g_{k}} ( p_{k} , r_{k}) + d_{g_{k}} (r_{k} ,p_{k}) - d_{g_{k}} (q_{k} ,r_{k}) - d_{g_{k}}(r_{k} ,q_{k})\right| + \frac{\delta}{2} < D_{\mathcal{M}_{k}} (p_{k} ,q_{k}) $$ for all $r_{k} \in \mathcal{M}_{k}$, which is impossible and therefore $D_{\widehat{\mathcal{M}}} = D$.    \\*
Hence, $(\widehat{\mathcal{M}},d)$ is a compact limit space in the strong topology since $\widehat{\mathcal{M}}$ is compact with respect to $D$ (it is a good exercise for the reader to check this).  I claim now that the $T_{0}$ quotient of $(\widehat{\mathcal{M}},d)$ is the desired limit space $(\mathcal{M}^{\textrm{str}} ,d)$.  This is an immediate consequence of the fact that the Gromov-Hausdorff distance between $(\mathcal{M}^{\textrm{str}} ,D)$ and $(\mathcal{M}_{k} , D_{\mathcal{M}_{k}})$ is less than $\frac{5}{2^{k+1}}$ and the inequality \label{pag83}
\begin{equation} \label{imp3} \left| d((p_{i})_{i \in \mathbb{N}} ,(q_{i})_{i \in \mathbb{N}}) - d_{g_{k}} (r_{k} ,s_{k}) \right| \leq \frac{3}{2^{k-2}} + D((p_{i})_{i \in \mathbb{N}} , r_{k}) + D((q_{i})_{i \in \mathbb{N}} , s_{k}) \end{equation}   $\square$ \\*
\\*
The uniqueness of the limit space is easily proven from theorem \ref{two}.  In fact, we have the following result.
\begin{theo} \label{four}
Let $(\mathcal{M}_{1},d_{1})$, $(\mathcal{M}_{2},d_{2})$ be two pairs, where $\mathcal{M}_{i}$ is a set with a Lorentz distance $d_{i}$ defined on it, such that $\mathcal{M}_{i}$ is compact in the strong metric topology defined by the metric $D_{\mathcal{M}_{i}}$ induced by $d_{i}$ for $i=1,2$.  Then $(\mathcal{M}_{1},d_{1})$ and  $(\mathcal{M}_{2},d_{2})$ cannot be distinguished by the Gromov-Hausdorff uniformity iff they are isometric w.r.t. the Lorentz distances.
\end{theo}
\textsl{Proof}: \\*
Let $\psi_{n} : \mathcal{M}_{1} \rightarrow \mathcal{M}_{2}$, $\zeta_{n} : \mathcal{M}_{2} \rightarrow \mathcal{M}_{1}$ such that $\psi_{n}$ and $\zeta_{n}$ make $(\mathcal{M}_{1},d_{1})$ and $(\mathcal{M}_{2},d_{2})$, $(\frac{1}{n} , \frac{1}{n})$ close.  Then, the inequalities (\ref{imp1}, \ref{imp2}) reveal that
$$ \left| D_{\mathcal{M}_{2}} ( \psi_{n} (p) , \psi_{n} (q)) - D_{\mathcal{M}_{1}} (p,q ) \right| < \frac{4}{n}, \, \forall p,q \in \mathcal{M}_{1} $$
and
$$ \left| D_{\mathcal{M}_{1}} ( \zeta_{n} (p) , \zeta_{n} (q)) - D_{\mathcal{M}_{2}} (p,q ) \right| < \frac{4}{n}, \, \forall p,q \in \mathcal{M}_{2}. $$
This, combined with,
$$ D_{\mathcal{M}_{1}} ( \zeta_{n} \circ \psi_{n} (p) , p), D_{\mathcal{M}_{2}}( \psi_{n} \circ \zeta_{n} (q) ,q) < \frac{1}{n}, \, \forall p \in \mathcal{M}_{1}, q \in \mathcal{M}_{2} $$
implies that we can find a subsequence $(n_{k})_{k \in \mathbb{N}}$ and an isometry $\psi$ w.r.t the Lorentz distances such that
$$\psi_{n_{k}} \stackrel{k \rightarrow \infty}{\rightarrow} \psi$$ and $$\zeta_{n_{k}} \stackrel{k \rightarrow \infty}{\rightarrow} \psi^{-1}$$ pointwise.  Choose $\mathcal{C}$ to be a countable dense subset of $\mathcal{M}_{1}$,
then by the usual diagonalisation argument we can find a
subsequence $(n_{k})_{k \in \mathbb{N}}$ such that
$\psi_{n_{k}}(p) \stackrel{k \rightarrow \infty}{\rightarrow} \psi
(p)$ for all $p \in \mathcal{C}$.  Clearly, $\psi$ preserves the
strong as well as the Lorentz metric.  Suppose $\psi ( \mathcal{C}
)$ is not dense in $\mathcal{M}_{2}$.  Then, choose a countable,
dense subset $\mathcal{D} \subset \mathcal{M}_{2}$ which contains
$\psi ( \mathcal{C} )$.  Taking a subsequence of $(n_{k})_{k \in
\mathbb{N}}$ (which we denote in the same way) if necessary, we
obtain that $\zeta_{n_{k}}(q) \stackrel{k \rightarrow
\infty}{\rightarrow} \zeta(q)$ for all $q \in \mathcal{D}$.
Obviously, $\zeta \circ \psi$ equals the identity on
$\mathcal{C}$.  Since $\psi ( \mathcal{C} )$ was supposed not to
be dense in $\mathcal{M}_{2}$, we can find an $\epsilon >0$ and a
point $q \in \mathcal{D}$ such that $D_{\mathcal{M}_{2}} (q
,\psi(p)) \geq \epsilon$ for all $p \in \mathcal{C}$, but this is
impossible since that would imply that $D_{\mathcal{M}_{1}}
(\zeta(q),r) \geq \epsilon$ for all $r \in \mathcal{C}$.  Hence,
we choose $\mathcal{D} = \psi( \mathcal{C})$ and $\psi \circ
\zeta$ equals the identity on $\psi ( \mathcal{C} )$.  I show now
that $\psi$ has a unique continuous extension.  Let $r \in
\mathcal{M}_{1} \setminus \mathcal{C}$ and choose a Cauchy
sequence $(r_{n})_{n \in \mathbb{N}}$ converging to $r$, then
$(\psi(r_{n}))_{n \in \mathbb{N}}$ is a Cauchy sequence converging
to, say, $\psi(r)$.  It is left as an easy exercise to the reader
to show that $\psi(r)$ is defined independently of the Cauchy
sequence.  I show now that $\psi_{n_{k}} (r) \stackrel{k
\rightarrow \infty }{\rightarrow} \psi(r)$.  Choose $\epsilon > 0$
and let $p \in \mathcal{C}$ such that $D_{\mathcal{M}_{1}} (p,r) <
\frac{\epsilon}{3}$.  Choose $k$ large enough that $n_{k} >
\frac{12}{\epsilon}$ and $\psi_{n_{k}} (p) \in
B_{D_{\mathcal{M}_{2}}}( \psi (r )) ,\frac{\epsilon}{3})$.  Then,
$D_{\mathcal{M}_{2}}( \psi_{n_{k}} (p) , \psi_{n_{k}} (r )) <
\frac{2 \epsilon}{3}$ and the triangle inequality implies that
$D_{\mathcal{M}_{2}}( \psi(r) , \psi_{n_{k}}(r)) < \epsilon$ which
completes the proof.  Clearly, $\psi$ must preserve the Lorentz distance.  $\square$

\subsection{Some first properties of the limit space} \label{prop}
Now, I shall study some first properties of the chronological relation in the limit space and start working towards a good definition of the causal relation $\prec$.  I will end this section by giving some conditions on the spacetimes $(\mathcal{M}_{i} ,g_{i})$ which imply that the timelike continuum of the corresponding limit space $(\mathcal{M}^{str},d)$ is ``as large as possible''.   To start with, I give an example, as a warm-up, of the phenomena we need to consider.
\begin{exie} \label{ex7}
\end{exie}
Choose $T, \epsilon >0$ and consider the cylinder $S^{1} \times \left[-T,T \right]$ with the usual coordinates $(\theta ,t)$.  Let $f$ be the function constructed in example \ref{ex4} and redefine $\beta_{\epsilon}$ as $\beta_{\epsilon} (t) = \epsilon + ( 1 - \epsilon) \chi( t - \frac{T}{6} )$.  Define $\rho_{\epsilon}$ as
$$ \rho_{\epsilon}(t,\theta) = \frac{\beta_{\epsilon}( \cdot +  \frac{T}{3})*\left( f( \cdot + \frac{T}{6})f(- \cdot + \frac{T}{6})\right)}{\int_{-\infty}^{+\infty}f(x+ \frac{T}{6})f(-x+\frac{T}{6})dx} (t)$$
$\rho_{\epsilon}$ is a function which equals $\epsilon$ for $t \leq - \frac{T}{3}$, $1$ for $t \geq 0$ and undergoes a smooth transition in the interval $\left[ - \frac{T}{3} , 0 \right]$.  Define metric tensors
$$ g_{\epsilon}(t, \theta) = -dt^{2} + \rho_{\epsilon}^{2}(t)d\theta^{2} $$
Denote by $d_{\epsilon}$ the associated Lorentz distances, put $\psi_{\delta}^{\epsilon}, \zeta_{\epsilon}^{\delta}$ equal to the identity on $S^{1} \times \left[ -T, T \right]$, and remember that $d$ is the ``limit distance'', $d = \lim_{\epsilon \rightarrow 0} d_{\epsilon}$.  What does the ($T_{0}$ quotient of the) limit space look like?   Let me give a dynamical picture of what happens: for any $\epsilon > 0$, the spacetime at hand is a tube of radius $\epsilon$ for $t \leq - \frac{T}{3}$ and $1$ for $t \geq 0$.  In the limit $\epsilon \rightarrow 0$, we are left with a one dimensional timelike line $\left[ -T ,-\frac{T}{3} \right]$ and a cylinder $S^{1} \times \left[0 , T \right]$ of radius one which are connected between $-\frac{T}{3}$ and $0$ by a tube. In this example, we close $\ll$ to $\prec$ by defining $J^{\pm}(p)$ as $J^{\pm}(p) = \overline{I^{\pm} (p)}$.  Obviously $J^{+}$ is the dual of $J^{-}$, and $\prec$ is a reflexive, antisymmetric and transitive relation.  Moreover, for $r \in \left(-T, -\frac{T}{3} \right)$, one has that $I^{+}(r) \neq \bigcap_{q \prec r, q \neq r} I^{+} (q)$ since $J^{\pm}(r) = I^{\pm}(r) \cup \left\{ r \right\}$.  $\square$
\\*
\\*
The following elementary properties are of immediate interest and examples are provided in examples \ref{ex8} and \ref{ex9}. \\*
\\*
\textbf{Elementary properties}  \\*
Define the timelike continuum, $\mathcal{TCON}$ of $\mathcal{M}^{\textrm{str}}$, as before.
\begin{itemize}
\item There exist limit spaces with elements $r \in \mathcal{TCON}$ such that $\overline{I^{+} (r)} \neq \bigcap_{s \ll r} I^{+} (s)$, i.e., $(\mathcal{M}^{str},d)$ is not causally continuous\footnote{By this I mean that $p \rightarrow I^{\pm}(p)$ is not outer continuous in the usual $C^{0}$ topology.}.  Hence, define the causal future $J^{+}(r)$ of $r \in \mathcal{TCON}$ as $J^{+}(r) = \bigcap_{s \ll r} I^{+} (s)$; the  causal past is defined dually.
\item One can construct limit spaces with two points $p,q$ such that $I^{-}(p) = I^{-}(q)$, $I^{+}(p) \subsetneq I^{+}(q)$, but $q$ cannot be in the causal past of $p$ without breaking inner continuity of $r \rightarrow J^{-}(r)$ in the usual $C^{0}$ topology.
\item The timelike continuum is, in general, not open.
\item On $\mathcal{TCON}$ the Alexandrov and strong topology coincide\footnote{We show that the strong topology is weaker as the Alexandrov topology on $\mathcal{TCON}$.  Choose $p \in \mathcal{TCON}$ and let $\epsilon > 0$, then we can find points $r,s \in B_{D_{\mathcal{M}}}(p, \epsilon)$ such that $r \ll p \ll s$ by definition of $\mathcal{TCON}$.  But then $I^{+}(r) \cap I^{-}(s) \subset B_{D_{\mathcal{M}}}(p, \epsilon)$ since the open balls in the strong topology are causally convex.}.
\end{itemize}  $\square$
\begin{exie} \label{ex8}
\end{exie}
Here, I construct an example of the first property.  The second one is considerably easier and is not treated\footnote{The reader who wants to know such example can look ahead to the next chapter.}.  Take the ``cylinder universe'' $\mathcal{CYL} = (S^{1} \times \left[0,1 \right], -dt^{2} + d\theta^{2})$ and define $(\mathcal{M}_{\epsilon}, -dt^{2} + d\theta^{2})$ by removing \\* $D^{+}(K^{+}((\pi , \frac{3}{4}))) \setminus K^{+}((\pi , \frac{3}{4}))$ from $\mathcal{CYL}$.  Recall that $K^{+}((\pi , \frac{3}{4}))$ is the future sphere of radius $\epsilon$ around $(\pi , \frac{3}{4})$ and $D^{+}(A)$ is the domain of dependence of a partial Cauchy surface $A$.  It is easy to see that $(\left(S^{1} \times \left[0,1\right]\right) \setminus I^{+}((\pi , \frac{3}{4})) , d)$ is a strong limit space of the sequence $(\mathcal{M}_{\epsilon}, -dt^{2} + d\theta^{2})_{\epsilon}$ where $d$ is the usual Lorentz distance associated to the metric tensor $-dt^{2} + d\theta^{2}$ on $\mathcal{M}^{str}$.  One can see that $(\pi - \frac{1}{4} , 1) \notin \overline{I^{+}((\pi + \frac{1}{4}, \frac{1}{2}))}$ but $(\pi - \frac{1}{4} , 1) \in \bigcap_{q \ll (\pi + \frac{1}{4}, \frac{1}{2})} I^{+}(q)$. $\square$
\begin{exie} \label{ex9}
\end{exie}
I give an example in which the timelike continuum is not open.  As in the previous example, I consider the cylinder universe.  For any $n \geq 6$, denote by $m(n)$ the largest even  integer (just for convenience) such that $\frac{1}{m(n)(m(n) + 1)} \geq \frac{3}{n}$.  Define a conformal factor $\Omega_{n}$ as follows:
\begin{itemize}
\item $\Omega_{n}(t) = 1$ for $t \geq \frac{1}{2} + \frac{1}{n}$
\item for all $0 < k < m(n)$, $k$ even, $\Omega_{n}(t) = \frac{1}{n}$ for $t \in \left[\frac{1}{k+1} + \frac{1}{n} ,  \frac{1}{k} - \frac{1}{n} \right]$ and $\Omega_{n}$ smoothly increases from $\frac{1}{n}$ to $1$ on the interval $\left[ \frac{1}{k} - \frac{1}{n} , \frac{1}{k} + \frac{1}{n} \right]$.
\item for all $1 < k < m(n)$, $k$ odd, $\Omega_{n}(t) = 1$ for $t \in \left[\frac{1}{k+1} + \frac{1}{n} ,  \frac{1}{k} - \frac{1}{n} \right]$ and $\Omega_{n}$ smoothly decreases from $1$ to $\frac{1}{n}$ on the interval $\left[ \frac{1}{k} -\frac{1}{n} , \frac{1}{k} + \frac{1}{n} \right]$.
\item for $t \leq \frac{1}{m(n)} + \frac{1}{n}$, $\Omega_{n} (t) = 1$
\end{itemize}
Consider the sequence of spacetimes $(S^{1} \times
\left[-1,1\right], \Omega_{n}^{2}(t) ( -dt^{2} + d\theta^{2}))_{n
\geq 6 }$, and the associated strong limit space
$(\mathcal{M}^{str} ,d)$ (w.r.t. mappings which equal the
identity) then any point $p$ with time coordinate $0$ belongs to
the timelike continuum.  However any neighbourhood of such $p$
contains points which do not belong to the timelike continuum.
$\square$
\\*
\\*
I show that $J^{+}(r)$ is closed for $r \in \mathcal{TCON}$ (the dual statement follows identically).  Remark first that for $q \ll r$, $\overline{I^{+}(r)} \subset I^{+}(q)$ which follows immediately from the continuity of $d$ and the reversed triangle inequality.  Hence, $\overline{I^{+}(r)} \subset J^{+}(r)$.  Let $(p^{i})_{i \in \mathbb{N}}$ be a sequence in $J^{+}(r)$ converging to an element $p$ and choose $q \ll r$.  Pick $s$ such that $q \ll s \ll r$, then $p \in \overline{I^{+}(s)}$ and hence, $p \in I^{+}(q)$, which finishes the proof. \\*
\\*
One might wonder whether two chronologically related points in a limit space can be joined by a timelike geodesic.  In the next example, I show that one can at most expect such geodesic to be causal.
\begin{exie} \label{ex10}
\end{exie}
I construct a limit space $\mathcal{M}^{\textrm{str}}$ in which the unique geodesic (causal curve with longest length) between two points $p \ll q$ is broken into a timelike and a null part.  The idea is to construct conformal factors $\Omega_{\epsilon}$ which equal $100$ in a specific area $A$ containing $q$, but not $p$, and $1$ outside $A$ apart from a small tube within a distance $\epsilon$ from $A$ (w.r.t. the obvious ``Riemannian'' metric) such that the geodesics from $p$ to $q$ bend towards the lightcone of $p$ as $\epsilon$ goes to zero, since they prefer to travel ``inside'' $A$ as much as possible.  We start from the cylinder universe $\mathcal{CYL}$ and let $p = (\pi ,\frac{1}{4})$, $q = (\pi , \frac{3}{4})$.  The past sphere of radius $\frac{1}{4}$ around $q$ intersects the null geodesic $t + \theta = \frac{1}{4} + \pi$ through $p$ in the point $r = ( \pi - \frac{3}{16} ,  \frac{7}{16})$.  Consider the null ray $t - \theta = \frac{5}{8} - \pi$ through $r$, then let $B \subset \left\{(\theta,t) | t-\theta \geq \frac{5}{8} - \pi \right\}$ be a closed rectangle containing $A(p,q) \cap \left\{(\theta,t) | t-\theta \geq \frac{5}{8} - \pi \right\}$ in its interior such that the distance of $B$ to the boundaries of $\mathcal{CYL}$, with respect to the usual Euclidian metric $\tilde{D}$ defined by $dt^{2} + d\theta^{2}$, is greater than zero.  $A$ is constructed by suitably rounding off the corners of $B$ outside $A(p,q)$.  We define now the outward $\epsilon$-tube $A^{\epsilon}$ of $A$  as the set of all $x \notin A$ which are a distance less or equal than $\epsilon$ apart from $A$, i.e.,
$$ A^{\epsilon} = \left\{ x \notin A | \exists a \in A : \tilde{D}(x,a) \leq \epsilon \right\} .$$
It is well known that for $\epsilon$ sufficiently small, $\partial
A^{\epsilon} \setminus \partial A$ is smooth (w.r.t. the usual differential structure) if $\partial A$ is.
For $\epsilon$ sufficiently small, let $\Omega_{\epsilon}$ be a
function which equals $100$ on $A$, $1$ in the complement of
$A^{\epsilon} \cup A$ and undergoes a transition in $A^{\epsilon}$
which is only dependent on, and decreasing in, the radial outward
coordinate.  As usual $g_{\epsilon} = \Omega^{2}_{\epsilon}(t,x) (
-dt^{2} + d\theta^{2} )$, $\psi^{\epsilon}_{\delta}$ and
$\zeta^{\delta}_{\epsilon}$ equal the identity, and $d =
\lim_{\epsilon \rightarrow 0} d_{\epsilon}$.  Define $\gamma_{1} :
\left[0,1\right] \rightarrow \mathcal{M}^{\textrm{str}}$ as
$\gamma_{1}(t) = (\pi - \frac{3 t}{16} , \frac{4 + 3t}{16})$.
Define $\gamma_{2} : \left[ 0 , 1 \right] \rightarrow
\mathcal{M}^{str}$ as the line running from $(\pi - \frac{3}{16} ,
\frac{7}{16})$ to $(\pi , \frac{3}{4})$ with uniform speed.  I
claim that $\gamma = \gamma_{2} \circ \gamma_{1}$ is (upon
reparametrisation) the unique distance maximizing causal curve
from $p$ to $q$ with length equal to $d(p,q)$.  With $\tilde{t} =
t - \frac{1}{4}$, $\tilde{\theta} = \theta - \pi$ we have that the
square of the length of any continuous causal curve which is
linear from $p$ to a point $s \in \left\{(\theta,t) | t-\theta =
\frac{5}{8} - \pi \right\}$ and from $s$ to $q$, equals
$$ 100^{2}\left[ \left(\tilde{t} - \frac{1}{2}\right)^{2} - \left(\tilde{t} - \frac{3}{8}\right)^{2}\right] + \left[ \tilde{t}^{2} - (\tilde{t} - \frac{3}{8})^{2} \right] = \frac{3 - 100^{2}}{4}\tilde{t} + \frac{7 \cdot 100^{2} - 9}{64}$$
for $\frac{3}{16} \leq \tilde{t} \leq \frac{7}{16}$, which proves the first part of the claim.  It follows from the above expression that the maximal length, $d(p,q)$, equals $25$.  $\square$                \\*
\\*
Intuitively, one would say in the last example that $ \mathcal{M}^{\textrm{str}} \setminus \mathcal{TCON} = \partial \mathcal{M}^{\textrm{str}}$.  Hence, it is time to give an intrinsic definition of the past and future boundary of a limit space of a sequence $(\mathcal{M}_{i},g_{i})_{i \in \mathbb{N}}$.
\begin{deffie}
A sequence $(p_{i})_{i \in \mathbb{N}} \in \mathcal{M}^{\textrm{str}}$ represents a point of the \emph{past} boundary, $\partial_{\textrm{P}} \mathcal{M}^{\textrm{str}}$, iff $\forall \epsilon > 0$, $\exists i_{0} \in \mathbb{N}:$  $\forall i \geq i_{0}$
$$ p_{i} \in \left( \partial_{\textrm{P}} \mathcal{M}_{i} \right)^{\epsilon} $$
where, now for all $A_{i} \subset \mathcal{M}_{i}$ $$A_{i}^{\epsilon} = \left\{ q_{i} \in \mathcal{M}_{i} | \exists a_{i} \in A_{i} : D_{\mathcal{M}_{i}} ( a_{i} , q_{i}) < \epsilon\right\}. $$
  The future boundary $\partial_{\textrm{F}} \mathcal{M}^{\textrm{str}}$ is defined dually.  $\square$
\end{deffie}
We have the following theorem.
\begin{theo}
The past and future boundaries of $\mathcal{M}^{\textrm{str}}$ are achronal sets.
\end{theo}
\textsl{Proof}: \\*
Suppose we can find $p,q \in \partial_{\textrm{P}} \mathcal{M}^{\textrm{str}}$ such that $d(p,q) > 0$.  Let $\epsilon = \frac{d(p,q)}{2}$ and choose $i$ sufficiently large such that
\begin{itemize}
\item $p_{i},q_{i} \in \left( \partial_{\textrm{P}} \mathcal{M}_{i} \right)^{\epsilon}$
\item $\left| d_{g_{i}} (p_{i} , q_{i}) - d(p,q) \right| < \frac{\epsilon}{2}$.
\end{itemize}
Then, by definition, there exist points $r_{i},s_{i} \in \partial_{\textrm{P}} \mathcal{M}_{i}$ such that $$D_{\mathcal{M}_{i}}(r_{i},p_{i}), D_{\mathcal{M}_{i}}(s_{i} ,q_{i}) < \epsilon$$ and, moreover, $d_{g_{i}}(p_{i},q_{i}) > \frac{3 \epsilon}{2}$.  Hence, since $\partial_{\textrm{P}} \mathcal{M}_{i}$ is spacelike, we have that
$$ \max_{t_{i} \in \partial_{\textrm{P}} \mathcal{M}_{i}} d_{g_{i}}(t_{i},p_{i}), \max_{t_{i} \in \partial_{\textrm{P}} \mathcal{M}_{i}} d_{g_{i}}(t_{i},q_{i}) < \epsilon $$
But, on the other hand, for $t_{i} \in I^{-}(p_{i}) \cap \partial_{\textrm{P}} \mathcal{M}_{i}$, we have that $d_{g_{i}} (t_{i}, q_{i}) > \frac{3 \epsilon}{2}$ which leads to a contradiction.  The proof for the future boundary is similar. $\square$
\\*
\\*
Using similar arguments, it is easy to see that there exists no point to the past of $\partial_{\textrm{P}} \mathcal{M}^{\textrm{str}}$, nor does there exist a point to the future of $\partial_{\textrm{F}} \mathcal{M}^{\textrm{str}}$.   \\*
\\*
As theorems \ref{two}, \ref{three} and \ref{four} show, the strong metric is an important object.  One might wonder whether for some spacetime $(\mathcal{M},g)$, the strong metric could be a path metric, since such property is stable in the Gromov-Hausdorff limit.  Moreover, a path metric reveals some interesting topological properties of the underlying space.  Unfortunately, we have the following result.
\begin{theo}
The strong metric $D_{\mathcal{M}}$ on a spacetime $(\mathcal{M},g)$ is \emph{never} a path metric.
\end{theo}
\textsl{Proof}: \\*
Suppose such a spacetime $(\mathcal{M},g)$ exists for which $D_{\mathcal{M}}$ is a path metric, and choose $p \in \partial_{\textrm{P}} \mathcal{M}$, $q \in \partial_{\textrm{F}} \mathcal{M}$ such that $d_{g}(p,q) = tdiam (\mathcal{M}) = D_{\mathcal{M}}(p,q) $.  I show that there exists no point $x$ such that $D_{\mathcal{M}} (p,x) = D_{\mathcal{M}}(x,q) = \frac{D_{\mathcal{M}}(p,q)}{2}$, which is a contradiction.  Let $r$ be a point such that $d_{g}(p,r) = d_{g}(r ,q) = \frac{d_{g}(p,q)}{2}$, then $$D_{\mathcal{M}} ( p,r ), D_{\mathcal{M}}(r,q) > \frac{d_{g}(p,q)}{2}$$  The first part is easily seen by noticing that for $s \in E^{+}(r) \setminus \left\{r \right\}$ $$\left| d_{g}(p,s) - d_{g}(r,s) \right| = d_{g}(p,s) > d_{g} (p,r).$$ The second part is proven similary.  Let $x \neq r$; then we have to distinguish two cases:
\begin{itemize}
\item $\max_{t \in \partial_{P} \mathcal{M}} d_{g}(t,x) \leq \frac{d_{g}(p,q)}{2}$
\item $\max_{t \in \partial_{P} \mathcal{M}} d_{g}(t,x) > \frac{d_{g}(p,q)}{2}$
\end{itemize}
Suppose the former is true, then there exists a point $s \in I^{-}(r)$ which is not in the causal past of $x$.\footnote{If this were not true then $r \prec x$ which would imply that $d_{g}(p,x) > \frac{d_{g}(p,q)}{2}$ which is a contradiction.}  Hence $\left| d_{g} (s,q) - d_{g}(s,x) \right| > \frac{d_{g}(p,q)}{2}$, which implies that $D_{\mathcal{M}}(x,q) > \frac{d_{g}(p,q)}{2}$.  \\*
Suppose the latter is satisfied; then $\max_{t \in \partial_{\textrm{F}} \mathcal{M}} d_{g}(x,t) < \frac{d_{g}(p,q)}{2}$ and this case is similar to the previous one. $\square$
\\*
\\*
It would be interesting to give a criterion which guarantees that $\mathcal{M}^{\textrm{str}} \setminus \partial \mathcal{M}^{\textrm{str}} = \mathcal{TCON}$ and every point of the boundary, which does not equal $\partial_{F} \mathcal{M} \cap \partial_{P} \mathcal{M}$, is the limit point of a timelike Cauchy sequence.  Particularly from the physical point of view, it is not entirely clear what degenerate regions would mean.  Moreover, it is far from easy to define a suitable causal relation between two points belonging to such a region as we shall see in section \ref{cau}.
\\*
\\*
\textbf{Intermezzo}: \\*
I suggest three, at first sight different, control mechanisms which prohibit the limit space from containing ``degenerate regions''.  Let $\alpha : \mathbb{R}^{+} \rightarrow  \mathbb{R}^{+}$ be a strictly increasing, continuous function such that $\alpha (x) \leq x$ for all $x \in \mathbb{R}^{+}$.  We say that $(\mathcal{M},g)$ has the $\mathcal{C}^{+}_{\alpha}, \mathcal{C}^{-}_{\alpha}$ or $\mathcal{C}_{\alpha}$ property iff for any $\epsilon$ with $tdiam(\mathcal{M}) \geq \epsilon > 0$ we have that, respectively:
\begin{itemize}
\item  $\alpha (\epsilon) \leq \min_{p \in \mathcal{M}^{\downarrow \epsilon}} \left[ \max_{r \in \overline{B_{D_{\mathcal{M}}} (p , \epsilon )}}  d_{g}(p,r)  \right] \leq \epsilon$
\item $\alpha (\epsilon) \leq \min_{p \in \mathcal{M}^{\uparrow \epsilon}} \left[ \max_{r \in \overline{B_{D_{\mathcal{M}}} (p , \epsilon )}} d_{g}(r,p) \right] \leq \epsilon$
\item $\alpha (\epsilon) \leq \min_{p \in \mathcal{M}} \left[ \max_{r \in \overline{B_{D_{\mathcal{M}}} (p , \epsilon )}} \left( d_{g}(r,p) + d_{g}(p,r) \right) \right] \leq \epsilon $
\end{itemize}
where $\mathcal{M}^{\downarrow \epsilon} = \left\{ p \in \mathcal{M} | p \notin \left(\partial_{\textrm{F}} \mathcal{M} \right)^{\epsilon} \right\}$ and $\mathcal{M}^{\uparrow \epsilon} = \left\{ p \in \mathcal{M} | p \notin \left(\partial_{\textrm{P}} \mathcal{M} \right)^{\epsilon} \right\}$. \\*
Clearly, not all functions $\alpha$ are meaningful.  In particular, they should satisfy $\alpha(x) + \alpha(y) \leq \alpha(x+y)$ for all $0 \leq x,y \leq x + y \leq tdiam (\mathcal{M})$.  This condition follows easily from the reverse triangle inequality satisfied by $d_{g}$ and the triangle inequality satisfied by $D_{\mathcal{M}}$.  Basic functions $\alpha_{n}^{K}$, $n >1$ and $K >0$, could be constructed by declaring that
$$ \alpha_{n}^{K} (x) = K \left(\frac{x}{K}\right)^{n} \textrm{ for } x \leq K, \textrm{ and } \alpha_{n}^{K}(x) = x \textrm{ otherwise.} $$
Obviously, $\alpha_{1} \leq \alpha_{2}$, implies that $\mathcal{C}^{\pm}_{\alpha_{2}} \subseteq \mathcal{C}^{\pm}_{\alpha_{1}}$.   Clearly, if both $\mathcal{C}^{+}_{\alpha}$ and $\mathcal{C}^{-}_{\alpha}$ hold, then $\mathcal{C}_{\alpha}$ is true, but $\mathcal{C}_{\alpha}$ implies neither of them.  The above expressions tell us there is a balance between local and global causal relations, in the sense that the local relations cannot become ``arbitrarily small'' while the global relations remain almost unaltered.  This perhaps needs a bit of explanation.  As an example, consider again the cylinder universe $\mathcal{CYL}$ and let $p = (\pi, \frac{2}{3})$.  Consider the ball $B(p , \frac{1}{100})$ of radius $\frac{1}{100}$ determined by the usual (observer dependent) Riemannian metric tensor $dt^{2} + d\theta^{2}$.  Construct a conformal factor $\Omega$ such that $\Omega(\theta ,t)$ equals $10^{-2003}$ for $(\theta ,t) \in B(p, \frac{1}{101})$, undergoes a smooth transition on the shell $\overline{B(p ,\frac{1}{100})} \setminus B(p, \frac{1}{101})$, and equals $1$ everywhere else.  It is easy to see that the strong metric determined by $(S^{1} \times \left[ 0 ,1 \right] , \Omega(\theta ,t)^{2} ( -dt^{2} + d\theta^{2} ))$ remains almost unchanged while the local Lorentz distance changes drastically!  So the idea behind the concept is, losely speaking,  to control the conformal factor with respect to some ``reference metric''.  The following result is of main interest.
\begin{theo}
For any $\alpha$ satisfying the above mentioned conditions, the $\mathcal{C}^{+}_{\alpha}, \mathcal{C}^{-}_{\alpha}$ and $\mathcal{C}_{\alpha}$ properties are stable under generalised, Gromov-Hausdorff convergence.  This means that if, say, $((\mathcal{M}_{i},g_{i}))_{i\in \mathbb{N}}$ is a generalised, Gromov-Hausdorff Cauchy sequence such that $(\mathcal{M}_{i},g_{i}) \in \mathcal{C}^{+}_{\alpha}$ for all $i \in \mathbb{N}$, then $(\mathcal{M}^{\textrm{str}}, d) \in \mathcal{C}^{+}_{\alpha}$ (and likewise for $\mathcal{C}^{-}_{\alpha}$ and $\mathcal{C}_{\alpha}$).
\end{theo}
\textsl{Proof}: \\*
Let $\epsilon > 0$, I show that $$\alpha (\epsilon) \leq \min_{p \in \mathcal{M}^{\textrm{str} \downarrow \epsilon}} \max_{r \in \overline{B_{D}(p , \epsilon )}} d(p,r).$$
Choose $(p_{i})_{i \in \mathbb{N}} \in \mathcal{M}^{\textrm{str} \downarrow \epsilon}$, $\alpha ( \epsilon ) > \delta  >0$ and $\delta > 4 \gamma > 0$ such that $\left| x - \epsilon \right| < \gamma$ implies that $\left| \alpha (x) - \alpha(\epsilon) \right| < \frac{\delta}{2}$.  Let $i$ be sufficiently large such that $\frac{1}{2^{i-3}} < \gamma$, $\left| D_{\mathcal{M}_{i}} (p_{i} , \partial_{\textrm{F}} \mathcal{M}_{i}) - \epsilon \right| < \frac{\gamma}{2}$ and $D((p_{j})_{j \in \mathbb{N}} , p_{i}) < \frac{3}{2^{i+1}}$, where $D$ denotes also the metric on the disjoint union $\bigsqcup_{i \in \mathbb{N}} \mathcal{M}_{i}$, constructed in the proof of theorem \ref{two}.  Then, there exists an $r_{i} \in B_{D_{\mathcal{M}_{i}}} (p_{i} , \epsilon - \frac{\gamma}{2})$ such that $d_{g_{i}} ( p_{i} ,r_{i}) > \alpha ( \epsilon ) - \frac{\delta}{2}$.   The final remarks of the same proof show that there exists a point $r \in \mathcal{M}^{\textrm{str}}$ such that $D(r,r_{i}) < \frac{5}{2^{i+1}}$.  Hence, $$D((p_{i})_{i \in \mathbb{N}} ,r ) < \epsilon - \frac{\gamma}{2} + \frac{3}{2^{i+1}} + \frac{5}{2^{i+1}} < \epsilon. $$   
Moreover, (\ref{imp3}) on page \pageref{pag83} implies that
$$ \left| d((p_{j})_{j \in \mathbb{N}} ,r ) - d_{g_{i}}(p_{i} , r_{i}) \right| < \frac{3}{2^{i-2}} + \frac{3}{2^{i+1}} + \frac{5}{2^{i+1}} = \frac{1}{2^{i-4}} < \frac{\delta}{2}. $$
Hence, $d((p_{j})_{j \in \mathbb{N}} ,r ) > \alpha ( \epsilon ) - \delta$.  This shows that for any such $\delta > 0$, we can find an $r(\delta) \in B_{D}((p_{i})_{i \in \mathbb{N}} , \epsilon)$ such that $d((p_{j})_{j \in \mathbb{N}} ,r ) > \alpha ( \epsilon ) - \delta$.  The compactness of the closed $\epsilon$-balls and the continuity of $d$ in the strong topology finish the proof. $\square$ \\* \\*
It is not difficult to see that $\mathcal{M} \setminus \partial \mathcal{M}^{\textrm{str}} = \mathcal{TCON}$ for a (limit) spacetime satisfying the $\mathcal{C}^{+}_{\alpha}$ and $\mathcal{C}^{-}_{\alpha}$ properties.  Moreover, every point of the boundary of such space, which does not equal $\partial_{F} \mathcal{M} \cap \partial_{P} \mathcal{M}$, is equivalent to a timelike Cauchy sequence.  There are a few serious questions which can be posed with respect to the above categories of objects. \\* \\*
\textbf{Questions}
\begin{itemize}
\item Can one find spacetimes such that only one of the properties $\mathcal{C}^{+}_{\alpha}, \mathcal{C}^{-}_{\alpha}$ or $\mathcal{C}_{\alpha}$ is satisfied?  If not, are some of them equivalent depending on $\alpha$?
\item Does the limit space of a $\mathcal{C}^{+}_{\alpha}$ sequence satisfy $\mathcal{M}^{\textrm{str}} \setminus \partial \mathcal{M}^{\textrm{str}} = \mathcal{TCON}$?
\item Same question for the limit space of a $\mathcal{C}^{-}_{\alpha}$ or $\mathcal{C}_{\alpha}$ sequence.
\end{itemize} $\square$
\\*
\\*
I shall give partial answers to these questions in the following chapter. $\square$

%% file: ls.tex
\chapter{Lorentz spaces} \label{lsp}
\section{Introduction}
In section \ref{construction}, we constructed the unique, limit space up to isometry (cfr. theorem \ref{four}), which is compact in the strong topology, of a sequence of compact, interpolating cobordisms.  The natural question which arises is whether one can find a maximal class of spaces, which is complete and Hausdorff separated in the natural extension of the GGH-uniformity.  The answer to this question will be affirmative, and such space will be called a Lorentz space.  \\* \\*
The main goal of chapter \ref{lsp} is to further study the properties of the moduli
space of isometry classes of Lorentz spaces.  In particular, in section \ref{GHvGGH}, we try to find out if the GH-metric and the GGH-uniformity are equivalent in the sense that
they have the same Cauchy sequences.  In section \ref{strD}, we touch upon the question whether the strong
metric determines the Lorentz distance uniquely up to time reversal.  This
would be particulary interesting since, if it were true, then Lorentzian interpolating spacetimes would be a subclass of ``Riemannian'', non-path metric, compact spaces modulo $\mathbb{Z}_2$.  Furthermore, in section \ref{cau}, we study the definition of a suitable causal relation and causal curves on limit spaces of compact, globally
hyperbolic interpolating spacetimes.  For example, if we knew how to define a causal relation
between two points in the ``degenerate area'' of a limit space, then one could
raise the question about the physical meaning of such ``causal relationships''.  Finally, we deal with the moduli space and some matters of precompactness in section \ref{comp}. \\*
\\*
For notational convenience, in the sequel, the subscript $\mathcal{M}$ will be dropped in the notation of the strong metric $D_{\mathcal{M}}$.

\section{Definition of Lorentz spaces}

The note following definition \ref{Str} and theorem \ref{four} strongly suggest the following definition of a Lorentz space.

\begin{deffie}

A Lorentz space is a pair $(\mathcal{M},d)$, where $\mathcal{M}$ is a set and $d$ is a Lorentz distance
on $\mathcal{M}$, such that $(\mathcal{M}, D)$ is a compact metric space (with $D$ the strong metric induced by $d$).

\end{deffie}
On the space of all Lorentz spaces $\aleph_{c}$, we can
introduce an equivalence relation $\sim$ by defining $(\mathcal{M}_{1},d_{1})
\sim (\mathcal{M}_{2},d_{2})$ iff there exists a bijection $\psi$ such that
$d_{2}(\psi(x), \psi(y)) = d_{1}(x,y)$ for all $x,y \in \mathcal{M}_{1}$.
Such a bijection is automatically a homeomorphism.
\begin{deffie}

The moduli space of all isometry classes of Lorentz spaces is the space
$\mathcal{LS} = \aleph _{c}/\sim$, equipped with the Hausdorff, quantitative,
GGH-uniformity.

\end{deffie}

\noindent Studying the proofs of theorems \ref{three} and \ref{four} (Ref. \cite{Noldus2}), the reader can see that $\mathcal{LS}$ is a complete, contractible space in which the finite spaces form a dense subset. It is also
easily seen that it is not a locally compact space.\\*
\\*
\textbf{Note}: The results in section \ref{GHvGGH}, in particular
theorem \ref{isom}, imply that the obvious extension of $d_{GH}$
to the moduli space of isometry classes is also a metric.  In the
above definition, we prefer to equip this space with the GGH
uniformity, since herewith $\mathcal{LS}$ is complete\footnote{The
author is unaware of any proof or counterexample of the fact that the moduli
space of isometry classes equipped with $d_{GH}$ would be complete.}.
$\square$  \\* \\* These results are surprisingly easy and
analogous to the ones obtained by Gromov for the metric case.  Let
me elaborate now a bit why this is so.  The main reason is that we
made a complete switch from the ``local'' (Alexandrov) viewpoint
to the more global perspective provided by the strong metric.  The
reader shall become even more aware of this after having read
section \ref{GHvGGH} in which, amongst other results, all results
proven in sections \ref{ld} and \ref{GHD} are generalised.

\section{$d_{\rm GH}$ versus the GGH-uniformity} \label{GHvGGH}

In this section, we examine the relationship between the GH-distance and the GGH-uniformity for Lorentz spaces $(\mathcal{M},d)$.  Along this study, some questions raised in section \ref{prop} will be solved.  Since the difference
between GH-closeness and GGH-closeness lies in the condition that the mappings
used in the definition be approximate inverses of each other, we find it
useful to introduce the concepts of $\epsilon$-isometry and $\epsilon$-surjection.

\begin{deffie}

Let $\epsilon >0$ and $(\mathcal{M},d)$ be a Lorentz space.  A mapping $f:\mathcal{M} \rightarrow \mathcal{M}$ is

\begin{itemize}

\item an $\epsilon$-isometry iff for all $x,y \in\mathcal{M}$
$$
       \left| d(f(x),f(y)) - d(x,y)  \right|
       < \epsilon \;.
$$
\item an $\epsilon$-surjection iff for all $p \in \mathcal{M}$ there
exists a $q \in \mathcal{M}$ such that
$$
       D(p, f(q)) < \epsilon \;.
$$

\end{itemize} \hfill$\square$

\end{deffie}
We start with the following theorem.
\begin{theo}

Let $(\mathcal{M},g)$ be a compact, globally hyperbolic cobordism.  Then for any $\eta > 0$, there exists an $\epsilon > 0$ such that for each
$\epsilon$-isometry $f$, there
exists an isometry $h$ of $\mathcal{M}$ for which the following holds:
$$
       D(f(x), h(x)) < \eta \quad \forall x \in \mathcal{M}\;.
$$
\end{theo}

\noindent\textsl{Proof:}\\*

\noindent Suppose that the statement is false.  Then there exists an $\eta > 0$ such that
for each $n \in \mathbb{N}_{0}$ there exists a $\frac{1}{n}$-isometry $f_{n}$, such
that for any isometry $h$ we can find a point $x(n,h)$ in $\mathcal{M}$ such that
$$
       D(f_{n}(x(n,h)), h(x(n,h))) \geq \eta\;.
$$
The proof of theorem \ref{GHD1} in section \ref{GHD} reveals that we can then find a subsequence $(f_{n_{k}})_{k \in \mathbb{N}}$ and an isometry $f$, such that $f_{n_{k}}
\stackrel{k \rightarrow \infty}{\rightarrow} f$ pointwise.  We now show that
this convergence is uniform in the strong metric, which provides the necessary
contradiction.  We restrict ourselves to proving that for any interior point
$p$ and $\epsilon >0$, there exists a $\delta >0$ such that $q \in B_{D}(p,
\delta)$ implies that $D(f(q),f_{n}(q)) < \epsilon$ for $n$ big enough.  The
rest of the statement is easy (but tedious) and is left as an exercise to the
courageous reader.  Choose $s,r \in B_{D}(p, \frac{\epsilon}{2})$ such that $s
\ll p \ll r$ with, say, $d(s,p)=d(p,r)$ as large as possible.  Let $\delta =
\frac{1}{3}d(p,r)$; then for $n > \frac{1}{\delta}$ such that $f_{n}(r) \in
B_{D}(f(r), \delta)$ and $f_{n}(s) \in B_{D}(f(s), \delta)$ we have that
$$
      f_{n}(B_{D}(p, \delta)) \subset A(f(r),f(s)) \subset B_{D}(f(p),
      {\textstyle\frac{\epsilon}{2}})\;.
$$
Hence,
$$D(f_{n}(q), f(q)) \leq \frac{\epsilon}{2} + \delta < \epsilon $$
for all $q \in B_{D}(p, \delta)$. \hfill$\square$
\\*
\\*
This result reveals that for any $\eta > 0$, there exists an $\epsilon >0$
such that any $\epsilon$-isometry is an $\eta$-surjection.  This, however, is
a fairly weak result and we would like to know if $\eta$ could be bounded by some universal
function of $\epsilon$, the timelike diameter and dimension of $\mathcal{M}$,
which goes to zero when either $\epsilon$ or the timelike diameter goes to
zero.  However, the following example shows that such a function cannot exist.
\begin{exie} \label{ex11}
\end{exie}
Consider the 2-dimensional flat cylinder $\mathcal{CYL} = ({\rm S}^1
\times  [0,1], -d t^{2} + d\theta^2)$ with the region $\mathcal{R}_\delta
= \{(\theta,t) \mid \theta \in [0,\pi],\ t > T(\theta)\}$ removed, where $T(\theta) \geq
1 - \delta$ for all $\theta \in [0,\pi]$, $T(0)=T(\pi)=1$ and
$\left|T'(\theta)\right| < 1$.
\begin{figure}[h]
    \begin{center}
        \scalebox{0.53}{\includegraphics{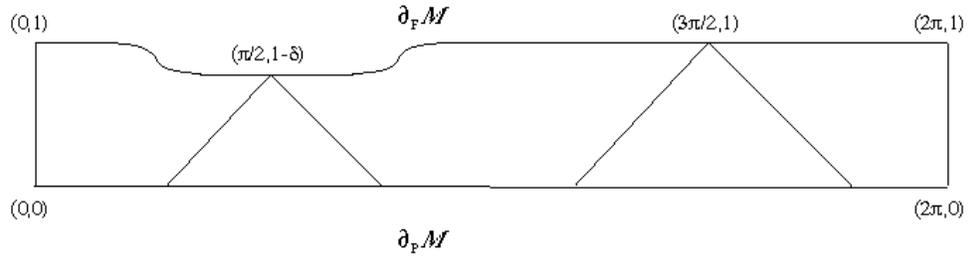}}
    \end{center}

    \caption{Illustration of example \ref{ex11}}
    \label{fig:grliph5-1}
\end{figure}
We will construct an approximate isometry $\psi$ which is far from
any isometry.  Since the only isometry is the identity, this
simplifies our analysis.  $\psi$ is constructed as the composition
of a rotation by $\pi$ times a retraction
$R_{\mathcal{R}_{\delta}}$ which maps a point $(\theta,t)$ to the
unique, closest point $(\theta,\tilde{t}) \in  \mathcal{R}_{\delta}^{c}$.
It is not difficult to check that $\psi$ is a $\sqrt{2
\delta}$-isometry.  However, the point $p$ defined by
$p=(\frac{3\pi}{2},1)$ gets mapped to a point which is strong
distance $1=tdiam(R_{\delta})$ apart.  This shows that for
$\delta$ arbitrarily small, one can construct universes which
allow $\delta$-isometries to be a distance $1$ apart from any
isometry.  $\square$
\begin{exie} \label{teg}
\end{exie}
In this example, we show that a near isometry can be arbitrarily far from
being a surjection.  The following picture shows a sequence of $N$ ``bumps'' with
a fixed width $L>1$.  Let $0 < \epsilon < \frac{1}{2}$ and consider the
function $g_{\epsilon} : \left[ -\epsilon , \epsilon \right] \rightarrow
\mathbb{R}^{+} : x \rightarrow \epsilon + x^{2}$.  Define a sequence of functions
$\Omega^{i}_{\epsilon} : \left[ (i-1)L , iL \right] \rightarrow \mathbb{R}^{+}$, $i=1
\ldots N$, which satisfy the following properties:
\begin{itemize}
\item $ 0 \leq \Omega^{i+1}_{\epsilon}(x+L) -
\Omega^{i}_{\epsilon}(x) \leq \frac{L}{\sqrt{2}N}$ for all $x \in
\left[ (i-1)L , iL \right]$ and $i:1 \ldots N-1$.
\item $\Omega^{i}_{\epsilon}$ is symmetric around $x = (i-\frac{1}{2})L$.
\item $\max_{x \in \left[ (i-1)L , iL \right] }
\Omega^{i}_{\epsilon}(x) = \frac{iL}{\sqrt{2}N}$
\item $\Omega^{i}_{\epsilon} (x) = g_{\epsilon}(x - (i-1)L)$ for $x
\in \left[ (i-1)L, (i-1)L + \epsilon \right]$
\item $ \left| \frac{d\Omega^{i}_{\epsilon} (x)}{dx} \right| < 1$ for
all $x \in \left[ (i-1)L , iL \right]$
\end{itemize}
Let $\Omega_{\epsilon}$ be the concatenation of all $\Omega^{i}_{\epsilon}$.
By identifying $0$ and $NL$, we obtain that $\Omega_{\epsilon}$ is a smooth function on
the circle of radius $\frac{NL}{2 \pi}$.  Define $\mathcal{A}$ as
$$
    \mathcal{A} = \left\{ (x,t) | x \in \left[0 , NL \right] \textrm{ and }
    t \in \left[ 0 , \Omega_{\epsilon} (x)\right] \right\}.
$$
Then, $(\mathcal{A}, -d t^{2} + d x^{2})$ is a globally hyperbolic
cobordism cut out of the cylinder universe with radius
$\frac{NL}{2 \pi}$. Define $\psi : \mathcal{A} \rightarrow
\mathcal{A}$ as the composition of a rotation to the left over an
angle of $\frac{2 \pi}{N}$ with a retraction $R_{\mathcal{A}}:
S^{1}_{NL} \times \mathbb{R} \rightarrow S^{1}_{NL} \times
\mathbb{R}$ which maps every point $(x,t)$ to the closest point
$(x,\tilde{t}) \in \mathcal{A}$.  Clearly, $\psi$ is a
$\frac{L}{\sqrt{N}}$-isometry which is not a
$\frac{(N-1)L}{\sqrt{2} N}$-surjection.   The figure is called the
carousel for obvious reasons. \hfill$\square$
\begin{figure}[]
    \scalebox{0.45}{\includegraphics{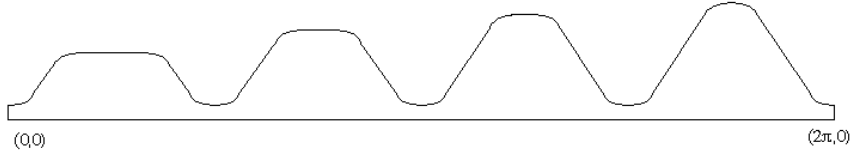}}

    \caption{The carousel}
    \label{fig:grliph5-2}
\end{figure}
\\*
\\*
Let $f,g : \mathcal{LS} \times \mathcal{LS} \times \mathbb{R}^{+}
\rightarrow \mathbb{R}^{+}$ be functions depending only upon the
timelike diameters of the respective Lorentz spaces, $f(x,y, 0) = g(x,y,0) = 0$ and $f,g$ are
continuous in the third element in $(x,y,0)$ for all $x,y \in
\mathcal{LS}$.  Do functions satisfying the above conditions exist
such that if $x$ and $y$ are $\epsilon$-close, then they are
$(f(x,y, \epsilon) , g(x,y, \epsilon))$-close?  It is not
difficult to prove that if there exist two mappings $\psi, \zeta$
which make $x$ and $y$ $\epsilon$-close, such that either $\psi$
or $\zeta$ is \emph{surjective}, then $x$ and $y$ are $(\epsilon,
2\epsilon)$ close.  Hence, if we want to find a counterexample
then we have to look for mappings which are ``far off'' being a
surjection.  The carousel hints the following counterexample.
\begin{exie} \label{teggie}
\end{exie}
Suppose $L=4m$, $m \in \mathbb{N} \setminus \left\{ 0,1 \right\}$ and let $\mathcal{P}^{L}_{1}, \mathcal{P}^{L}_{2}$ be causal sets given by the Hasse diagrams below.  In the pictures, it is understood that the fatter dots are identified.  On a locally finite causal set $\mathcal{P}$, the maximum number of links between two timelike related points $p \ll q$ determines a Lorentz distance.  Obviously, if $\mathcal{P}$ is finite, then it has a natural interpretation as a Lorentz space.  In Appendix E, the following theorem is proven.
\begin{theo} \label{conGHvGGH}
$\mathcal{P}^{L}_{1}, \mathcal{P}^{L}_{2}$ are $1$-close and for every pair of mappings $\psi : \mathcal{P}^{L}_{1} \rightarrow \mathcal{P}^{L}_{2}$, $\zeta :\mathcal{P}^{L}_{2} \rightarrow \mathcal{P}^{L}_{1}$ which make $\mathcal{P}^{L}_{1}, \mathcal{P}^{L}_{2}$ $k$-close, with $k < L/4$, there exists a $p \in \mathcal{P}^{L}_{2}$ such that $$D(p , \psi \circ \zeta (p) ) = L.$$ $\square$
\end{theo}
\begin{figure}[h]
\begin{center}
  \setlength{\unitlength}{0.7cm}
\begin{picture}(10,10)
\thicklines
\put(0,0){\line(0,1){1}}
\put(0,1){\line(1,-1){1}}
\put(1,0){\line(0,1){1}}
\put(1,1){\line(-1,-1){1}}
\put(1,1){\line(1,-1){1}}
\put(2,0){\line(0,1){}}
\put(2,1){\line(-1,-1){1}}
\put(2,1){\line(1,-1){1}}
\put(3,0){\line(0,1){1}}
\put(3,1){\line(-1,-1){1}}
\put(3,1){\line(1,-1){1}}
\put(3,0){\line(1,1){1}}
\put(5,0){\ldots}
\put(5,1){\ldots}
\put(7,0){\line(-1,1){1}}
\put(7,1){\line(-1,-1){1}}
\put(7,0){\line(1,1){1}}
\put(7,1){\line(1,-1){1}}
\put(8,0){\line(0,1){1}}
\put(8,0){\line(1,1){1}}
\put(8,1){\line(1,-1){1}}
\put(7,0){\line(0,1){1}}
\put(8,0){\line(0,1){1}}
\put(9,0){\line(0,1){1}}
\put(0,1){\line(0,1){1}}
\put(0,2){\line(0,1){1}}
\put(0,4){\vdots}
\put(0,5){\line(0,1){1}}
\put(0,6){\line(0,1){1}}
\put(0,7){\line(0,1){1}}
\put(0,8){\line(0,1){1}}
\put(1,1){\line(0,1){1}}
\put(1,2){\line(0,1){1}}
\put(1,4){\vdots}
\put(1,5){\line(0,1){1}}
\put(1,6){\line(0,1){1}}
\put(1,7){\line(0,1){1}}
\put(2,1){\line(0,1){1}}
\put(2,2){\line(0,1){1}}
\put(2,4){\vdots}
\put(2,5){\line(0,1){1}}
\put(2,6){\line(0,1){1}}
\put(3,1){\line(0,1){1}}
\put(3,2){\line(0,1){1}}
\put(3,4){\vdots}
\put(3,5){\line(0,1){1}}
\put(4,1){\line(0,1){1}}
\put(4,2){\line(0,1){1}}
\put(7,1){\line(0,1){1}}
\put(7,2){\line(0,1){1}}
\put(7,3){\line(0,1){1}}
\put(8,1){\line(0,1){1}}
\put(8,2){\line(0,1){1}}
\put(9,1){\line(0,1){1}}
\put(4,4){\vdots}
\put(0,0){\circle*{0.2}}
\put(0,1){\circle*{0.3}}
\put(1,1){\circle*{0.1}}
\put(1,0){\circle*{0.1}}
\put(2,0){\circle*{0.1}}
\put(2,1){\circle*{0.1}}
\put(3,0){\circle*{0.1}}
\put(3,1){\circle*{0.1}}
\put(7,0){\circle*{0.1}}
\put(7,1){\circle*{0.1}}
\put(8,0){\circle*{0.1}}
\put(8,1){\circle*{0.1}}
\put(9,0){\circle*{0.1}}
\put(9,1){\circle*{0.1}}
\put(0,2){\circle*{0.1}}
\put(0,3){\circle*{0.1}}
\put(0,5){\circle*{0.1}}
\put(0,6){\circle*{0.1}}
\put(0,7){\circle*{0.1}}
\put(0,8){\circle*{0.1}}
\put(0,9){\circle*{0.1}}
\put(1,2){\circle*{0.1}}
\put(1,3){\circle*{0.1}}
\put(1,5){\circle*{0.1}}
\put(1,6){\circle*{0.1}}
\put(1,7){\circle*{0.1}}
\put(1,8){\circle*{0.1}}
\put(2,2){\circle*{0.1}}
\put(2,3){\circle*{0.1}}
\put(2,5){\circle*{0.1}}
\put(2,6){\circle*{0.1}}
\put(2,7){\circle*{0.1}}
\put(3,2){\circle*{0.1}}
\put(3,3){\circle*{0.1}}
\put(3,5){\circle*{0.1}}
\put(3,6){\circle*{0.1}}
\put(7,2){\circle*{0.1}}
\put(7,3){\circle*{0.1}}
\put(7,4){\circle*{0.1}}
\put(8,2){\circle*{0.1}}
\put(8,3){\circle*{0.1}}
\put(9,2){\circle*{0.1}}
\put(9,1){\line(1,-1){1}}
\put(9,0){\line(1,1){1}}
\put(10,1){\circle*{0.3}}
\put(10,0){\circle*{0.2}}
\put(7,8){{\Large $ \mathcal{P}_2^{L}$}}
\put(0,10){$L+1$}
\put(1,9){$L$}
\put(2,8){$L-1$}
\put(3,7){$L-2$}
\put(7,5){$4$}
\put(8,4){$3$}
\put(9,3){$2$}
\end{picture}
\caption{Example \ref{teggie}, Lorentz spaces which are GH but not GGH close.}
\label{figure3}
\end{center}

\end{figure}
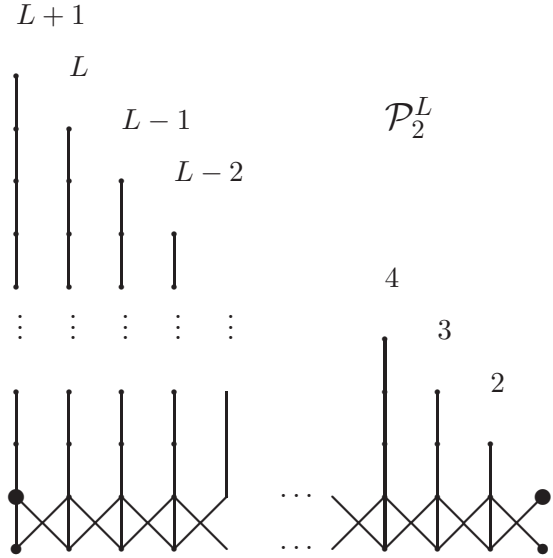
\begin{figure}[h]
\begin{center}
  \setlength{\unitlength}{0.7cm}
\begin{picture}(10,13)
\thicklines
\put(0,0){\line(0,1){1}}
\put(0,1){\line(1,-1){1}}
\put(1,0){\line(0,1){1}}
\put(1,1){\line(-1,-1){1}}
\put(1,1){\line(1,-1){1}}
\put(2,0){\line(0,1){}}
\put(2,1){\line(-1,-1){1}}
\put(2,1){\line(1,-1){1}}
\put(3,0){\line(0,1){1}}
\put(3,1){\line(-1,-1){1}}
\put(3,1){\line(1,-1){1}}
\put(3,0){\line(1,1){1}}
\put(5,0){\ldots}
\put(5,1){\ldots}
\put(7,0){\line(-1,1){1}}
\put(7,1){\line(-1,-1){1}}
\put(7,0){\line(1,1){1}}
\put(7,1){\line(1,-1){1}}
\put(8,0){\line(0,1){1}}
\put(8,0){\line(1,1){1}}
\put(8,1){\line(1,-1){1}}
\put(7,0){\line(0,1){1}}
\put(8,0){\line(0,1){1}}
\put(9,0){\line(0,1){1}}
\put(0,1){\line(0,1){1}}
\put(0,2){\line(0,1){1}}
\put(0,4){\vdots}
\put(0,5){\line(0,1){1}}
\put(1,1){\line(0,1){1}}
\put(1,2){\line(0,1){1}}
\put(1,4){\vdots}
\put(1,5){\line(0,1){1}}
\put(1,7){\vdots}
\put(1,8){\line(0,1){1}}
\put(1,9){\line(0,1){1}}
\put(1,10){\line(0,1){1}}
\put(1,11){\line(0,1){1}}
\put(2,1){\line(0,1){1}}
\put(2,2){\line(0,1){1}}
\put(2,4){\vdots}
\put(2,5){\line(0,1){1}}
\put(2,7){\vdots}
\put(2,8){\line(0,1){1}}
\put(2,9){\line(0,1){1}}
\put(3,1){\line(0,1){1}}
\put(3,2){\line(0,1){1}}
\put(3,4){\vdots}
\put(3,5){\line(0,1){1}}
\put(3,7){\vdots}
\put(3,8){\line(0,1){1}}
\put(4,1){\line(0,1){1}}
\put(4,2){\line(0,1){1}}
\put(4,4){\vdots}
\put(7,1){\line(0,1){1}}
\put(7,2){\line(0,1){1}}
\put(7,3){\line(0,1){1}}
\put(8,1){\line(0,1){1}}
\put(8,2){\line(0,1){1}}
\put(9,1){\line(0,1){1}}
\put(4,4){\vdots}
\put(0,0){\circle*{0.2}}
\put(0,1){\circle*{0.3}}
\put(1,1){\circle*{0.1}}
\put(1,0){\circle*{0.1}}
\put(2,0){\circle*{0.1}}
\put(2,1){\circle*{0.1}}
\put(3,0){\circle*{0.1}}
\put(3,1){\circle*{0.1}}
\put(7,0){\circle*{0.1}}
\put(7,1){\circle*{0.1}}
\put(8,0){\circle*{0.1}}
\put(8,1){\circle*{0.1}}
\put(9,0){\circle*{0.1}}
\put(9,1){\circle*{0.1}}
\put(0,2){\circle*{0.1}}
\put(0,3){\circle*{0.1}}
\put(0,5){\circle*{0.1}}
\put(0,6){\circle*{0.1}}
\put(1,2){\circle*{0.1}}
\put(1,3){\circle*{0.1}}
\put(1,5){\circle*{0.1}}
\put(1,6){\circle*{0.1}}
\put(1,8){\circle*{0.1}}
\put(1,9){\circle*{0.1}}
\put(1,10){\circle*{0.1}}
\put(1,11){\circle*{0.1}}
\put(1,12){\circle*{0.1}}
\put(2,2){\circle*{0.1}}
\put(2,3){\circle*{0.1}}
\put(2,5){\circle*{0.1}}
\put(2,6){\circle*{0.1}}
\put(2,8){\circle*{0.1}}
\put(2,9){\circle*{0.1}}
\put(2,10){\circle*{0.1}}
\put(3,2){\circle*{0.1}}
\put(3,3){\circle*{0.1}}
\put(3,5){\circle*{0.1}}
\put(3,6){\circle*{0.1}}
\put(7,2){\circle*{0.1}}
\put(7,3){\circle*{0.1}}
\put(7,4){\circle*{0.1}}
\put(8,2){\circle*{0.1}}
\put(8,3){\circle*{0.1}}
\put(9,2){\circle*{0.1}}
\put(9,1){\line(1,-1){1}}
\put(9,0){\line(1,1){1}}
\put(10,1){\circle*{0.3}}
\put(10,0){\circle*{0.2}}
\put(3,8){\circle*{0.1}}
\put(3,9){\circle*{0.1}}
\put(4,5){\line(0,1){1}}
\put(4,5){\circle*{0.1}}
\put(4,6){\circle*{0.1}}
\put(4,7){$L-3$}
\put(7,8){{\Large $ \mathcal{P}_1^{L}$}}
\put(0,7){$\frac{L}{2}$}
\put(1,13){$L+1$}
\put(2,11){$L-1$}
\put(3,10){$L-2$}
\put(7,5){$4$}
\put(8,4){$3$}
\put(9,3){$2$}
\end{picture}
\caption{Example \ref{teggie}, Lorentz spaces which are GH but not GGH close.}
\label{figure4}
\end{center}
\end{figure}
Choose $\alpha > 0$ and let $\epsilon_{m} = \frac{4\alpha}{4m + 1}$, $L_{m} = 4m$, $m \in \mathbb{N} \setminus \left\{ 0,1 \right\}$.  Define for $i=1,2$ the discrete Lorentz spaces $\mathcal{P}_{i}^{\epsilon_{m}, L_{m}}$ by the same Hasse diagrams, but now suppose that every link has length $\epsilon_{m}$.  The previous theorem teaches us that $\mathcal{P}_{1}^{\epsilon_{m},L_{m}}$ and $\mathcal{P}_{2}^{\epsilon_{m},L_{m}}$ are $\epsilon_{m}$-close, but for every pair of mappings $\psi_{m} : \mathcal{P}^{\epsilon_{m},L_{m}}_{1} \rightarrow \mathcal{P}^{\epsilon_{m},L_{m}}_{2}$, $\zeta_{m} :\mathcal{P}^{\epsilon_{m},L_{m}}_{2} \rightarrow \mathcal{P}^{\epsilon_{m},L_{m}}_{1}$ which make $\mathcal{P}^{\epsilon_{m},L_{m}}_{1}, \mathcal{P}^{\epsilon_{m},L_{m}}_{2}$ $\epsilon$- close, with $\epsilon < \frac{4m\alpha}{4m+1}$, there exists a $p \in \mathcal{P}^{\epsilon_{m},L_{m}}_{2}$ such that $$D_{m}(p , \psi_{m} \circ \zeta_{m} (p) ) = \frac{16m\alpha}{4m+1}.$$
However, $tdiam(\mathcal{P}_{i}^{\epsilon_{m}, L_{m}}) = 4 \alpha$ for $i=1,2$ and $m >1$.  This proves the claim that universal functions $f$ and $g$ satisfying the above conditions do not exist.  $\square$
\\*
\\*
Hence, the qualitative uniformities defined by GH and GGH are inequivalent.  However, this does not prove yet that there exist GH Cauchy sequences which are not GGH.  In fact, we show now that if a Lorentz space $(\mathcal{M},d)$ is a limit space of a GH Cauchy sequence $(\mathcal{M}_{i},d_{i})_{i \in \mathbb{N}}$, then this sequence is GGH Cauchy and converges to the same limit (up to isometry).  An intermediate result is the following.
\begin{theo} \label{bij}
Any isometry $\psi$ on a Lorentz space $(\mathcal{M},d)$ is a bijection.
\end{theo}
\textsl{Proof}: \\*
Evidently, $D(\psi(p), \psi(q)) \geq D(p,q)$ for all $p,q \in \mathcal{M}$.  Hence, we only have to show that $\psi$ is a surjection since, obviously, it is an injection.  Suppose we can find an open ball $B_{D}(r, \epsilon)$ which is not in $\psi(\mathcal{M})$, then $\psi^{k}(r) \notin B_{D}(\psi^{l}(r), \epsilon)$ for all $k > l$.  Since $\mathcal{M}$ is compact, we may, by passing to a subsequence if necessary, assume that $\psi^{l}(r) \stackrel{l \rightarrow \infty}{\rightarrow} \psi^{\infty}(r)$.  Hence, we arrive at the contradiction that $\psi^{\infty}(r) \notin B_{D}(\psi^{\infty}(r), \epsilon)$.  But then we have that $$D(\psi(p), \psi(q)) = \sup_{r \in \mathcal{M}} \left| d(\psi(r),\psi(p)) + d(\psi(p),\psi(r)) - d(\psi(r), \psi(q)) - d(\psi(q) , \psi(r)) \right|  $$
The rhs. of this equation equals $D(p,q)$.  This shows that $\psi$ is surjective since all isometries of compact metric spaces are.  $\square$     \\*
\\*
Before we prove the main result, we still need the following.
\begin{theo} \label{cov}
Let $\left\{\psi_{i} | i \in \mathbb{N}_{0} \right\}$ be a set of $\frac{1}{i}$-isometries on $\mathcal{M}$.  Then, there exists a subsequence $(\psi_{i_{n}})_{n \in \mathbb{N}}$ which uniformly converges in the strong sense to an isometry $\psi$.
\end{theo}
\textsl{Proof}:  \\*
As usual, let $\mathcal{C}$ be a countable dense subset of $\mathcal{M}$ and let $(\psi_{i_{n}})_{n \in \mathbb{N}}$ be a subsequence such that $\psi_{i_{n}}(p) \stackrel{n \rightarrow \infty}{\rightarrow} \psi(p)$ for all $p \in \mathcal{C}$.  It is easy to see that $\psi$ has a unique extension to a $D$-isometry (and $d$-isometry) using Theorem \ref{bij}.  The proof of Theorem \ref{bij} also implies that $\psi(\mathcal{C})$ is dense in $\mathcal{M}$.   As a consequence, we have that for any $\epsilon > 0$ there exists a $k(\epsilon) > 0$ such that $\psi_{i_{k}} ( \mathcal{C} )$ is $\epsilon$-dense in $\mathcal{M}$ for $k > k(\epsilon)$.   Hence, $$\left| D(\psi_{i_{k}}(p), \psi_{i_{k}}(q)) - D(p,q) \right| < \epsilon + \frac{2}{i_{k}}$$ for $k > k(\epsilon)$ and for all $p,q \in \mathcal{M}$.  This implies that
$$ D(\psi(r), \psi_{i_{k}}(r)) \leq  D(\psi(p) , \psi_{i_{k}}(p)) + 2D(p,r) + \frac{2}{i_{k}} + \epsilon $$ for $k > k(\epsilon)$ and $p \in \mathcal{C}$.  Since $\epsilon$ and $p$ can independently be chosen arbitrarly close to $0$ and $r$ respectively, the result follows. $\square$ \\*
\\*
We are now in position to prove the main result.
\begin{theo} \label{isom}
Let $(\mathcal{M}_{i},d_{i})_{i \in \mathbb{N}}$ be a GH Cauchy sequence of Lorentz spaces converging to a Lorentz space $(\mathcal{M},d)$, then this sequence is GGH Cauchy and converges to the same limit space.
\end{theo}
\textsl{Proof}:  \\*
Choose $\delta > 0$, then Theorem \ref{cov} implies that there exists a $\gamma >0$, such that if $f$ is a $\gamma$-isometry, then there exists an isometry $g$ such that $$D(f(x),g(x)) < \frac{\delta}{2} \, \forall x \in \mathcal{M}$$
Let $\psi_{i} : \mathcal{M}_{i} \rightarrow \mathcal{M}$ and $\zeta_{i} : \mathcal{M} \rightarrow \mathcal{M}_{i}$ which make $(\mathcal{M}_{i},d_{i})$ and $(\mathcal{M},d)$ $\epsilon_{i}$-close, where $\epsilon_{i} \stackrel{i \rightarrow \infty}{\rightarrow} 0$.  Then, the previous remark implies that for $i$ sufficiently large such that $2 \epsilon_{i} < \min \left\{\gamma , \frac{\delta}{2} \right\}$, there exists an isometry $\beta_{i}$ such that
$$D(\beta_{i}(x), \psi_{i} \circ \zeta_{i} (x)) < \frac{\delta}{2} \, \forall x \in \mathcal{M}$$
or,
$$D(x, \psi_{i} \circ \zeta_{i} \circ \beta_{i}^{-1}(x)) < \frac{\delta}{2} \, \forall x \in \mathcal{M}.$$
Hence,
$$ D_{i}(p_{i} , \zeta_{i} \circ \beta_{i}^{-1} \circ \psi_{i} ( p_{i})) \leq 2 \epsilon_{i} + D(\psi_{i}(p_{i}), \psi_{i} \circ \zeta_{i} \circ \beta_{i}^{-1} \circ \psi_{i} ( p_{i}))   $$
which implies that
$$ D_{i}(p_{i} , \zeta_{i} \circ \beta_{i}^{-1} \circ \psi_{i} ( p_{i})) \leq 2 \epsilon_{i} + \frac{\delta}{2} < \delta $$
Hence, for $i$ sufficiently large, $\psi_{i}$ and $\zeta_{i} \circ \beta_{i}^{-1}$ make $(\mathcal{M}_{i},d_{i})$ and $(\mathcal{M},d)$, $(\epsilon_{i}, \delta)$-close.  $\square$ \\*
\\*
This does not show yet that every GH Cauchy sequence is GGH, but a GH Cauchy sequence which is not GGH has either no sensible limit, or a limit which is not ``spatially compact''.  The last theorem has for consequence that the trivial extension of $d_{GH}$ to the moduli space of isometry classes of Lorentz spaces is a metric.
\section{The strong metric D} \label{strD}
In this section,  we study some properties of the strong metric.  Particular questions of interest are:
\begin{itemize}
\item What is the ``shape'' of the balls in the strong metric for spacetimes?
\item Does the strong metric determine the Lorentz metric up to time reversal?
\end{itemize}
We shall treat the first question in considerable detail, the second one is only answered partially.  \\*
\\*
In what follows, $\mathcal{M}$ is assumed to be a compact, interpolating spacetime.  To start with, we ``split'' the strong metric $D$ into two
pseudodistances $D^\pm$, which will be useful later on, and then study
properties of the open balls $B_D(p,\epsilon)$ of $D$-radius $\epsilon$
around $p$.  We start by defining
$$
       D^+( p,q) = \max_{r \in \mathcal{M}}
       \left| d_{g}(p,r) - d_{g}(q,r) \right|
$$
and
$$
       D^-( p,q)
       = \max_{r \in \mathcal{M}} \left| d_{g}(r,p) - d_{g}(r,q) \right|.
$$
Then $D$ can be recovered from $D^\pm$ as \cite{Noldus2}
$$
       D(p,q) = \max \left\{ D^+(p,q) , D^-(p,q) \right\},
$$
although, separately, $D^+$ and $D^-$ are \emph{pseudo} metrics ($D^{\pm}(p,q)
= 0$ does not necessarily imply that $p=q$).  However, this limitation of $D^+$
($D^-$) arises only for $p$ and $q$ both belonging to the future (past)
boundary of $\mathcal{M}$.  For example, clearly $D^+(p,q) = 0$ for all $p,q \in
\partial_{\rm F}\mathcal{M}$, but if $p \not\in \partial_{\rm F}\mathcal{M}$ and $q\in
\partial_{\rm F}\mathcal{M}$, any $r \in I^+(p)$ gives $d_g(p,r) = |d_g(p,r)-d_g(q,r)|
> 0$, and if both $p,q\not\in \partial_{\rm F}\mathcal{M}$, the same holds for any
$r \in I^+(p)\,\triangle\,I^+(q)$\footnote{For any two sets $A$
and $B$, $A\,\triangle\,B$ stands for the symmetric difference $(A
\setminus B) \cup (B \setminus A)$.}.  \\* \\* These remarks show
that both $D^\pm$ are true distances on the interior of
$\mathcal{M}$, and they also motivate us to try to locate the
``distance-maximising points'', i.e., points which realise the
maximum in the definition of both functions for given $p$ and
$q$.  \\* \\*

\noindent\textbf{Property}: {\em Given any two points $p$ and $q$ not both
belonging to $\partial_{\rm F} \mathcal{M}$, a point $r$ such that $D^+(p,q)
= |d_{g}( p,r) -  d_{g}(q,r)|$ is an element of $I^{+} (p) \,\triangle\,
I^{+}(q)$, and $I^+(r)
\subset I^+(p) \cap I^+(q)$.  A dual property holds for $D^-$.}\\*
\\*
\noindent\textsl{Proof:}\\* \noindent Obviously, the distance-maximising point $r$ belongs to $I^+(p)
\cup I^+(q)$. Suppose $r \in I^{+}(p) \cap I^{+}(q)$ and, without
loss of generality, assume that $d_{g}(p,r) > d_{g}(q,r)$.  Let
$\gamma$ be a distance maximising  geodesic from $p$ to $r$; then
$\gamma$ cuts $E^{+}(q)$ in a point $s$.  But then, the reverse
triangle inequality implies that
\begin{eqnarray*}
       d_g(p,r) - d_g(q,r) &=& d_g(p,s) + d_g(s,r) - d_g(q,r) \\
       &<& d_g(p,s)\ \ =\ \ d_g(p,s) - d_g(q,s)\;,
\end{eqnarray*}
which is a contradiction.  Hence, $r \in I^{+}(p)\,\triangle\,
I^{+}(q)$ and without loss of generality we may assume that $r \in I^+(p)
\setminus I^+(q)$, which means that $p \notin \partial_{F} \mathcal{M}$.  Either $r \in \partial_{F} \mathcal{M}$ which implies that $I^{+}(r) = \emptyset$ or $I^{+}(r) \neq \emptyset$.  The latter implies that $I^{+}(r) \subset I^{+}(p) \cap I^{+}(q)$ since otherwise there exists a point $s$ such that $$d_g(p,r) < d_g(p,s) = d_g(p,s) - d_g(q,s) $$ which is a contradiction.  \hfill$\square$\\*
\\*
Now, if $(\mathcal{M},g)$ contains no cut points\footnote{The definition given in \cite{Beem} is more general but reduces to the following in the globally hyperbolic case: a future oriented causal geodesic $\gamma$ starting at $p$ ($\gamma(0) = p$) has a future cut point $\gamma(t_{0})$ iff $\gamma$ is distance maximising between $p$ and $\gamma(t_{0})$, i.e.,  $d_{g}(\gamma(s),\gamma(t)) = L(\gamma_{\left[s,t\right]})$ for all $0 \leq t < s \leq t_{0}$ and $t_{0}$ is the largest (affine) parameter with this property; a past cut point for a (past oriented) geodesic is defined similarly.  A Lorentzian equivalent of an earlier result by Poincar\'e shows that a causal geodesic with initial endpoint $p$ has a cut point $\gamma(t_{0})$ iff there exists a second causal geodesic starting at $p$ which contains $\gamma(t_{0})$ or if $\gamma(t_{0})$ is \emph{conjugate} point for $\gamma$ (i.e., there exists a Jacobi field along $\gamma$ which vanishes in $p$ and $\gamma(t_{0})$).  A spacetime has no cut points iff any causal geodesic with initial (final) endpoint contains no future (past) cut points.}, then the distance-maximising point $r$ must
belong to $\partial_{\rm F} \mathcal{M}$.  Suppose that $r$ does not belong to
$\partial_{\rm F}\mathcal{M}$, then $I^{+}(r) \subset I^{+}(p)
\cap I^{+} (q)$ implies that $r$ belongs to $E^{+}(q)$. Let
$\gamma$ be the unique null geodesic from $q$ to $r$, then moving
$r$ to the future along this null geodesic up to $\partial_{\rm F}
\mathcal{M}$ keeps $r$ out of $I^{+}(q)$, otherwise the geodesic
would have a cut point, which is contrary to the assumption.  \\* \\*
The next theorem is also valid when there are cut points, but then the
statement can be made sharper.  As was remarked before in theorem \ref{amal}, the $\epsilon$-balls $B_D(p,\epsilon)$
are causally convex, in the sense that if $x,y \in B_D(p,\epsilon)$, then the Alexandrov set
$A(x,y) \subset B_D(p,\epsilon)$.  We now wish to find out more about those
sets. To begin with, notice that
$$
      B_D(p,\epsilon) = B_{D^+}(p,\epsilon) \cap B_{D^-}(p,\epsilon)\;.
$$
Then, we have:

\begin{theo} \label{ball}

Let $(\mathcal{M},g)$ be a spacetime with no cut points and choose a point
$p \in \mathcal{M} \setminus \partial_{\rm F} \mathcal{M}$ and an $\epsilon >
0$ such that $K^{+} ( p , \epsilon ) \neq \emptyset$.  Then the open ``sphere''
$B_{D^+} (p,\epsilon)$ of radius $\epsilon$, centered at $p$ with respect to
the  pseudometric $D^{+}$, satisfies
$$
      B_{D^+} (p , \epsilon) \subseteq \left[ \bigcap_{x \in \mathcal{H}^{+}
      (p)}  \left( \mathcal{O}^{-} (x, \epsilon ) \right)^{\rm c} \right]
      \bigcap \left[
      \bigcap_{x \in \mathcal{F}^{+} (p , \epsilon )} I^{-} (x) \right],
$$
where
\begin{itemize}

\item $K^{+} (x , \epsilon) = \left\{ y \in \mathcal{M} \mid d_{g}
(x,y) = \epsilon \right\}$, i.e. the future $\epsilon$-sphere centered at $x$,

\item $\mathcal{O}^{-} (x , \epsilon) = \left\{ y \in \mathcal{M}
\mid d_{g} (y,x) \geq \epsilon \right\}$, i.e. the closed outer past
$\epsilon$-ball around $x$,

\item $\mathcal{H}^{+}(p) = E^{+}(p) \cap \partial_{\rm F} \mathcal{M}$,

\item $\mathcal{F}^{+}(p , \epsilon) = K^{+}( p ,\epsilon) \cap
\partial_{\rm F} \mathcal{M}$.

\end{itemize}

\noindent The open sphere $B_{D^-} ( p, \epsilon )$ defined with respect to
the pseudometric $D^{-}$, satisfies a similar inclusion property with all
pasts and futures interchanged.
\end{theo}

\noindent\textsl{Proof:}\\*
\noindent Let $x \in B_{D^+} (p , \epsilon)$; then $x$ must be chronologically
connected to all points in $\mathcal{F}^{+}(p , \epsilon)$.  For, suppose there
exists a point $y \in \mathcal{F}^{+}(p , \epsilon)$ such that $x \notin
I^{-}(y)$ then $d_{g}(p,y) - d_{g}(x,y)  = \epsilon$, which is a contradiction.
On the other hand, $x$ cannot belong to $\mathcal{O}^-(z,\epsilon)$ for any $z
\in \mathcal{H}^+(p)$, since otherwise
$$
       d_{g} ( x , z ) - d_{g}(p,z) \geq \epsilon\;,
$$
which is impossible. \hfill$\square$
\\*

\begin{figure}[h]

    \begin{center}

        \scalebox{0.50}{\includegraphics{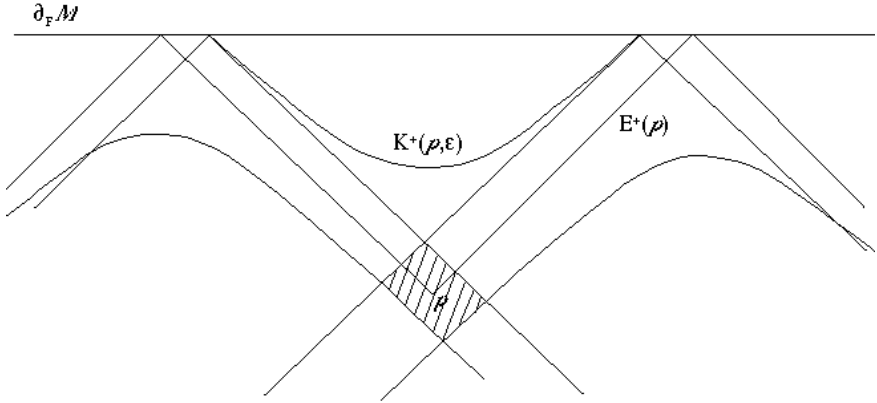}}

    \end{center}

\caption{Illustration of theorem \ref{ball} }

\label{figball}
\end{figure}

\noindent Figure \ref{figball} shows that the above inclusion can be an
equality.  The universe is $(S^{1} \times \left[ 0 , 1\right], - dt^{2} +
d\theta^{2})$  and the shaded area represents $B_{D^+} (p, \epsilon)$ for
$\epsilon$ sufficiently small.
Obviously, a dual statement holds for $B_{D^-}(p,\epsilon)$. For $B_D$, notice
that we can write down a simpler, but weaker bound
$$
      B_D(p,\epsilon) \subseteq
      \bigcap_{x\in\mathcal{F}^-(p,\epsilon),\,y\in\mathcal{F}^+(p,\epsilon)}
      A(x,y)\;.
$$
\\*
Concerning the second question posed at the beginning of this section, we notice that the strong metric $D$
does not determine the Lorentz distance $d$ up to time reversal for discrete Lorentz spaces.
\begin{exie}
\end{exie}
Consider the causal sets $\mathcal{P}_{1}$ and $\mathcal{P}_{2}$ defined by the Hasse diagrams below.
\begin{figure}[h]
\begin{center}
  \setlength{\unitlength}{0.7cm}
\begin{picture}(3.5,2)
\thicklines
\put(0,0){\line(0,1){1}}
\put(0,1){\line(1,-1){1}}
\put(1,0){\line(0,1){1}}

\put(0,0){\circle*{0.1}}
\put(0,1){\circle*{0.1}}
\put(1,1){\circle*{0.1}}
\put(1,0){\circle*{0.1}}
\put(2,0){\circle*{0.1}}

\put(1, 1.5){{\Large $ \mathcal{P}_1$}}
\end{picture}
\begin{picture}(2,2)
\thicklines
\put(0,0){\line(0,1){1}}
\put(0,1){\line(1,-1){1}}
\put(1,0){\line(0,1){1}}
\put(1,1){\line(1,-1){1}}

\put(0,0){\circle*{0.1}}
\put(0,1){\circle*{0.1}}
\put(1,1){\circle*{0.1}}
\put(1,0){\circle*{0.1}}
\put(2,0){\circle*{0.1}}

\put(1, 1.5){{\Large $\mathcal{P}_2$}}
\end{picture}
\end{center}
\caption{Hasse diagrams}
\label{fig7}
\end{figure}
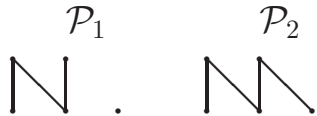
Clearly, $d_{GH}((\mathcal{P}_{1},d_{1}),(\mathcal{P}_{2},d_{2}))
= 1$ while
$d_{GH}((\mathcal{P}_{1},D_{1}),(\mathcal{P}_{2},D_{2})) = 0$.
Moreover, $D_{1}^{+} \neq D_{2}^{+}$ while $D_{1}^{-} =
D_{2}^{-}$.  This clearly shows that convergence in the strong
metric is not sufficient to guarantee convergence of the Lorentz
metrics, as far as discrete Lorentz spaces is concerned.
$\square$

\section{Causality} \label{cau}

\begin{figure}[h]
\begin{center}
  \setlength{\unitlength}{1.7cm}
\begin{picture}(8,3.5)

\put(1,1){\line(1,0){6}}
\put(1,1){\line(0,1){1.91}}
\put(1,2.91){\line(1,0){6}}
\put(7,1){\line(0,1){1.91}}
\put(1,0.5){$0$}
\put(7,0.5){$2 \pi$}
\put(0.5,3){$2$}
\put(3.53,1.96){\circle*{0.05}}
\put(4.45,1.96){\circle*{0.05}}
\thicklines
\put(3.53,1.96){\line(1,1){0.95}}
\put(3.53,1.96){\line(1,-1){0.95}}
\put(3.53,1.96){\line(-1,1){0.95}}
\put(3.53,1.96){\line(-1,-1){0.95}}
\put(4.45,1.96){\line(1,1){0.95}}
\put(4.45,1.96){\line(1,-1){0.95}}
\put(4.45,1.96){\line(-1,1){0.95}}
\put(4.45,1.96){\line(-1,-1){0.95}}

\thinlines
\put(3.53,2.06){\Blue{\line(1,-1){0.5}}}
\put(3.43,1.96){\Blue{\line(1,-1){0.5}}}
\put(3.38,1.91){\Blue{\line(1,-1){0.5}}}
\put(3.33,1.86){\Blue{\line(1,-1){0.5}}}
\put(3.28,1.81){\Blue{\line(1,-1){0.5}}}
\put(3.23,1.76){\Blue{\line(1,-1){0.5}}}
\put(3.18,1.71){\Blue{\line(1,-1){0.5}}}
\put(3.13,1.66){\Blue{\line(1,-1){0.5}}}
\put(3.08,1.61){\Blue{\line(1,-1){0.5}}}
\put(3.03,1.56){\Blue{\line(1,-1){0.5}}}
\put(2.98,1.51){\Blue{\line(1,-1){0.5}}}
\put(2.93,1.46){\Blue{\line(1,-1){0.45}}}
\put(2.88,1.41){\Blue{\line(1,-1){0.40}}}
\put(2.83,1.36){\Blue{\line(1,-1){0.35}}}
\put(2.78,1.31){\Blue{\line(1,-1){0.30}}}
\put(2.73,1.26){\Blue{\line(1,-1){0.25}}}
\put(2.68,1.21){\Blue{\line(1,-1){0.20}}}
\put(2.63,1.16){\Blue{\line(1,-1){0.15}}}
\put(2.58,1.11){\Blue{\line(1,-1){0.10}}}
\put(2.53,1.06){\Blue{\line(1,-1){0.10}}}
\put(2.48,1.01){\Blue{\line(1,-1){0.10}}}
\put(2.43,0.96){\Blue{\line(1,-1){0.10}}}

\put(3.93,2.36){\Blue{\line(1,-1){0.5}}}
\put(4.03,2.46){\Blue{\line(1,-1){0.5}}}
\put(4.08,2.51){\Blue{\line(1,-1){0.5}}}
\put(4.13,2.56){\Blue{\line(1,-1){0.5}}}
\put(4.18,2.61){\Blue{\line(1,-1){0.5}}}
\put(4.23,2.66){\Blue{\line(1,-1){0.5}}}
\put(4.28,2.71){\Blue{\line(1,-1){0.5}}}
\put(4.33,2.76){\Blue{\line(1,-1){0.5}}}
\put(4.38,2.81){\Blue{\line(1,-1){0.5}}}
\put(4.43,2.86){\Blue{\line(1,-1){0.5}}}
\put(4.48,2.91){\Blue{\line(1,-1){0.5}}}
\put(4.58,2.91){\Blue{\line(1,-1){0.45}}}
\put(4.68,2.91){\Blue{\line(1,-1){0.40}}}
\put(4.78,2.91){\Blue{\line(1,-1){0.35}}}
\put(4.88,2.91){\Blue{\line(1,-1){0.30}}}
\put(4.98,2.91){\Blue{\line(1,-1){0.25}}}
\put(5.08,2.91){\Blue{\line(1,-1){0.20}}}
\put(5.18,2.91){\Blue{\line(1,-1){0.15}}}
\put(5.28,2.91){\Blue{\line(1,-1){0.10}}}
\put(3.24,1.96){$p$}
\put(4.65,1.96){$q$}
\end{picture}
\caption{Example \ref{ex16}, cylinder universe with degenerate regions.}
\label{seven}
\end{center}
\end{figure}
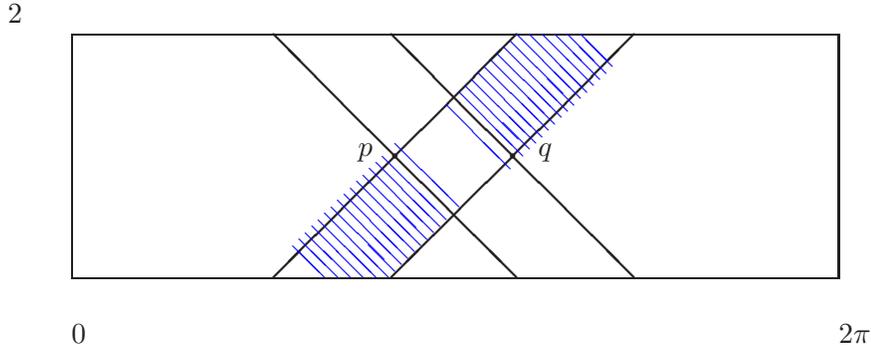

In this section, we study further causal properties of limit
spaces.  The most daunting problem is how to define a causal
relation on the degenerate regions.  First of all, one might ask
if this is physically meaningful in the sense that on may wonder whether
``particles'' could travel on such causal curves?  The answer to
this question \emph{appears} to be negative, since one would like particle motion in a spacetime region to
be determined by the distribution of matter and initial
data in the past region of the considered events, while any
definition of causality in the degenerate region \emph{will} be
based upon \emph{global (future)} data, such as specific differences
in chronological futures.  We cannot stress the word \emph{global}
enough, since, if there were a difference in the future
lightcones, then this can already be spotted by a tiny Alexandrov
neighbourhood which is ``boost equivalent'' to a neighbourhood
which might look more localised to a specific class of observers,
such as defined by the strong distance\footnote{Pick a point $p$
of the past boundary, note then that $x \rightarrow D^{-}(p,x)$ is
a time function.}.  Hence, it is not entirely clear whether
defining a causal relation is physically meaningful or not.  It is
for sure a quite interesting mathematical problem and it shall be
treated as such in the rest of this section. \\* \\* In the
sequel, we speak about a \emph{good} proposal for the causal
relation if, roughly speaking, it coincides on limit spaces of
arbitrary GGH Cauchy sequences of conformally equivalent
spacetimes with the causal relations defined by the elements of
these sequences.  To start with, we give an example which shows
that the closure of the space of discrete Lorentz spaces contains
spaces, whose causal behaviour differs significantly from the kind
of limit spaces already considered before\footnote{A similar kind
of limit space was communicated to me by R. Sorkin.}.
\begin{exie} \label{ex15}
\end{exie}
Suppose $L > 2$ and let $(\mathcal{P}^{L}_{n},d_{n})$ be a discrete Lorentz space defined by the set $\mathcal{P}_{n}^{L} = \left\{ \frac{iL}{n} | i=0 \ldots n \right\}$ and $d_{n}(p,q) = \max \left\{ 0 , q - p - 1 \right\}$.  It is easy to check that $d_{n}$ defines a Lorentz distance.  Moreover, $(\mathcal{P}^{L}_{n},d_{n}) \stackrel{n \rightarrow \infty}{\rightarrow} (\left[0,L \right], d)$ where, evidently, $d(p,q) = \max \left\{ 0 , q - p - 1 \right\}$.  Obviously, we want the causal relation to be the ordinary order relation on $\left[0,L\right]$.  Hence, any pair of timelike related points can \emph{only} be connected by a causal curve $\gamma$, which is nowhere timelike in the sense that for any $\gamma(t)$, $0 < \gamma(s) - \gamma(t) < 1$ implies that $d(\gamma(t),\gamma(s)) = 0$.  The strong metric $D(t,s)$ between points $t<s$ equals $s-t$ unless $0 < t < 1$ and $L-1 < s < L$, then it equals $\max \left\{L-t,s\right\} -1$.  Hence, locally, $D$ is the path metric defined by the the standard line element $dt^{2}$.  Conclusion: although the limit space has a manifold structure, the Lorentz distance is far from being derived from a tensor. $\square$ \\* \\*
In the previous example, we have defined the causal relation using our intuition.  Since we are looking for a general prescription for the causal relation, we might postulate something like: $p \prec q$ iff $I^{+}(q) \subset I^{+}(p)$ and $I^{-}(p) \subset I^{-}(q)$, although, as mentioned in section \ref{prop}, this is not sufficient.  Therefore, let us start by defining the causal relation on this subset of $\mathcal{M}$ which we are most familiar with namely $\overline{\mathcal{TCON}}$.  A good candidate for $\prec$ on $\overline{\mathcal{TCON}}$ is the $K^{+}$-causal relation defined by Sorkin and Woolgar
\cite{Sorkin2}, i.e.,

\begin{deffie}
$K^{+}$ is the smallest, topologically closed, partial order in
$\mathcal{M} \times
\mathcal{M}$ containing $I^{+}$. $\square$
\end{deffie}
\textbf{Remarks:}
As mentioned in \cite{Sorkin2}, $K^{+}$ can be build by transfinite induction as follows:
\begin{itemize}
\item $\prec^{0} = I^{+}$
\item $\prec^{\alpha} = \bigcup_{\beta < \alpha} \prec^{\beta}$ if $\alpha$ is a limit ordinal
\item $\prec^{\beta + 1}$ is constructed from $\prec^{\beta}$ by adding pairs which are implied either by transitivity or closure.
\end{itemize}
Since $\mathcal{M} \times \mathcal{M}$ has at most
$2^{\aleph_{0}}$ elements,the procedure has to terminate at an
ordinal\footnote{For more information about ordinals and
transitive induction see \cite{Kelley}.} with cardinality less or equal to $2^{\aleph_{0}}$.  The
following example illustrates that the procedure can run up to an
ordinal with cardinality $\aleph_{0}$. $\square$
\begin{exie}
\end{exie}
Let $\mathbb{N} = \aleph_{0}$ and construct the discrete Lorentz space $(\mathcal{P},d)$ as follows:
\begin{itemize}
\item the set of points is $\mathcal{P} = \left\{w^{i}, x^{j}_{k}, y^{j}_{k}, z^{j} | i \in \mathbb{N}+1 \textrm{ and } j,k \in \mathbb{N} \right\}$.
\item the Lorentz distance if given by $d(w^{i},z^{j}) = d(x^{i}_{k},z^{j}) = \frac{1}{(j+1)^{2}}$ and $d(y^{i}_{k}, z^{j+1}) = \frac{1}{(j+2)^{2}}$ for all $k$ and $i \leq j$ in $\mathbb{N}$.  Moreover, $d(x^{i}_{j},y^{i}_{j}) = \frac{1}{(j+1)(i+1)^{2}}$.  All other distances are calculated from these values by taking the maximum over all ``timelike'' chains.
\item from these data, it is easy to calculate the strong distance: \\* $D(w^{i},x^{i}_{j}),D(w^{i+1},y^{i}_{j}) = \frac{1}{(j+1)(i+1)^{2}}$ and $D(x^{i}_{j},y^{i}_{j}) = \frac{1}{(i+1)^{2}}$.
\end{itemize}
We now start our program: $\prec^{1}$ is the closure of $I^{+}$.  Obviously, the new relations induced by this procedure are $w^{i} \prec^{1} w^{i+1}$.  $\prec^{2}$ is constructed from $\prec^{1}$ by adding pairs which are implied by transitivity and closure: this results in $w^{i} \prec^{2} w^{i+2}$ for all $i \in \mathbb{N}$ and $w^{\mathbb{N}}  \prec^{2} w^{\mathbb{N}}$.  The reader may easily check that at stage $n>2$ : $\prec^{n} = \prec^{n-1} \cup \left\{(w^{i}, w^{j}) | i+2^{n-1} < j \leq i + 2^{n} \in \mathbb{N} \right\}$.  Hence, $$ \prec^{\mathbb{N}} = I^{+} \cup \left\{(w^{i}, w^{j}) | i < j \in \mathbb{N} \right \} \cup \left\{(w^{\mathbb{N}} ,w^{\mathbb{N}}) \right\}.$$  But this relation is not closed yet and
$$ \prec^{\mathbb{N}+1} = I^{+} \cup \left\{(w^{i}, w^{j}) | i < j \in \mathbb{N}+1 \right \} \cup \left\{(w^{\mathbb{N}} ,w^{\mathbb{N}}) \right\}.$$    So the procedure stops at the $\mathbb{N}+1$'th step and the cardinality of $\mathbb{N}+1$ is $\aleph_{0}$.  $\square$ \\* \\*
Since $K^{+}$ gives in general more information\footnote{One can construct Lorentz spaces, where $K^{+}(q) \subseteq K^{+}(p)$ and $K^{-}(p) \subseteq K^{-} (q)$ do \emph{not} imply that $I^{+}(q) \subseteq I^{+}(p)$ and $I^{-}(p) \subseteq I^{-}(q)$ and vice versa. However, $K^{+}(q) \subseteq K^{+}(p)$ and $K^{-}(p) \subseteq K^{-} (q)$ does imply that $I^{+}(q) \subseteq I^{+}(p)$ and $I^{-}(p) \subseteq I^{-}(q)$ for Lorentz spaces $(\mathcal{M},d)$ satisfying the following division property:
$$\forall p \ll q, \exists r : p \ll r \ll q$$} than $I^{+}$, we might hope that adding the conditions $K^{+}(q) \subset K^{+}(p)$ and $K^{-}(p) \subset K^{-}(q)$ in order for $p \prec q$ leads to a satisfying definition.  Unfortunately, it does not, as illustrated in the following example.
\begin{exie} \label{ex16}
\end{exie}
To simplify the discussion, define a relation $\mathcal{R}$ between $p,q \in \mathcal{M} \setminus \overline{\mathcal{TCON}}$ as follows:
$$ p\mathcal{R}q \Leftrightarrow K^{+}(q) \subseteq K^{+}(p), K^{-}(p) \subseteq K^{-} (q), I^{+}(q) \subseteq I^{+}(p) \textrm{ and } I^{-}(p) \subseteq I^{-}(q). $$
Picture \ref{seven} shows a Lorentz space with points $p$ and $q$ such that $p\mathcal{R}q$ holds.  Clearly, we do not want that $p \prec q$.  However, there exists no curve $\gamma$ between $p$ and $q$ satisfying the condition that $\gamma(t) \mathcal{R} \gamma (s)$ for all $t \leq s$.  The picture shows a part of the cylinder universe with degenerate regions, which are indicated by the shading.   $\square$
\\* \\*
The above example is very unfortunate in the sense that it shows
that using all imaginable relations between points derived from
the chronological partial order, using zero dimensional objects
only, is not sufficient for obtaining a satisfactory causal
relation.  However, it also suggests that the following definition
might have more success.
\begin{deffie}
Define a partial order $\mathcal{P}$ on a Lorentz space $(\mathcal{M},d)$ by putting $p\mathcal{P}q$ iff there exists a continuous curve $\gamma : \left[0,1\right] \rightarrow \mathcal{M}$ from $p$ to $q$ such that $\gamma(t)\mathcal{R}\gamma(s)$ for all $0 \leq t\leq s \leq 1$.
Finally, define $\prec_{d}$ on $\mathcal{M} \setminus \overline{\mathcal{TCON}}$ as the smallest topologically closed transitive relationship containing $\mathcal{P}$ and $I^{+}$.  It is easy to see that $\prec_{d}$ is compatible with $K^{+}$, i.e., $p \prec_{d} q$ and $q \in K^{-}(r)$ imply that $p \in K^{-}(r)$ and vice versa. $\square$
\end{deffie}
The following example in three dimensions shows that also this definition has its limitations.  However, in two dimensions, it does work as is proven in Theorem \ref{causr}.
\begin{exie} \label{ex17}
\end{exie}
Consider the three dimensional cylinder universe $(S^{2} \times
\left[-1,1\right], -dt^{2} + d\theta^{2} + \sin^{2}\theta
d\phi^{2})$.  Consider the spacelike geodesic
$\gamma:\left[0,\frac{1}{4} \right] \rightarrow S^{2} \times
\left[-1,1\right]:s \rightarrow \gamma(s)=( \theta_{0}+s
,\phi_{0}, 0)$.  Take the limit $(S^{2} \times
\left[-1,1\right],d)$ over a suitable sequence of conformally
equivalent metrics, with conformal factors which converge to zero
on thin, specific, (see picture \ref{eight} below) open
neighbourhoods of $J^{+}(\gamma(s))\setminus J^{+}(\gamma(t)) $
and $J^{-}(\gamma(t)) \setminus J^{-}(\gamma(s))$, which are
subsets of $J^{+}(\gamma(t))^{c}$ and $J^{-}(\gamma(s))^{c}$
respectively for all $t<s$\footnote{The relation $J^{+}$ denotes
here the usual causal relation defined by the metric tensor
$-dt^{2} + d\theta^{2} + \sin^{2}\theta d\phi^{2}$.}.
\begin{figure}[h]
\begin{center}
  \setlength{\unitlength}{1.7cm}
\begin{picture}(8,3.5)

\put(1,1){\line(1,0){6}}
\put(1,1){\line(0,1){1.91}}
\put(1,2.91){\line(1,0){6}}
\put(7,1){\line(0,1){1.91}}
\put(1,0.5){$0$}
\put(7,0.5){$2 \pi$}
\put(3.53,1.96){\circle*{0.05}}
\put(4.45,1.96){\circle*{0.05}}
\thicklines
\put(3.53,1.96){\line(1,0){0.94}}
\put(3.53,1.96){\circle{1}}
\put(4.45,1.96){\circle{1}}
\put(3.53,2.37){\line(1,0){0.94}}
\put(3.53,1.54){\line(1,0){0.94}}
\thinlines
\put(3.53,2.42){\Blue{\line(1,0){0.99}}}
\put(3.64,2.38){\Blue{\line(1,0){0.99}}}
\put(3.75,2.32){\Blue{\line(1,0){0.99}}}
\put(3.80,2.27){\Blue{\line(1,0){0.99}}}
\put(3.85,2.22){\Blue{\line(1,0){0.99}}}
\put(3.89,2.17){\Blue{\line(1,0){0.99}}}
\put(3.91,2.12){\Blue{\line(1,0){0.99}}}
\put(3.92,2.07){\Blue{\line(1,0){0.99}}}
\put(3.94,2.02){\Blue{\line(1,0){0.99}}}
\put(3.95,1.97){\Blue{\line(1,0){0.99}}}
\put(3.94,1.92){\Blue{\line(1,0){0.99}}}
\put(3.92,1.87){\Blue{\line(1,0){0.99}}}
\put(3.91,1.82){\Blue{\line(1,0){0.99}}}
\put(3.89,1.77){\Blue{\line(1,0){0.99}}}
\put(3.87,1.72){\Blue{\line(1,0){0.99}}}
\put(3.83,1.67){\Blue{\line(1,0){0.99}}}
\put(3.79,1.62){\Blue{\line(1,0){0.99}}}
\put(3.73,1.57){\Blue{\line(1,0){0.99}}}
\put(3.63,1.52){\Blue{\line(1,0){0.99}}}
\put(3.52,1.47){\Blue{\line(1,0){0.99}}}
\put(3.24,1.96){$p$}
\put(4.65,1.96){$q$}
\end{picture}
\caption{Example \ref{ex17}, intersection of the lightcones with a constant time hypersurface.  The shading indicates the degenerate regions.}
\label{eight}
\end{center}
\end{figure}
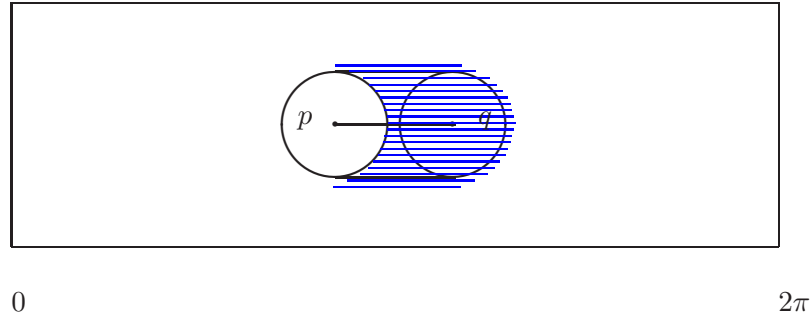
It is not difficult to see that for any point $q$ belonging to the degenerate region, either $I_{d}^{+}(q) \cap I_{d}^{+}(\gamma(0)) \neq \emptyset$ or $I_{d}^{-}(q) \cap I_{d}^{-}(\gamma(1/4)) \neq \emptyset$.  Using this, it is easy to see that for any two points $p$ and $q$ belonging to the degenerate region, we have that $I_{d}^{-}(p) \neq I_{d}^{-}(q)$ or $I_{d}^{+}(p) \neq I_{d}^{+}(q)$. $\square$  \\*

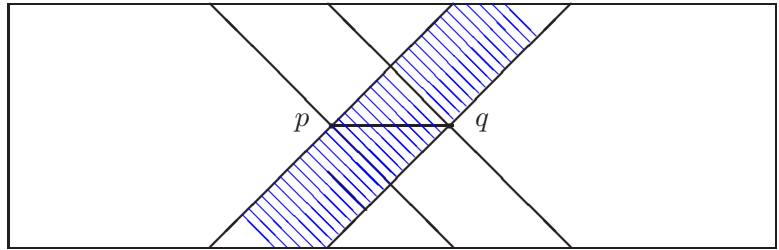
\begin{figure}[h]
\begin{center}
  \setlength{\unitlength}{1.7cm}
\begin{picture}(7,3)
\put(1,1){\line(1,0){6}}
\put(1,1){\line(0,1){1.91}}
\put(1,2.91){\line(1,0){6}}
\put(7,1){\line(0,1){1.91}}
\put(3.53,1.96){\circle*{0.05}}
\put(4.45,1.96){\circle*{0.05}}
\thicklines
\put(3.53,1.96){\line(1,1){0.95}}
\put(3.53,1.96){\line(1,-1){0.95}}
\put(3.53,1.96){\line(-1,1){0.95}}
\put(3.53,1.96){\line(-1,-1){0.95}}
\put(4.45,1.96){\line(1,1){0.95}}
\put(4.45,1.96){\line(1,-1){0.95}}
\put(4.45,1.96){\line(-1,1){0.95}}
\put(4.45,1.96){\line(-1,-1){0.95}}
\put(3.53,1.96){\line(1,0){0.95}}
\put(3.8,1.3){\line(-1,1){0.3}}
\thinlines
\put(3.58,2.01){\Blue{\line(1,-1){0.45}}}
\put(3.48,1.91){\Blue{\line(1,-1){0.45}}}
\put(3.43,1.86){\Blue{\line(1,-1){0.45}}}
\put(3.38,1.81){\Blue{\line(1,-1){0.45}}}
\put(3.33,1.76){\Blue{\line(1,-1){0.45}}}
\put(3.28,1.71){\Blue{\line(1,-1){0.45}}}
\put(3.23,1.66){\Blue{\line(1,-1){0.45}}}
\put(3.18,1.61){\Blue{\line(1,-1){0.45}}}
\put(3.13,1.56){\Blue{\line(1,-1){0.45}}}
\put(3.08,1.51){\Blue{\line(1,-1){0.45}}}
\put(3.03,1.46){\Blue{\line(1,-1){0.45}}}
\put(2.98,1.41){\Blue{\line(1,-1){0.4}}}
\put(2.93,1.36){\Blue{\line(1,-1){0.35}}}
\put(2.88,1.31){\Blue{\line(1,-1){0.30}}}
\put(2.83,1.26){\Blue{\line(1,-1){0.25}}}
\put(2.78,1.21){\Blue{\line(1,-1){0.20}}}
\put(2.73,1.16){\Blue{\line(1,-1){0.15}}}
\put(2.68,1.11){\Blue{\line(1,-1){0.10}}}
\put(2.63,1.06){\Blue{\line(1,-1){0.10}}}
\put(2.58,1.01){\Blue{\line(1,-1){0.10}}}

\put(3.48,1.91){\Blue{\line(1,-1){0.45}}}
\put(3.53,1.96){\Blue{\line(1,-1){0.45}}}
\put(3.58,2.01){\Blue{\line(1,-1){0.45}}}
\put(3.63,2.06){\Blue{\line(1,-1){0.45}}}
\put(3.68,2.11){\Blue{\line(1,-1){0.45}}}
\put(3.73,2.16){\Blue{\line(1,-1){0.45}}}
\put(3.78,2.21){\Blue{\line(1,-1){0.45}}}
\put(3.83,2.26){\Blue{\line(1,-1){0.45}}}
\put(3.88,2.31){\Blue{\line(1,-1){0.45}}}
\put(3.93,2.36){\Blue{\line(1,-1){0.45}}}
\put(4.03,2.46){\Blue{\line(1,-1){0.45}}}
\put(4.08,2.51){\Blue{\line(1,-1){0.45}}}
\put(4.13,2.56){\Blue{\line(1,-1){0.45}}}
\put(4.18,2.61){\Blue{\line(1,-1){0.45}}}
\put(4.23,2.66){\Blue{\line(1,-1){0.45}}}
\put(4.28,2.71){\Blue{\line(1,-1){0.45}}}
\put(4.33,2.76){\Blue{\line(1,-1){0.45}}}
\put(4.38,2.81){\Blue{\line(1,-1){0.45}}}
\put(4.43,2.86){\Blue{\line(1,-1){0.45}}}
\put(4.48,2.91){\Blue{\line(1,-1){0.45}}}
\put(4.58,2.91){\Blue{\line(1,-1){0.40}}}
\put(4.68,2.91){\Blue{\line(1,-1){0.35}}}
\put(4.78,2.91){\Blue{\line(1,-1){0.30}}}
\put(4.88,2.91){\Blue{\line(1,-1){0.25}}}
\put(4.98,2.91){\Blue{\line(1,-1){0.20}}}
\put(5.08,2.91){\Blue{\line(1,-1){0.15}}}
\put(5.18,2.91){\Blue{\line(1,-1){0.10}}}
\put(5.28,2.91){\Blue{\line(1,-1){0.10}}}
\put(3.24,1.96){$p$}
\put(4.65,1.96){$q$}
\end{picture}
\caption{Proof of theorem \ref{causr}, conflict with the $T_{0}$ property.}
\label{nine}
\end{center}
\end{figure}

\begin{theo} \label{causr} Let $(\mathcal{M},g)$ be a \emph{two dimensional}, globally hyperbolic, interpolating spacetime which is isometrically embeddable in the interior of an interpolating spacetime \emph{without cut points} and suppose $\Omega_{i}$ is a sequence of positive $C^{\infty}$ functions on $\mathcal{M}$ such that $\left| d_{\Omega_{i}^{2}g}(p,q) - d_{\Omega_{j}^{2}g}(p,q) \right| < \frac{1}{i}$ for all $j > i > 0$ and $p,q \in \mathcal{M}$.  Denote by $(\mathcal{N},d)$ the GGH limit space.  Suppose that $\mathcal{M}=\mathcal{N}$, i.e., no points get identified.  Then, one has that $p \prec_{g} q$ iff $p \prec_{d} q$ for all $p,q  \in \mathcal{M}$.
\end{theo}
\textbf{Note}: If $(\mathcal{M},g)$ were
allowed to have cut points, then the theorem would not be valid
anymore.  It is possible to construct a counterexample by making a
drawing on the two dimensional cylinder universe. \\* \\*
\textsl{Proof}: \\* First notice that the strong topology defined
by $D$ coincides with the manifold topology\footnote{Clearly, $D$
is continuous in the manifold topology on $\mathcal{M} \times
\mathcal{M}$ since it is the uniform limit of a sequence of
continuous functions.  On the other hand, let $p \in \mathcal{M}$
and $p \in \mathcal{V} \subset \overline{\mathcal{V}} \subset
\mathcal{U}$ where $\mathcal{U}$ and $\mathcal{V}$ are open
neighbourhoods of $p$ in the manifold topology.  Suppose,
moreover, that there exists a sequence $(p_{n})_{n \in
\mathbb{N}}$ such that $p_{n} \notin \mathcal{U}$ and $D(p,p_{n})
< \frac{1}{n}$.  By passing to a subsequence if necessary, we may
assume that $p_{n} \stackrel{n \rightarrow \infty}{\rightarrow} q
\in \overline{V}^{c}$ in the manifold topology.  Hence, $D(p,q)=0$
which contradicts $\mathcal{M} = \mathcal{N}$.}. \\*
$\Rightarrow)$ We show that \emph{any} $g$ causal curve $\gamma$
is a $\mathcal{R}$ causal curve.  Clearly, $I^{+}_{d}(\gamma(s))
\subseteq I^{+}_{d}(\gamma(t))$ and $I^{-}_{d}(\gamma(t))
\subseteq I^{-}_{d}(\gamma(s))$ for all $t < s$ which proves the
basis of induction.  Let $\alpha = \beta + 1$ and suppose that $t
< s$ implies that $\gamma(s) \prec^{\beta} y \Longrightarrow
\gamma(t) \prec^{\beta} y$.  Obviously, if $\gamma(s)
\prec^{\beta} y \prec^{\beta} z$ then $\gamma(t) \prec^{\beta} y
\prec^{\beta} z$.  So suppose that there exist sequences
$(q_{n})_{n \in \mathbb{N}}$, $(y_{n})_{n \in \mathbb{N}}$
converging to $\gamma(s)$ and $y$ respectively such that $q_{n}
\prec^{\beta} y_{n}$ for any $n$, then there exists a sequence $(p_{n})_{n \in
\mathbb{N}}$ converging to $\gamma(t)$ with $p_{n} \prec_{g}
q_{n}$.  The induction hypothesis then implies that $p_{n}
\prec^{\beta} y_{n}$ for all $n$ which proves the claim.  \\*
$\Leftarrow)$ We have to show that $g$-spacelike events cannot be
connected by an $\mathcal{R}$-causal curve.  Suppose $p$ and $q$
are such events, and suppose $\gamma: \left[0,1\right] \rightarrow
\mathcal{M}$ is an $\mathcal{R}$-causal curve connecting them.
Without loss of generality, we may assume that $\gamma$ is
spacelike to $p$ and $q$ in the sense that $\gamma(t)$, $p$ and
$q$ are $g$-spacelike events for all $t \in
\left(0,1\right)$\footnote{Note that $\gamma$ cannot intersect
$E^{-}(p)$, nor $E^{+}(q)$, because this would violate the
assumption that the limit space equals $\mathcal{M}$.  By
continuity, there exists a $t$, such that $\gamma(t) \in E^{+}(p)$
but $\gamma(u) \notin E^{+}(p)$ for all $u > t$, and a minimal
$s>t$ such that $\gamma(s) \in E^{-}(q)$.}.  Moreover, we may
assume that $\gamma$ is a subset of a convex open neighborhood
$\mathcal{U}$ on which $g$ is conformally flat.  By using the
nonexistence of cut points, one can deduce that the set
$\mathcal{S} \subset \mathcal{M}$ bounded by the two right $g$
null geodesics containing $\gamma$ is entirely degenerate as is
shown in picture \ref{nine}\footnote{Choose $\gamma(t), t \in
\left(0,1\right)$.  Then, there exists an open neighborhood
$\mathcal{O}$ of $\gamma(t)$ such that for all $r \in \mathcal{O}$
which are $g$-spacelike to the left or in the $g$ chronological
past of $\gamma(t)$, we have that the left, future oriented, null
geodesic starting at $r$ does \emph{not} intersect the future
oriented, right null geodesic starting at $\gamma(t)$.  Otherwise,
$\gamma(t)$ would have a cut point in any extension of
$(\mathcal{M},g)$.  Hence, for all $s < t$ such that $\gamma(s)
\in \mathcal{O}$ is such point $r$, we have that any point in
$J^{+}(\gamma(t))$ to the right of the right null geodesic
emanating from $\gamma(s)$ belongs to the degenerate area.  A
similar argument is valid for the past, with left and right
switched.  Using this for all $t$ leads to picture \ref{nine}.}.
Any two points in $\mathcal{S} \cap \mathcal{U}$ belonging to any
left $g$ null geodesic have the same chronological relations.
$\square$
\\* \\* The results of example \ref{ex17} and theorem \ref{causr} are quite discouraging, since any good definition of a causal relation seems to depend upon some notion of \emph{dimension} of the Lorentz space.  One could try to make the definition more restrictive, so that it would be possible to reproduce a result analogous to theorem \ref{causr} in all dimensions.  It seems to me that ``local''\footnote{``Local'' in the sense that one studies properties of local congruences of curves between neighbourhoods of points.  One such idea would be to construct the following kind of definition: define the relation $p\mathcal{Q}q$ iff there exist neighbourhoods $\mathcal{U}$, $\mathcal{V}$ of $p$ and $q$ respectively and a mapping $\psi : \mathcal{U} \times \left[0,1 \right] \rightarrow \mathcal{M}$ such that
\begin{itemize}
\item $\psi_{t} : \mathcal{U} \rightarrow \mathcal{M}: r \rightarrow \psi(r,t)$ is a homeomorphism for any $t$ and $\psi_{1} (\mathcal{U}) = \mathcal{V}$.
\item $\psi_{r}: \left[0,1\right] \rightarrow \mathcal{M} : t \rightarrow \psi(r,t)$ defines a $\mathcal{R}$-causal curve for any $r \in \mathcal{U}$.
\end{itemize}
$\prec_{d}$ is then defined as the smallest topologically closed transitive relation encompassing $\mathcal{Q}$ and $I^{+}$.   Again it is not difficult to construct a counterexample similar to example \ref{ex17}.} ideas won't work. \\* \\*
The rest of this paragraph is devoted to proving that the limit space $(\mathcal{M},d)$ of a $\mathcal{C}^{+}_{\alpha}$ and $\mathcal{C}^{-}_{\alpha}$ sequence $(\mathcal{M}_{i},d_{i})_{i \in \mathbb{N}}$ of path metric\footnote{For a precise definition of a path metric Lorentz space, see definition \ref{path}.} Lorentz spaces is a path metric Lorentz space.  Strictly speaking, we should still define the $\mathcal{C}^{\pm}_{\alpha}$ properties for general Lorentz spaces $(\mathcal{M},d)$.  Looking at the definition in the intermezzo of section \ref{prop}, the reader can see that this boils down to defining the future and past boundaries of $(\mathcal{M},d)$.  Obviously, the past boundary $\partial_{P}\mathcal{M}$ is the set of points $p$ such that $I^{-}(p) = \emptyset$, the future boundary $\partial_{F}\mathcal{M}$ is defined dually.  \\* \\*
\textbf{Property}: The $\mathcal{C}^{+}_{\alpha}$ property implies that the interior of $\partial_{P} \mathcal{M}$ is empty and, likewise, the $\mathcal{C}^{-}_{\alpha}$ property implies that the interior of $\partial_{F} \mathcal{M}$ is empty. \\*
\\*
\textsl{Proof}: \\*
I shall only prove the former.  First, note that $\partial_{P} \mathcal{M} \cap \partial_{F} \mathcal{M}$ contains at most one point.  Let $p \in \partial_{P} \mathcal{M} \setminus \partial_{F}  \mathcal{M}$ and $\epsilon > 0$ be such that $B_{D}(p, \epsilon) \subset \partial_{P} \mathcal{M} \setminus \partial_{F}  \mathcal{M}$.  Then, $d(p,r) = 0$ for all $r \in B_{D}(p, \epsilon)$ which is impossible by the $\mathcal{C}^{+}_{\alpha}$ property.  $\square$ \\* \\*
As a consequence, we have that for a Lorentz space $(\mathcal{M},d)$ satisfying the $\mathcal{C}^{+}_{\alpha}$ and $\mathcal{C}^{-}_{\alpha}$ property, $\overline{\mathcal{TCON}} \cup \left( \partial_{P} \mathcal{M} \cap \partial_{F} \mathcal{M} \right) = \mathcal{M}$\footnote{For example, let $p \in \partial_{P} \mathcal{M} \setminus \partial_{F}  \mathcal{M}$, then for $\epsilon > 0$ sufficiently small, we have that $B_{D}(p , \epsilon) \cap \partial_{F} \mathcal{M} = \emptyset$.  By the $\mathcal{C}^{+}_{\alpha}$ property, there exists an $r \in \overline{B_{D}(p, \frac{\epsilon}{2})}$ such that $d(p,r) = \alpha(\frac{\epsilon}{2})$.  Hence $$D(r, \partial_{P} \mathcal{M}), D(r, \partial_{F} \mathcal{M}) \geq \alpha \left(\frac{\epsilon}{2} \right)$$ which implies, by the $\mathcal{C}^{+}_{\alpha}$ and $\mathcal{C}^{-}_{\alpha}$ properties, that $r \in \mathcal{TCON}$.}.  Note that the second term on the left-hand side of this equality only needs to be accounted for iff $\partial_{P} \mathcal{M} \cap \partial_{F} \mathcal{M}$ is an \emph{isolated} point.  Hence, the causal relation on such space is the $K^{+}$ relation.   \\* \\*
First, we give an example of spacetime with the $\mathcal{C}^{\pm}_{x^{2}/2}$ property.
\begin{exie} \label{ex18}
\end{exie}
Consider again the cylinder universe $\mathcal{CYL} = ({\rm S}^1
\times  [0,1], -d t^{2} + d\theta^2)$.  We argue that $\mathcal{CYL}$
belongs to the  category defined by $\alpha : \mathbb{R}^{+} \rightarrow \mathbb{R}^{+} : x
\rightarrow \frac{x^{2}}{2}$.  Since SO(2) is the isometry group of $d_g$, it is
sufficient to prove the assertion for  points with a fixed spatial coordinate,
say, $\theta = \pi$.  Let $1 \geq \tilde{t} > t \geq 0$; then it is easy to prove that
$$
       D((\pi , t) ,(\pi ,\tilde{t})) = \sqrt{\tilde{t} - t} \,
       \max \left\{\sqrt{\tilde{t} + t} , \sqrt{2 - (t + \tilde{t})} \right\},
$$
where $D$ is the strong metric on $\mathcal{CYL}$.  Hence
$$
       d_g ((\pi,t) , (\pi, \tilde{t})) \leq D((\pi , t) ,(\pi ,\tilde{t}))^{2}
       \leq 2\, d_g ((\pi,t) , (\pi, \tilde{t}))\;,
$$
which proves the assertion. \hfill$\square$ \\* \\*
Before we proceed, we should define causal curves $\gamma$ and lengths thereof\footnote{The reader can find similar definitions in \cite{Busemann}.}.
\begin{deffie}
Let $(\mathcal{M},d)$ be a Lorentz space.  Assume $a<b$ and let $\gamma : [a, b] \rightarrow \mathcal{M}$ be a continuous (w.r.t. the strong topology) mapping such that for all $a \leq t < s \leq b$ : $ \gamma (t) \prec \gamma
(s)$ ($ \gamma (t) \ll \gamma (s)$); then $\gamma$ is a basic, causal (timelike)
curve.  Let $a < b$, $c < d$ and $\gamma_{1}: [a, b] \rightarrow \mathcal{M}$,
$\gamma_{2}: [c ,d] \rightarrow \mathcal{M}$ be basic causal curves such that
$\gamma_{2}(c) = \gamma_{1}(b)$.  We define the concatenation $\gamma_{2}
\circ \gamma_{1}$ of $\gamma_{2}$ with $\gamma_{1}$ as the basic causal  curve
$\gamma_{2} \circ \gamma_{1}: [a, b + d - c] \rightarrow \mathcal{M}$ such that
$$
       \gamma_{2} \circ \gamma_{1} (t)
       = \begin{cases} \gamma_{1} (t) & {\rm if } \quad a \leq t \leq b \cr
       \gamma_{2} (t + c -b) & {\rm if } \quad b \leq t \leq b + d - c. \end{cases}
$$
A (countably infinite) concatenation of basic, causal curves is a causal
curve. \\*
The length $L(\gamma)$ of a basic, causal curve $\gamma : [a , b] \rightarrow
\mathcal{M}$ is defined as
$$
       L(\gamma) = \inf_{ \Delta } \sum_{i = 0}^{ \left| \Delta \right| - 1}
       d(\gamma(t_{i}) , \gamma(t_{i+1}))\;,
$$
where, $\Delta = \left\{t_{i} | a = t_{0} < t_{1} < \ldots < t_{n-1} < t_{n}
= b \right\}$ is a partition of $\left[a,b \right]$.  Obviously,
$$
       L(\gamma_{2} \circ \gamma_{1}) = L(\gamma_{1}) + L(\gamma_{2})\;.
$$
\hfill$\square$

\end{deffie}
Now, we are able to give the definition of a path metric Lorentz space.
\begin{deffie} \label{path}
$(\mathcal{M},d)$ is a path metric Lorentz space iff for any $p \prec q$, there exists a causal curve $\gamma$ from $p$ to $q$, such that $L(\gamma) = d(p,q)$.
\end{deffie}
In case the causal relation coincides with the $K^{+}$ relation, I will prove that $(\mathcal{M},d)$ is a path metric space iff for any $p \ll q$, there exists a distance realising ($K^{+}$) causal curve from $p$ to $q$.  We need to introduce the Vietoris topology on the set $2^{(\mathcal{M},D)}$ of all closed, non-empty subsets of $(\mathcal{M},D)$ for which a \emph{sub-basis} is given by the sets $\mathcal{B}(\mathcal{M},\mathcal{O})$ and $\mathcal{B}(\mathcal{O}, \mathcal{M})$.  The former are sets with as members closed sets which meet the open set $\mathcal{O}$, the latter consists of the closed subsets of $\mathcal{O}$.  It is known that $2^{(\mathcal{M},D)}$ equipped with the Vietoris topology is compact \cite{Sorkin2}.   Also, it is proven in this paper that the Vietoris limit of a sequence of $K^{+}$ causal curves is a $K^{+}$ causal curve using topological arguments only.
\begin{theo} \label{theo27} Let $(\mathcal{M},d)$ be a Lorentz space, then $(\mathcal{M},d)$ is a path metric space with respect to the $K^{+}$ relation iff for any $p \ll q$, there exists a distance realising $K^{+}$ causal curve from $p$ to $q$.
\end{theo}
\textsl{Proof}: \\* We only have to prove that the latter implies
the former, the other way around being obvious.  We shall once
more proceed by transfinite induction with as induction hypothesis
$H_{\alpha}$ the statement that $p \prec^{\alpha} q$ implies that
there exists a distance realising $K^{+}$ causal curve from $p$ to
$q$.  The basis of induction is nothing else but our assumption.
Hence, let $\alpha = \beta + 1$ and assume $H_{\beta}$ is valid.
If $p \prec^{\beta} q \prec^{\beta} r$ then there exists a $K^{+}$
causal curve from $p$ to $r$, by the induction hypothesis and
concatenation, which is obviously distance maximising.  So assume
that there exist sequences $(p_{n})_{n \in \mathbb{N}}$,
$(q_{n})_{n \in \mathbb{N}}$ converging to $p$ and $q$
respectively and that $p_{n} \prec^{\beta} q_{n}$ for all $n \in
\mathbb{N}$.  Then, $H_{\beta}$ implies that there exist $K^{+}$
causal curves $\gamma_{n}$ from $p_{n}$ to $q_{n}$.  We may assume
that, by passing to a subsequence if necessary, $(\gamma_{n})_{n
\in \mathbb{N}}$ converges in the Vietoris topology to a (distance
maximising) $K^{+}$ causal curve connecting $p$ with $q$.
$\square$ \\* \\* Before I prove the main result, it is useful to
study some properties of causal curves with respect to the strong
metric $D$.
\begin{exie} \label{ex19}
\end{exie}
We show that the $D$-length of a compact, basic causal curve is in general
infinite. Obviously, the way to define the $D$-length, $DL( \gamma )$, of a
basic causal curve $\gamma: [a , b] \rightarrow \mathcal{M}$ is
$$
       DL( \gamma ) = \sup_{\Delta} \sum_{i = 0}^{\left| \Delta \right| - 1 }
       D(\gamma (t_{i}) , \gamma(t_{i+1}))\;,
$$
where, as before, $\Delta = \left\{t_{i} \mid a = t_{0} < t_{1} <
\ldots < t_{n-1} < t_{n} = b \right\}$ is a partition of $[a,b]$.
Returning to example \ref{ex18}, we prove that the length of the interval $\left\{
(\pi , t) \mid 0 \leq t \leq 1 \right\}$ equals $\infty$.  Choose $\Delta_{n}
= \left\{ 1 - \frac{1}{k} \mid k  = 1 \ldots n \right\} \cup \{1\}$, then
$$
       \sum_{k = 0}^{ n } D((\pi , t_{k}) , (\pi , t_{k+1}))
       > \sum_{k = 2}^{n+2} \frac{1}{k}\;,
$$
which proves the claim. \hfill$\square$
\\*
\\*
Since example \ref{ex19} shows that the $D$-length of a causal
curve is a meaningless concept, we have to come up with some other
way to divide a causal curve into smaller pieces.  As a starter,
we mention the following result.
\begin{theo}
Let $(\mathcal{M},g)$ be an interpolating spacetime and let $\gamma : \left[a , b \right] \rightarrow \mathcal{M}$ be any
basic, causal curve.  Then, there exists no $t \in \left( a , b
\right)$ such that $D(\gamma(a), \gamma(t)) = D(\gamma(t) ,\gamma(b))
= \frac{D(\gamma(a), \gamma(b))}{2}$, i.e., $\gamma$ has no
$D$-midpoint.
\end{theo}
\textsl{Proof:} \\* We show that for all $t \in \left( a ,b
\right)$: $D(\gamma(a), \gamma(b)) < D(\gamma(a) , \gamma(t)) +
D(\gamma(t), \gamma(b))$. Assume that the point $r$, which
realises $D(\gamma(a), \gamma(b))$, belongs to $I^{+} (\gamma(a))
\setminus I^{+} (\gamma(b))$.  The case where $r$ belongs to
$I^{-}(\gamma(b)) \setminus I^{-} (\gamma(a))$ is identical and is
left as an exercise to the reader.  Then,
$$ D(\gamma(a), \gamma(b)) =  d(\gamma(a),r) = \left( d(\gamma(a),r)
- d(\gamma(t),r) \right) + d(\gamma(t),r) $$ Both terms on the
rhs. are nonnegative and bounded by $D(\gamma(a) , \gamma(t))$ and
$D(\gamma(t) , \gamma(b))$ respectively.  The first term can only
realise $D(\gamma(a), \gamma(t))$ if $r \in I^{+}(\gamma(a))
\setminus I^{+} (\gamma(t))$.  But, in that case $d(\gamma(t),r) =
0$ and this concludes the proof.   $\square$ \\*
\\*
Hence, we define the following concept of division of a causal curve.
\begin{deffie}
Let $\gamma : \left[ a ,b \right] \rightarrow \mathcal{M}$ be a basic
causal curve and
denote $\delta = D (\gamma(a) , \gamma(b))$.  The division,
$\gamma^{1/2}$, of $\gamma$ is defined as the set of points $\left\{p_{i} | i=0 \ldots k \right\}$ such
that $\gamma (a) = p_{0} \prec p_{1} \prec \ldots \prec p_{k-1} \prec p_{k} = \gamma (b)$, $D (p_{i} , p_{i+1}) = \textstyle{\frac{\delta}{2}},\; \forall i: 0 \ldots k-2$, $\textstyle{\frac{\delta}{4}} \leq D(p_{k-1}, p_{k}) < {\textstyle{\frac{3\delta}{4}}}$ and there exists no point $q$ such that $p_{k-1} \prec q \prec \gamma(b)$ with $D(p_{k-1},q) =\textstyle{\frac{\delta}{2}}$ and $\textstyle{\frac{\delta}{4}} \leq D(q, \gamma(b)) < {\textstyle{\frac{3\delta}{4}}}$.  Such finite number $2 \leq k = \left|\gamma^{1/2} \right| - 1$ exists, since $\gamma$ is continuous with respect to the strong topology.
\end{deffie}
This new concept facilitates the proof of the final theorem.
\begin{theo}
The limit space $(\mathcal{M},d)$ of a GGH $\mathcal{C}^{+}_{\alpha}$ and
$\mathcal{C}^{-}_{\alpha}$ Cauchy sequence $(\mathcal{M}_{i},d_{i})_{i \in
\mathbb{N}}$ of path metric Lorentz spaces, is a path metric Lorentz space.
\end{theo}
\noindent\textsl{Proof}:
According to theorem \ref{theo27} and arguments preceding it, we only have to prove that $p \ll q$, $p,q \in \mathcal{M}$, implies that there exists a $K^{+}$ causal curve connecting $p$ with $q$. \\*
Let $\psi_{i}:\mathcal{M}_{i} \rightarrow \mathcal{M}$ and $\zeta_{i}: \mathcal{M} \rightarrow \mathcal{M}_{i}$ be mappings which make $(\mathcal{M}_{i},d_{i})$ and $(\mathcal{M},d)$ $(\epsilon_{i},\epsilon_{i})$-close where $\epsilon_{i} \stackrel{i \rightarrow \infty}{\rightarrow} 0$.  Choose $p,q \in \mathcal{M}$ such that $d(p,q) > 0$.  Let $\epsilon < \frac{d(p,q)}{8} $ and choose $i$ sufficiently large such that $\epsilon_{i} < \alpha(\epsilon)$.  Hence, $\left| d_{i}(\zeta_{i}(p), \zeta_{i}(q)) - d(p,q) \right| < \epsilon_{i}$ and $\left| D_{i}(\zeta_{i}(p), \zeta_{i}(q)) - D(p,q)  \right| < 4 \epsilon_{i}$.  Let $\gamma_{i}$, be a geodesic from $\zeta_{i}(p)$ to $\zeta_{i}(q)$ and consider $\gamma_{i}^{\frac{1}{2}} = \left\{p^{i}_{s} | s=0 \ldots k_{i} \right\}$.  Assume that $s^{i}$ is the largest number such that $d_{i}(p^{i}_{s^{i}+1}, \zeta_{i}(q)) > \frac{d_{i}(\zeta_{i}(p),\zeta_{i}(q))}{2}$.  Then, for all $s \leq s^{i}$, pick $r^{i}_{s+1}$ such that $\zeta_{i}(q) \gg_i r^{i}_{s+1} \gg_i p^{i}_{s+1}$, $D_{i}(p^{i}_{s+1},r^{i}_{s+1}) \leq \epsilon$ and $d_{i}(p^{i}_{s+1},r^{i}_{s+1}) = \alpha (\epsilon )$.  This is possible, since the $\mathcal{C}^{+}_{\alpha}$ property is valid and since $\frac{d_{i}(\zeta_{i}(p),\zeta_{i}(q))}{2} > \frac{d(p,q)-\alpha(\epsilon)}{2} > \frac{7d(p,q)}{16}$.  If $d_{i}(p^{i}_{s^{i}+1},p^{i}_{s^{i}+2}) < \alpha(\epsilon)$, then construct in a similar way $r^{i}_{s^{i}+2}$; this is possible since $\frac{5d(p,q)}{16} > \epsilon$.  For all $s > s^{i}+1$, define $t^{i}_{s} \ll_i p^{i}_{s}$ such that $D_{i}(t^{i}_{s},p^{i}_{s+1}) \leq \epsilon$ and $d_{i}(t^{i}_{s},p^{i}_{s+1}) = \alpha (\epsilon)$.  Obviously, $d_{i}(\zeta_{i}(p),t^{i}_{s}),d_{i}(r^{i}_{s},\zeta_{i}(q)) > \frac{3d(p,q)}{16}$ and $d_{i}(p^{i}_{s},r^{i}_{s+1}), d_{i}(t^{i}_{s}, p^{i}_{s+1}) \geq \alpha \left( \epsilon \right)$.  Hence, one can uniquely define sequences of the types $$\left\{\zeta_{i}(p),r^{i}_{1},p^{i}_{1}, r^{i}_{2}, p^{i}_{2}, \ldots , p^{i}_{s^{i}}, r^{i}_{s^{i}+1}, p^{i}_{s^{i}+1}, r^{i}_{s^{i}+2}, t^{i}_{s^{i}+2}, p^{i}_{s^{i}+3}, t^{i}_{s^{i}+3}, \ldots , t^{i}_{k_{i}-1}, \zeta_{i}(q)   \right\}$$ and $$\left\{\zeta_{i}(p),r^{i}_{1},p^{i}_{1}, r^{i}_{2}, p^{i}_{2}, \ldots , p^{i}_{s^{i}}, r^{i}_{s^{i}+1}, p^{i}_{s^{i}+1}, p^{i}_{s^{i}+2}, t^{i}_{s^{i}+2}, p^{i}_{s^{i}+3}, t^{i}_{s^{i}+3}, \ldots , t^{i}_{k_{i}-1}, \zeta_{i}(q)   \right\}$$ depending on the fact if $d_{i}(p^{i}_{s^{i}+1},p^{i}_{s^{i}+2}) < \alpha(\epsilon)$ or $d_{i}(p^{i}_{s^{i}+1},p^{i}_{s^{i}+2}) \geq \alpha(\epsilon)$ respectively.  In general, we have constructed a sequence of the form $(z^{i}_{s})_{s=0}^{2k_{i}-1}$, with the following useful properties:
\begin{itemize}
\item $z^{i}_{0} = \zeta_{i}(p)$ and $z^{i}_{2k_{i}-1} = \zeta_{i}(q)$
\item $\frac{D_{i}(\zeta_{i}(p),\zeta_{i}(q))}{2} \leq D_{i}(z^{i}_{2s}, z^{i}_{2s+1}) \leq \frac{D_{i}(\zeta_{i}(p),\zeta_{i}(q))}{2} + \epsilon$ for $s \leq k^{i}-2$,\\* $\frac{D_{i}(\zeta_{i}(p),\zeta_{i}(q))}{4} \leq D_{i}(z^{i}_{2k^{i}-2}, z^{i}_{2k^{i}-1}) < \frac{3D_{i}(\zeta_{i}(p),\zeta_{i}(q))}{4} + \epsilon$ and \\* $d_{i}(z^{i}_{2s}, z^{i}_{2s+1}) \geq \alpha(\epsilon)$ for all $s \leq k^{i}-1$.
\item $D_{i}(z^{i}_{2s-1}, z^{i}_{2s}) < 2 \epsilon$
\end{itemize}
Hence, the sequence $(\psi_{i}(z^{i}_{s}))_{s=0}^{2k_{i}-1}$ satisfies:
\begin{itemize}
\item $D(\psi_{i}(z^{i}_{0}),p), D(\psi_{i}(z^{i}_{2k_{i}-1}),q) < \alpha(\epsilon)$
\item $\frac{D(p,q)}{2} - 7\epsilon \leq D(\psi_{i}(z^{i}_{2s}), \psi_{i}(z^{i}_{2s+1})) \leq \frac{D(p,q)}{2} + 8\epsilon$ for $s \leq k^{i}-2$, \\* $\frac{D(p,q)}{4} -6 \epsilon < D(\psi_{i}(z^{i}_{2k^{i}-2}), \psi_{i}(z^{i}_{2k^{i}-1})) < \frac{3D(p,q)}{4} + 10\epsilon$ and \\* $d_{i}(\psi_{i}(z^{i}_{2s}), \psi_{i}(z^{i}_{2s+1})) > 0$ for all $s \leq k^{i}-1$
\item $D(\psi_{i}(z^{i}_{2s-1}), \psi_{i}(z^{i}_{2s})) < 6 \epsilon$
\end{itemize}
Hence, for every $n$ such that $\epsilon_{n} < \alpha(d(p,q)/8)$ we can find a
sequence $(\alpha^{n}_{s})_{s=0}^{2k_{n}-1}$ in $\mathcal{M}$
satisfying the above properties\footnote{In the sequel, the reader should keep in
mind that these finite sequences can be extended to infinite ones
by putting everything after $2k_{n}-1$ equal to $q$.}.  By using a
diagonalisation argument, we can find a subsequence (which we
label with the same index) such that $k_{n+1} \geq k_{n}$ for all
$n \in \mathbb{N}$ and a sequence
$(\alpha_{s})_{s=0}^{2\sup_{n}k_{n} -1}$ such that $\alpha^{n}_{s}
\stackrel{n \rightarrow \infty}{\rightarrow} \alpha_{s}$ for all
$s \leq 2\sup_{n}k_{n} -1$.  Obviously $\sup_{n}k_{n}$ must be
finite, since otherwise we found an infinite sequence of points
which are all a distance greater or equal than $\frac{D(p,q)}{2}$
apart, which is impossible by compactness\footnote{To see this,
notice that the strong distance is increasing along
causal paths.}. Hence, we have found a finite sequence of points
$\beta_{s} \leq \beta_{s+1}$, $s=0 \ldots k$, such that:
\begin{itemize}
\item $\beta_{0}=p$ and $\beta_{k}=q$
\item $\sum_{s=0}^{k-1}d(\beta_{s}, \beta_{s+1}) = d(p,q)$
\item $D(\beta_{s},\beta_{s+1}) = \frac{D(p,q)}{2}$, $s \leq k-2$ and \\* $\frac{D(p,q)}{4} \leq D(\beta_{k-1},q) \leq \frac{3D(p,q)}{4}$.
\end{itemize}
It is possible that for some $s$, $d(\beta_{s}, \beta_{s+1}) = 0$ but these are limits of timelike intervals as follows from the construction.  Subdividing each of these approximating timelike intervals and using a compactness argument together with the continuity of $K^{+}$ in the strong topology, one obtains that every two timelike related points are connected by a causal geodesic.  $\square$ \\* \\*
This result is, in the author's viewpoint, very encouraging since it shows that all concepts fit nicely together.  Notice also that the proof is considerably more difficult than the one in the metric case, where it suffices to use the existence of a midpoint for path metrics.
\section{Compactness of classes of Lorentz spaces.} \label{comp}
To end this chapter, we give some criteria for a collection of Lorentz spaces to
be precompact with respect to the GGH-uniformity.
The ideas presented here can be traced back to Gromov and proofs of the
results at hand can be found in Petersen \cite{Petersen}.  Let $(\mathcal{M},d)$
be a Lorentz space, and  define (as in Gromov \cite{Gromov})

\begin{itemize}

\item $\textrm{Cap}_{\mathcal{M}} ( \epsilon) = $ the maximum number
of disjoint $\frac{\epsilon}{2}$-balls in $(\mathcal{M},D)$.

\item $\textrm{Cov}_{\mathcal{M}} (\epsilon) = $ the minimum number of
$\epsilon$-balls needed to cover $\mathcal{M}$.

\end{itemize}
Clearly, $\textrm{Cov}_{\mathcal{M}} (\epsilon ) \leq
\textrm{Cap}_{\mathcal{M}}
(\epsilon)$ and both are decreasing functions of $\epsilon$.  What do these
definitions mean?  $\textrm{Cov}_{\mathcal{M}} (\epsilon)$ tells us that one
can choose $\textrm{Cov}_{\mathcal{M}} (\epsilon)$ points $p_{i}$ in
$\mathcal{M}$ such that the pair $\left( \left\{ p_{i} \mid i = 1 \ldots
\textrm{Cov}_{\mathcal{M}} (\epsilon)\right\},d\right)$ is $(2 \epsilon,
\epsilon)$-close in the Gromov-Hausdorff metric to $(\mathcal{M},d)$.  On the
other hand, suppose that $(\mathcal{M}_{1},d_{1})$ and $(\mathcal{M}_{2} ,
d_{2})$ are $(\epsilon, \delta)$ GGH-close, then we know that
$$
       d_{\rm GH} ( (\mathcal{M}_{1},D_{1}),
       (\mathcal{M}_{2},D_{2})) \leq \epsilon + {\textstyle\frac{3\delta}{2}}\;,
$$
and therefore one obtains from the triangle inequality that
$$
       \textrm{Cov}_{\mathcal{M}_{1}} ( \gamma + 2 \epsilon + 3 \delta)
       \leq  \textrm{Cov}_{\mathcal{M}_{2}} ( \gamma )
$$
and
$$
       \textrm{Cap}_{\mathcal{M}_{1}} (\gamma )
       \geq \textrm{Cap}_{\mathcal{M}_{2}} ( \gamma + 4 \epsilon + 6 \delta )
$$
for all $\gamma > 0$.  Since we have a quantitative Hausdorff uniformity on the moduli space
$\mathcal{LS}$ with a countable basis around every point, the following two
criteria for compactness are equivalent:

\begin{itemize}

\item Every open cover has a finite subcover.

\item Every sequence has a subsequence which converges to a limit point.

\end{itemize}

\begin{theo} For a class $\mathcal{C} \subset \mathcal{LS}$, the
following statements
are equivalent:

\begin{enumerate}

\item $\mathcal{C}$ is precompact in $\mathcal{LS}$, i.e., every sequence
has a subsequence that is convergent in $\mathcal{LS}$

\item There is a function $N: (0 ,\alpha ) \rightarrow
(0, \infty)$ such that $\textrm{Cap}_{\mathcal{M}} ( \epsilon ) \leq N(
\epsilon )$ for all $(\mathcal{M},d) \in \mathcal{C}$.

\item There is a function $N: (0 ,\alpha ) \rightarrow
(0, \infty)$ such that $\textrm{Cov}_{\mathcal{M}} ( \epsilon ) \leq N(
\epsilon )$ for all $(\mathcal{M},d) \in \mathcal{C}$.

\end{enumerate}

\end{theo}

\noindent\textsl{Proof}: \\*
\noindent $1 \Rightarrow 2)$ If $\mathcal{C}$ is precompact, then
for any $\epsilon > 0$ there exist points $(\mathcal{M}_{1},d_{1})$, ...,
$(\mathcal{M}_{k},d_{k}) \in \mathcal{C}$ such that any $(\mathcal{M},d)$ is
$(\frac{\epsilon}{16}, \frac{\epsilon}{24})$ close to some
$(\mathcal{M}_{i},d_{i})$.  Hence, $\textrm{Cap}_{\mathcal{M}}(\epsilon) \leq
\textrm{Cap}_{\mathcal{M}_{i}} (\frac{\epsilon}{2}) \leq \max_{j}
\textrm{Cap}_{\mathcal{M}_{j}} (\frac{\epsilon}{2})$, which clearly proves a
bound for $\textrm{Cap}_{\mathcal{M}}(\epsilon)$ for any $\epsilon > 0$. \\*

\noindent $2 \Rightarrow 3)$ is obvious. \\*

\noindent $3 \Rightarrow 1)$ Because of the generalised triangle
inequality, it suffices to show that for any $\epsilon >0$, there
exists a finite collection $\mathcal{A}$ of spaces in
$\mathcal{LS}$ such that any pair $(\mathcal{M},d) \in
\mathcal{C}$ is $(\epsilon, \epsilon)$-close to one of the
elements in $\mathcal{A}$.  Observe that for any $(\mathcal{M},d)$
and $\delta > 0$: $tdiam(\mathcal{M}) \leq 2 \delta\,
\textrm{Cov}_{\mathcal{M}} (\delta)$, since $D_{\mathcal{M}}(p,q)
\geq d(p,q)$ for all $p,q \in \mathcal{M}$. The hypothesis implies
the existence of a  function $N(\epsilon)$ such that
$\textrm{Cov}_{\mathcal{M}}(\frac{\epsilon}{8}) \leq
N(\frac{\epsilon}{8})$.  Hence, every space in $\mathcal{C}$ is
$(\frac{\epsilon}{4}, \frac{\epsilon}{8})$-close to a finite space
with $N(\frac{\epsilon}{8})$ elements, such that the timelike
distance between any two points does not exceed the value
$\frac{\epsilon}{4}\, N(\frac{\epsilon}{8})$.  The Lorentz metric
on such a finite space  consists of a square matrix $(d_{ij})_{1
\leq i,j \leq  N(\epsilon/8)}$ such that $0 \leq  d_{ij} \leq
\frac{\epsilon}{4}\, N(\frac{\epsilon}{8})$.  Obviously, one can
find a finite collection $\mathcal{A}$ of Lorentz spaces with
$N(\frac{\epsilon}{8})$ elements such that any of the $(d_{ij})_{1
\leq i,j \leq  N(\epsilon/8)}$ is $(\frac{\epsilon}{4}, 0)$-close
to some element of $\mathcal{A}$.  Hence, all spaces
$(\mathcal{M},d) \in \mathcal{C}$ are $(\frac{\epsilon}{2},
\frac{5\epsilon}{8})$-close to some element of $\mathcal{A}$,
which concludes the proof. \hfill$\square$ \\* \\* We show that
the covering property with covering function $N$ is stable under
GGH-convergence provided that $N$ is continuous (cfr.\ the
$\mathcal{C}^{\pm}_{\alpha}$ properties in section \ref{prop}).

\begin{theo}

Let $\mathcal{C}(N(\epsilon))$ be the collection of pairs $(\mathcal{M},d) \in
\mathcal{LS}$ such that $\textrm{Cov}_{\mathcal{M}}(\epsilon) \leq
N(\epsilon)$ for all $\epsilon > 0$; suppose $N$ is continuous. Then,
$\mathcal{C}(N(\epsilon))$ is compact.

\end{theo}

\noindent\textsl{Proof}:\\*
We already know that $\mathcal{C}(N(\epsilon))$ is precompact, hence
suppose $(\mathcal{M}_{i},d_{i}) \stackrel{i \rightarrow \infty}{\rightarrow}
(\mathcal{M},d)$ in the GGH-uniformity, then with
$\alpha_{i} \stackrel{i \rightarrow \infty}{\rightarrow} 0$ such that
$(\mathcal{M}_{i},d_{i})$ and $(\mathcal{M},d)$ are $(\alpha_{i},
\alpha_{i})$-close, we obtain that
$$
       \textrm{Cov}_{ \mathcal{M}}(\epsilon) \leq \textrm{Cov}_{\mathcal{M}_{i}}
       ( \epsilon - 5 \alpha_{i} )  \leq N( \epsilon - 5 \alpha_{i})\;.
$$
The continuity of $N$ concludes the proof. \hfill$\square$

%% file: conclusions.tex
\chapter{Criticisms and perspectives}
This thesis dealt with an introduction to a Lorentzian Gromov Hausdorff theory of convergence and its possible applications in physics.  Taking into account the succes the geometric and topological control theory of Gromov, Petersen et al had in the field of Euclidian dynamical triangulations \cite{Carfora}, it might be expected that this Lorentzian Gromov Hausdorff theory will have a similar effect in fields as the Lorentzian dynamical triangulations approach to quantum gravity and causal set theory.  This thesis is hopefully just the beginning of a long journey in which, without any doubt, better geometers than myself could make substantial progress.  In this last chapter, I wish to advocate some ideas about future work and criticisms about the basic concepts introduced in this thesis, which should be taken seriously.  \\* \\*            
I am very aware of the fact that the definition of the strong metric is not optimal from the physical point of view and the use of the pseudo metric $D^{-}$ defined by
$$ D^{-}(p,q) = \sup_{r \in \mathcal{M}} \left| d(r,p) - d(r,q) \right| $$
which depends only upon the difference in the \emph{past} sets, would have been desirable.  In that case however, we would encounter several technical problems:
\begin{itemize} 
\item points of the past boundary cannot be distinguished. 
\item if we would only use $D^{-}$ for defining the metric topology on points not belonging to the past boundary, then the Lorentz distance $d$ could be discontinuous, hence the continuity of $d$ would be an extra assumption.
\end{itemize}
Notice first that for an element $p$ of the past boundary, the continuous function $q \rightarrow D^{-}(p,q)$ determines a \emph{physical} time which is only non differentiable (on an interpolating spacetime) when caustics occur (i.e., on a set of Borel measure zero)\footnote{Note that this function does not depend of the point $p$.}.  The (Lipschitz) continuous surfaces of constant time provide a \emph{physical} ``slicing'' even in a non differentiable setting and it might be interesting to study the differentiability breakdown of $D^{-}$ in case caustics occur.  In the field of Lorentzian dynamical triangulations for example, this time function is (apart from an overall scale) integer valued and it counts the \emph{physical} ``spacelike hypersurfaces''\footnote{This remark was communicated to me by R. Loll}.  Concerning the above objections, our aim is to extract enough ``meaningful'', \emph{local}\footnote{``Local'' with respect to the physical time function.} information of the past boundary.  Local means that the extracted information belongs to the intersection over all $\epsilon>0$ of the $D^{-}$ $\epsilon$-tubes around $\partial_{P} \mathcal{M}$.  It is also natural to ask that the extracted information is robust enough so that it is stable under convergence.  It is here that we let ourselves be guided by our physical motivations: basically we are only interested in discrete models and continuous nondegenerate spacetimes and the discreteness implies that we introduce a fundamental scale $\mathcal{S}$.  Hence, we \emph{demand} that for any $p \in \mathcal{M}$ either $I^{\pm}(p) \cap B_{D^{-}}(p, \mathcal{S}) = \emptyset$ or $I^{\pm}(p) \cap B_{D^{-}}(p ,\epsilon)$ is nonempty for any $\epsilon > 0$.  The reader has learned in chapter $3$ that in order for such a property to be stable under convergence, we need to introduce something similar to the $\mathcal{C}^{+}_{\alpha}$ and $\mathcal{C}^{-}_{\alpha}$ properties defined in section \ref{prop}.  Also, the continuity of $d$ is only stable under convergence iff $d$ is \emph{Lipschitz continuous} (with a \emph{fixed} Lipschitz constant $\gamma > 1$) with respect to $D^{-}$, i.e., $$\left| d(p,q) - d(r,s) \right| \leq \gamma \left( D^{-}(p,r) + D^{-}(q,s) \right) \quad \forall p,q,r,s \in \mathcal{M}.$$
However, this is even not sufficient since the above would mean that for any $p,r \in \partial_{P} \mathcal{M}$ and $q \in \mathcal{M}$ we have $d(p,q) = d(r,q)$.  We solve this matter by demanding that the above formula is true for points $p,r$ belonging to the interior of $\mathcal{M}^{\uparrow \mathcal{S}}$.  For points $p$ or $r$ belonging to the closed $\mathcal{S}$-tube around $\partial_{P} \mathcal{M}$, we demand that the above is only true in case $p \ll r$ or $r \ll p$.  These remarks could lead to the following alternative definition of a physical Lorentz space:
\begin{deffie}
Let $\alpha: \mathbb{R}^{+} \rightarrow \mathbb{R}^{+}$ be a continuous, strictly increasing function such that $\alpha(x) + \alpha(y) \leq \alpha(x+y) \leq x + y$ for all $x,y \in \mathbb{R}^{+}$, let $\mathcal{S} > 0$ be a fixed fundamental scale and suppose $\gamma > 1$.  Let $\mathcal{M}$ be a set with a Lorentz distance $d$ defined on it such that $(\mathcal{M},D^{-})$ is a compact topological space and $D^{-}$ is a metric on $\mathcal{M} \setminus \partial_{P} \mathcal{M}$.  Such space is said to have the $\mathcal{D}^{+}_{\alpha}$ property iff for any $p \in \mathcal{M}$ either $I^{+}(p) \cap B_{D^{-}}(p, \mathcal{S}) = \emptyset$ or for any $\epsilon$ with $\min \left\{ D^{-}(p, \partial_{F}\mathcal{M}), \mathcal{S} \right\} \geq \epsilon > 0$:
$$ \alpha(\epsilon) \leq \max_{r \in \overline{B_{D^{-}}(p, \epsilon)}} d(p,r). $$
Similarly, $(\mathcal{M},d)$ satisfies the $\mathcal{D}^{-}_{\alpha}$ property iff for any $p \in \mathcal{M}$ either $I^{-}(p) \cap B_{D^{-}}(p, \mathcal{S}) = \emptyset$ or for any $\epsilon$ with $\min \left\{ D^{-}(p, \partial_{P}\mathcal{M}), \mathcal{S} \right\} \geq \epsilon > 0$:
$$ \alpha(\epsilon) \leq \max_{r \in \overline{B_{D^{-}}(p, \epsilon)}} d(r,p). $$ 
Define the \emph{discrete} past boundary $\partial_{P}^{d} \mathcal{M}$ as the set of all $p \in \partial_{P}\mathcal{M}$ such that $I^{+}(p) \cap B_{D^{-}}(p, \mathcal{S}) = \emptyset$ and the continuum past boundary $\partial_{P}^{c} \mathcal{M}$ as the complement of the discrete past boundary in $\partial_{P}\mathcal{M}$ and assume that the cardinality of $\partial_{P}^{d} \mathcal{M}$ is at most countable.  A space satisfying the previous properties is called a \emph{physical} Lorentz space of the type $(\alpha, \mathcal{S}, \gamma)$ iff 
$$\left| d(p,q) - d(r,s) \right| \leq \gamma \left( D^{-}(p,r) + D^{-}(q,s) \right) $$
for all $q,s  \in \mathcal{M}$; $r,p \in \overset{\circ}{\mathcal{M}^{\uparrow \mathcal{S}}}$, $r \ll p$ or $p \ll r$.
\end{deffie}
$\partial_{P}^{c} \mathcal{M}$ can still look quite exotic but we can safely disregard it, since deletion would not influence the value of $D^{-}$ on $\mathcal{M} \setminus \partial_{P} \mathcal{M}$.   Define the future reduction of a physical Lorentz space $(\mathcal{M},d)$ as the space which can be obtained from $\mathcal{M}\setminus \partial_{P}^{c} \mathcal{M}$ by identifying those points of $\partial_{P}^{d} \mathcal{M}$ which have the same future.   We declare two physical Lorentz spaces $(\mathcal{M}_{1},d_{1})$ and $(\mathcal{M}_{2},d_{2})$ to be \emph{physically} equivalent iff the future reductions are isometrically equivalent.  I did not further investigate implications of this definition, however I believe that most results in chapters $3$ and $4$ could be reproduced without too much difficulty. \\* \\*
Looking at the near future, I believe it is important to study the influence of sectional curvature and volume bounds on convergence, which might be interesting from the physical point of view since one then could study perturbations in geometry arising from fluctuations in matter sources from a global and observer independent (purely Lorentzian) point of view.  This might not only be attractive for quantum gravity but possibly also have some importance for (classical) cosmology.  A study of the influence of sectional curvature and volume bounds on convergence has already been made in the Riemannian case by a.o. Gromov, Cheeger and Anderson. 

%% file: AppendixA.tex
\chapter {Causality conditions}

\begin{verse}
\hspace{13pt} \textsl{We give a list of causality conditions which
are standard in the literature.  We discuss important cases, where
some of these properties coincide and provide the reader with
equivalent characterisations, which have some technical advantage.
Also, a clear point of view is taken concerning the problem which
causality requirement is regarded as minimal in order for the
spacetime to be physical.}
\end{verse}
\begin{enumerate}
\item The \emph{chronology} condition forbids the existence of
closed timelike curves. \item The \emph{causality} condition
excludes closed causal curves. \item A spacetime model $(\mathcal{M},g)$ is
\emph{future distinguishing} in a point $p$ if
and only if every neighbourhood of $p$ contains a neighbourhood of
$p$ which no future directed curve from $p$ intersects more
than once.  This is equivalent to $I^{+}(p) = I^{+}(q)
\Rightarrow p=q$ and the spacetime is causal.  Suppose $I^{+}(p) = I^{+}(q)$ for $p \neq q$
and the future distinguishing condition holds at $p$.  Then, for
any $s \in I^{+}(p)$, we can find a future directed causal curve
starting in $p$, passing trough an arbitrary small neighbourhood
of $q$, and ending at $s$ which clearly gives a contradiction.  On
the other hand, suppose one has that the second characterisation
is satisfied, but not the first.  Let $B_{n}$ be a local basis for
the topology at $p$ ordered by inclusion, $B_{n} \subset B_{m}$ for
$n > m$, such that for any $n$ there exists a future oriented,
inextensible, timelike curve $\lambda_{n}$ starting at $p$ and
coming back to $B_{n}$.  Let $\mathcal{U}$ be a convex normal
neighbourhood of $p$, $p$ is the origin of the coordinate frame,
$x^{4} = 0$ is a spacelike hypersurface and let $\mathcal{V} =
B^{3}(0,\epsilon) \times \left[- \epsilon , - \frac{\epsilon}{2}
\right]$ be a compact subset of $\mathcal{U}$ for $\epsilon$ small
enough.  For $n$ big enough $\lambda_{n}$ intersects $\mathcal{V}$
just before intersecting $B_{n}$.  Since $\mathcal{V}$ is compact,
we can find a point in $\mathcal{V}$ and a subsequence
$\lambda_{n_{k}}$ such that $r$ is a limit point of the
$\lambda_{n_{k}}$.  Now it is obvious that $r \in
E_{\mathcal{U}}^{-}(p)$ otherwise it would be impossible for the
$\lambda_{n_{k}}$ to intersect the $B_{n_{k}}$ in the limit for $k
\rightarrow \infty$, or, one would obtain a closed causal curve
which is in contradiction with the causality requirement.  So one
clearly has that $I^{+}(p) = I^{+}(r)$ which is impossible. 
\item The past distinguishing condition is defined similarly and is satisfied iff the spacetime is causal and $I^{-}(p) = I^{-}(q)$ implies that $p=q$.
\item
The \emph{strong causality} condition is satisfied in a point $p
\in \mathcal{M}$ if every neighbourhood of $p$ contains a
neighbourhood of $p$, which no causal curve intersects more than
once. \item The \emph{stable causality} condition needs a bit of
introduction.  Let $\mathcal{M}$ be fixed and ${T_{S}}^{0}_{2}
(\mathcal{M})$ the bundle of symmetric tensors of type $(0,2)$
over $\mathcal{M}$.  A Lorentz metric $g$ is a section of
$T^{0}_{S2} (\mathcal{M}) \stackrel{\pi}{\rightarrow}
\mathcal{M}$.  One can topologise the space of $C^{r}$ sections by
defining a basis.  Let $\mathcal{U}$ be any open set in
$T^{0}_{S2}$ such that $\pi ( \mathcal{U}) = \mathcal{M}$ and
define $O(\mathcal{U})$ to be the set of all sections $g$ such
that $g( \mathcal{M}) \subset \mathcal{U}$.  We say that $g$ is
stably causal iff there exists a neighbourhood in the $C^{0}$
topology such that every $\tilde{g}$ in this neighbourhood
satisfies the chronology condition. \item A model
$(\mathcal{M},g)$ is \emph{globally hyperbolic} iff the strong
causality condition is satisfied everywhere and if $\forall p,q
\in \mathcal{M}$ one has that $A(p,q)$ is compact.
\end{enumerate}
There needs to be said a bit more about these causality conditions
since they are quite important for the rest of this thesis.  Under
reasonable physical conditions the causality and chronology
condition are the same.  More precise, if $R_{ab}K^{a}K^{b} \geq
0$ for every null vector $\textbf{K}$ and every null geodesic
contains a point at which the tidal force $$K_{ [ a } R_{ b ] c d
[ e } K_{f ]} K^{c} K^{d}$$ is nonzero and if the chronology condition holds on
$\mathcal{M}$, then the causality condition holds on
$\mathcal{M}$.  If, in addition, $\mathcal{M}$ is geodesically
complete, then the strong causality holds everywhere on
$\mathcal{M}$.  Besides the fact that condition 3 does not allow the existence of
closed causal curves, it guarantees that if it holds on a compact
set $\mathcal{S}$ then no future inextensible causal curve can be
totally future imprisoned in $\mathcal{S}$.  This means that every
``signal'', which has no future endpoint in $\mathcal{S}$,
eventually has to escape from $\mathcal{S}$.  In the literature,
one usually requires at least that condition 3 and 4 be
satisfied.  On the other hand it is quite daring to extrapolate
our local observations which seem to confirm that there exist no closed timelike
curves, to observations which should be valid for the whole
universe!  There is good evidence that closed timelike curves cannot be manufactured by human beings, however the possibility that they occur spontaneously should not be omitted.  Unfortunately, spacetimes with closed timelike curves are difficult, if not impossible, to manage globally.  Hence, in this thesis, we shall take the pragmatic point of view that physical spacetimes have to be stably causal since on one hand this is a very physical condition and on the other hand, as we shall see, the space of all compact, stably causal interpolating spacetimes can be well controlled from the technical point of view.  \\* The stable
causality condition is equivalent to the existence a ``cosmic
time''.  More precise,  $(\mathcal{M},g)$ satisfies the stable
causality condition iff there exists a function $f$ on
$\mathcal{M}$ whose gradient is everywhere (future) timelike.  The
spacelike hypersurfaces $\{ f = \text{constant} \}$ are not
necessarily diffeomorphism equivalent to each other unless they
are all compact. \\* \\* The definition of global hyperbolicity given here
is one out of a set of equivalent characterisations:
\begin{itemize}
\item $(\mathcal{M},g)$ is strongly causal and $A(p,q)$ is compact
$\forall p,q \in \mathcal{M}$ \item $(\mathcal{M},g)$ is strongly
causal and $\mathcal{C}(p,q)$ is compact $\forall p,q \in
\mathcal{M}$ \item there exists a spacelike hypersurface $\Sigma$,
which every inextensible causal curve intersects exactly once,
i.e., $D(\Sigma) = \mathcal{M}$; $\Sigma$ is called a Cauchy
surface.  One has moreover that $ \mathcal{M}  \sim \mathbb{R}
\times \Sigma $ and every $\{a\} \times \Sigma$ is a Cauchy
surface.
\end{itemize}
So, in contrast to stable causality, in a globally hyperbolic spacetime, all surfaces of constant
cosmic time are \emph{a priori} diffeomorphism equivalent. Hence,
in a globally hyperbolic spacetimes no topology change can occur.  All the previous implies that for a compact interpolating spacetime the conditions of global hyperbolicity and stable causality are equivalent.  

%% file: appendB.tex
\chapter{Isometries}
The next theorem is a \emph{slight} generalisation of proposition $4.21$
in \cite{Beem} and is necessary to prove that every map with strictly positive timelike dilatation and co-dilatation is a conformal isometry.
\begin{theo}
If $(\mathcal{M},g)$ is strongly causal, then every onto map $f: \mathcal{M} \rightarrow \mathcal{N}$ with finite, strictly positive timelike dilatation and co-dilatation is a homeomorphism
\end{theo}
\textsl{Proof}: \\* Observe first that for all $p,q \in
\mathcal{M}$, one has that $d(f(p),f(q)) > 0$ iff $d(p,q)>0$.
Hence, $f(I^{ \pm}(x)) = I^{ \pm}(f(x))$ (since $f$ is onto) and
$f(I^{+}(p) \cap I^{-}(q)) = I^{+}(f(p)) \cap I^{-} (f(q))$. Since
$(\mathcal{M},g)$ is strongly causal, the Alexandrov topology
coincides with the manifold topology.  Hence, $f$ is an open
mapping.  $f$ is also injective, since if $p \neq q$ and $f(p) =
f(q)$, we arrive to the following contradiction.  Let
$\mathcal{U}$ be a locally convex neighbourhood of $p$ which does
not contain $q$ and satisfies the condition that every causal
curve intersects $\mathcal{U}$ exactly once.  Take then $r \ll p
\ll s$ with $r,s \in \mathcal{U}$ then $I^{+} (r) \cap I^{-} (s)
\subset \mathcal{U}$.  A fortiori $f(r) \ll f(p) = f(q) \ll f(s)$
which is a contradiction since $q \notin I^{+} (r) \cap I^{-}
(s)$.  We are done if we prove that $f^{-1}$ is open. For this it
is sufficient to prove that $(\mathcal{N},h)$ is strongly causal.
Suppose that strong causality is not satisfied at $f(p)$.  First,
choose a locally convex neighbourhood $\mathcal{U}$ of $f(p)$ such
that $(\mathcal{U},h_{ | \mathcal{U} })$ is globally hyperbolic.
Let $\mathcal{W}$ be a neighbourhood of $f(p)$ of compact closure
in $\mathcal{U}$.  If strong causality is not satisfied at $f(p)$
then there exist points $q_{n} \ll f(p) \ll  r_{n}$ in
$\mathcal{W}$ such that $q_{n},r_{n} \stackrel{n \rightarrow
\infty}{\rightarrow} f(p)$ and causal curves $\lambda_{n}$ from
$p_{n}$ to $q_{n}$ which leave $\mathcal{U}$.  Denote by $z_{n}$
the first intersection with $\partial \mathcal{W}$ of
$\lambda_{n}$.  Then there exists a subsequence $z_{n_{k}}$ such
that $z_{n_{k}} \stackrel{k \rightarrow \infty}{\rightarrow} z$.
Obviously, $p = f^{-1} (z)$ otherwise the continuity of $f^{-1}$
would contradict the strong causality of $(\mathcal{M},g)$.  But
on the other hand $p = f^{-1} (z)$ contradicts the injectivity of
$f$.  $\square$   \\* We show now that $f$ takes null geodesics to
null geodesics.  Take a small enough convex, normal neighbourhood
$\mathcal{U}$ of $p$ which no causal curve intersects more than
once and such that $(\mathcal{U} , g_{| \mathcal{U} } )$ is
globally hyperbolic.  Moreover, we assume that the closure of
$f(\mathcal{U})$ belongs to a convex, normal neighbourhood
$\mathcal{V}$ of $f(p)$ which no causal curve intersects more than
once, with $(\mathcal{V}, h_{| \mathcal{V}})$ globally hyperbolic.
Let $\alpha(q,r)$ be a null geodesic in $\mathcal{U}$ and take
sequences $q_{n} \rightarrow q$, $r_{n} \rightarrow r$ with $q_{n}
\ll r_{n}$ for all $n$.  $f$ takes timelike geodesics
$\alpha(q_{n},r_{n})$ with length $d(q_{n} , r_{n} )$ to timelike
curves $\gamma(f(q_{n}),f(r_{n}) )$  with length at most $\beta
d(q_{n} ,r_{n})$.  Moreover, $f(q_{n}) \rightarrow f(q)$ and
$f(r_{n}) \rightarrow f(r)$.  The geodesics $\alpha(q_{n} ,r_{n}
)$ converge to the null geodesic $\alpha(q,r)$.  Because of the
global hyperbolicity of $(\mathcal{V}, h_{| \mathcal{V}})$ a
subsequence of the timelike curves $\gamma(f(q_{n}),f(r_{n}))$
converges to a causal curve from $q$ to $r$.  This causal curve
need to be an unbroken null geodesic $\alpha(f(q),f(r))$ since
$d(f(q),f(r)) = 0$.  In fact, it is easy to see that the whole
sequence $\gamma(f(q_{n}) ,f( r_{n}))$ converges in the $C^{0}$
topology of curves to $\alpha(f(q),f(r))$, which concludes the
proof.  \\*  It is easy to check that if $(\mathcal{M},g)$ is a
strongly causal spacetime with spacelike boundary, then the above
results are still valid, i.e., the homeomorphism extends to the
boundary. A well known result of Hawking, King and McCarthy
\cite{Hawking2}, which is the Lorentzian equivalent of an earlier
theorem by Palais, states that every homeomorphism which maps null
geodesics to null geodesics must be a conformal isometry.

%% file: appendF.tex
\chapter{Uniformities}
For the convenience of the reader, I collect here some results on uniformities.  More results can be found in \cite{Kelley}.  Let $(X,d)$ be a topological space where $d$ is a (pseudo)
distance and denote by $ \tau $ the corresponding locally compact
topology. It is an elementary fact that the open balls $B_{1/n } (
p )$ with radius $1/n : n \in \mathbb{N}_{0} $ around $p$ define a
countable basis for $ \tau $ in $p$.  In this appendix $I,J$ will
denote index sets.  A $(X, \tau)$ cover $C$ is defined as follows:
$$ C = \{ A_{i} | A_{i} \in \tau, i \in I \} $$
such that $$ \bigcup_{ i \in I } A_{i} = X .$$ If $C =  \{ A_{i} |
A_{i} \in \tau, i \in I \} , D = \{ B_{j} | B_{j} \in \tau, j \in
J \} $ are $(X, \tau)$ covers then we say that $C$ is finer than
or is a refinement of $D$, $ C < D$ if and only if $$ \forall i
\in I \quad \exists j \in J : A_{i} \subset B_{j} .$$  Next we
define a few operations on the set of covers $C(X, \tau )$:
\\
\textbf{Operations on covers}
\begin{itemize}
\item Let $C,D$ be as before,  $$ C \wedge D  = \{ A_{i} \cap B_{j} | A_{i}, B_{j} \in \tau \quad
i \in I, j \in J \}$$ $C \wedge D$ is obviously a cover, moreover
the doublet $ C(X ,\tau),\wedge $ is a commutative semigroup.
\item For $ A \subset X$ the star of $A$ with respect to $C$ is
defined as follows:
$$ St(A,C) = \cup_{ A_{i} \in C: A \cap A_{i} \neq \emptyset}
A_{i} $$
\item The star of $C, C^{*}$ is then defined as:
$$ C^{*} = \{ St(A_{i},C) | A_{i} \in C \} $$
Remark that $ C < C^{*} < C^{**} \ldots $ and that if $I $ is
finite then there exists a $n \in \mathbb{N} $ such that after $n$
star operations $C$ has become the trivial cover.
\end{itemize}
Using the topological basis of open balls, we can define
elementary covers $C_{n} \quad n \in \mathbb{N}_{0} $ as follows:
$$ C_{n} = \{ B_{1/n } ( p ) | p \in X \} $$
These elementary covers now define a subset $U$ of $C(X, \tau )$ :
$$ U = \{ C \in C(X, \tau )| \exists C_{n} : C_{n} < C \} $$ The set $U$
satisfies the following obvious properties:
\begin{enumerate}
\item If $ C \in U $ and $ C < D $ then $ D \in U $
\item If $C,D \in U $ then $ C \wedge D \in U $
\item If $ C \in U $ then $ \exists D \in U : D^{*} < C $
\end{enumerate}
From now on we take the above properties as a \textbf{definition}
for a \textbf{uniformity}:
\begin{deffie}
Let $X$ be a set, a cover $C $ is defined as:
$$ C = \{ A_{i} | A_{i} \subset X, i \in I \} $$
such that $$ \bigcup_{ i \in I } A_{i} = X $$ A collection of
covers $U$ is called a \textbf{uniformity} for $X$ if and only if
\begin{enumerate}
\item If $ C \in U $ and $ C < D $ then $ D \in U $
\item If $C,D \in U $ then $ C \wedge D \in U $
\item If $ C \in U $ then $ \exists D \in U : D^{*} < C $
\end{enumerate}
where all definitions of $ < , \wedge $ and $^{*}$ are independent
of $ \tau $.
\end{deffie}
It has been proven that any uniformity can be generated by a
family of pseudodistances \cite{Page}.  This indicates that a
uniformity defines a topology.  For our applications we need a
different ingredient.
\begin{deffie}
Let $I$ be a directed net, and suppose $B_{i}(x) \subset X$
satisfy the following properties:
\begin{enumerate}
\item $x \in B_{i} ( x ) \quad \forall x \in X, i \in I $
\item If $ i \leq j $ then $ B_{i} (x) \subset B_{j} (x) \quad
\forall x \in X $
\item $ \forall i \in I , \exists j \in I$  such that $ \forall y \in
B_{j}(x): x \in B_{i} (y)$
\item $ \forall i \in I, \exists j \in I$ such that if $ z \in
B_{j} (y), y \in B_{j}(x) $ then $ z \in B_{i} (x) $.
\end{enumerate}
then we call the family of all $B_{i}(x)$ \textbf{a uniform
neighbourhood system}.
\end{deffie}
Now it has been proven that if $ \{ B_{i}(x) | x \in X, i \in I
\}$ is a uniform neighbourhood system then the family of covers:
$$ C_{i} = \{ B_{i} (x) | x \in X \} $$
$i \in I$ is a basis for a uniformity on $X$.  On the other hand
every uniformity can be constructed from a uniform neighbourhood
system.
\\
The topology $ \tau_{U} $ defined by a uniformity $ U $, the
\textbf{uniform topology}, is constructed as follows:
$$ O(x) \in \tau_{U} \Longleftrightarrow \exists C \in U : St(x,C)
\subset O(x) $$ so $ \{ St(x,C) | x \in X , C \in U \} $ defines a
basis for the topology. The topology is Hausdorff if and only if $
\bigcap_{O(x) \in \tau_{U}} O(x) = \{x\} $, but it is not difficult
to see that this is equivalent with: $$\bigcap_{i \in I } B_{i}(x)
= \{x\}
$$ where $ \{ B_{i} (x) | i \in I, x \in X \} $ is the uniform
neighbourhood system which generates $U$.  \\* \\* We state a few facts
about \textbf{quotient uniformities}. \\* \textbf{Terminology}
\begin{itemize}
\item Let $(X,U)$ and $(Y,V)$ be uniform spaces, a map $f : X
\rightarrow Y $ is \textbf{uniformly continuous} if and only if
$$ \forall C \in V : f^{ -1} ( C ) \in U $$
where for $ C = \{ A_{i} | i \in I \} $ , $ f^{-1} (C) = \{ f^{-1}
(A_{i}) | i \in I \} $.
\item A uniformity $ \tilde{U} $ on $X$ is finer than $U$ if and
only if every cover in $U$ belongs to $ \tilde{U} $.
\item Let $ \pi : X \rightarrow \tilde{X} $ be a surjective map
and $(X,U)$ a uniform space, the \textbf{quotient uniformity} $
\tilde{U} $ on $ \tilde{X} $ is the finest uniformity which makes
$ \pi $ uniformly continuous.
\end{itemize}
Notice that the existence of a quotient uniformity is guaranteed
by the lemma of Zorn, the uniqueness is immediate.  The obvious
question now is if $ \tau_{ \tilde{U}} $ is equal to the quotient
topology of $ \tau_{U} $.  The answer is in general no, but under
some special circumstances it works.
\begin{deffie}
A uniform neighbourhood system $ \{ B_{i} (x) | x \in X, i \in I
\}$ is \textbf{compatible} with an equivalence relation on $X$ if
and only if
$$ \forall i \in I, x' \sim x \text{ and } y \in B_{i}(x)
\quad \exists y' \sim y : y' \in B_{i} ( x' )
$$
\end{deffie}
As envisaged, compatibility implies that $ \tau_{ \tilde{U}} $ is
equal to the quotient topology of $ \tau_{U} $.
\\
\begin{theo}
If $U$ is generated by $ \{ B_{i}(x) | i \in I, x \in X \} $ which
is compatible with $ \sim $  which is for example defined by a
surjective map, then the quotient uniformity $\tilde{U} $ on $
\tilde{X} = X / \sim $ is generated by the uniform neighbourhood
system defined by:
$$ \tilde{B}_{i} ( \tilde{x} ) = \{ \tilde{y} | \exists x \in
\tilde{x} \text{ and } y \in \tilde{y} : y \in B_{i} ( x)\} $$ $
\forall \tilde{x} \in \tilde{X}, i \in I$.  Moreover $ \tau_{
\tilde{U}} $ is equal to the quotient topology of $ \tau_{U} $ and
a basis of neighbourhoods of $ \tilde{x} \in \tilde{X} $ is $ \{
\tilde{B}_{i} ( \tilde{x} ) | i \in I \} $
\end{theo}
As mentioned, every uniformity can be generated by a family of
pseudodistances. In the case that the uniformity is generated by a
countable uniform neighbourhood system, the topology is defined by
one pseudodistance, which is a distance when the uniformity is
Hausdorff.  Suppose $ C_{n} = \{ B_{n} (x) | x \in X \} $, $ n \in
\mathbb{N} $ , is a countable basis for a uniformity $U$, then we
can find a subsequence $ (n_{k})_{k} $ such that:
$$ \forall k, \quad w \in B_{n_{k}} ( z),  z \in B_{n_{k}} (y),  y \in B_{n_{k}} (x
) \Rightarrow  w \in B_{n_{k-1}} (x)$$ Assume $C_{n}$ is such a
basis.
\\
\begin{theo}
Let $ C_{n} $ be a countable basis of $U$, then with
$$ \rho (x,y) = \inf_{ \{ n \geq 0, y \in B_{n} (x) \} }
2^{-n} $$ the function
$$ d(x,y) = \inf_{ K \in \mathbb{N},   x_{k}} \sum_{ k = 1}^{K} \frac{1}{2} (
\rho (x_{k-1} , x_{k} ) + \rho ( x_{k} , x_{k-1} ) ) $$ is a
pseudodistance which generates $U$.  $ \{ x_{0} , \ldots , x_{K}
\} $ with $ x_{0} = x, x_{K} = y$ is a path in $X$. If $U$ is
Hausdorff then $d$ is a distance.
\end{theo}
Note that the function $d$ depends on the choice of basis $C_{n}$
and is therefore not canonical.

%% file: appendG.tex
\chapter{Continuity of $d$ on $\mathcal{TCON}$.}
The aim of this Appendix is to prove the claims made on page \pageref{page64}, namely that $d$ is continuous on $\mathcal{TCON}$ in $\overline{T_{0}\mathcal{S}}$ by using ``local'' techniques only.  After having read the following pages, the reader shall be very aware of the advantages the ``global'' approach brings with respect to the Alexandrov technique.  \\* \\*
\textbf{Technical preliminaries 1} \\*
Let $\mathcal{C}_{i}$ be countable dense subsets of $\mathcal{M}_{i}$; then I prove that $$ \mathcal{C} = \bigcup_{i \in \mathbb{N}} \bigcup_{x_{i} \in \mathcal{C}_{i}} \left((\psi_{l}^{i} (x_{i}))_{i < l \in \mathbb{N}}\right)$$
is a countable dense subset of $T_{0} \mathcal{S}$ in the strong topology (hence $\mathcal{C}$ is dense in $\overline{T_{0} \mathcal{S}}$ in the Alexandrov topology).  Let $p \in T_{0} \mathcal{S}$ and choose $\epsilon > 0$, we will find a $q \in \mathcal{C}$ such that $D_{T_{0} \mathcal{S}} ( p ,q) < \epsilon $.  The result is a consequence of the following estimate \\*
\\*
$\left| d(r,p) - d(r,q) \right| \leq $
\begin{eqnarray*} & & \left| d(r,p) - d_{g_{l}} ( r_{l} , \psi^{k}_{l} (p_{k})) \right| + \left| d_{g_{l}} (r_{l} , \psi^{k}_{l} (p_{k})) - d_{g_{k}} ( \zeta^{l}_{k} (r_{l}) , \zeta^{l}_{k} \circ \psi^{k}_{l} (p_{k})) \right| + \\ & &
\left| d_{g_{k}} ( \zeta^{l}_{k} (r_{l}) , \zeta^{l}_{k} \circ \psi^{k}_{l} (p_{k})) - d_{g_{k}} ( \zeta^{l}_{k}(r_{l}) , p_{k}) \right| + \left| d_{g_{k}} (\zeta^{l}_{k}(r_{l}) , p_{k}) - d_{g_{k}} ( \zeta^{l}_{k} (r_{l}) , q_{k} ) \right| \\ & & + \left| d_{g_{k}} ( \zeta^{l}_{k} (r_{l}) , \zeta^{l}_{k} \circ \psi^{k}_{l} (q_{k})) - d_{g_{k}} ( \zeta^{l}_{k}(r_{l}) , q_{k}) \right| + \left| d(r,q) - d_{g_{l}} ( r_{l} , \psi^{k}_{l} (q_{k})) \right| + \\ & & \left| d_{g_{l}} (r_{l} , \psi^{k}_{l} (q_{k})) - d_{g_{k}} ( \zeta^{l}_{k} (r_{l}) , \zeta^{l}_{k} \circ \psi^{k}_{l} (q_{k})) \right| .
\end{eqnarray*}
Choose $k, \delta > 0$ such that $\frac{8}{2^{k-1}} + \delta < \epsilon$ and $p = (\psi^{k}_{l} ( p_{k}))_{k < l \in \mathbb{N}}$.  For $q_{k} \in \mathcal{C}_{k}$ such that $d_{\mathcal{M}_{k}}(p_{k} ,q_{k}) < \delta$ and for $l$ large enough, it is easily seen that
$$ \left| d(r,p) - d(r,q) \right| < \epsilon.  $$ \\*
\textbf{Technical preliminaries 2}  \\*
Define $\psi_{\infty}^{k} : \mathcal{M}_{k} \rightarrow \overline{T_{0} \mathcal{S}} : x_{k} \rightarrow (\psi_{l}^{k} (x_{k}))_{k < l \in \mathbb{N}}$; then obviously $$\left| d(\psi_{\infty}^{k} (x_{k}) , \psi_{\infty}^{k} (y_{k})) - d_{g_{k}} (x_{k} , y_{k}) \right| \leq \frac{1}{2^{k-1}}.$$
Constructing $\zeta^{\infty}_{k}: \overline{T_{0} \mathcal{S}} \rightarrow \mathcal{M}_{k}$ is a bit more involved and by no means canonical.  Let $\mathcal{C}$ be the dense subset of $\overline{T_{0} \mathcal{S}}$ in the Alexandrov topology constructed above.  Furthermore, we assume that every $T_{0}$ equivalence class in $\overline{T_{0} \mathcal{S}}$ has at most one representative in $\mathcal{C}$. The reason for this shall become clear soon.
For every $x_{\infty} = (\psi_{l}^{i} (x_{i}) )_{i < l \in \mathbb{N}} \in \mathcal{C}$, the sequence
$$ ( \zeta_{k}^{l} \circ \psi_{l}^{i} (x_{i}) )_{ i , k < l \in \mathbb{N}} $$
in $\mathcal{M}_{k}$ has a subsequence which converges to an
element which we label as $\zeta_{k}^{\infty} (x_{\infty} )$.  In
fact, by using a diagonalisation argument, we may assume that
there exists \emph{one} subsequence $(\zeta_{k}^{l_{n}})_{n \in
\mathbb{N}}$ such that for all $x_{\infty} \in \mathcal{C}$
$$\zeta_{k}^{l_{n}} \circ \psi_{l_{n}}^{i} (x_{i}) \stackrel{n
\rightarrow \infty}{\rightarrow} \zeta_{k}^{\infty} (x_{\infty})
\in \mathcal{M}_{k}$$  whenever the above expressions are defined.
Obviously, we have that for all $x_{\infty} , y_{\infty} \in
\mathcal{C}$
\begin{eqnarray*}
\left| d(x_{\infty} , y_{\infty} ) - d_{g_{k}} ( \zeta_{k}^{\infty} (x_{\infty}) , \zeta_{k}^{ \infty} (y_{\infty}) ) \right| & \leq & \left| d_{g_{l_{n}}} ( x_{l_{n}} , y_{l_{n}} ) - d(x_{\infty} , y_{\infty} ) \right|  \end{eqnarray*}
\begin{eqnarray*}  & & \left| d_{g_{k}} ( \zeta_{k}^{\infty} (x_{\infty}) , \zeta_{k}^{ \infty} (y_{\infty}) ) - d_{g_{k}} ( \zeta_{k}^{l_{n}} (x_{l_{n}}) , \zeta_{k}^{l_{n}} (y_{l_{n}}) ) \right| + \\ & & \left| d_{g_{k}} (\zeta_{k}^{l_{n}} x_{l_{n}} , \zeta_{k}^{l_{n}} y_{l_{n}} ) - d_{g_{l_{n}}} ( x_{l_{n}} , y_{l_{n}} ) \right|
\end{eqnarray*}
which implies that, \begin{equation} \label{equ} \left| d(x_{\infty} , y_{\infty} ) - d_{g_{k}} ( \zeta_{k}^{\infty} (x_{\infty}) , \zeta_{k}^{ \infty} (y_{\infty}) ) \right| \leq \frac{1}{2^{k-1}} . \end{equation}  However, suppose $y_{\infty} \sim x_{\infty} \in \mathcal{C}$, then it is not guaranteed that $\zeta_{k}^{l_{n}} (y_{l_{n}})$ converges in $\mathcal{M}_{k}$ and even if it does, the limit points do not necessarily coincide.  Hence, the map $\zeta_{k}^{\infty}$ is defined on the $T_{0}$ equivalence classes by picking out particular representatives and subsequences, which justifies our comment made earlier.  Note that in the previous construction, I did not use ``convergence to invertibility''.  By making some elementary estimates, the reader can easily see that this property does not make the sequence $(\zeta_{k}^{n} \circ \psi_{n}^{i} (x_{i}))_{n \in \mathbb{N}}$ convergent\footnote{It would however only guarantee that there exists a sphere (in the strong topology) of radius $\frac{1}{2^{k-3}}$ which contains an infinite number of elements of this sequence.}.  We now prove that $\psi^{k}_{\infty} \circ \zeta_{k}^{\infty}$ converges to the identity on $\mathcal{C}$ in the limit $k \rightarrow \infty$.  The following estimate is crucial:
\begin{eqnarray*}
 \left| d( r , \psi^{k}_{\infty} \circ \zeta^{\infty}_{k} (x_{\infty}) ) - d(r , x_{\infty}) \right| & \leq & \left| d(r , \psi_{\infty }^{k} \circ \zeta_{k}^{\infty} ( x_{\infty} )) - d_{g_{l_{n}}} ( r_{l_{n}} , \psi_{l_{n}}^{k} \circ \zeta^{\infty}_{k} ( x_{\infty} )) \right| +
 \end{eqnarray*}
\begin{eqnarray*} & & \left| d_{g_{l_{n}}} ( r_{l_{n}} , \psi^{k}_{l_{n}} \circ \zeta^{\infty}_{k} ( x_{\infty} )) - d_{g_{k}} ( \zeta^{l_{n}}_{k} (r_{l_{n}} ) , \zeta^{l_{n}}_{k} \circ \psi^{k}_{l_{n}} \circ \zeta^{\infty}_{k} ( x_{\infty} ) ) \right|  + \\ & &
\left| d_{g_{k}} ( \zeta^{l_{n}}_{k} (r_{l_{n}}) , \zeta_{k}^{l_{n}} \circ \psi_{l_{n}}^{k} \circ \zeta^{\infty}_{k} ( x_{ \infty} ) )  - d_{g_{k}} ( \zeta^{l_{n}}_{k} (r_{l_{n}}) , \zeta^{\infty}_{k} ( x_{\infty} )) \right| +  \\ & &
\left| d_{g_{k}} ( \zeta^{l_{n}}_{k} ( r_{l_{n}} ) , \zeta^{l_{n}}_{k} ( x_{l_{n}} ) ) - d_{g_{k}} ( \zeta^{l_{n}}_{k} (r_{l_{n}}) , \zeta^{\infty}_{k} ( x_{\infty} ) ) \right| + \\ & & \left| d_{g_{k}} ( \zeta^{l_{n}}_{k} (r_{l_{n}}) , \zeta^{l_{n}}_{k} (x_{l_{n}})) - d_{g_{l_{n}}} (r_{l_{n}} , x_{l_{n}} ) \right| + \left| d_{g_{l_{n}}} (r_{l_{n}} , x_{l_{n}} ) - d(r, x_{\infty}) \right|
\end{eqnarray*}
for all $x_{\infty} \in \mathcal{C}$ and $r \in T_{0} \mathcal{S}$.  Choose $\delta > 0$, and $n$ sufficiently large such that the first, fourth, and last term on the rhs. are smaller than $\frac{\delta}{3}$.  For the fourth term this is possible, since for $n$ sufficiently large $D_{\mathcal{M}_{k}}(\zeta^{l_{n}}_{k} ( x_{l_{n}} ) , \zeta^{\infty}_{k} (x_{\infty})) < \frac{\delta}{3}$.\footnote{Remember that for globally hyperbolic spacetimes the strong topology coincides with the manifold topology.}  The second and the fifth term are smaller than $\frac{1}{2^{k-1}}$ and finally the third one is smaller than $\frac{3}{2^{k-1}}$.  Hence,
$$ \left| d( r , \psi^{k}_{\infty} \circ \zeta^{\infty}_{k} (x_{\infty}) ) - d(r , x_{\infty}) \right| \leq \frac{5}{2^{k-1}} + \delta, $$
which proves the claim.  $\square$
\\*
\\*
Now, we are in position to prove the following theorem.
\begin{theo}
$d$ has a unique continuous (in the Alexandrov topology) extension on the timelike continuum of $\overline{T_{0} \mathcal{S}}$
\end{theo}
\textsl{Proof}:
Let $(x^{i})_{i \in \mathbb{N}}$ be a future Cauchy sequence which determines an element of $\overline{T_{0} \mathcal{S}}$ (the past case is identical) and let $y \in T_{0} \mathcal{S}$.  Then we define
$$ d(y,(x^{i})_{i \in \mathbb{N}}) = \lim_{i \rightarrow \infty} d(y, x^{i}). $$
Note that the limit on the rhs. is well defined since the sequence $(d(y, x^{i}))_{i \in \mathbb{N}}$ is increasing and bounded by $\sup_{i \in \mathbb{N}} \textrm{tdiam} ( \mathcal{M}_{i} ) < \infty$.  $d((x^{i})_{i \in \mathbb{N}} , y)$ is obviously defined as
$$ d((x^{i})_{i \in \mathbb{N}} , y) = \lim_{i \rightarrow \infty} d(x^{i} , y), $$
in which case the existence of the limit is guaranteed since the sequence $(d(x^{i} ,y))_{i \in \mathbb{N}}$ is decreasing and bounded by zero.  We now prove that the limit is the same when we replace the sequence $(x^{i})_{i \in \mathbb{N}}$ by an equivalent one, say, $(z^{i})_{i \in \mathbb{N}}$.  If both sequences are future timelike equivalent, then this statement is obvious.  Therefore, let $(z^{i})_{i \in \mathbb{N}}$ be past timelike Cauchy (such a sequence exists since $(x^{i})_{i \in \mathbb{N}}$ belongs to the timelike continuum) and moreover, assume that $\lim_{i \rightarrow \infty} d(y,z^{i}) > \lim_{i \rightarrow \infty} d(y,x^{i}) + \delta$ for a certain $\delta >0$.  The other possibility (the position of $y$ and $z^{i}$,$x^{i}$ as arguments of $d$ swapped and the inequality reversed) is similar and therefore we only bother proving this one.  Moreover, without loss of generality, we may assume that all elements $x^{i} , z^{i}$ belong to $\mathcal{C}$.  Choose $k$ sufficiently large such that $\frac{1}{2^{k-1}} < \frac{\delta}{60}$ and $y_{k} = \psi_{k}^{i_{0}} (y_{i_{0}})$ for some $i_{0} < k$, then the following estimate
\begin{eqnarray*} \left| d_{g_{k}} ( y_{k} , \zeta^{\infty}_{k} (z^{m} ) ) - d( y, z^{m} ) \right| & \leq & \left|d_{g_{k}} ( y_{k} , \zeta^{\infty}_{k} (z^{m}) ) - d(y , \psi^{k}_{\infty} \circ \zeta^{\infty}_{k} ( z^{m} )) \right| + \\ & & \left| d(y , \psi^{k}_{\infty} \circ \zeta^{\infty}_{k} ( z^{m} ) ) - d( y ,z^{m} ) \right|  \\ & < & \frac{\delta}{60} + \frac{5 \delta }{60} \leq \frac{\delta}{10}
\end{eqnarray*}
reveals that
$$ \frac{8 \delta}{10} + d_{g_{k}} ( y_{k} , \zeta^{\infty}_{k} (x^{l}) ) < d_{g_{k}} ( y_{k} , \zeta^{\infty}_{k} (z^{m}) ) $$
for all $l,m \in \mathbb{N}$.  Since $\mathcal{M}_{k}$ is compact,
we can find subsequences $(\zeta^{\infty}_{k} (x^{n}))_{n \in
\mathbb{N}}$, $(\zeta^{\infty}_{k} (z^{n}))_{n \in \mathbb{N}}$
such that $\lim_{n \rightarrow \infty} \zeta^{\infty}_{k} (x^{n})
= \tilde{x}_{k}$ and $\lim_{n \rightarrow \infty}
\zeta^{\infty}_{k} (z^{n}) = \tilde{z}_{k}$.  Global hyperbolicity
then implies that $$d_{g_{k}} ( y_{k} , \tilde{x}_{k} ) + \frac{ 8
\delta }{10} \leq d_{g_{k}} ( y_{k} , \tilde{z}_{k} ).$$ Let
$\gamma_{k}$ be a distance maximising geodesic in
$\mathcal{M}_{k}$ from $y_{k}$ to $\tilde{z}_{k}$, and define
points $q^{1}_{k}, q^{2}_{k} , q^{3}_{k}$ on $\gamma_{k}$ by
$$d_{g_{k}}( q^{1}_{k} , q^{2}_{k} ) = d_{g_{k}} ( q^{2}_{k} ,
q^{3}_{k}) = d_{g_{k}} (q^{3}_{k} , \tilde{z}_{k} ) = \frac{
\delta }{5};$$ then for $n$ large enough $d_{g_{k}} ( q^{1}_{k} ,
\zeta^{\infty}_{k} (x^{n})) = 0$ and $d_{g_{k}} ( q^{3}_{k} ,
\zeta^{\infty}_{k} (z^{n}) ) > \frac{\delta}{6}$.  Hence with
$q^{j} = (\psi_{s}^{k} (q^{j}_{k}) )_{k < s \in \mathbb{N}}$, one
obtains for $n$ sufficiently large
\begin{itemize}
\item $d(q^{1} , x^{n}) < \frac{ \delta}{10}$
\item $d(q^{1}, q^{2}) , d(q^{2} , q^{3}) > \frac{\delta}{6}$
\item $d(q^{3}, z^{n} ) > \frac{\delta}{15}$
\end{itemize}
Hence, $q^{2}$ is not in the chronological past of \emph{all} $x^{n}$ (it is for sure valid for the subsequence and since the original sequence is \emph{future} Cauchy, it is valid for every element), and $q^{3}$ is in the chronological past of \emph{all} $z^{n}$ (again it is obviously valid for the subsequence, and since the whole sequence is \emph{past} Cauchy, the result follows).  Hence, $(x^{i})_{i \in \mathbb{N}}$ is not equivalent to $(z^{i})_{i \in \mathbb{N}}$, which gives the necessary contradiction.  The result we have obtained so far is that $d$ is timelike continuous in one variable.  We now prove that it is continuous in both variables.  Again we will only prove one case, the others being similar.  Let $(x^{i})_{i \in \mathbb{N}}$, $(y^{i})_{i \in \mathbb{N}}$ be two inequivalent future timelike Cauchy sequences in $\mathcal{C}$.  We show that \begin{equation} \label{limmie}  \lim_{i \rightarrow \infty} \lim_{j \rightarrow \infty} d(x^{i}, y^{j}) \end{equation} exists.  For sure we can find a subsequence $(x^{i_{n}} , y^{i_{n}})_{n \in \mathbb{N}}$ such that $$ \lim_{n \rightarrow \infty} d(x^{i_{n}} , y^{i_{n}} ) = \lim_{i \rightarrow \infty} d(x^{i} , (y^{j})_{j \in \mathbb{N}} ) = \alpha .$$
The limit on the rhs. exists since $(d(x^{i} ,(y^{j})_{j \in \mathbb{N}})_{i \in \mathbb{N}}$ is a decreasing sequence.  Suppose the limit (\ref{limmie}) does not exist; then we can find another subsequence $(x^{j_{m}} , y^{j_{m}})_{m \in \mathbb{N}}$ such that
$$ \lim_{m \rightarrow \infty} d(x^{j_{m}} , y^{j_{m}} ) + \delta <  \alpha $$
for some $\delta > 0$.  Choose $k$ sufficiently large such that $\frac{1}{2^{k-1}} < \frac{\delta}{3}$; by eventually passing to a subsequence we may assume that $\zeta^{\infty}_{k} (x^{j_{m}}) \stackrel{m \rightarrow \infty}{\rightarrow} \tilde{x}_{k}$ and $\zeta^{\infty}_{k} (y^{j_{m}}) \stackrel{m \rightarrow \infty}{\rightarrow} \tilde{y}_{k}$.  Obviously, $d_{g_{k}} ( \tilde{x}_{k} , \tilde{y_{k}} ) + \frac{ 2 \delta }{3} < \alpha $.  Moreover, since $d_{g_{k}}$ is continuous in both arguments, there exists an $n_{0}$ such that for $n,m \geq n_{0}$  $$d_{g_{k}} ( \zeta^{\infty}_{k}(x^{j_{m}}) , \zeta^{\infty}_{k} (y^{j_{n}} ) ) + \frac{2 \delta}{3} < \alpha .$$
Hence, for $n,m \geq n_{0}$ we have that
$$ d( x^{j_{m}} , y^{j_{n}} ) + \frac{ \delta}{3} < \alpha .$$
Choose $r$ large enough such that
\begin{itemize}
\item $d(x^{i_{r}} , y^{i_{r}}) > \alpha - \frac{\delta}{3}$
\item there exist $n,m \geq n_{0}$ with $x^{j_{n}} \ll x^{i_{r}}$ and $y^{i_{r}} \ll y^{j_{m}}$
\end{itemize}
we obtain that
$$\alpha - \frac{\delta}{3} > d(x^{j_{n}} , y^{j_{m}}) > d(x^{i_{r}} , y^{i_{r}} ),$$ which is the necessary contradiction.  We should still prove that this limit is independent of the sequences representing both equivalence classes.  Again I will do this for one case, the rest is left as an exercise to the reader.  As before, assume that $(x^{i})_{i \in \mathbb{N}}$ and $(y^{i})_{i \in \mathbb{N}}$ are inequivalent future timelike Cauchy sequences and let $(u^{i})_{i \in \mathbb{N}} \sim (x^{i})_{i \in \mathbb{N}}$ and $(v^{i})_{i \in \mathbb{N}} \sim (y^{i})_{i \in \mathbb{N}}$ be past timelike Cauchy.  Then, we know that $d((u^{i})_{i \in \mathbb{N}} , (v^{i})_{i \in \mathbb{N}})$ is well defined and equals $\lim_{i \rightarrow \infty} d(u^{i} , (v^{j})_{j \in \mathbb{N}})$.   The first part of the proof tells us that $d(u^{i} , (v^{j})_{j \in \mathbb{N}}) = d( u^{i} , (y^{j})_{j \in \mathbb{N}})$, hence $$d((u^{i})_{i \in \mathbb{N}} , (v^{i})_{i \in \mathbb{N}} ) \leq d((x^{i})_{i \in \mathbb{N}} , (y^{i})_{i \in \mathbb{N}} ). $$   Similarly, $d((u^{i})_{i \in \mathbb{N}} , v^{j} ) = d((x^{i})_{i \in \mathbb{N}} , v^{j}) > d((x^{i})_{i \in \mathbb{N}} , y^{j})$, hence
$$ d((u^{i})_{i \in \mathbb{N}} , (v^{i})_{i \in \mathbb{N}} ) \geq d((x^{i})_{i \in \mathbb{N}} , (y^{i})_{i \in \mathbb{N}}) ,$$
which proves the equality.    $\square$
\\*
\\*
We now prove that $\overline{T_{0} \mathcal{S}}$ has the $T_{2}$ property on the points belonging to the timelike continuum.   \begin{theo}
$\overline{T_{0} \mathcal{S}}$ has the $T_{2}$ property on the timelike continuum.
\end{theo}
\textsl{Proof}: \\*
Let $x,y \in \overline{T_{0} \mathcal{S}}$ be two $T_{0}$ separated points belonging to the timelike continuum.  Hence, we may assume there exist points $r,s$ such that, say, $r \ll x \ll s$ but $y \notin I^{+} (r) \cap I^{-} (s)$.  This implies that either $d(r,y)=0$ or $d(y,s) = 0$.  Without loss of generality, we may assume the former.  Pick $q \in \overline{T_{0} \mathcal{S}}$ such that $r \ll q \ll x$ and $d(r,q) > \frac{d(r,x)}{2}$.  Let $(p^{i})_{i \in \mathbb{N}}$ be a past timelike Cauchy sequence which is $T_{0}$ equivalent to $y$.  Then, since $d$ is continuous, we have that for $i$ sufficiently large $d(r , p^{i} ) < \frac{d(r,q)}{2}$.  Hence, $d(q,p^{i}) = 0$ and $y$ and $x$ are separated by $I^{-} (p^{i})$ and $I^{+}(q) \cap I^{-}(s)$ respectively. $\square$

%% file: appendI.tex
\chapter{Counterexample GH versus GGH}
In this Appendix, we prove theorem \ref{conGHvGGH}.  First, we
introduce some notational conventions.  Denote by
$\alpha^{i}_{j,k}$ the $k$'th element of the $j$'th column
$K^{i}_{j}$ in the causal set $\mathcal{P}^{L}_{i}$.  The
labelling of elements in a column starts from zero.  For example:
the maximal element in $K^{1}_{2}$ is $\alpha^{1}_{2,L+1}$.  For
notational simplicity, we agree that $\alpha^{i}_{j,0} \equiv
b^{i}_{j}$ and the top element of the column $K^{i}_{j}$ is
denoted by $t^{i}_{j}$.  In $\mathcal{P}_{2}^{L}$, this results in
$t^{2}_{j} = \alpha^{2}_{j, L+2-j}$.  The idea of the proof is to
determine how the bottom and top elements shift under the maps
$\psi$ and $\zeta$.  The following Lemma is crucial.
\newtheorem{lem}{Lemma}
\begin{lem}
Let $r < \frac{L}{4}+5$ ($1 < r <\frac{L}{4}+5$), and suppose $\zeta ( t^{2}_{r} ) \in  K^{1}_{s} $ ($\psi ( t^{1}_{r} ) \in  K^{2}_{s}$) then $\zeta ( b^{2}_{r+j}) \in \left\{ b^{1}_{s-1}, b^{1}_{s}, b^{1}_{s+1} \right\}$ ($\psi ( b^{1}_{r}) \in \left\{ b^{2}_{s-1}, b^{2}_{s}, b^{2}_{s+1} \right\}$) where $j=0,1$ all the indices have to be taken modulo $L+1$.
\end{lem}
\textsl{Proof}:  \\*
Remark first that $\zeta ( t^{2}_{r} ) \in K^{1}_{s}$ with $s$ a natural number between $2$ ($1$) and $r+k$ if $r < \frac{L}{2} + 1 - k$ ($r \geq \frac{L}{2} + 1 - k$).  Obviously, $\zeta ( t^{2}_{r} ) \geq \alpha^{1}_{s, L+2-r-k}$ where $\geq$ means ``in the causal future of''.  Suppose $\zeta ( b^{2}_{r+j}) \notin \left\{ b^{1}_{s-1}, b^{1}_{s}, b^{1}_{s+1} \right\}$ for some $j=0,1$, then $\zeta ( b^{2}_{r+j}) =  \alpha^{1}_{r,q}$ with $q \geq 1$ since,
$$d_{2}(b^{2}_{r+j}, t^{2}_{r}) - k \geq L + 2 - (\frac{L}{4} +4) - (\frac{L}{4}-1) = \frac{L}{2} - 1 > 0.$$
But in this case $\zeta (t^{2}_{r+1}) \geq \alpha^{1}_{s,L+1-k-r+q}$.  The above calculation reveals that $d_{2}(b^{2}_{r+2},t^{2}_{r+1})-k \geq \frac{L}{2}-2 > 0$ and since, moreover, $d_{2}(b^{2}_{r+j}, b^{2}_{r+2}) = 0$, we obtain that $\zeta ( b^{2}_{r+2}) \leq \alpha^{1}_{s, q + j}$.  Hence, $$d_{1}(\zeta ( b^{2}_{r+2}),\zeta ( t^{2}_{r})) \geq L + 2 - k - r + q - (k + q) \geq \frac{L}{4},$$ which is impossible since $d_{2}(b^{2}_{r+2}, t^{2}_{r}) = 0$.  The result for $\psi$ is obvious. $\square$
\\*
\\*
We shall further construct $\zeta$ and state similar properties of $\psi$ later on.
\begin{lem}
$\zeta (b^{2}_{r}) = b^{1}_{s}$ if $\zeta(t^{2}_{r}) \in K^{1}_{s}$ with $r$ between $1$ and $\frac{L}{4}+3$.
\end{lem}
\textsl{Proof}: \\* According to Lemma $1$, we have that
$\zeta(b^{2}_{r}) \in \left\{b^{1}_{s-1}, b^{1}_{s} , b^{1}_{s+1}
\right\}$.  Suppose that $\zeta(b^{2}_{r}) = b^{1}_{s+1}$, then we
show that $\zeta(b^{2}_{r+1}) \notin \left\{b^{1}_{s-1}, b^{1}_{s}
, b^{1}_{s+1} \right\}$, which is impossible by the same Lemma.
The arguments for $\zeta(b^{2}_{r}) = b^{1}_{s-1}$ are identical.
Suppose $\zeta(b^{2}_{r+1}) = b^{1}_{s+1}$, then $d_{1}(
\zeta(b^{2}_{r}) , \zeta(t^{2}_{r+2})) \geq L+2-(r+2)-k \geq
\frac{L}{2} - 2 \geq \frac{L}{4}$, which is impossible since
$d_{2} ( b^{2}_{r} , t^{2}_{r+2}) = 0$. Hence, suppose that $\zeta
(b^{2}_{r+1}) = b^{1}_{s-1}$, then $\zeta (t^{2}_{r+1}) \in
K^{1}_{s}$.  Hence, $\zeta(b^{2}_{r+2}) \in \left\{ b^{1}_{s-1},
b^{1}_{s} , b^{1}_{s+1} \right\}$, which is impossible since then
$d_{1}(\zeta(b^{2}_{r+2}) , \zeta(t^{2}_{r}) \geq L+2-r-k \geq
\frac{L}{2}$. So, we are only left with $\zeta(b^{2}_{r+1}) =
b^{1}_{s}$.  Obviously, $\zeta (t^{2}_{r+1}) \in K^{1}_{s+1}$,
since otherwise $\zeta (t^{2}_{r+1}) \in K^{1}_{s}$, which was
proven impossible before.  Hence, $\zeta (b^{2}_{r+2}) \in \left\{
b^{1}_{s}, b^{1}_{s+1}, b^{1}_{s+2} \right\}$.  Previous arguments
show that $\zeta (b^{2}_{r+2}) = b^{1}_{s+2}$, but then $\zeta
(t^{2}_{r+2}) \geq \alpha^{1}_{s+1, L-r-k}$, which implies that
$d_{1}( \zeta(b^{2}_{r}), \zeta(t^{2}_{r+2})) \geq \frac{L}{2}-2
\geq \frac{L}{4}$.  $\square$
\\*
\\*
Obviously, the same theorem applies to $\psi$ for $1<r<\frac{L}{4}+4$.
The following Lemma almost gives the necessary result.
\begin{lem}
If $\zeta ( t^{2}_{1}) \in K^{1}_{s}$ with $s$ ranging between $2$ and $k+1 \leq \frac{L}{4}$ then $\zeta (b^{2}_{i}) = b^{1}_{s+i-1}$ and $\zeta(t^{2}_{i}) \geq \alpha^{1}_{s+i-1, L+2-i-k}$ for $i \leq \frac{L}{4}+3$.
\end{lem}
\textsl{Proof}: \\*
As a consequence of Lemma 2, we have only two possibilities.  Either $\zeta (b^{2}_{i}) = b^{1}_{s+i-1}$ and $\zeta(t^{2}_{i}) \geq \alpha^{1}_{s+i-1, L+2-i-k}$, or, $\zeta (b^{2}_{i}) = b^{1}_{s-i-1}$ and $\zeta(t^{2}_{i}) \geq \alpha^{1}_{s-i-1, L+2-i-k}$ for $i \leq \frac{L}{4}+3$.  If the latter were true then $\zeta(t^{2}_{s+1}) \geq \alpha^{1}_{l,L+2-s-k}$ since $s+1 \leq \frac{L}{4} +1$, which is impossible since $ L+2-s-k > 2$. $\square$
\\*
\\*
First of all, it is easy to see that if $\psi ( t^{1}_{2} ) \in K^{2}_{\tilde{s}}$ with $\tilde{s}$ between $2$ and $k+1$.  Since, suppose $\psi ( t^{1}_{2} ) \in K^{2}_{1}$, then $\psi(b^{1}_{1}) = b^{2}_{L}$, since otherwise or $\psi(b^{1}_{1}) \geq \alpha^{2}_{1,1}$, or $\psi(b^{1}_{1}) \in \left\{ b^{2}_{1}, b^{2}_{2} \right\}$.  The former is impossible, since then $d_{2}( \psi (b^{1}_{3}), \psi (t^{1}_{1}) ) \geq \frac{L}{2} + 2 - k > \frac{L}{4}$.  The latter would imply that $\psi(b^{1}_{1}), \psi(t^{1}_{3})) \geq L-1-k$, which is also impossible.  But, then $\psi(t^{1}_{1}) \in K^{2}_{L-1} \cup K^{2}_{L} \cup K^{2}_{1}$, which is impossible for $L \geq 8$ (which was the assumption).  By a reasoning analogous to the one in Lemma 3,  we obtain that $\psi (b^{1}_{i}) = b^{2}_{\tilde{s}+i-2}$ and $\psi(t^{1}_{i}) \geq \alpha^{2}_{\tilde{s}+i-2, L+2-i-k}$ for $1< i \leq \frac{L}{4}+3$.  Moreover, $\psi(b^{1}_{1}) = b^{2}_{\tilde{s}-1}$.  \\*
\\*
We finish the proof by remarking that $d_{1}(b^{1}_{s+j}, \zeta( t^{2}_{1})) \geq L+1-k$ for $j=-1,0,1$.  Since $1 \geq s-1 <s+1 \leq \frac{L}{4} + 1$, we have that $\psi (b^{1}_{s+j}) = b^{2}_{s + \tilde{s} - 2 + j}$.  Since $L+1-2k \geq \frac{L}{2} + 3$ we obtain that $\psi \circ \zeta (t^{2}_{1}) \in K^{2}_{s + \tilde{s} - 2}$.  Hence, $D_{2}(t^{2}_{1}, \psi \circ \zeta (t^{2}_{1})) = L$.  $\square$

%% file: appendJ.tex
\chapter{Paper one: A new topology on the space of Lorentzian metrics on a fixed manifold}

%% file: hoofd.bbl
\begin{thebibliography}{99}

\bibitem{Yosida} K. Yosida, Functional analysis, Springer 1978
\bibitem{Keller} H. H. Keller, Differential Calculus in Locally
Convex Space, Lecture Notes in Mathematics 417, Springer 1974
\bibitem{Dubinsky} E. Dubinsky, The Structure of Nuclear Fr\'echet
Spaces, Lecture Notes in Mathematics 720, Springer 1979
\bibitem{Greenleaf} F. P. Greenleaf, Invariant means on
topological groups, Van Nostrand Mathematical Studies 16, 1969
\bibitem{Pier} J.P. Pier, Amenable locally compact groups,
John Wiley and sons, 1984
\bibitem{Higgins} P.J. Higgins, An introduction to topological
groups, London Mathematical Society Lecture Note Series $15$,
Cambridge university press, 1974
\bibitem{Kriele} M. Kriele, Spacetime, Springer 1999
\bibitem{Hawking1} S.W. Hawking, G.F.R. Ellis, The large scale
structure of space-time, Cambrige monographs on mathematical
physics, $1973$
\bibitem{Jost} J. Jost, Riemannian Geometry and Geometric
Analysis, Springer, 1998
\bibitem{Hawking2} S.W. Hawking, A.R King, P.J. McCarthy, A new topology for curved
space-time which incorporates the causal, differential and
conformal structures Journal of Mathematical Physics volume 17,
page 174, 1976
\bibitem{Bombelli} L. Bombelli, R.D. Sorkin, When are two
Lorentzian metrics close?, Preliminary version, May 1991
\bibitem{Bombelli2} L. Bombelli, D. A. Meyer 1989, The origin of
Lorentzian geometry Phys. Lett. A1414 226-228
\bibitem{Bombelli3} L. Bombelli, 2000, Statistical Lorentzian
geometry and the closeness of Lorentzian manifolds J. Math. Phys
41 6944-6958 and gr-qc/0002053
\bibitem{Carfora}  J. Ambj\o rn, M. Carfora, A. Marzuoli, The
geometry of Dynamical Triangulations, Springer 1997
\bibitem{Donaldson} S. Donaldson, Connections, cohomology and the
intersection forms of 4 - manifolds, Ibidem 24 (1986) also An
application of gauge theory to the topology of 4 - manifolds, J.
of Differential Geometry. 8 (1983)
\bibitem{Sorkin} A. Borde, H.F Dowker, R.S. Garcia, R.D. Sorkin,
S.Surya, Causal Continuity in degenerate spacetimes, arXiv:
gr-qc/9901063 v3, 1999
\bibitem{Abbati}  M.C. Abbati, A. Mania,
On differential structure for projective limits of manifolds. J.
Geom. and Phys. 29 (1999) 35-63.
\bibitem{Balanzat} M. M. Balanzat La differentielle d'Hadamard-Fr\'echet dans
les espaces vectoriels topologiques C. R. Acad. Sci. Sci. Paris
251 (1960) 2459-2461.
\bibitem{Kriegl1}  A. Fr\"ohlicher, A. Kriegl, Linear Spaces and Differentiation Theory
, J. Wiley and Sons 1988
\bibitem{Kriegl2}  A. Kriegl, P. W. Michor The convenient Setting of Global Analysis,
American Mathematical Society Math. Surv. and Mon. 53 (1997)
\bibitem{Michor}  P. W. Michor, Manifolds of differentiable mappings, Shiva, Orpington 1980
\bibitem{Michor1} P.W. Michor, Manifolds of smooth maps, II: the Lie group of diffeomorphisms of a non-compact smooth manifold, Cah. Top. G\'eom. Diff., 1980 ; Manifolds of smooth maps III: the prinicipal bundle of embeddings of a non compact smooth manifold, Cah. Top. G\'eom. Diff.,1980 ; Manifolds of smooth maps IV: theorem of de Rham, Cah. Top. G\'eom. Diff., 1983
\bibitem{Kelley} J.L. Kelley, General topology, Van Nostrand, 1955
\bibitem{Szabados} L. B. Szabados, Causal measurability in
chronological spaces, Journal of General Relativity and
Gravitation 19, 1091-100, 1987
\bibitem{Page} W. Page, 1978, Topological uniform structures, New
York, Wiley
\bibitem{Wald} R. M. Wald, 1984, General Relativity, Chicago University press
\bibitem{Thiemann} T. Thiemann, Introduction to Modern Canonical Quantum General Relativity, preprint gr-qc/011034v1 
\bibitem{Nakahara} M. Nakahara, Geometry, Topology and Physics, Adam Hilger, Bristol and New York, 1990
\bibitem{Choquet}  Y. Choquet - Bruhat, R. Geroch, Global Aspects of the Cauchy Problem in General Relativity, \textsl{Commun. math. Phys.} \textbf{14}, 329-335, 1969
\bibitem{Dixon} W. G. Dixon, 1974, Dynamics of Extended Bodies in General Relativity. III Equations of Motion, \textsl{Phil. Trans. Roy. Soc. Lond.} \textbf{A277}, 59-119 
\bibitem{Fock} V. A. Fock, 1939, Sur le Mouvement des masses finies d'apres la th\'eorie de gravitation Einsteinienne, \textsl{J. Phys. U.S.S.R.}, \textbf{1}, 81-116
\bibitem{Chandrasekhar} S. Chandrasekhar, The Mathematical Theory of Black Holes, Clarendon-Press, Oxford, 1985
\bibitem{Steenrod} N. Steenrod, The topology of fibre bundles, Princeton University Press, 1951
\bibitem{Townsend} Townsend, Black holes, preprint gr-qc/9707012 
\bibitem{Hawking3} J.M. Bardeen, B. Carter, S.W. Hawking, The four laws of black hole mechanics, \textsl{Commun. math. Phys} \textbf{31}, 161-170 (1973)
\bibitem{Wald1} R. Wald, The thermodynamics of black holes, \textsl{preprint  gr-qc/9912119v2} 
\bibitem{Wald2} R. M. Wald and Vivek Iyer, Some properties of the Noether charge and a proposal for dynamical black hole entropy, \textsl{Phys. Rev.} \textbf{D}, 846 - 864, 1994
\bibitem{Wald3} R. M. Wald, Quantum Field theory in curved spacetime and black hole thermodynamics, Chicago university press, 1994
\bibitem{Wald4} R. M. Wald, On particle creation by black holes,  \textsl{Commun. math. Phys.} \textbf{45}, 9 - 34 (1975)
\bibitem{Hawking4} S. W. Hawking, Particle creation by black holes, \textsl{Commun. math. Phys.} \textbf{43},  199-220, 1975
\bibitem{Bra} O. Bratteli, D.W. Robinson, Operator Algebras and Quantum Statistical Mechanics, Texts and Monographs in physics, Springer
\bibitem{Ash} A. Ashtekar, A. Magnon, Quantum fields in curved space-times, \textsl{Proc. R. Soc. Lond. A.} \textbf{346}, 375-394 (1975)
\bibitem{Borde} A. Borde, Topology change in classical general relativity, \textsl{gr-qc/9406053 v1}, 30 jun. 1994
\bibitem{Dowker} F. Dowker, Topology change in quantum gravity, \textsl{gr-qc/0206020 v1}, 7 Jun 2002
\bibitem{gra} J. Grabowski, Derivative of the exponential mapping for infinite dimensional Lie groups, Ann. Global. Anal. Geom. \textbf{11} (1993), 213-220
\bibitem{abba} M.C. Abbati, R. Cirelli, S. DeSantis, E. Ruffini The Second Noether Theorem in the formalism of jet-bundles: Symmetries and degeneration, \textsl{J.Geom. and Phys.} \textbf{17} (1995)
\bibitem{mars} J. E. Marsden, P.R. Chernoff,  Properties of infinite dimensional Hamiltonian systems, Lecture notes in mathematics, Springer-Verlag, Berlin 1974.
\bibitem{Beem} J.K. Beem, P.E. Ehrlich, K.L. Easley, Global Lorentzian Geometry, Marcel Dekker, 1996
\bibitem{Petersen} P. Petersen, Riemannian Geometry, Springer, 1998 
\bibitem{Noldus} J. Noldus, A new topology on the space of Lorentzian metrics on a fixed manifold, Class. Quantum Grav. \textbf{19}, 2002, 6075-6107
\bibitem{Defrise} L. Defrise-Carter,  Conformal groups and conformally equivalent isometry groups, Commun. math. Phys, 40, 273-282, (1975)
\bibitem{Sorkin1} R.D. Sorkin, Causal sets: discrete gravity, notes for the Valdivia summer school, Jan. 2002 
\bibitem{Noldus1} J. Noldus, A Lorentzian Gromov-Hausdorff notion of distance, Class. Quant. Grav. \textbf{21}, 839-850 Jan 2004 
\bibitem{Sorkin2} R. Sorkin, E. Woolgar, A causal order for spacetimes with $C^{0}$ Lorentzian metrics: proof of compactness of the space of causal curves, gr-qc/9508018 v3, 24 Jun 1999
\bibitem{Busemann} H. Busemann, Timelike spaces, Dissertationes Mathematicae, Warszawa 1967
\bibitem{Meyer} D. Meyer, A metric space construction for the boundary of space-time, J. Math. Phys. 27, 1986  
\bibitem{Kriele1} M. Kriele, Comment on ``A metric space construction for the boundary of spacetime'' by D.A. Meyer, J. Math. Phys. 30, 1989   
\bibitem{Gromov} M. Gromov, Metric structures for Riemannian and Non-Riemannian spaces, Birkh\"auser, 1997
\bibitem{Noldus2} L. Bombelli, J. Noldus, 2004, The moduli space of isometry classes of globally hyperbolic spacetimes, submitted for publication in CQG 
\bibitem{Noldus3} J. Noldus, The limit space of a Cauchy sequence of globally hyperbolic spacetimes, Class. Quant. Grav. \textbf{21}, 851-874, Jan 2004 
\bibitem{Wein} S. Weinberg, Gravitation and cosmology, Wiley and Sons, 1972
\bibitem{Isham1} C.J. Isham, K.V. Kucha\v{r}, Representations of Spacetime Diffeomorphisms I: Canonical Parametrized Field Theories, Annals of Physics, \textbf{164}, 288-315 (1985)
\bibitem{Isham2} C.J. Isham, K.V. Kucha\v{r}, Representations of Spacetime Diffeomorphisms II : Canonical Geometrodynamics, Annals of Physics, \textbf{164}, 316-333 (1985)
\bibitem{Sav1} K. Savvidou, General relativity histories theory I: The spacetime character of the canonical description, gr-qc/0306034v1, 9 Jun 2003
\bibitem{Sav2} K. Savvidou, General relativity histories theory II: Invariance groups, gr-qc/0306036v1, 9 Jun 2003
\bibitem{Synge} J.L. Synge, Relativity:The General Theory, North Holland Publishing Company, 1960
\bibitem{Sorkin3} R.D. Sorkin, Forks in the road, on the way to quantum gravity, gr-qc/9706002
\bibitem{Isham} C.J. Isham: Quantum Logic and the Histories approach to quantum theory, gr-qc/9308006 ; Quantum temporal logic and decoherence functionals in the histories approach to generalized quantum theory, gr-qc/9405029 ; The classification of decoherence functionals : an analogue of Gleason's theorem, gr-qc/9406015 ; Continuous histories and the history group in generalized quantum theory, gr-qc/9503063
\bibitem{Rideout} D. Rideout, A classical sequential growth dynamics for causal sets, Phys.Rev.D61:024002,2000 and  gr-qc/9904062 
\bibitem{Sorkin4} G. Brightwell, H. F. Dowker, R. S. Garcia, J. Henson, R. D. Sorkin, Observables in causal set cosmology, Phys.Rev.D67:084031,2003 and gr-qc/0210061  
\bibitem{Lovelock} D. Lovelock, J. Math. Phys. 13, 874 (1972)
\end{thebibliography}
